\documentclass[12pt, aps, floatfix, nofootinbib,  superscriptaddress]{revtex4-1}
\usepackage{latexsym}
\usepackage{amsmath}
\usepackage{amssymb}
\usepackage{amsfonts}
\usepackage{bm}

\usepackage[vcentermath]{youngtab}
\usepackage{mathtools}
\usepackage[squaren]{SIunits}

\usepackage[margin=2cm]{geometry}
\usepackage{makecell}

\usepackage{color}

\usepackage{longtable}
\usepackage{multirow}
\usepackage{supertabular}
\usepackage{graphicx}

\usepackage[mathscr,scaled=1.15]{urwchancal}
\DeclareFontFamily{OT1}{pzc}{}
\DeclareFontShape{OT1}{pzc}{m}{it}%
{<-> s * [1.15] pzcmi7t}{}
\DeclareMathAlphabet{\mathpzc}{OT1}{pzc}{m}{it}

\definecolor{purple}{rgb}{0.5,0,0.5}
\definecolor{blue}{rgb}{0.0,0,0.9}
\definecolor{prdblue}{rgb}{0.133,0.118,0.498}
\usepackage[colorlinks=true, pdfstartview=FitV, linkcolor=prdblue, citecolor= prdblue, urlcolor=prdblue]{hyperref}

\def\tstrut{\vrule height2.25ex depth0pt width0pt} 
\def\mystrut{\rule[-4pt]{0sp}{13pt}}

\newlength{\Wfockuud}

\newcommand*{\arrowpalette}[2]{\raisebox{\depth}{\resizebox*{!}{\heightof{$#1X$}}{$#1#2$}}}
\newcommand*{\gooduparrow}{{\mathpalette{\arrowpalette}{\uparrow}}}
\newcommand*{\gooddownarrow}{{\mathpalette{\arrowpalette}{\downarrow}}}

\newcommand*{\VmA}{V\mathord{-}A}%
\newcommand*{\BnoL}{B\mathord{\smash{\neq}}\Lambda}
\newcommand{\overbar}[1]{\mkern 1.5mu\overline{\mkern-1.5mu#1\mkern-1.5mu}\mkern 1.5mu}
\newcommand{\MSbar}{\overbar{\text{MS}}}

\begin{document}


\title{$\,$\\[-7ex]\hspace*{\fill}{\normalsize{\sf\emph{Preprint no}. NJU-INP 024/20}}\\[1ex]
\vspace{1cm}Diquark Correlations in Hadron Physics: \\ Origin, Impact and Evidence}


\author{M. Yu. Barabanov}
\affiliation{Joint Institute for Nuclear Research, Dubna, 141980, Russia}

\author{M. A. Bedolla}
\affiliation{Mesoamerican Centre for Theoretical Physics, Universidad Aut\'onoma de Chiapas, 
Tuxtla Guti\'errez  29040, Chiapas, M\'exico}

\author{W. K. Brooks}
\affiliation{Departamento de F\'isica, Universidad T\'ecnica Federico Santa Mar\'ia, Valpara\'iso, Chile.}

\author{G. D. Cates}
\affiliation{University of Virginia, Charlottesville, VA 22904, USA}

\author{C. Chen}
\affiliation{Institut f\"ur Theoretische Physik, Justus-Liebig-Universit\"at  Gie\ss en, D-35392 Gie\ss en, Germany}

\author{Y. Chen}
\affiliation{Institute of High Energy Physics, Chinese Academy of Sciences, Beijing 100049, China}
\affiliation{School of Physics, University of Chinese Academy of Sciences, Beijing 100049, China}

\author{\mbox{E. Cisbani}}
\affiliation{Istituto Superiore di Sanit\`a, I-00161 Rome, Italy}

\author{M. Ding}
\affiliation{European Centre for Theoretical Studies in Nuclear Physics and Related Areas (ECT$^\ast$) and Fondazione Bruno Kessler, Villa Tambosi,  Strada delle Tabarelle 286, I-38123 Villazzano (TN) Italy}

\author{G. Eichmann}
\affiliation{LIP Lisboa, Av. Prof. Gama Pinto 2, P-1649-003 Lisboa, Portugal}
\affiliation{Departamento de F\'isica, Instituto Superior T\'ecnico, P-1049-001 Lisboa, Portugal}

\author{R. Ent}
\affiliation{Thomas Jefferson National Accelerator Facility, Newport News, Virginia 23606, USA}

\author{J. Ferretti}
\email{jacopo.j.ferretti@jyu.fi}
\affiliation{Department of Physics, University of Jyv\"askyl\"a, P.O. Box 35 (YFL), 40014 Jyv\"askyl\"a, Finland}

\author{R. W. Gothe}
\affiliation{Department of Physics and Astronomy, University of South Carolina, Columbia, SC 29208, USA}

\author{\mbox{T. Horn}}
\affiliation{Catholic University of America, Washington, D.C. 20064, USA}
\affiliation{Thomas Jefferson National Accelerator Facility, Newport News, Virginia 23606, USA}

\author{S. Liuti}
\affiliation{University of Virginia, Charlottesville, VA 22904, USA}

\author{C. Mezrag}
\affiliation{IRFU, CEA, Universit\'e Paris-Saclay, F-91191 Gif-sur-Yvette, France}

\author{A. Pilloni}
\affiliation{European Centre for Theoretical Studies in Nuclear Physics and Related Areas (ECT$^\ast$) and Fondazione Bruno Kessler, Villa Tambosi,  Strada delle Tabarelle 286, I-38123 Villazzano (TN) Italy}

\author{A. J. R. Puckett}
\affiliation{University of Connecticut, Storrs, Connecticut 06269, USA}

\author{C. D. Roberts}
\email{cdroberts@nju.edu.cn}
\affiliation{School of Physics, Nanjing University, Nanjing, Jiangsu 210093, China}
\affiliation{Institute for Nonperturbative Physics, Nanjing University, Nanjing, Jiangsu 210093, China}

\author{P. Rossi}
\affiliation{Thomas Jefferson National Accelerator Facility, Newport News, Virginia 23606, USA}
\affiliation{INFN, Laboratori Nazionali di Frascati, I-00044 Frascati, Italy}

\author{G. Salm\'e}
\affiliation{Istituto Nazionale di Fisica Nucleare (INFN), Sezione di Roma, I-00185 Rome, Italy}

\author{E. Santopinto}
\email{Elena.Santopinto@ge.infn.it}
\affiliation{Istituto Nazionale di Fisica Nucleare (INFN), Sezione di Genova, via Dodecaneso 33, I-16146 Genova, Italy}

\author{J. Segovia}
\email[]{jsegovia@upo.es}
\affiliation{Departamento de Sistemas F\'isicos, Qu\'imicos y Naturales, \\ Universidad Pablo de Olavide, E-41013 Sevilla, Spain}
\affiliation{Institute for Nonperturbative Physics, Nanjing University, Nanjing, Jiangsu 210093, China}

\author{S. N. Syritsyn}
\affiliation{RIKEN BNL Research Center, Brookhaven National Laboratory, Upton, NY 11973, USA}
\affiliation{Department of Physics and Astronomy, Stony Brook University, Stony Brook, NY 11794, USA}

\author{\mbox{M. Takizawa}}
\affiliation{Showa Pharmaceutical University, Machida, Tokyo 194-8543, Japan}
\affiliation{J-PARC Branch, KEK Theory Center, IPNS, KEK, Tokai, Ibaraki, 319-1106, Japan}
\affiliation{Meson Science Laboratory, Cluster for Pioneering Research, RIKEN,  Wako, Saitama 351-0198, Japan}

\author{E. Tomasi-Gustafsson}
\affiliation{IRFU, CEA, Universit\'e Paris-Saclay, F-91191 Gif-sur-Yvette, France}

\author{P. Wein}
\affiliation{Institut f\"ur Theoretische Physik, Universit\"at Regensburg, Universit\"atsstra\ss e 31, D-93040 Regensburg, Germany}

\author{B. B. Wojtsekhowski}
\email{bogdanw@jlab.org}
\affiliation{Thomas Jefferson National Accelerator Facility, Newport News, Virginia 23606, USA}

\date{2020 August 07}

\begin{abstract}
\centerline{\bf Abstract}
The last decade has seen a marked shift in how the internal structure of hadrons is understood.  Modern experimental facilities, new theoretical techniques for the continuum bound-state problem and progress with lattice-regularised QCD have provided strong indications that soft quark+quark (diquark) correlations play a crucial role in hadron physics. For example, theory indicates that the appearance of such correlations is a necessary consequence of dynamical chiral symmetry breaking, \emph{viz}.\ a corollary of emergent hadronic mass that is responsible for almost all visible mass in the universe; experiment has uncovered signals for such correlations in the flavour-separation of the proton's electromagnetic form factors; and phenomenology suggests that diquark correlations might be critical to the formation of exotic tetra- and penta-quark hadrons.  A broad spectrum of such information is evaluated herein, with a view to consolidating the facts and therefrom moving toward a coherent, unified picture of hadron structure and the role that diquark correlations might play.
\end{abstract}

\maketitle




\section*{Contents}
\unskip
\begin{itemize}
\item Sec.\,\ref{sec:Introduction} -- Introduction \dotfill\ \pageref{sec:Introduction}
\item Sec.\,\ref{sec:Theory} -- Diquarks in Theory \dotfill\ \pageref{sec:Theory}
\begin{itemize}
\item Sec.\,\ref{sec:quark-models} -- Quark Models \dotfill\ \pageref{sec:quark-models}
\item Sec.\,\ref{sec:DSEs} -- Continuum Schwinger Function Methods \dotfill \pageref{sec:DSEs}
\item Sec.\,\ref{sec:lattice} -- Lattice-regularised QCD \dotfill\ \pageref{sec:lattice}
\end{itemize}
\item Sec.\,\ref{sec:Experiment}  -- Observation of Diquarks \dotfill\ \pageref{sec:Experiment}
\begin{itemize}
\item Sec.\,\ref{SLNFFS} -- Space-like Nucleon Form Factors \dotfill\ \pageref{SLNFFS}
\item Sec.\,\ref{TLNFFS} -- Time-like Nucleon Form Factors \dotfill\ \pageref{TLNFFS}
\item Sec.\,\ref{secNRTFFs} -- Nucleon to Resonance Transition Form Factors \dotfill\ \pageref{secNRTFFs}
\item Sec.\,\ref{MDSoBs} -- Multidimensional Structure of Baryons \dotfill\ \pageref{MDSoBs}
\item Sec.\,\ref{MSaaWoDS} -- Meson Structure as a Window onto Diquark Structure \dotfill\ \pageref{MSaaWoDS}
\item Sec.\,\ref{EHatCtDs} -- Exotic Hadrons and their Connection to Diquarks \dotfill\ \pageref{EHatCtDs}
\end{itemize}
\item Sec.\,\ref{sec:Future} -- Prospects for a Diquark Future  \dotfill\ \pageref{sec:Future}
\begin{itemize}
\item Sec.\,\ref{SBSSLNFSS} -- SBS Programme on High-$Q^2$ Space-like Nucleon Form Factors \dotfill \pageref{SBSSLNFSS}
\item Sec.\,\ref{CPHF} -- Colour Propagation and Hadron Formation \dotfill \pageref{CPHF}
\item Sec.\,\ref{PCsBB} -- Production Cross-sections of Baryons at Belle \dotfill \pageref{PCsBB}
\item Sec.\,\ref{sec:DSEsFuture} -- Next Steps for Continuum Schwinger Function Methods \dotfill \pageref{sec:DSEsFuture}
\item Sec.\,\ref{SAnilQCD} -- Selected Advances (needed) in lattice-QCD \dotfill \pageref{SAnilQCD}
\end{itemize}
\item Sec.\,\ref{sec:Epilogue} -- Epilogue \dotfill \pageref{sec:Epilogue}
\item Abbreviations \dotfill\ \pageref{abbreviations}
\item References \dotfill\ \pageref{references}
\end{itemize}
\newpage

\pagenumbering{arabic}
\setcounter{page}{1}

\section{Introduction}
\label{sec:Introduction}

More than one century of fundamental research in atomic and nuclear physics has shown that all matter is corpuscular, with the atoms that comprise us, themselves containing a dense nuclear core. This core is composed of protons and neutrons, referred to collectively as nucleons, which are members of a broader class of fm-scale particles, called hadrons. In working towards an understanding of hadrons, it has been found that they are complicated bound-states of gluons and quarks whose interactions are described by a Poincar\'e-invariant quantum non-Abelian gauge field theory; namely, quantum chromodynamics (QCD).

QCD is fundamentally different from other pieces of the Standard Model of Particle Physics (SM): whilst perturbation theory is a powerful tool when used in connection with high-energy processes, this technique is powerless when it comes to developing an understanding of observable low-energy characteristics of QCD. The body of experimental and theoretical methods used to probe and map QCD's infrared domain can be called strong-QCD \cite{Pennington:1996dy} and they must deal with emergent nonperturbative phenomena, such as confinement of gluons and quarks and dynamical chiral symmetry breaking (DCSB).

The QCD running coupling lies at the heart of many attempts to define and understand confinement because almost immediately following the demonstration of asymptotic freedom \cite{Politzer:2005kc, Wilczek:2005az, Gross:2005kv} the associated appearance of an infrared Landau pole in the perturbative expression for the running coupling spawned the idea of infrared slavery, \emph{viz}.\ confinement expressed through a far-infrared divergence in the running coupling.  In the absence of a nonperturbative definition of a unique running coupling, this idea is not more than a conjecture; but recent studies \cite{Binosi:2014aea, Binosi:2016nme, Cui:2019dwv} support a conclusion that the Landau pole is screened (eliminated) in QCD by the dynamical generation of a gluon mass-scale and the theory possesses an infrared stable fixed point.

In numerical simulations of lattice-regularised QCD (lQCD) that use static sources to represent the valence-quarks of, for instance, a proton, a ``Y-junction" flux-tube  picture of nucleon structure is drawn, \emph{e.g}.\ Refs.\,\cite{Bissey:2006bz, Bissey:2009gw}.  Such results and notions could suggest an important role for the three-gluon vertex, which is a signature of the non-Abelian character of QCD and the source of asymptotic freedom, in quark (and gluon) confinement inside the hadron.  That is, if the static-quark picture were equally valid in real-world QCD.  In dynamical QCD, however, wherein active light quarks are ubiquitous, it is not; so a different explanation of binding within the nucleon, and most generally within any hadron, must be found.

Based on an accumulated body of evidence, it appears likely that confinement, defined via the violation of reflection positivity by coloured Schwinger functions (see, \emph{e.g}.\ Refs.\,\cite{Gribov:1999ui, Munczek:1983dx, Stingl:1983pt, Cahill:1985mh, Stingl:1985hx, Cahill:1988zi, Krein:1990sf, Burden:1991gd, Hawes:1993ef, Roberts:1994dr, Roberts:2000aa, Alkofer:2000wg, Roberts:2007ji, Brodsky:2012ku, Strauss:2012dg, Gao:2015kea, Papavassiliou:2015aga, Binosi:2019ecz} and citations therein and thereof)  and DCSB have a common origin in the SM; but this does not mean that confinement and DCSB must necessarily appear together.  Models can readily be built that express one without the other, \emph{e.g}.\ numerous constituent quark models express confinement through potentials that rise rapidly with interparticle separation, yet possess no ready definition of a chiral limit \cite{Segovia:2008zza, Yang:2017qan}; and models of the Nambu--Jona-Lasinio type typically express DCSB but not confinement \cite{Nambu:1961tp, JonaLasinio:1964cw, Nambu:2009zza}.

DCSB ensures the existence of nearly-massless pseudo-Nambu-Goldstone (NG) modes (pions), each constituted from a valence-quark and -antiquark whose individual Lagrangian current-quark masses are $<1$\% of the proton mass \cite{Maris:1997hd}. In the presence of these modes, no flux tube between a static colour source and sink can have a measurable existence. To verify this statement, consider such a tube being stretched between a source and sink. The potential energy accumulated within the tube may increase only until it reaches that required to produce a particle-antiparticle pair of the theory's pseudo-NG modes. Simulations of lQCD show \cite{Bali:2005fu, Prkacin:2005dc} that the flux tube then disappears instantaneously along its entire length, leaving two isolated colour-singlet systems. The length-scale associated with this effect in QCD is $\simeq 1/3\,\text{fm}$.  Hence, if any such string forms, it would dissolve well within hadron interiors.

Another equally important consequence of DCSB is less well known.  Namely, any interaction capable of creating pseudo-NG modes as bound-states of a light dressed-quark and -antiquark, and reproducing the measured value of their leptonic decay constants, will necessarily also generate strong colour-antitriplet correlations between any two dressed quarks contained within a hadron. Although a rigorous proof within QCD is not known, this assertion is based upon an accumulated body of evidence, gathered in three decades of studying two- and three-body bound-state problems in hadron physics, \emph{e.g}.\ Refs. \cite{Cahill:1987qr, Cahill:1988dx, Oettel:1998bk, Bloch:1999vk, Bender:2002as, Eichmann:2008ef, Roberts:2011cf, Segovia:2015ufa, Eichmann:2016yit}. No realistic counter examples are known; and the existence of such quark+quark (diquark) correlations is also supported by simulations of lQCD \cite{Hess:1998sd, Orginos:2005vr, Alexandrou:2006cq, DeGrand:2007vu, Babich:2007ah, Green:2010vc, Bi:2015ifa}.

It is worth remarking here that in a dynamical theory based on SU$(2)$-colour, diquarks are colour-singlets. They would thus exist as asymptotic states and form mass-degenerate multiplets with mesons composed from like-flavoured quarks. (These properties are a manifestation of Pauli-G\"ursey symmetry \cite{Pauli:1957, Gursey:1958}.) Consequently, the isoscalar-scalar, $[ud]_{0^+}$, diquark would be massless in the presence of DCSB, matching the pion, and the isovector-pseudovector, $\{ud\}_{1^+}$, diquark would be degenerate with the theory's $\rho$-meson.  Such identities are lost in changing the gauge group to SU$(3)$-colour [SU$_{\rm c}(3)$]; but clear and instructive similarities between mesons and diquarks nevertheless remain, such as \cite{Cahill:1987qr, Roberts:2000aa, Maris:2002yu, Maris:2004bp, Eichmann:2008ef, Aznauryan:2012ba, Segovia:2015hra, Roberts:2015lja, Eichmann:2016hgl, Burkert:2017djo, Lu:2017cln, Chen:2017pse, Chen:2019fzn, Yin:2019bxe, Lu:2019bjs}: (\emph{i}) isoscalar-scalar and isovector-pseudovector diquark correlations are the strongest, but others could appear inside a hadron so long as their quantum numbers are allowed by Fermi-Dirac statistics; (\emph{ii}) the associated diquark mass-scales express the strength and range of the correlation and are each bounded below by the partnered meson's mass; and (\emph{iii}) realistic diquark correlations are soft, \emph{i.e}.\ they possess an electromagnetic size that is bounded below by that of the analogous mesonic system.

It is important to appreciate that these fully dynamical diquark correlations are different from the static, pointlike diquarks which featured in early attempts \cite{Anselmino:1992vg} to understand the baryon spectrum and to explain the so-called missing resonance problem \cite{Ripani:2002ss, Burkert:2012ee, Kamano:2013iva}. Modern diquarks are fully dynamical inside hadrons: no valence quark holds a special place because each one participates in all diquarks to the fullest extent allowed by the quantum numbers of the quark, the diquark and the hadron in hand. The continual rearrangement of the quarks guarantees a hadron spectrum as rich as that found experimentally and that obtained in modern constituent quark models \cite{Yang:2017qan} and lQCD calculations \cite{Edwards:2011jj}.

Evidently, the notion of diquark correlations is spread widely across modern nuclear and high-energy physics; for example, experiment has uncovered signals for such correlations in the flavour-separation of the proton's electromagnetic form factors \cite{Cates:2011pz, Wojtsekhowski:2020tlo}; and phenomenology suggests that diquark correlations might play a material role in the formation of exotic tetra- and penta-quark hadrons \cite{Chen:2016qju, Esposito:2016noz, Ali:2017jda, Olsen:2017bmm, Guo:2017jvc, Liu:2019zoy, Guo:2019twa}. At issue, however, is whether all these things called diquarks are the same; and if there are dissimilarities, can they be understood and reconciled so that experiment can properly search for clean observable signals.

Herein, therefore, a critical review of existing information is undertaken in order to consolidate available facts and identify a path toward a consistent description of diquark correlations inside hadrons that answers the following basic questions:
\begin{itemize}
\item[(i)] How firmly founded are continuum theoretical predictions of diquark correlations in hadrons?
\item[(ii)] What does lQCD have to say about the existence and character of diquark correlations in baryons and multiquark systems?
\item[(iii)] Are there strategies for combining continuum and lattice methods in pursuit of an insightful understanding of hadron structure?
\item[(iv)] Can theory identify experimental observables that would constitute unambiguous measurable signals for the presence of diquark correlations?
\item[(v)] Is there a traceable connection between the so-called diquarks used to build phenomenological models of high-energy processes and the correlations predicted by contemporary theory; and if so, how can such models be improved therefrom?
\item[(vi)] Are diquarks the only type of two-body correlations that play a role in hadron structure?
\item[(vii)] Which new experiments, facilities and analysis tools are best suited to test the emerging picture of two-body correlations in hadrons?
\end{itemize}
Note, too, that the last millennium saw publications which treat the diquark concept explicitly or implicitly. It is not our intention to recapitulate that work. Interested parties may consult other documents that supply additional material, \emph{e.g}.\ Refs.\,\cite{Anselmino:1992vg, Szczekowski:1988pq}, the proceedings of some workshops in the 1990s \cite{Anselmino:1989vt, Anselmino:1994uj, Anselmino:1998rf}, and a compilation of references to articles on diquarks \cite{Skytt:1991rm}.

Before proceeding further, it is worth remarking that this perspective supplies a wide-ranging view of the diquark concept, providing a discussion of many variations on the theme.  There are some occasions in which different approaches might appear to be mutually inconsistent.  In such cases, the reader should understand that in science there is room for constructive disagreement on the road of progress.

The manuscript is arranged as follows. In Sec.~\ref{sec:Theory} we revisit the theoretical concept of diquark correlations inside hadrons; review the latest advances on this topic using phenomenological quark models, continuum Schwinger functional methods and lattice-regularised QCD techniques; and highlight some examples of their most relevant results compared with experimental data. Section~\ref{sec:Experiment} is devoted to an experimental overview of the most prominent signals of diquark correlations inside hadrons, either conventional or unconventional. We dedicate Sec.~\ref{sec:Future} to discuss possible theoretical and experimental pathways, which have not yet been explored and can consolidate the concept of diquark correlations. We finish with a summary and perspective in Sec.~\ref{sec:Epilogue}. 
\section{Diquarks in Theory}
\label{sec:Theory}
\subsection{Phenomenological Quark Models}
\label{sec:quark-models}
The notion of diquarks dates back to the foundations of the quark model (QM) itself \cite{GellMann:1964nj, Zweig:1981pd}. Its introduction had the purpose to provide an alternative description of baryons as bound states of a constituent-quark and -diquark \cite{Ida:1966ev, Lichtenberg:1967zz, Lichtenberg:1981pp}. Later, phenomenological indications for the emergence of diquark-like correlations were given. They included the $\Delta I = \frac{1}{2}$ rule in weak non-leptonic decays \cite{Neubert:1989fp}; some regularities in parton distribution functions (DFs) and spin-dependent structure functions \cite{Close:1988br}; $\Lambda(1116)$ and $\Lambda(1520)$ fragmentation functions \cite{Jaffe:2004ph, Wilczek:2004im, Selem:2006nd}; the Regge behaviour of hadrons, namely the fact that baryons and mesons can be accommodated on Regge trajectories with approximately the same slope \cite{Martin:1985hw, Johnson:1975sg, Santopinto:2004hw, Wilczek:2004im, Selem:2006nd}; the absence from the baryon spectrum of the $\Lambda\,\frac32^+$ baryon state \cite{Jaffe:2004ph} and, more generally, the problem of missing baryon resonances \cite{Santopinto:2004hw, Galata:2012xt}.

The concept of diquarks as effective degrees-of-freedom in QMs has proven useful in the calculation of baryon spectra, \emph{e.g}.\ SU(3) light-quark baryons \cite{Lichtenberg:1979de, Lichtenberg:1982jp, Santopinto:2004hw, Ferretti:2011zz, Galata:2012xt, Santopinto:2014opa, Gutierrez:2014qpa, DeSanctis:2014ria} and also heavy-light systems \cite{Roncaglia:1995az, Ebert:1996ec, Gershtein:2000nx, Ebert:2002ig, Giannuzzi:2009gh, Ali:2017wsf, Santopinto:2018ljf}.
As discussed in Refs.~\cite{Santopinto:2004hw, Galata:2012xt, Santopinto:2016fay}, the introduction of hard diquark correlations in light baryon spectroscopy could also provide a solution to the old problem of missing baryon resonances, which affects all the three-quark model predictions for baryon masses \cite{Isgur:1978xj, Capstick:1986bm, Capstick:1992th, Bijker:1994yr, Glozman:1995fu, Loring:2001kx, Giannini:2001kb}.  However, it is no longer certain that such a problem exists because modern data and recent analyses have reduced the number of missing resonances \cite{Ripani:2002ss, Burkert:2012ee, Kamano:2013iva, Crede:2013sze, Mokeev:2015moa, Anisovich:2017pmi}.
Diquark degrees-of-freedom within the framework of a quark model were also applied to baryon structure; some examples are nucleon electromagnetic form factors \cite{Jakob:1993th, Keiner:1995bu, Ma:2002ir, Santopinto:2004hw, DeSanctis:2011zz}, baryon magnetic moments \cite{Lipkin:1981qb, Weiss:1993kv, DeSanctis:2014ria}, electromagnetic transition helicity amplitudes or form factors \cite{Santopinto:2004hw, Ramalho:2008ra}, and in the study of transversity problems and fragmentation functions \cite{Jakob:1997wg, Brodsky:2002cx, Gamberg:2003ey}. Moreover, in the case of the ratio of electric and magnetic form factors of the proton as a function of photon momentum, the quark+diquark model predicts a zero \cite{DeSanctis:2011zz, Santopinto:2016fay}.

Diquark degrees-of-freedom may even play an important role in the context of the spectroscopy and structure of multiquark states. Such systems are hadrons that cannot be described solely in terms of three valence quark, $qqq$, three valence antiquark, $\bar q\bar q\bar q$, or quark+antiquark, $q \bar q$, degrees-of-freedom. They include $XYZ$ states (suspected tetraquarks), such as the $X(3872)$ [now denoted $\chi_{\text{c1}}(3872)$] \cite{Choi:2003ue, Aubert:2004ns, Acosta:2003zx, Abazov:2004kp} and the $X(4274)$ [$\chi_{\text{c1}}(4274)$] \cite{Aaltonen:2011at, Aaij:2016iza}; and the $P_{\text c}$  pentaquark candidates recently discovered by the LHCb Collaboration in $\Lambda_{\rm b}\to J/\psi \Lambda^\ast$ and $\Lambda_{\rm b} \to P_{\rm c}^+ K^- \to (J/\psi p) K^-$ decays \cite{Aaij:2015tga, Aaij:2019vzc}. In addition to heavy+light multiquark configurations, such as $qQ \bar q \bar Q$ and $Q\bar Q qqq$ (with $Q = c$ or $b$), one may also expect the emergence of fully heavy $QQ \bar Q \bar Q$ systems \cite{Berezhnoy:2011xn, Chen:2016jxd, Karliner:2016zzc, Wang:2017jtz, Anwar:2017toa, Esposito:2018cwh, Bedolla:2019zwg, Wu:2016vtq, Liu:2019zuc}. It has been argued \cite{Eichten:2017ual} that if stable $QQ \bar Q \bar Q$ tetraquarks exist, they may be observable at LHC.  However, the empirical status is uncertain \cite{Aaij:2018zrb, Aaij:2020fnh}.

The possible existence of diquark+antidiquark bound states was suggested long ago \cite{Jaffe:1976ig}. Even though they have never been clearly identified experimentally, compact diquark+anti\-diquark configurations might provide an explanation of the properties of hidden-charm/bottom $XYZ$ exotic mesons \cite{Lichtenberg:1979de, Lichtenberg:1996fi, Brink:1998as, Maiani:2004vq, Ebert:2008wm, Deng:2014gqa, Zhao:2014qva, Bicudo:2015vta, Lu:2016cwr, Anwar:2018sol, Yang:2019itm}. On the pentaquark side, diquarks may also play an important role by providing a description of the properties of $P_{\rm c}$ states as diquark+diquark+antiquark configurations \cite{Lichtenberg:1998dm, Cheung:2003de, Shuryak:2003zi, Stewart:2004pd, Maiani:2015vwa, Lebed:2015tna, Li:2015gta}. It is important to note here that multiquark candidates for the exotic $XYZ$ states can alternatively be interpreted as meson+meson molecules, hadro-quarkonium states, and kinematic or threshold effects caused by virtual particles \cite{Chen:2016qju, Esposito:2016noz, Ali:2017jda, Olsen:2017bmm, Guo:2017jvc, Liu:2019zoy, Guo:2019twa}.

In summary, the concept of quark+quark effective degrees-of-freedom is very helpful within the QM phenomenological approach to simplify the description of either conventional or exotic hadrons. This applies not only to spectroscopy but also to structure properties. However, whether these hard diquarks should be understood only as mathematical artifices or as ``physical'' degrees-of-freedom in the hadron's wave function is still a matter of study and debate. To understand their role in three-quark and multiquark bound-state systems, one should compare the predictions of the diquark model with those obtained using explicit quark degrees-of-freedom.


\subsubsection{Diquark wave functions}

A diquark's colour wave function is a superposition of two different SU$_{\rm c}(3)$ configurations,
\begin{equation}
\label{eqn:psi-cD}
\left| \psi_{{\rm c},D} \right\rangle = \alpha \left| ({\bf 3}_{\rm c1},{\bf 3}_{\rm c2}) \bar {\bf 3}_{\rm c12} \right\rangle
+ \beta \left| ({\bf 3}_{\rm c1},{\bf 3}_{\rm c2}) {\bf 6}_{\rm c12} \right\rangle  \,,
\end{equation}	
where ${\bf 3}_{{\rm c}i}$ (with $i = 1$ or 2) are fundamental representations of SU$_{\rm c}(3)$, corresponding to the quark constituents of the diquark, and the coefficients $\alpha$ and $\beta$ satisfy $\alpha^2 + \beta^2$ $=$$ 1$.
In compact tetraquark (diquark+antidiquark) states, the diquark colour wave function of Eq.\,\eqref{eqn:psi-cD} must be combined with that of the antidiquark to obtain a colour-singlet wave function; \emph{i.e}.\ the tetraquark colour wave function is obtained by superposing the $\left| \bar {\bf 3}_{\rm c12}, {\bf 3}_{\rm c34}; {\bf 1}_{\rm c1234} \right\rangle$ and $\left| {\bf 6}_{\rm c12}, \bar {\bf 6}_{\rm c34}; {\bf 1}_{\rm c1234} \right\rangle$ colour-singlet components. In the baryon case, the diquark must be in the $\bar {\bf 3}_{\rm c}$ representation of SU$_{\rm c}(3)$ to satisfy the requirement of a colourless baryon. The baryon colour wave function is then given by $\left| \bar {\bf 3}_{\rm c12}, {\bf 3}_{\rm c3}; {\bf 1}_{\rm c123} \right\rangle$, where ${\bf 3}_{\rm c3}$ is the colour wave function of the third quark inside the baryon.

The QM procedure to construct diquark spin-flavour wave functions is straightforward.  For simplicity, the illustration is restricted to light diquarks, namely those composed of a pair drawn from the set $ \{u, d, s\}$. The extension to heavy+light and fully-heavy diquarks is straightforward and can be found, \emph{e.g}.\ in Refs.\,\cite{Maiani:2004vq, Anwar:2018sol}.

The SU$_{\text{sf}}(6)$ (spin-flavour) diquark wave functions can be constructed using Young diagrams \cite{Hamermesh} by combining two fundamental representations of SU$_{\text{sf}}(6)$, {\bf 6}$_{\rm sf}$, which correspond to the quark constituents of the diquark. One has
\begin{align}
\yng(1) \quad \otimes \quad \yng(1) \quad&=\quad \yng(1,1) \quad \oplus \quad \yng(2) \nonumber \\[0.5ex]
{\bf 6}_{\rm sf} \quad \otimes \quad {\bf 6}_{\rm sf} \quad &= \quad {\bf 15}_{\rm sf} \quad \oplus \quad {\bf 21}_{\rm sf} \,,
\label{eqn:young-diq}
\end{align}
where ${\bf 15}_{\rm sf}$ and ${\bf 21}_{\rm sf}$ are, respectively, the completely antisymmetric and symmetric diquark spin-flavour states.

The diquark total wave function,
\begin{equation}
\psi_D = \psi_{{\rm c},D} \otimes \psi_{{\rm sf},D} \otimes \psi_{{\rm sp},D} \,,
\label{eq:psiD}
\end{equation}
must be completely antisymmetric in order to satisfy the Pauli principle. Here, $\psi_{{\rm c},D}$, $\psi_{{\rm sf},D}$ and $\psi_{{\rm sp},D}$ are, respectively, its colour, spin-flavour, and spatial parts.

Focusing on light baryons with masses below $2.5\,\text{GeV}$, their diquark constituents can be regarded as $S$-wave configurations; namely, with no internal spatial excitations. Therefore, the diquark's colour and spatial wave functions are, respectively, completely antisymmetric and symmetric; and then the diquark spin-flavour wave function has to be completely symmetric. The diquark ${\bf 15}_{\rm sf}$ representation of Eq.\,\eqref{eqn:young-diq} is thus forbidden in the case of low-lying SU(3)-flavour [SU$_{\rm f}(3)$] baryon resonances \cite{Jaffe:2004ph, Wilczek:2004im, Santopinto:2004hw}.
By decomposing the ${\bf 21}_{\rm sf}$ diquark wave function of Eq.~\eqref{eqn:young-diq} in terms of SU$_{\rm s}$(2) $\otimes$ SU$_{\rm f}$(3), one gets two different diquark configurations, the scalar diquark, with flavour $\bar {\bf 3}_{\rm f}$ and spin $S = 0$, and the axial-vector diquark, with flavour ${\bf 6}_{\rm f}$ and spin $S = 1$. (They have been called ``good'' and ``bad'', respectively; but since both appear crucial to the structure of all baryons, that terminology is not employed herein because it is misleading.)  By means of a one-gluon-exchange interaction between the two quarks, one can show that the scalar diquark is $\sim 20$\% lighter; hence, should be the dominant configuration in low-lying baryon states \cite{DeRujula:1975qlm, DeGrand:1975cf, Jaffe:2004ph, Wilczek:2004im, Santopinto:2004hw, Santopinto:2014opa}.

The baryon spin-flavour states are obtained by combining the two-quark SU$_{\rm sf}$(6) representations of Eq.\,\eqref{eqn:young-diq} with a ${\bf 6}_{\rm sf}$ representation, which corresponds to the third constituent quark within the baryon.
One has
\begin{subequations}
\label{eqn:young-bar}
\begin{equation}
\label{eqn:young-bar-A}
\left. \begin{array}{ccccccc}
\yng(1,1) & \otimes & \yng(1) & = & \yng(1,1,1) & \oplus & \yng(2,1)  \\
&&&&&&\\
{\bf 15}_{\rm sf} & \otimes & {\bf 6}_{\rm sf} & = & {\bf 20}_{\rm sf}  & \oplus & {\bf 70}_{\rm sf}
\end{array} \right. \,,
\end{equation}
and
\begin{equation}
\label{eqn:young-bar-S}
\left. \begin{array}{ccccccc}
\yng(2) & \otimes & \yng(1) & = & \yng(3) & \oplus & \yng(2,1) \\
&&&&&&\\
{\bf 21}_{\rm sf} & \otimes & {\bf 6}_{\rm sf} & = & {\bf 56}_{\rm sf} & \oplus & {\bf 70}_{\rm sf}
\end{array} \right. \,.
\end{equation}
\end{subequations}
In the three-quark model, all spin-flavour states in Eqs.\,\eqref{eqn:young-bar} are achievable. Conversely, in the quark+diquark model only those of Eq.\,\eqref{eqn:young-bar-S} are accessible. Therefore, in the quark+diquark model the number of states is much reduced with respect to the three-quark model. This argument \cite{Santopinto:2004hw, Galata:2012xt} offers a solution to the missing baryon resonance problem, if it exists.  

The missing baryon resonances are states predicted by QMs, with (as yet) no corresponding experimentally observed counterparts.  One may argue that there could be baryon states very weakly coupled to the single pion, but with higher probabilities of decaying into two or more pions or into other mesons \cite{Capstick:1992th, Bijker:1994yr, Ferretti:2015ada}.
The detection of such resonances is further complicated by the problem of separating experimental data from backgrounds and by the expansion of the differential cross section into many partial waves.
Alternately, it is possible to consider models that are characterised by a smaller number of effective degrees of freedom with respect to the three-constituent-quark models and to assume that some of the missing states, not yet observed experimentally, simply do not exist. This is the case for the quark+diquark models discussed in Ref.\,\cite{Galata:2012xt, Santopinto:2004hw, Santopinto:2014opa}, in particular Ref.\,\cite[Table\,III]{Galata:2012xt}.  At the same time, it should be kept in mind that quark+diquark models \cite{Galata:2012xt, Santopinto:2014opa, Santopinto:2016zkl} also have missing baryon states, but only fewer than three-quark models.

The construction of light and heavy+light tetraquarks as diquark+antidiquark states can be found in, for instance, Refs.\,\cite{Maiani:2004vq, Maiani:2004uc, Santopinto:2006my, Anwar:2018sol}; for the construction of pentaquark wave functions as diquark+diquark+antiquark states, see \emph{e.g}.\ Ref.\,\cite{Maiani:2015vwa}.


\subsubsection{Diquark masses}

There are three standard ways to estimate diquark masses in QMs: they can be considered as model parameters to be fitted to experimental data \cite{Santopinto:2004hw, Galata:2012xt, Ferretti:2011zz, Santopinto:2014opa}; they can be estimated via phenomenological considerations \cite{Selem:2006nd, Jaffe:2004ph}; or they can be calculated by binding two quarks via one-gluon-exchange interaction \cite{Anwar:2017toa, Anwar:2018sol, Bedolla:2019zwg, Ferretti:2019zyh} plus a spin-spin contribution \cite{Maiani:2004vq}.

Ref.\,\cite{Jaffe:2004ph} highlighted that in heavy+light baryons an elementary scalar diquark, $[q_1, q_2]_{0^+}$, has no spin interaction with the spectator heavy quark, $Q$, while the kindred axial-vector diquark, $\{q_1,q_2\}_{1^+}$, does. One has
\begin{equation}
\label{eqn:jaffe-ss}
H(Q,\{q_1,q_2\}_{1^+}) = K(Q,\{q_1,q_2\}_{1^+}) \, 2 \, {\bf S}_{\{q_1,q_2\}_{1^+}} \cdot {\bf S}_Q \,,
\end{equation}	
where ${\bf S}_{\{q_1,q_2\}_{1^+}}$ and ${\bf S}_Q$ are the spins of the light axial-vector diquark and heavy quark, respectively; and the coefficient $K(Q,\{q_1,q_2\})$ depends on the quark masses. To estimate the difference between scalar and axial-vector diquark masses, it is necessary to take linear combinations of baryon (and meson) masses that eliminate the spin-dependent interaction of Eq.~\eqref{eqn:jaffe-ss}. For example, one has: $M_{ud}^{\rm av} - M_{ud}^{\rm sc} = \frac{1}{3} \left(2 M(\Sigma_Q^*) + M(\Sigma_Q) \right)- M(\Lambda_Q)$. This leads to the following results for the scalar--axial-vector diquark mass differences \cite{Jaffe:2004ph}: $M_{ud}^{\rm av} - M_{ud}^{\rm sc} \simeq 210$ MeV, $M_{ud}^{\rm sc} - M_u \simeq 315$ MeV, $M_{us}^{\rm av} - M_{us}^{\rm sc} = 152$ MeV, and $M_{us}^{\rm sc} - M_s = 498$ MeV.

A similar idea was used in Ref.\,\cite{Maiani:2004vq}, wherein the diquark masses were estimated by first extracting the strength of the quark-quark spin-spin interaction in a colour antitriplet state, $(\kappa_{qq})_{\bar {\bf 3}}$, from several baryon masses, like that of the $\Lambda$ (to evaluate the scalar diquark mass) and that of the $\Sigma$ (to estimate the axial-vector diquark mass). By plugging the previous $\kappa$ estimates into an algebraic mass formula with spin-spin interactions for tetraquarks, light diquark masses were inferred by fitting their values to the $a_0(980)$ and $\sigma(480)$ experimental levels: $M_{ud}^{\rm sc} = 395$ MeV and $M_{sq}^{\rm sc} = 590$ MeV (with $q = u$ or $d$).  Using the same approach to fit the $X(3872)$ tetraquark mass, then $M_{cq}^{\rm sc} = 1933$ MeV.
(Such low values for the scalar and axial-vector diquark masses are inconsistent with many calculations; \emph{e.g}.\ herein see: Table~\ref{tab:diquark-masses-OGE}; Fig.\,\ref{fig:meson-diquark-masses} and Eq.\,\eqref{diquarkmassescalculated}; and Table~\ref{tab:mass}.  Moreover, continuum Schwinger function methods (CSMs) applied to QCD suggest that $\sigma$, $a_0$ are dominated by meson+meson, not diquark+antidiquark, channels; and the $X(3872)$ is primarily a molecule-like $DD^\ast$ system.  More on this in Sec.\,\ref{sec:diquarks-dse}.)

Ref.\,\cite{Selem:2006nd} approached the probem by generalizing the Chew-Frautschi formula, $M^2 = a + \sigma L$, which describes the Regge trajectories of resonances with the same internal quantum numbers but different values of $J^P$.  Here, $\sigma$ is a constant ($\simeq 1.1$ GeV$^2$), $a$ depends on the quantum numbers and $L$ is the orbital angular momentum. By considering two masses, $m_1$ and $m_2$, connected by a relativistic string with angular momentum $L$ and constant tension $T$, and in the limit of small $m_{1,2}$, the following expression was obtained
\begin{equation}
E \simeq \sqrt{\sigma L} + \kappa L^{-1/4} \mu^{3/2},
\end{equation}
where $\kappa \simeq 1.15$ GeV$^{-1/2}$ and $\mu^{3/2} = m_1^{3/2} + m_2^{3/2}$.
Using a simple picture in which baryons contain only one type of diquark, then comparing those with scalar diquarks and those containing axial-vector diquarks, inferences were made regarding the mass difference between diquarks, \emph{e.g}.\  $M_{ud}^{\rm av} > M_{us}^{\rm sc} > M_s > M_{ud}^{\rm sc}$ and  $(M_{ud}^{\rm av})^{3/2} - (M_{ud}^{\rm sc})^{3/2} = 0.28$ GeV$^{3/2}$.  If $M_{ud}^{\rm sc}$ varies from 100 to 500 MeV, then $M_{ud}^{\rm av} - M_{ud}^{\rm sc}$ ranges from 360 to 240 MeV.

\begin{table}[!t]
\caption{\label{tab:diquark-masses-OGE} Scalar and axial-vector diquark masses, $M^{\rm sc}$ and $M^{\rm av}$, respectively, computed by means of the relativised QM Hamiltonian of Refs.~\cite{Godfrey:1985xj, Capstick:1986bm}. Notation: $q$ indicates light, $u$ or $d$, quarks. These results were previously reported in Ref.\,\cite[Table 1]{Ferretti:2019zyh}.}
\centering
\begin{ruledtabular}
\begin{tabular}{ccc}
Flavour content & $M^{\rm sc}$ (MeV) & $M^{\rm av}$ (MeV) \\
\hline
$qq$ & 691  & 840 \\
$qs$ & 886  & 992 \\
$ss$ & --   & 1136 \\
$qc$ & 2099 & 2138 \\
$sc$ & 2229 & 2264 \\
$cc$ & --   & 3329 \\
$qb$ & 5451 & 5465 \\
$sb$ & 5572 & 5585 \\
$cb$ & 6599 & 6611 \\
$bb$ & --   & 9845 \\
\end{tabular}
\end{ruledtabular}
\end{table}

The remaining approach is exemplified in Refs.~\cite{Anwar:2017toa, Anwar:2018sol, Bedolla:2019zwg, Ferretti:2019zyh}, wherein a relativised QM Hamiltonian \cite{Godfrey:1985xj, Capstick:1986bm} was used to bind a quark+quark pair. To do that, one needs a relation between quark-quark and quark-antiquark colour Casimirs, $\langle {\bf F}_q \cdot {\bf F}_{\bar q}\rangle = - \frac{4}{3} = 2 \langle {\bf F}_q \cdot {\bf F}_q\rangle$~\cite[Eqs. (3, 4, 8)]{Godfrey:1985xj}, where the ${\bf F}$'s are related to the Gell-Mann colour matrices by ${\bf F}^a = \frac{\lambda^a}{2}$. The results are shown in Table~\ref{tab:diquark-masses-OGE}.


\subsubsection{Light and heavy-light baryons in the diquark model}

The description of baryons as quark+diquark bound states has important consequences. The main one is that the internal dynamics among quark+diquark constituents can be described by a single relative coordinate, ${\bf r}_{\rm rel}$, instead of the usual ${\bm \rho}$ and $\bm \lambda$ Jacobi coordinates of a three-quark system. As a result, one obtains a spectrum characterised by a smaller number of states with respect to the one predicted by three-quark models, as  discussed in Ref.~\cite{Santopinto:2004hw} and below.

There are several quark+diquark models for baryon spectroscopy. Some of them are potential models, like the interacting quark+diquark model of Refs.\,\cite{Santopinto:2004hw, Ferretti:2011zz, Santopinto:2014opa, DeSanctis:2014ria}, the relativised quark+diquark models of Refs.\,\cite{Lichtenberg:1982jp, Ebert:1996ec, Ebert:2002ig}, and the nonrelativistic potential model of Ref.~\cite{Gershtein:2000nx}. Others are simple algebraic models, such as the quark+diquark model of Ref.~\cite{Galata:2012xt}.

Refs.\,\cite{Ebert:1996ec, Ebert:2002ig} report a spectrum of doubly-heavy baryons computed using a relativised quark+di\-quark model. In particular, the result for the ground-state mass of the $\Xi_{\rm cc}$ with $J^P = \frac{1}{2}^+$, 3620 MeV, is compatible with the experimental mass of the $\Xi_{\rm cc}^{++}$ resonance listed recently by the PDG \cite{Tanabashi:2018oca}: $3621.2\pm0.7$ MeV, even though the experimental quantum numbers are still unknown. The theoretical predictions for the ground-state masses of the $\Xi_{\rm bb}$, $\Omega_{\rm bb}$, and $\Omega_{\rm cc}$ configurations are, respectively, 10202 MeV, 10359 MeV, and 3778 MeV.
(Complete spectra, obtained using CSMs and exploiting all possible dynamical diquark configurations, are drawn in Fig.\,\ref{figPeiLin}.)

In the interacting quark+diquark model of Refs.\,\cite{Santopinto:2004hw, Ferretti:2011zz, Santopinto:2014opa, DeSanctis:2014ria}, the quark-diquark interaction is the sum of a Coulomb-like + linear-confining potential, $V_{\rm conf} = - \frac{\alpha}{r} + \beta r$, $\alpha$ and $\beta$ being free parameters, plus an exchange interaction,
\begin{align}
M_{\rm ex}(r) &= \left(-1 \right)^{L + 1} \mbox{ } e^{-\sigma r}  \left[ A_{\rm S} \mbox{ } {\bf s}_1
\cdot {\bf s}_2  + A_{\rm F} \mbox{ } {\bm \lambda}_1^{\rm f} \cdot  {\bm \lambda}_2^{\rm f} \mbox{ } + A_{\rm I} \mbox{ } {\bf t}_1 \cdot {\bf t}_2  \right] \,,
\label{eqn:Vexch-strange}
\end{align}
which depends on the quantum numbers of the quark and diquark: their relative orbital angular momentum ($L$), their spins (${\bf s}_i$, with $i = 1,\,2$), isospins (${\bf t}_i$), and flavour representations [the SU$_{\rm f}$(3) Gell-Mann matrices ${\bm \lambda}_i^{\rm f}$]; $A_{\rm S}$, $A_{\rm F}$, $A_{\rm I}$, and $\sigma$ are model parameters, fitted to the experimental data. This model was applied to both nonstrange \cite{Santopinto:2004hw, Ferretti:2011zz, DeSanctis:2014ria} and strange \cite{Santopinto:2014opa} baryon spectroscopy. In the nonstrange sector, the spectrum of the model shows no missing baryon resonances up to an energy of 2\,GeV; the calculated spectrum of hyperons is also reasonably reproduced.


\subsubsection{Compact tetraquarks in the diquark model}

The diquark model was also used in the context of compact (diquark+antidiquark) tetraquark spectroscopy. In particular, it was applied to the study of light \cite{Lichtenberg:1996fi, Santopinto:2006my, Maiani:2004uc} and heavy+light \cite{Brink:1998as, Ebert:2008wm, Lichtenberg:1996fi, Deng:2014gqa, Bicudo:2015vta, Lu:2016cwr, Zhao:2014qva, Maiani:2004vq, Anwar:2018sol, Yang:2019itm} tetraquarks. The study of compact heavy+light tetraquark configurations might provide an explanation of the properties of some hidden-charm/bottom $XYZ$ exotic mesons \cite{Chen:2016qju, Ali:2017jda, Olsen:2017bmm, Guo:2017jvc}.

\begin{figure}[!t]
\begin{center}
\includegraphics[width=0.6\textwidth]{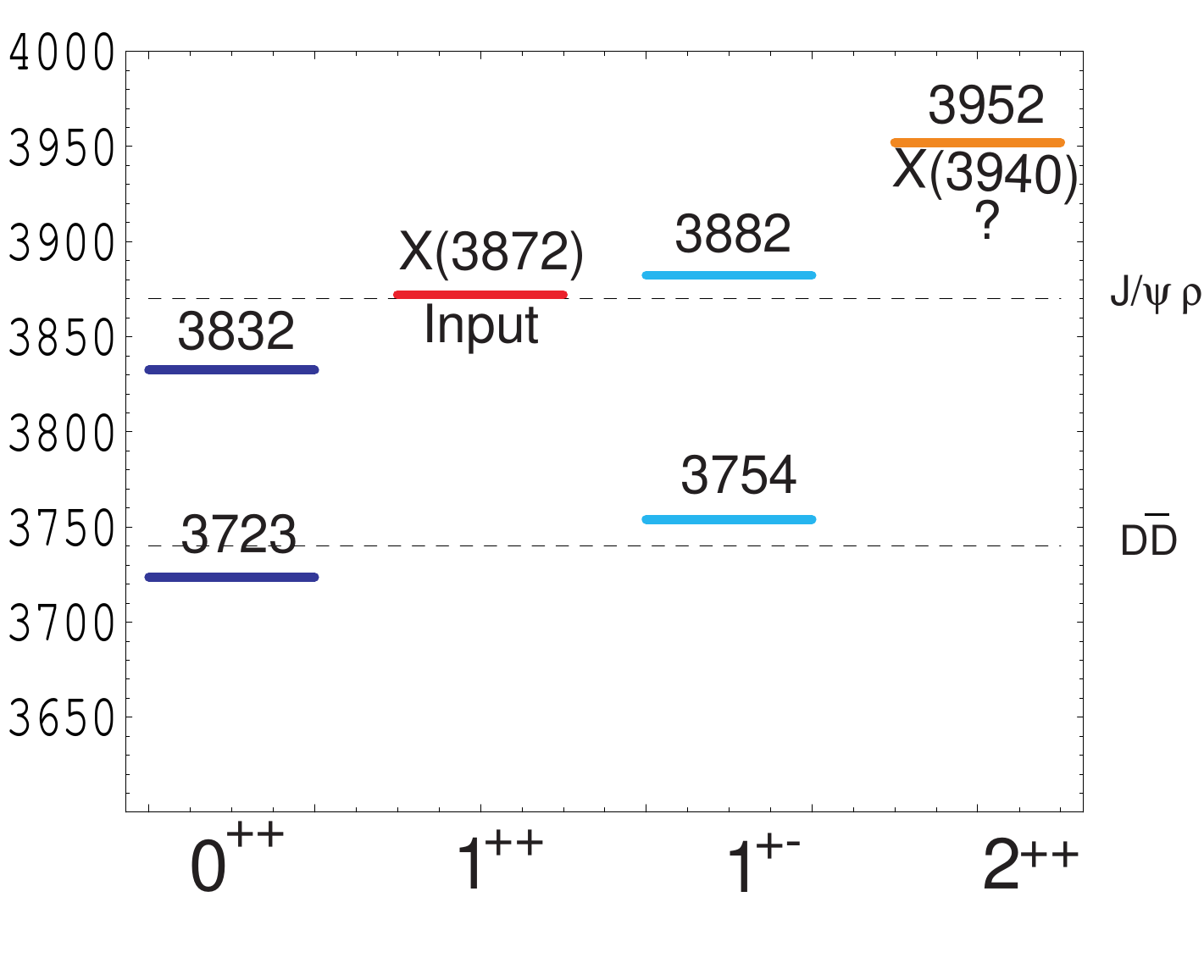}\vspace*{-5ex}
\end{center}
\caption{\label{fig:X(3872)}
Spectrum of the $X(3872)$-containing multiplet from Ref.\,\cite{Maiani:2004vq}. (Masses in MeV.)}
\end{figure}

Ref.\,\cite{Maiani:2004vq} discussed the possible appearance of heavy-light tetraquarks within an algebraic model, proposing the following mass formula:
\begin{align}
H &= 2 M_{qc}^{\rm sc} \nonumber \\
& + 2(\kappa_{cq})_{\bar {\bf 3}} \left[ {\bf S}_c \cdot {\bf S}_q + {\bf S}_{\bar c} \cdot {\bf S}_{\bar q'}\right] +2 \kappa_{q \bar q} \left( {\bf S}_c \cdot {\bf S}_{\bar q'} \right)
+ 2 \kappa_{c \bar q} 	\left[ {\bf S}_c \cdot {\bf S}_{\bar q'} + {\bf S}_{\bar c} \cdot {\bf S}_q\right] + 2 \kappa_{c \bar c} \left( {\bf S}_c \cdot {\bf S}_{\bar c} \right) \,, \label{MaianiTI}
\end{align}	
where the $\kappa$ parameters are flavour-dependent strengths of the spin-spin interaction, fitted to light and heavy+light baryon mass differences.  After fitting the $M_{qc}^{\rm sc}$ parameter to the mass of the $X(3872)$, Ref.\,\cite{Maiani:2004vq} computed the spectrum of tetraquarks belonging to the $X(3872)$ multiplet, with the result drawn in Fig.\,\ref{fig:X(3872)}.  (See also the discussion of Fig.\,\ref{spettroXf}.)

Ref.\,\cite{Ebert:2008wm} calculated the heavy+light tetraquark spectrum using a relativistic diquark+antidiquark model with one-gluon exchange and long-range vector and scalar linear-confinement potentials. The interpretation therein of the $X(3872)$ as a $qc \bar q \bar c$ state is the same as Ref.\,\cite{Maiani:2004vq}.

Ref.\,\cite{Anwar:2018sol} computed the spectrum of hidden-charm ($q c \bar q \bar c$ and $s c \bar s \bar c$) tetraquarks by means of a relativised potential model with linear-confinement and one-gluon exchange (OGE) interactions. In particular, it was shown that $13$ charmonium-like observed states can be accommodated in the tetraquark picture, with the exception of the $X(4274)$.
Ref.\,\cite{Lu:2016cwr} used a similar model to study the $s \bar s c \bar c$ sector and discussed possible assignments for the $X(4140)$, $X(4274)$, $X(4500)$, and $X(4700)$.  As in Ref.\,\cite{Anwar:2018sol}, the $X(4274)$ could not be accommodated in this tetraquark picture.


\subsubsection{Compact pentaquarks in the diquark model}

The potential hidden-charm pentaquark signals, $P_{\rm c}$, were observed by the LHCb Collaboration in $\Lambda_{\rm b} \rightarrow J/\psi \Lambda^*$ and $\Lambda_{\rm b} \rightarrow P_{\rm c}^+ K^- \rightarrow (J/\psi p) K^-$ decays~\cite{Aaij:2015tga, Aaij:2019vzc}. They carry one unit of baryon number and show the peculiar quark structure $P_{\rm c}^+ = uud c \bar c$, whence the name pentaquarks. The mass difference between the observed pentaquarks, $P_{\rm c}(4312)^+$ on one side, $P_{\rm c}(4440)^+$ and $P_{\rm c}(4457)^+$ on the other, is of the order of $\Delta M = 140$ MeV. This is much smaller than the energy associated with an orbital excitation, $\mathcal O (300)$ MeV, as \emph{e.g}.\ in the case $M_{\Lambda(1405)} - M_{\Lambda(1116)} \simeq 290$ MeV.

In Ref.\,\cite{Maiani:2015vwa}, the splitting $\Delta M$ was explained in the context of the pentaquark model by considering 5-quark states characterised by different diquark contents. In particular, two possible valence quark structures were proposed:
\begin{equation}
P_{\rm c,u} = \epsilon^{\alpha\beta\gamma} \bar c_\alpha [cu]_{\beta; S =0,1} [ud]_{\gamma; S = 0,1} \,,
\quad
P_{\rm c,d} = \epsilon^{\alpha\beta\gamma} \bar c_\alpha [cd]_{\beta; S =0,1} [uu]_{\gamma; S = 1} \,,
\end{equation}
where Greek letters are colour indices and the diquarks are in the colour anti-triplet, $\bar {\bf 3}_{\rm c}$, configuration.

\begin{figure}[!t]
\begin{center}
\includegraphics[width=0.75\textwidth]{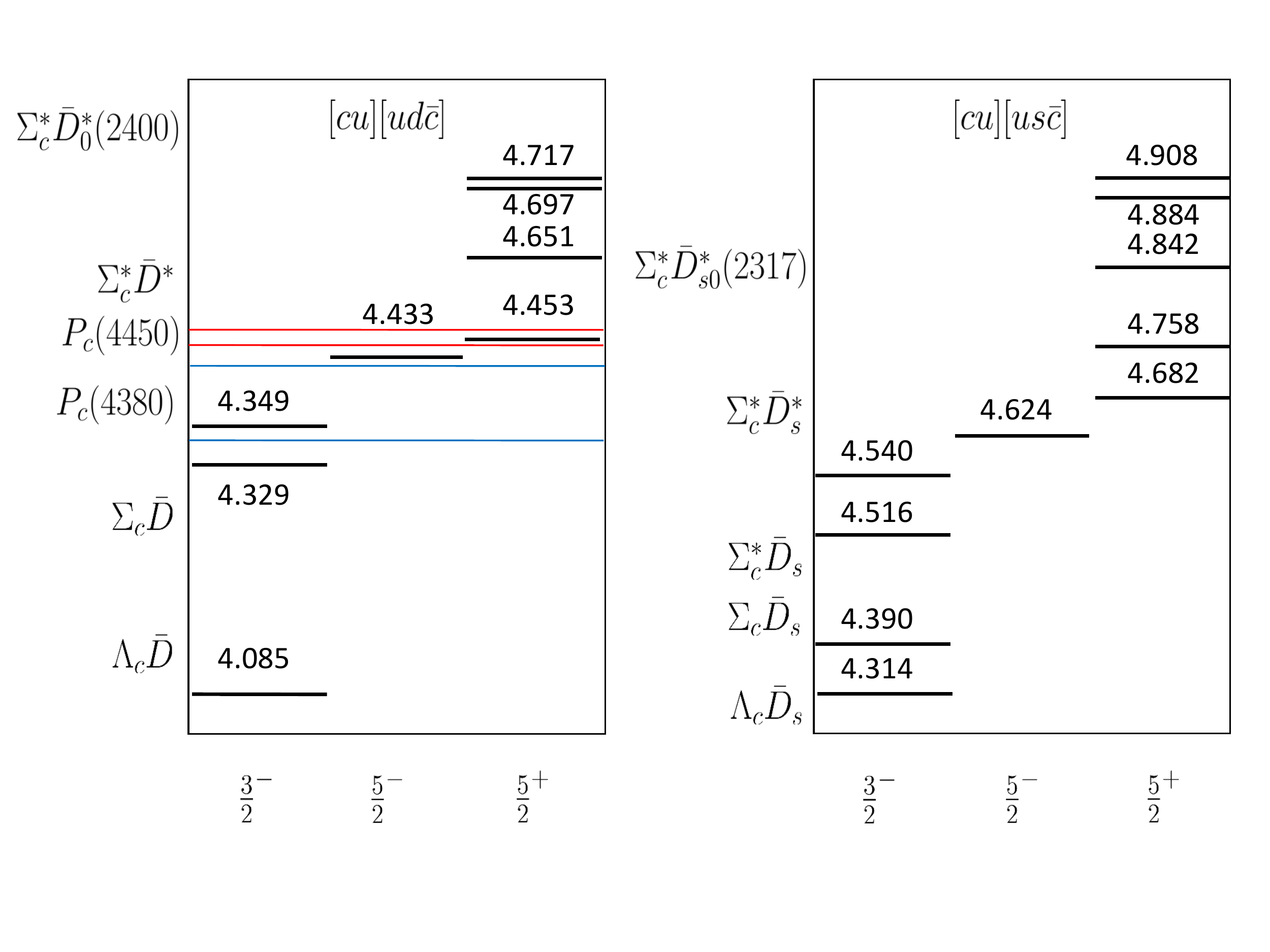}\vspace*{-10ex}
\end{center}
\caption{Predicted masses, in MeV, for hidden-charm pentaquarks \cite{Zhu:2015bba} (thick black lines) compared with experimental data \cite{Tanabashi:2018oca} (thin coloured lines).}
\label{fig:Pc-penta}
\end{figure}

The properties and quantum numbers of $P_{\rm c}$ pentaquarks were also studied in the context of the diquark model in Refs.\,\cite{Li:2015gta, Zhu:2015bba, Ali:2016dkf, Ghosh:2017fwg, Giannuzzi:2019esi}.
Ref.\,\cite{Zhu:2015bba} interpreted the LHCb hidden-charm pentaquarks as diquark, $q_1 q_2$, and triquark, $q_3 q_4 \bar q_5$, bound states. The colour structure of the diquark constituent is the same as Eq.\,\eqref{eqn:psi-cD}, namely ${\bf 3}_{{\rm c}1} \otimes {\bf 3}_{{\rm c}2} = \bar {\bf 3}_{{\rm c}12} + {\bf 6}_{{\rm c}12}$; in the triquark case, one has ${\bf 3}_{\rm c3} \otimes {\bf 3}_{\rm c4} \otimes \bar {\bf 3}_{\rm c5} = {\bf 3}_{\rm c345} \oplus \bar {\bf 6}_{\rm c345} \oplus {\bf 3}_{\rm c345} \oplus {\bf 15}_{\rm c345}$. The colour-singlet pentaquark wave function, ${\bf 1}_{\rm c12345}$, was obtained by combining a diquark in the $\bar {\bf 3}_{{\rm c}12}$ configuration and a triquark in ${\bf 3}_{\rm c345}$. The masses of the $P_{\rm c}$ pentaquark were also computed by means of an algebraic mass formula, characterised by spin-spin and spin-orbit interactions, with the results shown in Fig.\,\ref{fig:Pc-penta}. A similar mass formula was used in Ref.\,\cite{Ali:2016dkf}, assuming a diquark+diquark+antiquark description of $P_{\rm c}$ states.

The masses of $qqqQ \bar Q$ pentaquark configurations (with $Q = c$ or $b$) were computed in Ref.\,\cite{Giannuzzi:2019esi} using a potential model inspired by an AdS/QCD model. The interaction is very similar to that typically described as the Cornell potential; and the results are $100-200$ MeV above the corresponding experimental data.



\begin{figure}[!t]
\begin{center}
\includegraphics[width=0.60\textwidth]{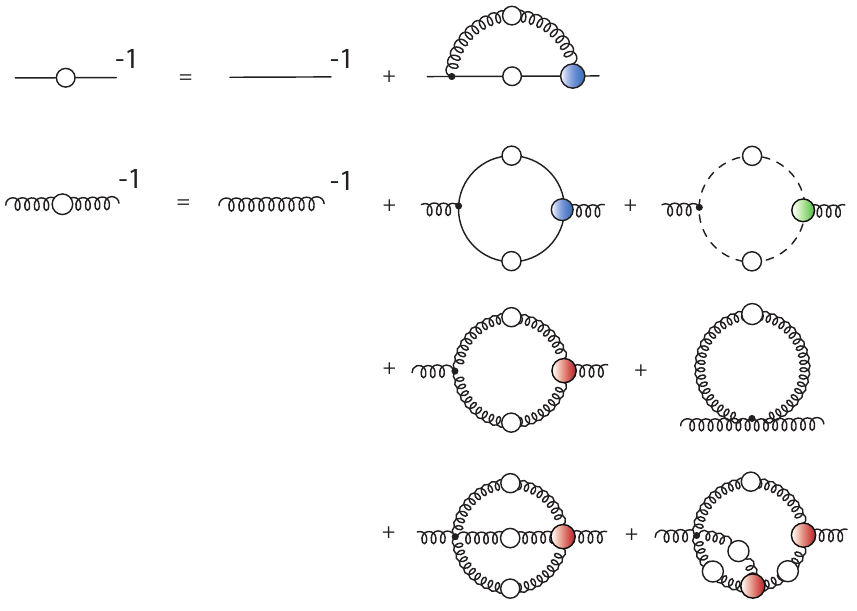}
\end{center}
\caption{\label{DSEs} DSEs for the quark two-point Schwinger function (propagator) (top) and the gluon two-point function (bottom). Solid, curly and dashed lines represent quarks, gluons and ghosts, respectively.}
\end{figure}

\subsection{Continuum Schwinger Function Methods}
\label{sec:DSEs}
The role of diquark correlations inside hadrons has also long been emphasised in studies using CSMs, such as the Dyson-Schwinger equations (DSEs); see, \emph{e.g}.\ Refs.\,\cite{Roberts:1994dr, Roberts:2000aa, Alkofer:2000wg, Chang:2011vu, Bashir:2012fs, Aznauryan:2012ba, Roberts:2015lja, Eichmann:2016yit, Burkert:2017djo, Fischer:2018sdj} for reviews on their applications to hadron physics.  As a quantum field theory equivalent of the Euler-Lagrange equations, the DSEs are a system of integral equations whose solutions deliver QCD's $n$-point Schwinger functions, \emph{i.e}.\ the same quantities computed in numerical simulations of lQCD.  The simplest DSEs are illustrated in Fig.\,\ref{DSEs}, \emph{viz}.\ the gap equations for the quark and gluon.  These equations provide the keys to understanding the emergence of hadronic mass in the SM, \emph{e.g}.\ \emph{\`a la} Nambu \cite{Nambu:2009zza}, a nonzero dressed-quark mass-function emerges in solving the quark gap equation even in the absence of couplings to the Higgs boson.  This is the basic signature of DCSB; namely, the emergence of \emph{mass from nothing}, and there is a firm theoretical position from which one can argue that DCSB is responsible for more than 98\% of the visible mass in the Universe \cite{Roberts:2016vyn}.

\begin{figure}[!t]
\begin{center}
\includegraphics[width=0.45\textwidth] {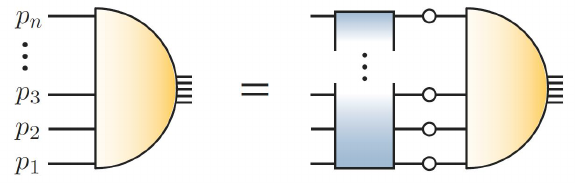}
\end{center}
\caption{\label{fig:BSE} Generic form of the homogeneous integral equation for an $n$-valence-body bound state, which is described herein as a (generalised) Bethe-Salpeter equation (BSE). The lines with circles are dressed quark propagators and the kernel is the sum of all irreducible two-, three-, \ldots, $n$-body contributions.}
\end{figure}

At the next level of complexity are the Poincar\'e-covariant bound-state equations, Bethe-Salpeter \cite{Salpeter:1951sz}, Faddeev \cite{Faddeev:1960su}, \emph{etc}., a generic form of which is illustrated in Fig.\,\ref{fig:BSE}.  The bound-state kernel, indicated by the shaded box, is the sum of all possible irreducible two-, three-, \ldots, $n$-body contributions.  The solution of such an equation yields the mass (pole-position) and bound-state amplitude for a bound-state (resonance) seeded by a total of $n$ valence quarks and/or antiquarks.  This information provides the foundation for computing all properties of the associated hadron.  Moreover, with the external legs reattached to the bound-state amplitude, one obtains a Poincar\'e covariant wave function that, under certain limiting conditions, possesses a mathematical connection to the wave functions typical of quantum mechanics.

As noted, the kernels depend on an array of QCD's $n$-point functions, sound information about which is therefore important in developing the solutions.  Here, the past two decades have seen substantial progress, with results provided by DSE studies \cite{Binosi:2014aea, Eichmann:2014xya, Williams:2014iea, Cyrol:2014kca, Aguilar:2015nqa, Williams:2015cvx, Athenodorou:2016oyh, Binosi:2016wcx, Binosi:2016nme, Aguilar:2019uob, Huber:2020keu}, functional renormalisation-group equations \cite{Mitter:2014wpa, Cyrol:2016tym, Cyrol:2017ewj} and lQCD \cite{Cucchieri:2007md, Cucchieri:2008qm, Bogolubsky:2009dc, Maas:2011se, Ayala:2012pb, Duarte:2016iko, Binosi:2016xxu, Gao:2017uox, Oliveira:2018lln, Boucaud:2018xup, Cui:2019dwv}.  Notably, where fair comparisons can be drawn, these three approaches agree; hence, the results provide a robust foundation from which to develop predictions for hadron observables.  (Landau gauge is typically employed because it is a fixed point of the renormalisation group and that gauge most readily implemented in lQCD.)

Furthermore, extensive progress has been made in developing symmetry-preserving schemes for combining QCD's $n$-point functions into Bethe-Salpeter kernels that guarantee all Ward-Green-Takahashi identities (WGTIs) are satisfied in the study of hadron observables.  For instance, the axial-vector WGTI is crucial to ensuring that DCSB is both a necessary and sufficient condition for the pion's emergence as a NG mode; and proving this and insightfully expressing its wide-ranging impact on hadron observables has been a distinguishing success of the DSE approach for more than twenty years.  The systematic, symmetry preserving truncation schemes that have been developed for this purpose can be traced from Refs.\,\cite{Chang:2009zb, Chang:2010hb, Chang:2011ei, Chang:2011vu, Bashir:2012fs, Heupel:2014ina, Sanchis-Alepuz:2015qra, Williams:2015cvx, Roberts:2015lja, Binosi:2016rxz, Eichmann:2016yit}

The leading order in such a scheme is the rainbow-ladder (RL) truncation, where the $q\bar{q}$ and $qq$ kernels in mesons and baryons are expressed by gluon exchanges with a momentum-dependent effective interaction that is provided by \emph{Ansatz}. This leaves the quark propagator to be solved from its DSE with information on other relevant $n$-point functions implicit in the interaction \emph{Ansatz}. RL truncation has been successful in a range of applications, including the properties of (isovector) pseudoscalar and vector ground-state mesons as well as $J^P = 1/2^+$ octet baryons and $J=3/2^+$ decuplet baryons. Its deficiencies in other meson and baryon channels and in the heavy+light meson sector are also well documented, see \emph{e.g}. Refs.\,\cite{Chang:2011vu, Bashir:2012fs, Roberts:2015lja, Eichmann:2016yit, Yao:2020vef} and references therein.

It is clear from Fig.\,\ref{DSEs} that improving upon RL truncation involves a substantial increase in complexity since it requires explicit information about the gluon propagator, quark-gluon vertex and other $n$-point functions. So far, kernels beyond RL have mostly been employed only for light mesons, where they improve the spectrum significantly \cite{Chang:2009zb, Chang:2010hb, Chang:2011ei, Williams:2015cvx}. For baryons, some exploratory calculations beyond RL are available \cite{Sanchis-Alepuz:2014wea, Sanchis-Alepuz:2015qra, Chen:2017pse, Chen:2019fzn}. This is also the point where connections to an underlying soft-diquark structure can be made and profitably exploited.


\subsubsection{Diquarks}
\label{sec:diquarks-dse}
There are many reasons to anticipate a role for diquark correlations within baryons.  For instance:
quark+quark scattering in the colour-antitriplet ($\bar 3_c$) channel is attractive;
the theory of superconductivity reveals that fermions pair even in the presence of an arbitrarily small attractive interaction;
phase space factors materially enhance two-body interactions over $n\geq 3$ body interactions within a baryon;
and the primary three-body force, produced by a three-gluon vertex attaching once, and only once, to each valence quark, vanishes when projected into the colour-singlet channel:
\begin{equation}
\label{primafacie}
\begin{array}{cccc}
 \mbox{final\,state} & \mbox{three\,gluon\,vertex} & \mbox{initial\,state} & \\[-1ex]
\mbox{colour\,wave function} &  & \mbox{colour\,wave function} & \\
 \varepsilon_{f_1 f_2 f_3} & f^{abc} [\lambda^a]_{f_1 i_1} \, [\lambda^b]_{f_2 i_2} \, [\lambda^c]_{f_3 i_3} & \varepsilon_{i_1 i_2 i_3} & = 0\,,
\end{array}
\end{equation}
where $\varepsilon_{ijk}$ is the Levi-Civita tensor, $\{\lambda^a\}$ are SU$_{\rm  c}(3)$ Gell-Mann matrices, and $f^{abc}$ is the structure tensor of SU$_{\rm c}(3)$.  Consequently, the leading role for the three-gluon vertex interaction within a baryon is the strengthening of quark+quark correlations by attaching twice to one of the valence quarks and additionally to one of the others.


A mathematical link between mesons and diquarks is forged by their Bethe-Salpeter (BS) amplitudes, whose tensors only differ by inclusion of the charge conjugation matrix. Denoting this matrix by $C$, a pseudoscalar meson ($\gamma_5$) is linked to a scalar diquark ($\gamma_5 C$), a vector meson ($\gamma^\mu$) to an axial-vector diquark ($\gamma^\mu C$), \emph{etc}. Diquarks are subject to the Pauli principle, which in turn determines their isospin. The full colour-spinor-flavour amplitude of a diquark must be antisymmetric under quark exchange; the colour part $\varepsilon_{ijk}$ is antisymmetric by itself and $\gamma^5 C$ is an antisymmetric Dirac matrix; hence, a scalar diquark made of light quarks must have an antisymmetric flavour wave function $[ud] \sim ud - du$ with $I=0$. In this way, the non-exotic meson channels with
\begin{equation}
J^{PC} = 0^{-+}, \;\; 1^{--}, \;\; 0^{++}, \;\; 1^{++}, \;\; 1^{+-}
\end{equation}
have the following diquark partners:
\begin{equation}
\label{diquarkpartners}
I(J^P) = 0(0^+), \;\; 1(1^+), \;\; 0(0^-), \;\; 0(1^-), \;\; 1(1^-).
\end{equation}

\begin{figure}[!t]
\begin{center}
\includegraphics[width=0.75\columnwidth]{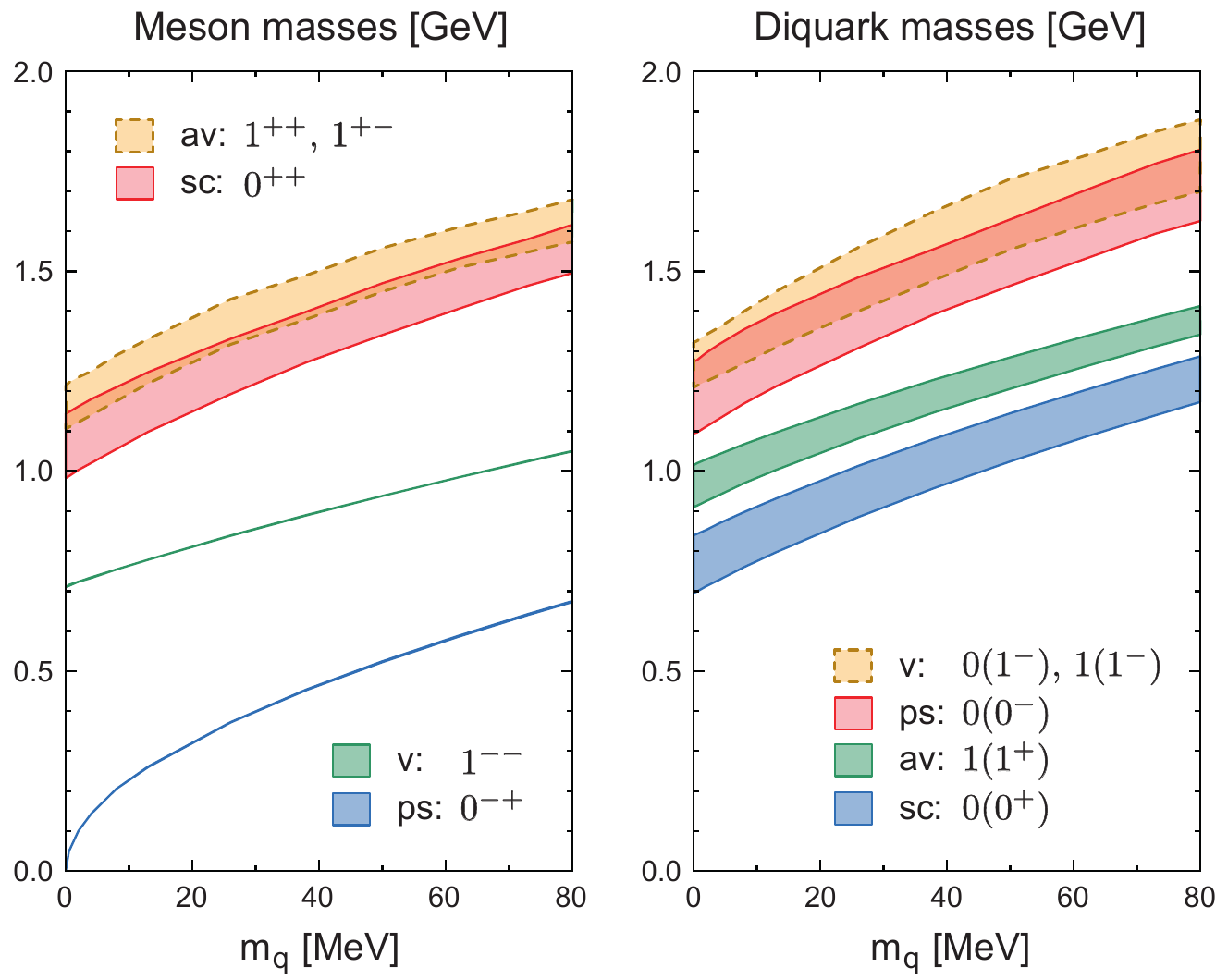}
\end{center}
\caption{\label{fig:meson-diquark-masses} Meson and diquark masses from their BSEs plotted versus current-quark mass. The bands express a RL interaction uncertainty \cite{Eichmann:2016hgl}. For the scalar and axial-vector mesons, following Ref.\,\cite{Roberts:2011cf}, the coupling strength in the BSE has been reduced by a common prefactor to simulate effects beyond rainbow-ladder, which pushes their masses into fair agreement with experiment. The pseudoscalar and vector diquark masses inflate accordingly.}
\end{figure}

This connection is explicit in the RL truncation, where the gluon exchange in both $q\bar{q}$ and $qq$ systems is identical, except for an extra factor of $1/2$ in the $qq$ channel deriving from differences in the colour structure. Thus, when calculating mesons from their BSEs, one simultaneously obtains the respective diquark properties. In Fig.\,\ref{fig:meson-diquark-masses}, the resulting meson and diquark masses are plotted against the current-quark mass, which enters in the quark DSE and is varied from the chiral limit up to the strange-quark mass \cite{Eichmann:2016hgl}. For light up/down quarks, the masses are (in GeV)
\begin{equation}
\label{diquarkmassescalculated}
\begin{array}{cccc}
0^+ & 1^+ & 0^- & 1^- \\
0.80(7) & 0.99(5) & 1.22(9) & 1.30(6)
\end{array}
\end{equation}
Such masses and splittings are similar to those obtained in quark models, the symmetry-preserving treatment of a vector$\,\otimes\,$vector contact interaction (SCI) and QCD-kindred DSE frameworks, and lQCD (see Secs.\,\ref{sec:quark-models}, \ref{sec:lattice}).

It is worth noting that
\begin{equation}
\label{CSMavsc}
\delta_{{1^+0^+}} = m_{1^+} - m_{0^+} = 0.19(2)\,{\rm GeV},
\end{equation}
which is significantly less than the empirically known splitting between the $\Delta$-baryon and nucleon, $\delta_{\Delta N} \approx 0.27\,$GeV.  The associated Faddeev equations nevertheless produce masses for the nucleon and $\Delta$ in fair agreement with experiment, as discussed in connection with Fig.\,\ref{SAPV} below.  Naturally, $\delta_{{1^+0^+}}$ is partly responsible for $\delta_{\Delta N}$; and neglecting meson cloud effects, there is a linear relationship between them, \emph{e.g}.\ see Ref.\,\cite[Fig.\,1]{Roberts:2011cf}.  However, the net result for $\delta_{\Delta N}$ is also contingent upon other effects.  For instance, the nucleon and $\Delta$-baryon possess intrinsic deformation \cite{Roberts:2019wov}, so spin-orbit interactions play a role; and meson cloud effects can increase the splitting by $0.05$-$0.10\,$GeV, depending on the formulation \cite{Hecht:2002ej}.

The stability of RL studies of pseudoscalar and vector mesons provides another indication that their scalar and axial-vector diquark partners should play an important role in baryons: irrespective of interaction details, they always appear much the same.  On the other hand, positive parity mesons are distinguished by the presence of significant orbital angular momentum.  RL truncation underestimates associated repulsive effects; hence, produces scalar and axial-vector mesons that are too light. Consequently, RL estimates of the masses of their diquark partners are probably also too low.  This and associated deficiencies are remedied in beyond-RL calculations \cite{Chang:2009zb, Chang:2010hb, Chang:2011ei, Williams:2015cvx}.  The corrections can be mimicked by introducing a repulsion factor into the BSEs for scalar and axial-vector mesons and their diquark partners \cite{Roberts:2011cf} and this expedient was used in the calculations that produced Fig.\,\ref{fig:meson-diquark-masses}.

\begin{figure}[!t]
\begin{center}
\includegraphics[width=0.75\columnwidth]{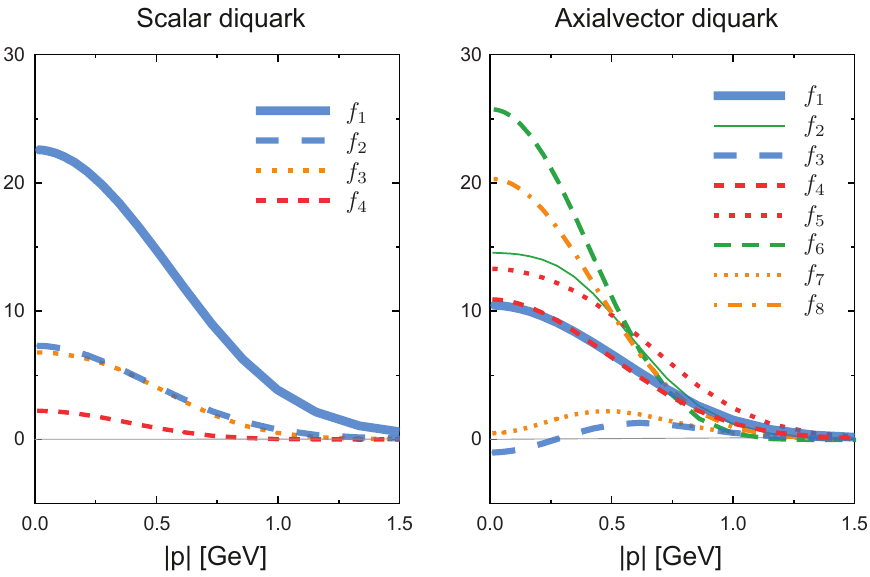}
\end{center}
\caption{\label{fig:diquark-amps} Dimensionless dressing functions in the correlation amplitudes of the light-quark scalar- (left) and axial-vector-diquark (right). 
}
\end{figure}

The diquarks calculated in the RL truncation are not pointlike objects.  Far from it: their BS amplitudes carry a rich tensorial structure that depends on the relative and total momentum, with four tensors for $J=0$ and eight for $J=1$ diquarks.  This structure is illustrated in Fig.\,\ref{fig:diquark-amps}, which depicts oft used projections of the Poincar\'e-covariant scalar dressing functions associated with the various tensor structures characterising scalar and pseudovector diquarks.  In both cases, $f_1(|p|)$ is associated with the leading tensor, \emph{i.e}.\ $\gamma_5 C$ and $\gamma_\mu C$, respectively.  These functions are dominant.  Whilst others are larger in Fig.\,\ref{fig:diquark-amps}, the associated Dirac-matrix tensors suppress their contributions to physical quantities.

It is worth reiterating that diquark correlations are coloured and it is only in connection with the partnering quark that a colour singlet system is obtained.  This means that diquarks are confined.  That is not true if RL truncation is used alone to define the quark+quark scattering problem \cite{Maris:2002yu}.  However, corrections to this leading-order truncation have been examined using the infrared-dominant interaction in Ref.\,\cite{Munczek:1983dx}; and in fully self-consistent symmetry-preserving studies, such corrections eliminate bound-state poles from the quark+quark scattering matrix, but preserve the strong correlations \cite{Bender:1996bb, Bender:2002as, Bhagwat:2004hn}.  These studies indicated that as coloured systems, like gluons and quarks, diquark propagation is described by a compound two-point function whose analytic structure is not that of an asymptotic state \cite{Gribov:1999ui, Munczek:1983dx, Stingl:1983pt, Cahill:1985mh, Stingl:1985hx, Cahill:1988zi, Krein:1990sf, Burden:1991gd, Hawes:1993ef, Roberts:1994dr, Roberts:2007ji, Brodsky:2012ku, Gao:2015kea, Papavassiliou:2015aga}; but which is nevertheless characterised by a mass-scale commensurate with that obtained in a RL analysis.

\begin{figure}[!t]
\centerline{%
\includegraphics[clip, width=0.6\textwidth] {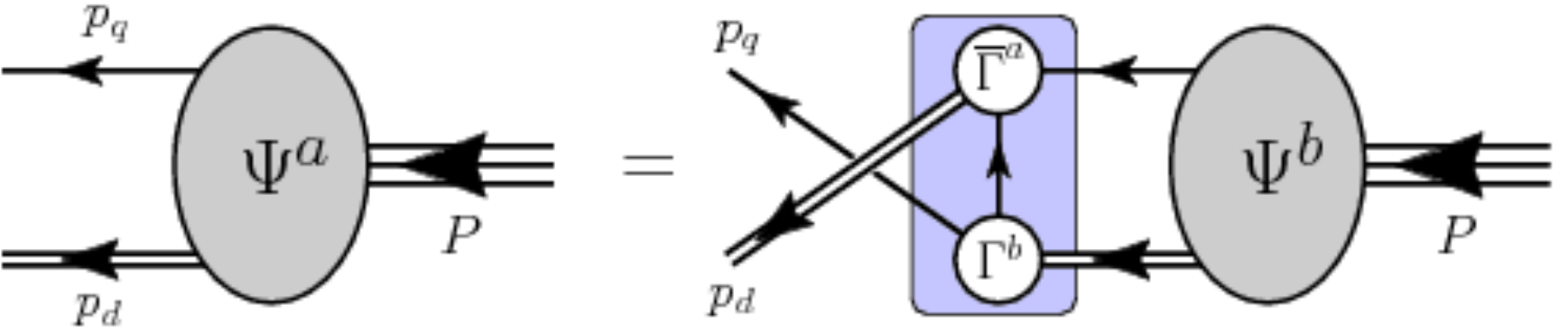}}
\caption{\label{figFaddeev}
Poincar\'e covariant quark+diquark Faddeev equation: a linear integral equation for the matrix-valued function $\Psi$, being the Faddeev amplitude for a baryon of total momentum $P= p_q + p_d$, which expresses the relative momentum correlation between the dressed-quarks and -nonpointlike-diquarks within the baryon.  The shaded rectangle demarcates the kernel of the Faddeev equation:
\emph{single line}, dressed-quark propagator; $\Gamma$,  diquark correlation amplitude; and \emph{double line}, diquark propagator.}
\end{figure}

In order to study the effect of diquark correlations on baryon structure and properties, the three-body version of Fig.\,\ref{fig:BSE} must be reformulated to make these correlations explicit.  This was first accomplished to produce a Poincar\'e-covariant baryon bound-state equation in Refs.\,\cite{Cahill:1988dx, Reinhardt:1989rw, Efimov:1990uz}, with the result illustrated in Fig.\,\ref{figFaddeev}.  The derivation involves resummation of all quark+quark interactions into quark+quark scattering matrices, $M$, subsequently approximated as follows:
\begin{equation}
M_{qq}(k,q;K) = \sum_{J^P=0^+,1^+,\ldots} \bar\Gamma^{J^P}\!(k;-K)\, \Delta^{J^P}\!(K) \, \Gamma^{J^P}\!(q;K)\,, \label{AnsatzMqq}
\end{equation}
where $\{\Gamma^{J^P}\!(q;K)\}$ are amplitudes describing the diquark correlations and $\{\Delta^{J^P}\!(K) \}$ are the associated propagators.  A \emph{prima facie} case in favour of this approximation was given in connection with Eq.\,\eqref{primafacie}.  Further validation is subsequently to be sought through comparison of resulting predictions with experiment.

In a baryon described by Fig.\,\ref{figFaddeev}, the binding has two contributions.  One part is expressed in the formation of tight diquark correlations; and that is augmented by attraction generated through the quark exchange depicted in the shaded area of Fig.\,\ref{figFaddeev}.  This exchange ensures that diquark correlations within the baryon are fully dynamical: no quark holds a special place because each one participates in all diquarks to the fullest extent allowed by its quantum numbers. The continual rearrangement of the quarks guarantees, \emph{inter} \emph{alia}, that the nucleon's dressed-quark wave function complies with Pauli statistics.  Gluons do not appear explicitly in Fig\,\ref{figFaddeev} because their effects are sublimated, being expressed in the properties of the elements in the Faddeev kernel.

Early attempts to use the Faddeev equation in Fig.\,\ref{figFaddeev} as a tool for studying baryons are described in Refs.\,\cite{Cahill:1988dx, Burden:1988dt, Keiner:1995bu, Hellstern:1995ri, Hellstern:1997pg, Oettel:1998bk, Oettel:1999gc}.  Hereafter, selected highlights from activities in the current millennium are described.  (A recent attempt to solve a quark+diquark BSE in Minkowski space using a ladder approximation is described in Ref.\,\cite{AlvarengaNogueira:2019zcs}.)

In closing this section, it is worth reiterating the result displayed in Fig.\,\ref{fig:meson-diquark-masses}; namely, a given meson is always lighter than its diquark partner.  It follows that if a system can form both internal meson and diquark correlations, the former will be dominant.  This is indeed what has been seen in four-body ($qq\bar{q}\bar{q}$) calculations of tetraquark systems based on the RL truncation \cite{Heupel:2012ua, Eichmann:2015cra, Wallbott:2019dng, Wallbott:2020jzh}. For example, it turns out that the ``light scalar mesons" such as the $\sigma$, $\kappa$, $a_0$ and $f_0$, when solved as four-quark systems, are dominated by meson+meson and not diquark+antidiquark channels \cite{Heupel:2012ua, Eichmann:2015cra}. Since the dominant mesons are the pseudoscalar NG bosons, the resulting four-quark states turn out to be especially light. These studies also indicate that the $X(3872)$ is dominated by molecule-like $DD^\ast$ components \cite{Wallbott:2019dng}.  The same is found for other states with $cq\bar{q}\bar{c}$ quark content; whereas for $cc\bar{q}\bar{q}$ systems, diquarks also play a role \cite{Wallbott:2020jzh}.  Regarding light-quark hybrid systems, a potentially important role is also played by different two-body correlations; namely, glue+quark and glue+antiquark \cite{Xu:2019sns}.

\subsubsection{Insights from a contact interaction}
As remarked above, DSEs provide a natural framework for the symmetry-preserving treatment of hadron bound states in quantum field theory.  The starting point in the matter sector is knowledge of the quark-quark interaction, which is now known with some certainty \cite{Binosi:2014aea, Binosi:2016nme, Cui:2019dwv}, as are its consequences: whilst the effective charge, and gluon and quark masses run with momentum, $k^2$, they all saturate at infrared momenta, each changing by $\lesssim20$\% on $0\lesssim \surd k^2 \lesssim m_0 \approx m_p/2$, where $m_0$ is a renormalisation-group-invariant gluon mass-scale and $m_p$ is the proton mass.  It follows that, employed judiciously, the symmetry-preserving treatment of a vector$\,\otimes\,$vector contact interaction (SCI) can provide insights and useful results for those hadron observables whose measurement involves probe momenta less than $m_0$, \emph{e.g}.\ hadron masses and form factors on $|Q^2|\lesssim M^2$, where is $M$ an infrared value of the dressed-quark mass and $M\lesssim m_0$ \cite{Chen:2012txa}.

The SCI formulation of the coupled two- and three-valence-body bound-state problems was introduced in Refs.\,\cite{Roberts:2011cf, Wilson:2011aa, Chen:2012qr}. It is based upon RL truncation and uses
\begin{equation}
\label{defineSCI}
g^2D_{\mu\nu}(p-q) = \delta_{\mu\nu}\frac{4\pi\alpha_{\rm IR}}{m_G^2}
\end{equation}
to represent the quark-quark interaction kernel, where $D_{\mu\nu}$ is the gluon propagator, $m_G \sim m_0$ is the gluon mass-scale, and the fitted parameter, $\alpha_{\rm IR}$, is commensurate with contemporary estimates of the zero-momentum value of the QCD effective charge \cite{Binosi:2016nme, Cui:2019dwv}.  Additionally, in the treatment of baryons, a variant of the ``static approximation'' \cite{Buck:1992wz} is employed, \emph{i.e}.\ the quark exchange interaction in Fig.\,\ref{figFaddeev} is treated as momentum-independent.  This has the virtue of ensuring that both the diquark correlation amplitudes in the Faddeev kernel and the baryon Faddeev amplitude produced by that kernel are momentum independent.  (Eliminating this static approximation increases computational effort, obscures insights, and does not bring material improvement in results \cite{Xu:2015kta}.)

As noted in connection with Eq.\,\eqref{diquarkpartners}, accounting for Fermi-Dirac statistics, five types of diquark correlation are possible in a $J=1/2$ bound-state: isoscalar-scalar ($I=0$, $J^P=0^+$), isovector-pseudo\-vector, isoscalar-pseudo\-scalar, iso\-scalar-vector, and iso\-vector-vector.  A $J=3/2$ bound-state may only contain isovector-pseudo\-vector and iso\-vector-vector diquarks. The SCI does not support an iso\-vector-vector diquark \cite{Roberts:2011wy}.

Ref.\,\cite{Roberts:2011cf} used the SCI to solve the Faddeev equations of the nucleon and $\Delta(1232)$-resonance, their parity partners, and the first radial excited states of these systems. Ref.\,\cite{Chen:2012qr} extended that work to all octet and decuplet baryons. These studies assumed that baryons are constituted solely from diquarks with the same parity, \emph{i.e}.\ positive-parity baryons only contain positive-parity diquarks, and negative parity baryons consist solely of negative-parity diquarks.

\begin{figure*}[!t]
\centering
\includegraphics[clip, width=0.75\textwidth] {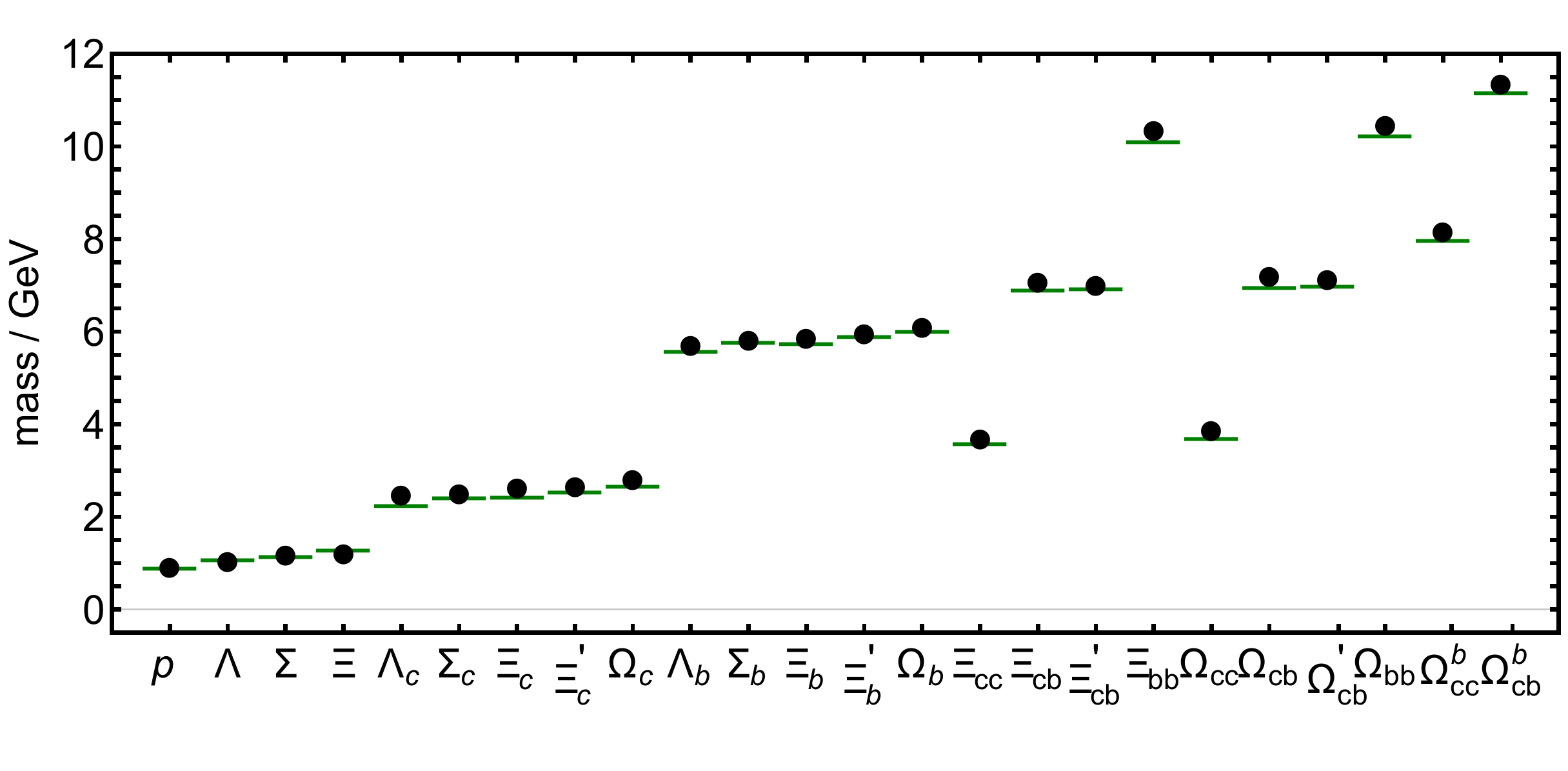} \\[2ex]
\vspace*{-0.80cm}
\includegraphics[clip, width=0.75\textwidth] {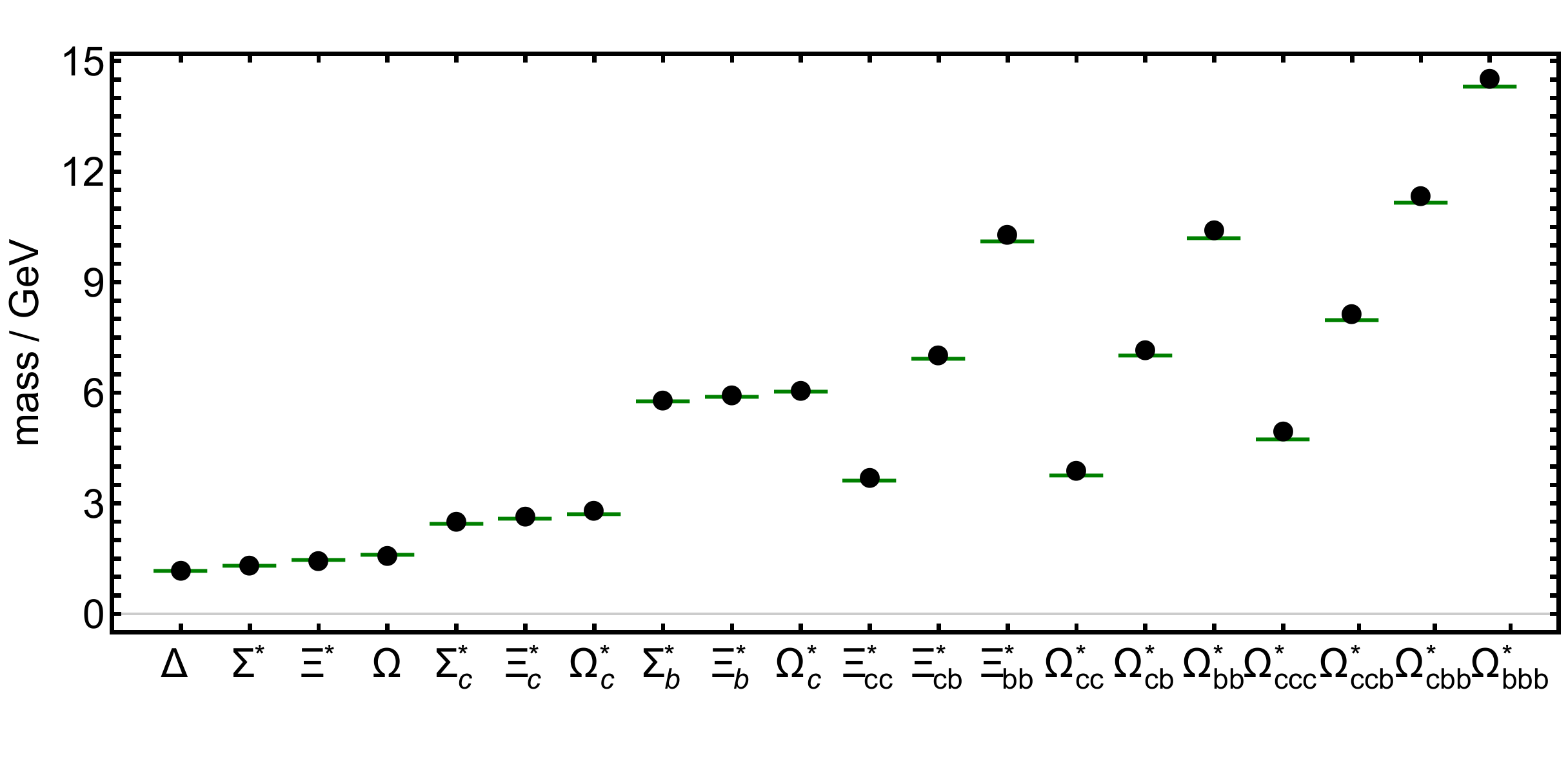}
\caption{\label{figPeiLin}
Comparison between SCI computed masses (black circles)  of ground-state flavour-SU(5) $J^P= 1/2^+$ (top) and $J^P=3/2^+$ (bottom) baryons and either experiment \cite{Tanabashi:2018oca} or lQCD \cite{Brown:2014ena, Mathur:2018epb} (green lines).  (Adapted from Ref.\,\cite{Yin:2019bxe}.)}
\end{figure*}

Ref.\,\cite{Lu:2017cln} eliminated the like-parity restriction and found that ground-state, even-parity baryons are indeed constituted, almost exclusively, from like-parity diquarks.  On the other hand, odd-parity baryons, in which quark+diquark orbital angular momentum plays a larger role, contain a measurable even-parity diquark component even though odd-parity diquarks are dominant.  Capitalizing on this information, the spectra of $J^P=1/2^+, 3/2^+$ $(fgh)$ baryons, with $f,g,h \in \{u,d,s,c,b\}$, were computed in Refs.\,\cite{Yin:2019bxe, Gutierrez-Guerrero:2019uwa}.  The strength of the simple SCI approach is highlighted by Fig.\,\ref{figPeiLin}.  Notably, Ref.\,\cite{Yin:2019bxe} predicts that diquark correlations are an important component of all baryons; and owing to the dynamical character of the diquarks, it is typically the lightest allowed diquark correlation which defines the most important component of a baryon's Faddeev amplitude.

As mentioned above, the SCI can also be used profitably to study hadron properties characterised by small momentum transfer, $|Q^2|\lesssim M^2$.  Ref.\,\cite{Wilson:2011aa} used the SCI to compute nucleon and Roper electromagnetic elastic and transition form factors, concluding that in the description of the nucleon and its first radial excitation, both scalar and pseudovector diquarks play an important role, and obtaining some qualitatively instructive results for the form factors.  The elastic and transition form factors of the $\Delta(1232)$ were computed in Refs.\,\cite{Segovia:2013rca, Segovia:2013uga}, solving a longstanding puzzle surrounding the $Q^2$ dependence of the magnetic transition form factor.  The nucleon $\sigma$-term and tensor charges were computed in Refs.\,\cite{Xu:2015kta, Pitschmann:2014jxa}, anticipating results obtained later using a more realistic interaction \cite{Wang:2018kto}.


\subsubsection{QCD-kindred formulation}
\label{QCDkindred}
The SCI is simple, algebraically solvable, and often delivers valuable insights.  It was introduced for these reasons and also to demonstrate conclusively that experiments are sensitive to the momentum-dependence of QCD's effective charge and its diverse expressions in observables \cite{Roberts:2018hpf}.  In working toward realistic QCD-connected predictions, one can adapt the pattern used for mesons; namely, solve gap equations for the dressed-quark propagators and BSEs for the diquark correlation amplitudes, build the Faddeev kernels therewith, and solve for baryon masses and Faddeev amplitudes.  As discussed in Sec.\,\ref{sec:dse-ab-initio}, this \emph{ab initio} approach has delivered successes, but it is computationally cumbersome and limited in reach by existing algorithms.  An alternative \cite{Hecht:2002ej} is to construct a QCD-kindred framework, in which all elements of the Faddeev kernels and interaction currents are momentum dependent and consistent with QCD scaling laws.

A successful QCD-kindred framework is described and employed in Refs.\,\cite{Segovia:2014aza, Segovia:2015hra, Segovia:2016zyc, Chen:2017pse, Chen:2018nsg, Chen:2019fzn, Lu:2019bjs, Cui:2020rmu}.  It uses an efficacious algebraic parametrisation for the dressed light-quark propagator, unchanged for two decades \cite{Hecht:2000xa}, yet consistent with contemporary numerical results \cite{Chen:2017pse}; expresses confinement and DCSB; retains the leading diquark amplitudes discussed in connection with Fig.\,\ref{fig:diquark-amps}; and describes diquark propagation in a manner consistent with colour confinement and asymptotic freedom.  The formulation has two parameters, \emph{viz}.\ mass-scales connected with the scalar and pseudovector diquark correlations.  They were fitted to obtain desired masses for the nucleon and $\Delta$-baryon.  The fitted values are consistent with those described in connection with Fig.\,\ref{fig:meson-diquark-masses} and that means with all existing complementary studies, continuum and lattice.

\begin{figure*}[!t]
\centering
\includegraphics[clip,width=0.9\linewidth]{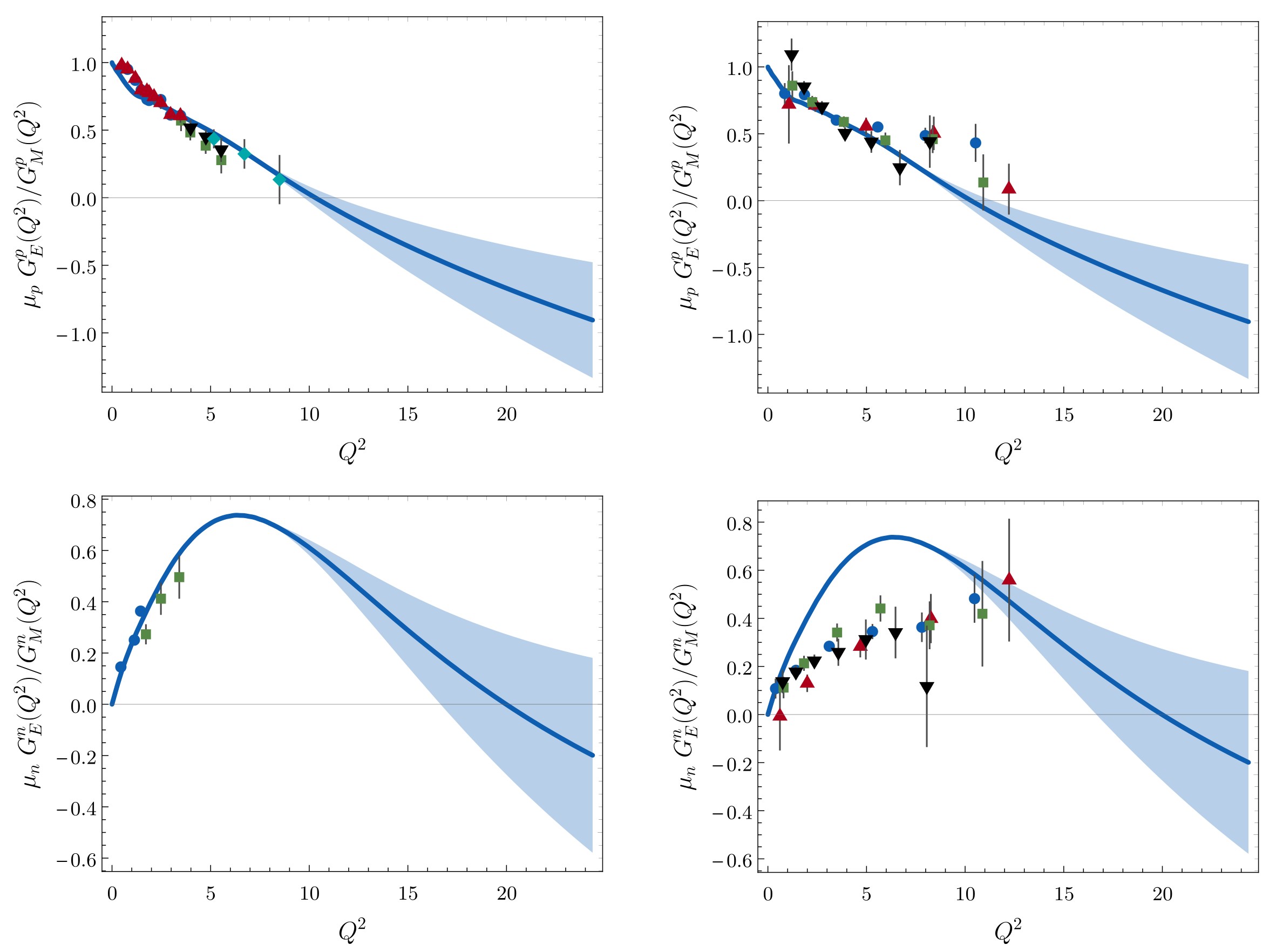}
\caption{\label{GEonGM}
Ratios of Sachs form factors, $\mu_N G_E^N(x)/G_M^N(x)$.
\emph{Upper panels} --  Proton.  \underline{Left}, prediction in Ref.\,\cite{Cui:2020rmu} compared with data
(red up-triangles \cite{Jones:1999rz};
green squares \cite{Gayou:2001qd};
blue circles \cite{Punjabi:2005wq};
black down-triangles \cite{Puckett:2010ac};
and cyan diamonds \cite{Puckett:2011xg});
\underline{right}, compared with available lQCD results, drawn from Ref.\,\cite{Syritsyn:2017jrc, Kallidonis:2018cas}.
\emph{Lower panels} -- Neutron.  \underline{Left}, comparison with data (blue circles \cite{Madey:2003av} and green squares \cite{Riordan:2010id}); \underline{right}, with available lQCD results, drawn from Ref.\,\cite{Kallidonis:2018cas}.
Ref.\,\cite{Cui:2020rmu} exploited a statistical implementation of the Schlessinger point method (SPM)  \cite{Schlessinger:1966zz, PhysRev.167.1411, Tripolt:2016cya, Chen:2018nsg, Binosi:2018rht, Binosi:2019ecz} for the interpolation and extrapolation of smooth functions in order to deliver predictions for form factors on $x>9$; and in all panels, the $1\sigma$ band for the SPM approximants is shaded in light blue.
}
\end{figure*}

This framework was first used to study the form factors of the simplest baryons: the nucleon and the $\Delta(1232)$. The nucleon's elastic electromagnetic form factors were calculated in \cite{Alkofer:2004yf, Cloet:2008re, Cloet:2013gva}; and the elastic and transition form factors of the $\Delta(1232)$ were computed in \cite{Segovia:2014aza}.  Today, predictions for nucleon form factors have been delivered on the entire domain of momentum transfers accessible at the upgraded JLab facility, \emph{i.e}.\ $0\leq Q^2\leq 18 \,m_N^2$ ($m_N$ is the nucleon mass) \cite{Cui:2020rmu}.  The results expose features of the form factors and the role of diquark correlations in the nucleon that can be tested in new generation experiments at existing facilities, \emph{e.g}.\ a zero in $G_E^p/G_M^p$ and a maximum in $G_E^n/G_M^n$ (see Fig.\,\ref{GEonGM}); and a zero in the proton's $d$-quark Dirac form factor, $F_1^d$.
(Aspects of the forthcoming JLab programme are discussed in Sec.\,\ref{SBSSLNFSS}.)
Additionally, examination of the associated light-front-transverse number and anomalous magnetisation densities reveals, \emph{inter alia}: a marked excess of valence $u$-quarks in the neighbourhood of the proton's centre of transverse momentum; and that the valence $d$-quark is markedly more active magnetically than either of the valence $u$-quarks.  Additional revelations about nucleon structure in Ref.\,\cite{Cui:2020rmu} cannot be tested at JLab, but could be validated using a high-luminosity accelerator capable of delivering higher beam energies than are currently available, \emph{e.g}.\ EIC and EicC.

Another important feature of this QCD-kindred framework is that it can be used to study the radial excitations of baryons and the associated electroproduction form factors. For instance, Refs.\,\cite{Segovia:2015hra, Segovia:2016zyc} computed the nucleon-to-Roper electromagnetic transition form factors, thereby making a profound contribution to a solution to the fifty-year puzzle of the Roper resonance \cite{Burkert:2017djo}.  The analysis indicates that the Roper-resonance is, at heart, the first radial excitation of the nucleon, consisting of a well-defined dressed-quark core augmented by a meson cloud.  (See also Sec.\,\ref{lQCDBSWF} below.)
In anticipation of new generation experiments at JLab, the nucleon-to-Roper electromagnetic transition form factors at large momentum transfers were computed in Ref.\,\cite{Chen:2018nsg}.   Likewise, Ref.\,\cite{Lu:2019bjs} supplied predictions for the $\gamma^\ast p \to \Delta^+(1232), \Delta^+(1600)$ transition form factors, providing the information necessary to test the conjecture that the $\Delta(1600)$ is an analogue of the Roper resonance, \emph{i.e}.\ the simplest radial excitation of the $\Delta(1232)$.  Notably, precise measurements of the $\gamma^\ast p \to \Delta^+(1232)$ transition already exist on $0 \leq Q^2 \lesssim 8\,$GeV$^2$ and the calculated results compare favourably with the data outside the meson-cloud domain.  The predictions for the $\gamma^\ast p \to \Delta^+(1600)$ are currently being compared with JLab data \cite{Trivedi:2018rgo, Burkert:2019opk}.

\begin{figure}[!t]
\begin{tabular}{ccc}
\includegraphics[clip, width=0.45\textwidth]{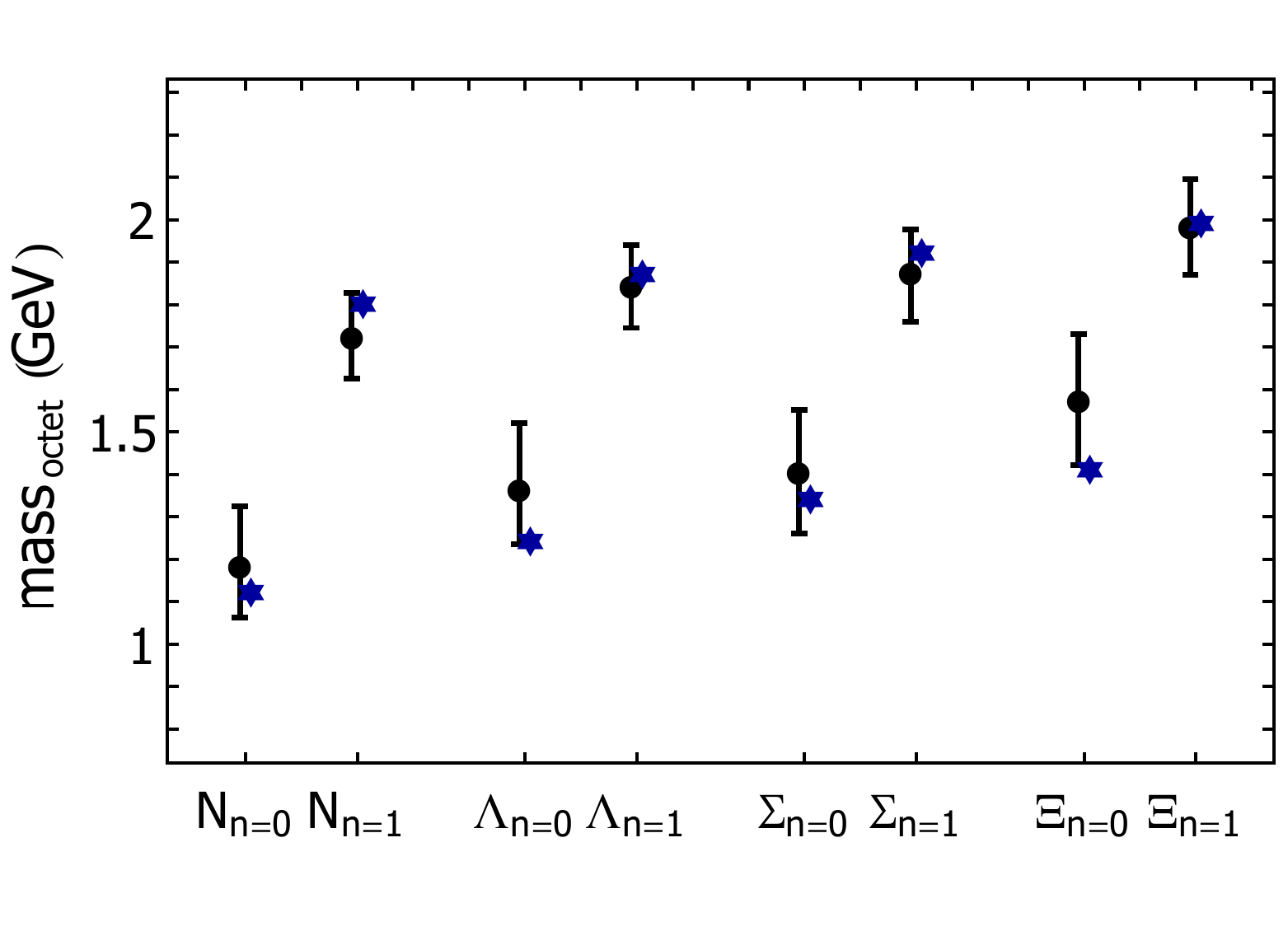} & \hspace*{2em} &
\includegraphics[clip, width=0.45\textwidth]{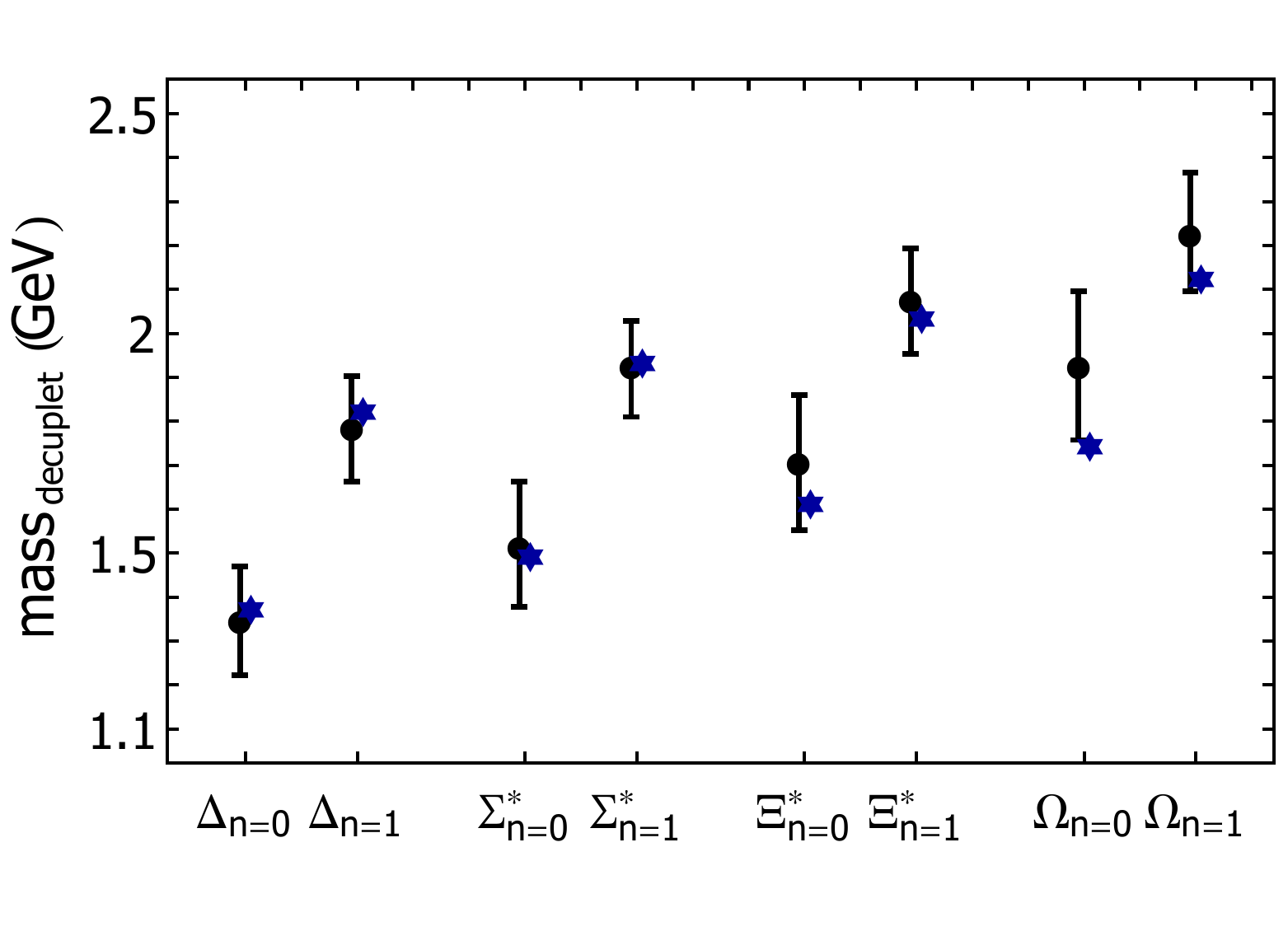}
\end{tabular}
\caption{\label{CompareYa}
Comparison between the masses computed using Faddeev equation kernels built with dressed-quarks and diquarks described by QCD-like momentum-dependent propagators and amplitudes and those obtained using a symmetry-preserving treatment of a vector$\,\otimes\,$vector contact-interaction (blue stars) \cite{Lu:2017cln}.
\emph{Left panel}: octet states.
\emph{Right panel}: decuplet states.
The vertical riser indicates the response of the Ref.\,\cite{Chen:2019fzn} results to a coherent $\pm 5$\% change in the mass-scales associated with the diquarks and dressed-quarks.
The horizontal axis lists a particle name with a subscript that indicates whether it is ground-state ($n=0$) or first positive-parity excitation ($n=1$).
}
\end{figure}

The QCD-kindred framework has also been used recently to perform a comparative study of the four lightest $(I=1/2,J^P=1/2^\pm)$ baryon isospin doublets \cite{Chen:2017pse}. This study indicates that in these doublets, isoscalar-scalar, isovector-pseudovector, isoscalar-pseudoscalar, and vector diquarks can all play a role. In the two lightest $(1/2,1/2^+)$ doublets, however, scalar and pseudovector diquarks are overwhelmingly dominant. The associated rest-frame wave functions are largely $S$-wave in nature; and the first excited state in this $1/2^+$ channel has the appearance of a radial excitation of the ground state. The two lightest $(1/2,1/2^-)$ doublets fit a different picture: accurate estimates of their masses are obtained by retaining only pseudovector diquarks; in their rest frames, the amplitudes describing their dressed-quark cores contain roughly equal fractions of even- and odd-parity diquarks; and the associated wave functions are predominantly $P$-wave in nature, yet possess measurable $S$-wave components. Moreover, the first excited state in each negative-parity channel has little of the appearance of a radial excitation. This analysis confirms the SCI prediction that one can safely ignore negative-parity diquarks in positive-parity baryons. However, ignoring positive-parity diquarks in negative-parity baryons is a poor approximation. Benefiting from such guidance, Ref.\,\cite{Chen:2019fzn} computed the spectrum and Poincar\'e-covariant wave functions for all SU$_{\rm f}(3)$ positive-parity octet and decuplet baryons and their first excitations.   A comparison of the QCD-kindred spectra with those obtained using the SCI is shown in Fig.\,\ref{CompareYa}.  Amongst other things, it highlights the response of baryon masses to changes in those of the dressed-quarks and -diquarks; and the usefulness of SCI analyses of infrared-dominated observables.

\subsubsection{\emph{Ab initio} approach}
\label{sec:dse-ab-initio}
Ideally, an {\em ab-initio} DSE approach should follow the program outlined at the beginning of Sec.\,\ref{sec:DSEs}: one settles on a truncation, which specifies an interaction kernel depending on QCD's $n$-point functions, and calculates all subsequent hadron properties without further approximations. In this way, the current-quark masses and the scale $\Lambda_\text{QCD}$ would remain the only parameters in all calculations and one could study the calculated observables as functions of the pion mass $m_\pi$. Although progress in this direction has been made, it is still at an early stage owing to the underlying complexity -- most {\em ab-initio} baryon calculations to date are based on the RL truncation.

\begin{figure*}[!t]
\begin{center}
\begin{tabular}{c}
\includegraphics[width=0.95\textwidth]{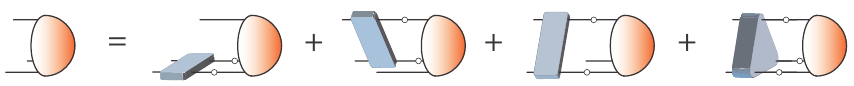}
\end{tabular}\vspace*{-5ex}
\end{center}
\caption{\label{qqq} Three-quark Faddeev equation. \emph{Solid line with open circle}, dressed quark propagator.}
\end{figure*}

If one views the BSE kernel as a \textit{black box}, there are two possible paths to proceed when studying baryons. The first is to solve the three-body Faddeev equation in Fig.\,\ref{qqq} directly. While this demands substantial numerical efforts, it is also conceptually simple because it only involves quarks and gluons; \emph{e.g}.\ the equation does not know about diquarks. Details and examples of the approach can be followed from Refs.\,\cite{Eichmann:2009zx, Sanchis-Alepuz:2017jjd, Qin:2019hgk}.  (Note that the explicit three-body interaction kernel is normally neglected, again supported by the discussion around Eq.\,\eqref{primafacie}.) The second strategy is to solve the quark+diquark Faddeev equation in Fig.\,\ref{figFaddeev} with all quark and diquark elements calculated from their own equations, \emph{i.e}.\ one solves the BSEs of mesons and diquarks, the Dyson equations for the diquark propagators, and finally the baryons' Faddeev equations.  (See Refs.~\cite{Eichmann:2007nn, Eichmann:2008ef, Eichmann:2009zx} for details.)


In both strategies, only the two-body kernel enters in the equations, either directly as in Fig.\,\ref{qqq} or indirectly through the diquark BSEs, producing the diquark amplitudes and propagators that appear in Fig.\,\ref{figFaddeev}. The RL kernel in particular depends \cite{Maris:1997tm, Maris:1999bh, Qin:2011dd} on a mass-scale parameter, which is usually fixed to the experimental pion decay constant, and a width parameter, generating \emph{e.g}.\ the bands in Fig.~\ref{fig:meson-diquark-masses}. Therefore, aspects of the goal outlined above are realised: mesons and baryons can be studied in the same approach, with only a few input parameters (the current-quark masses, a scale, and a shape parameter), and one can calculate the dependence of observables on the current-quark mass, as in Fig.~\ref{fig:meson-diquark-masses}.

In both cases one needs to solve the quark DSE in the complex momentum plane to obtain numerical solutions for the quark propagator. These solutions typically have complex conjugate poles, which pose an obstacle because they produce upper limits for the possible on-shell hadron masses that can be obtained when using straightforward algorithms. In this case, for three light quarks the largest baryon mass one can reach directly is $\sim$ 1.5 GeV.  Above that value, extrapolations are commonly used, see \emph{e.g}.\  Refs.\,\cite{Eichmann:2016hgl, Eichmann:2018adq}. In stepping beyond RL truncation, one must also take care of the singularity structure in other correlation functions.  In principle this problem can be overcome using contour deformations \cite{Maris:1995ns, Strauss:2012dg, Windisch:2012zd, Windisch:2012sz, Pawlowski:2017gxj, Weil:2017knt, Williams:2018adr, Miramontes:2019mco, Eichmann:2019dts}.  Alternatively, perturbation theory integral representations \cite{Nakanishi:1969ph} can be used in the manner exploited successfully for mesons \cite{Chang:2013nia}.

The first {\em ab-initio} quark+diquark study in the RL truncation is described in Ref~\cite{Eichmann:2007nn}, where the nucleon mass and its electromagnetic form factors were calculated as functions of the current-quark mass. Ref.\,\cite{Eichmann:2008ef} discussed the simultaneous prediction of meson and baryon observables; these results are in qualitative agreement with the corresponding ones in the QCD-kindred framework \cite{Cloet:2008re, Segovia:2014aza}. The mass of the $\Delta$ resonance was calculated in Ref.\,\cite{Nicmorus:2008vb}, its electromagnetic form factors in Ref.\,\cite{Nicmorus:2010sd} and the $N\to \gamma^\ast \Delta$ transition form factors in Ref.\,\cite{Eichmann:2011aa}.

In Ref.\,\cite{Eichmann:2009qa}, the nucleon's three-body Faddeev equation was solved for the first time, using the RL truncation. The resulting current-mass evolution of the nucleon mass compares well with lQCD results and deviates by only $\sim5\%$ from the quark+diquark result. The approach was later extended to $\Delta$ and $\Omega$ baryons \cite{Sanchis-Alepuz:2014sca}, the full octet and decuplet ground-state spectrum \cite{Sanchis-Alepuz:2014sca}, and baryons involving heavy quarks \cite{Qin:2018dqp}. In Ref.\,\cite{Qin:2019hgk}, the calculated ground states and first excitations of baryons with $J={1/2}^+$ and ${3/2}^+$, and with quark content from light to bottom, were found to reproduce the known spectrum of 39 states with an accuracy of $\sim 3\%$.

The three-body approach has also been applied to compute structure observables, such as form factors, including the electromagnetic form factors of the nucleon \cite{Eichmann:2011vu}, its axial and pseudoscalar form factors \cite{Eichmann:2011pv}, the electromagnetic form factors of ground-state octet and decuplet baryons (including those with strangeness) \cite{Sanchis-Alepuz:2015fcg}, and the electromagnetic transition form factors between octet and decuplet baryons \cite{Sanchis-Alepuz:2014sca}. Overall, the results are in good agreement with available experimental data, except at low $Q^2$, where discrepancies can be attributed to meson-cloud effects (which a RL kernel does not incorporate). In Ref.~\cite{Wang:2018kto}, the proton's tensor charges were computed, presenting a favorable comparison with lQCD results.

\begin{figure*}[!t]
\begin{center}
\begin{tabular}{c}
\includegraphics[width=0.9\textwidth]{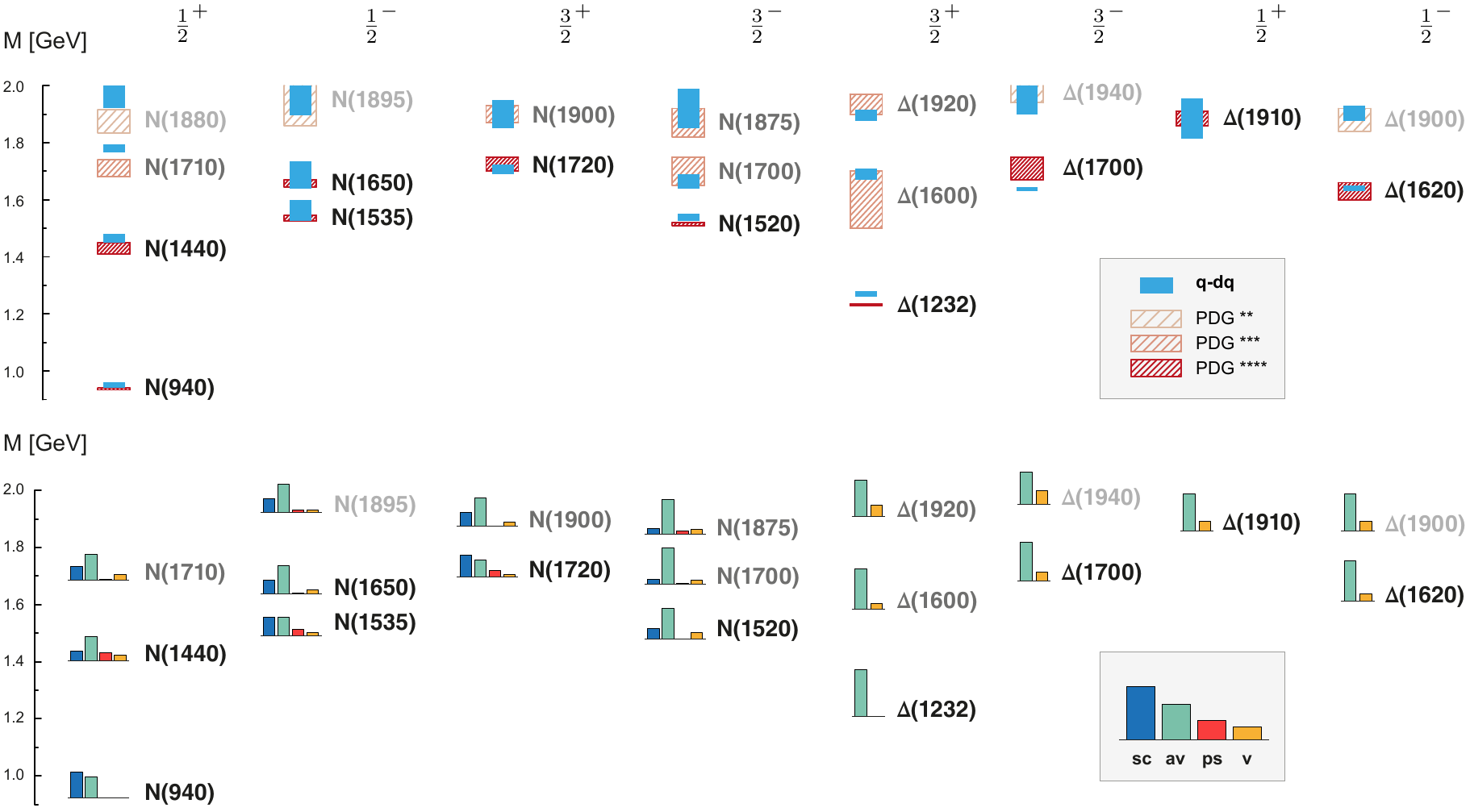}
\end{tabular} \vspace*{-3ex}
\end{center}
\caption{\label{SAPV} Light-baryon spectrum for nucleon and $\Delta$ states with $J^P=1/2^\pm$ and $3/2^\pm$ obtained from the quark-diquark Faddeev calculation (top) and their individual diquark contributions (bottom) \cite{Eichmann:2016hgl}.
}
\end{figure*}

Returning to the question of diquarks and their impact on the baryon spectrum, Ref.\,\cite{Eichmann:2016hgl} calculated the ground and excited states of light octet and decuplet baryons, both in the three-body Faddeev framework and the quark+diquark approximation. Scalar, axial-vector, pseudoscalar and vector diquarks were included because they can all contribute to the nucleon channels, whereas the ($I=3/2$) $\Delta$ baryons only permit axial-vector and vector diquarks with $I=1$. The two approaches were found to be mutually consistent; a similar conclusion was also made in Ref.\,\cite{Eichmann:2018adq} for the strange-baryon sector. Since both approaches employ the same RL interaction, this confirms that a quark+diquark picture is a good approximation and underlines the role of diquark correlations in the baryon spectrum.

Of course, it should be noted that while the $N(1/2^+)$ and $\Delta(3/2^+)$ masses calculated in this direct approach agree well with experiment, the remaining spin-parity channels come out too light (see, e.g. Ref.\,\cite[Fig.\,3]{Eichmann:2016hgl}). Recalling the analogous situation for mesons discussed in connection with Fig.\,\ref{fig:meson-diquark-masses}, the spectrum shown in Fig.\,\ref{SAPV} was obtained by reducing the strength of the pseudoscalar and vector diquarks in the quark+diquark Faddeev equation by a multiplicative factor $c=0.35$ to simulate beyond-RL contributions. As a result, the masses in the problematic channels are increased and one achieves overall agreement with the empirical spectrum.

\begin{figure*}[!t]
\begin{center}
\begin{tabular}{c}
\includegraphics[width=0.95\textwidth]{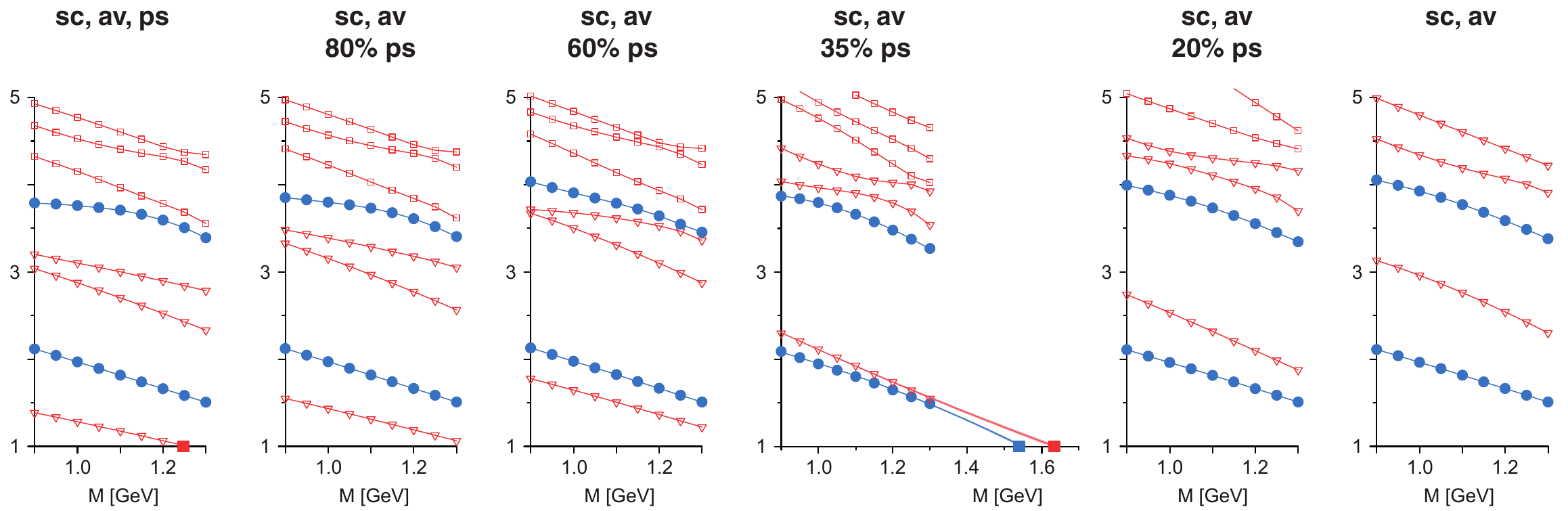}
\end{tabular} \vspace*{-3ex}
\end{center}
\caption{\label{BSE-EVs-12-} Eigenvalues of the baryon's quark-diquark equation in the $N(1/2^-)$ channel plotted over the baryon mass~\cite{Eichmann:2016hgl}. As the strength of the pseudoscalar diquark is reduced, the lowest eigenvalue moves up in the spectrum. At $35\%$ reduction, which corresponds to Fig.~\ref{SAPV}, one obtains two nearby states as also seen in experiment.
}
\end{figure*}

An especially interesting case is the $N(1/2^-)$ channel, where one experimentally finds two nearby states: the $N(1535)$, which is the parity partner of the nucleon, and the $N(1650)$. (As discussed elsewhere \cite{Burkert:2017djo}, the level ordering between the $N(1535)$ and the Roper resonance $N(1440)$ has been a longstanding issue in quark models \cite{Isgur:1978xj, Capstick:1986bm, Bijker:1994yr, Glozman:1995fu, Loring:2001kx, Capstick:1992th, Giannini:2001kb}.)
In the RL truncation, both Faddeev calculations produce a low-lying state around $\sim 1.2$ GeV in the $J=1/2^-$ channel; hence, the wrong level ordering. This can be seen in Fig.\,\ref{BSE-EVs-12-}, which shows the eigenvalues of the quark+diquark BSE; each eigenvalue can produce a bound state if $\lambda_i(M) = 1$. When scalar, axial-vector and pseudoscalar diquarks are included, one finds a low-lying ground state (like in the three-body calculation) which is dominated by the pseudoscalar diquark. As the strength of the pseudoscalar diquark is gradually turned off, two of the eigenvalues (filled symbols) are insensitive, whereas others (open symbols) strongly react to this change: the ground state moves up in the spectrum and eventually even switches its role with the first excitation. At $c=0.35$, which corresponds to the spectrum in Fig.\,\ref{SAPV}, this results in two nearby states which produce masses in the experimental neighborhood. Apparently, the heavier odd-parity diquarks contaminate the baryon spectrum; and, as with their meson partners, beyond-RL effects should be expected to have a net repulsive effect in these channels, thereby reducing their importance.

The lower panel in Fig.\,\ref{SAPV} shows a calculation of the diquark contributions to the Bethe-Salpeter norm of each calculated state. (An analogous breakdown into partial-wave contributions can be found in Ref.\,\cite{Eichmann:2016nsu}.) For the $N$ and $\Delta$ ground states, only the scalar and pseudovector diquarks play a role, whereas the higher-lying diquarks provide small but relevant contributions in all other cases. Note also that the axial-vector diquark is significant in many channels.

The two measures used in Refs.\,\cite{Segovia:2015hra, Chen:2017pse} to evaluate a baryon's diquark content are different from that used to produce Fig.\,\ref{BSE-EVs-12-}: one focuses on the Faddeev wave function and the other on the contribution of each diquark type to the bound-state's mass.  Of these, the former is similar to that used for Fig.\,\ref{BSE-EVs-12-}; whereas the latter samples effects very differently, delivering results which emphasise that in the computation of an observable quantity, there is significant interference between the distinct diquark components in a baryon's Faddeev amplitude.  Notwithstanding these things, a basic fact remains: the nucleon and Roper possess very similar diquark content.  One learns from these analyses that comparisons between diquark fractions computed for different baryons using the same indicator are easily interpreted, whereas that is not always the case for comparisons between results obtained for the same baryon using different schemes.

These remarks reemphasise that if one chooses to take a diquark perspective seriously, then a full understanding of hadrons requires the careful consideration of all physically allowed quark+quark correlations, \emph{e.g}.\ Eq.\,\eqref{AnsatzMqq}.  Failing that, one is liable to arrive at a simplistic approximation to quark+quark scattering within the compound system under study.

It is worth adding a final comment on the agreement between theory and experiment in Fig.\,\ref{SAPV}, which might seem puzzling because meson-cloud effects introduce mass shifts \cite{Suzuki:2009nj} and, more generally, all states except the proton are resonances that decay hadronically.
Namely, in choosing the mass-scale parameter in the RL interaction so as to describe $f_\pi$, some influences of the meson cloud are implicitly incorporated \cite{Eichmann:2008ae}: after all, a match with experiment has been required.  The operating conjecture for RL truncation is that the impact of meson cloud effects on a resonance's Breit-Wigner mass is captured by the choice of interaction scale, even though a width is not generated.  This should be reasonable for states whose width is a small fraction of their mass; and in practice, as already illustrated herein and in many other studies, the supposition appears to be correct.
Explicit studies aimed at exploring this conjecture, with explicit implementation of hadronic decay channels in BSEs are described elsewhere \cite{Pichowsky:1999mu, Williams:2018adr, Miramontes:2019mco}.

At present, few {\em ab-initio} Faddeev studies employ a beyond-RL interaction kernel. A calculation within a 2PI truncation \cite{Sanchis-Alepuz:2015qra} did not significantly improve the spectrum.  A 3PI calculation has so far has only been employed for light mesons \cite{Williams:2015cvx}. The effect of pion-cloud contributions on $N$ and $\Delta$ masses was explored in Ref.\,\cite{Sanchis-Alepuz:2014wea}, where the terms responsible for feedback of the pion onto the quark were resolved.  This leads to rainbow-ladder-like pion-cloud effects in bound states.  In Refs.\cite{Bender:1996bb, Bender:2002as, Bhagwat:2004hn}, the diquark correlations were studied in a truncation scheme that systematically extends the RL approximation and ensures that, in the chiral limit, the isovector, pseudoscalar meson remains massless. It was found that diquarks are removed from the observable spectrum by repulsive contributions that only appear at higher order in the Bethe-Salpeter kernel, whereas the net effects of higher order terms on meson bound-state masses are small.


\subsubsection{Baryon distribution amplitudes}
\label{CedricMezrag}
An important way of exposing the impact of strong diquark correlations within baryons is to compute their parton distribution amplitudes \cite{Lepage:1979zb, Efremov:1979qk, Lepage:1980fj}.  These quantities provide for a probability interpretation, like wave functions in quantum mechanics, and feature in the scattering formulae that describe hard exclusive processes in QCD.  In the case of mesons, the PDAs have been studied extensively \cite{Chang:2013pq, Chang:2013epa, Segovia:2013eca, Gao:2014bca, Mezrag:2014jka, Shi:2015esa, Ding:2015rkn, Li:2016dzv, Li:2016mah, Ding:2018xwy, Binosi:2018rht, Cui:2020dlm, Cui:2020piK}; and analogous calculations for diquarks are possible.  However, calculations of baryon PDAs are much more difficult because of their three-body complexity.

\begin{figure*}[!t]
\begin{center}
\begin{tabular}{ccc}
\parbox[c]{0.30\linewidth}{\includegraphics[clip, width=\linewidth]{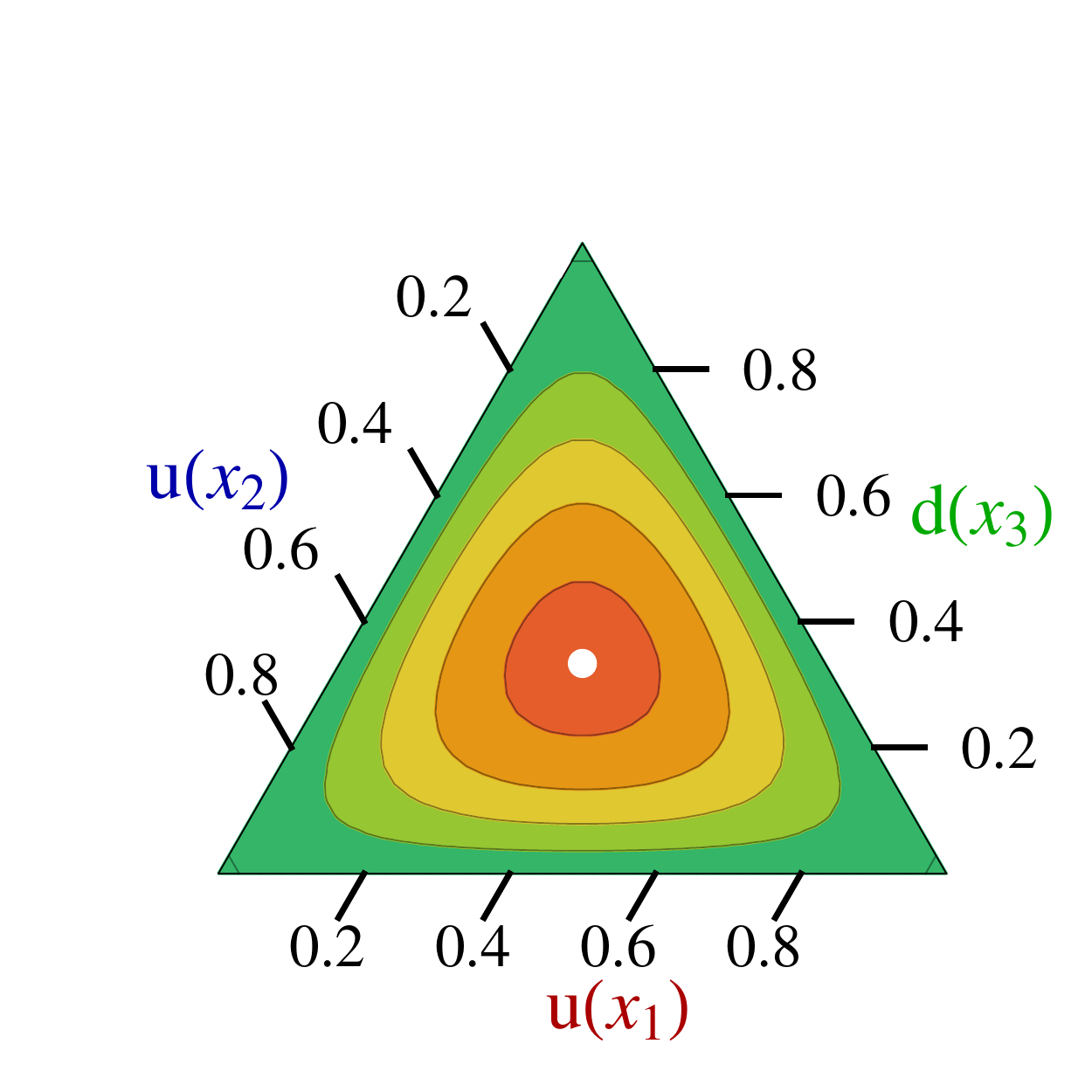}} &
\parbox[c]{0.30\linewidth}{\includegraphics[clip, width=\linewidth]{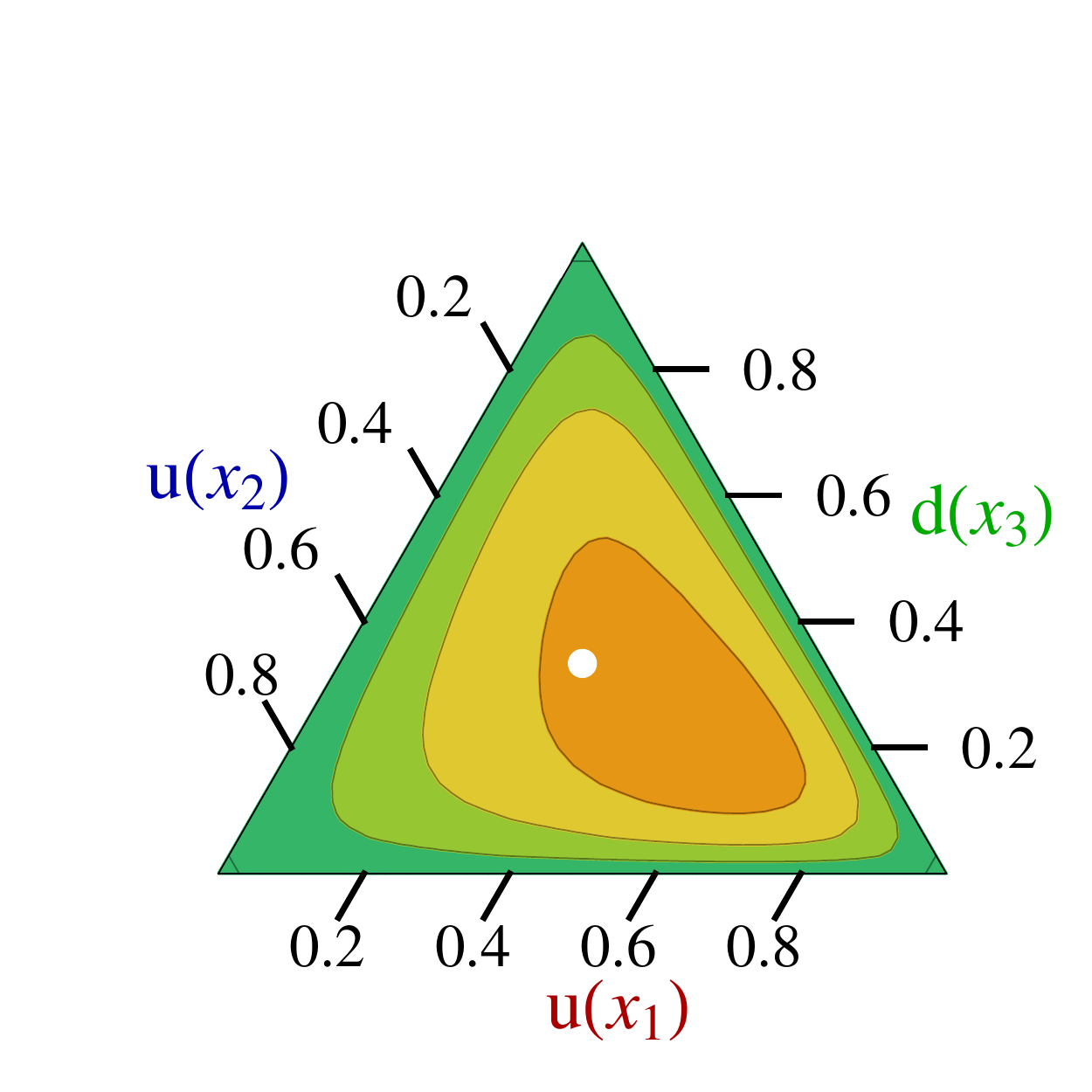}} &
\parbox[c]{0.36\linewidth}{\includegraphics[clip, width=\linewidth]{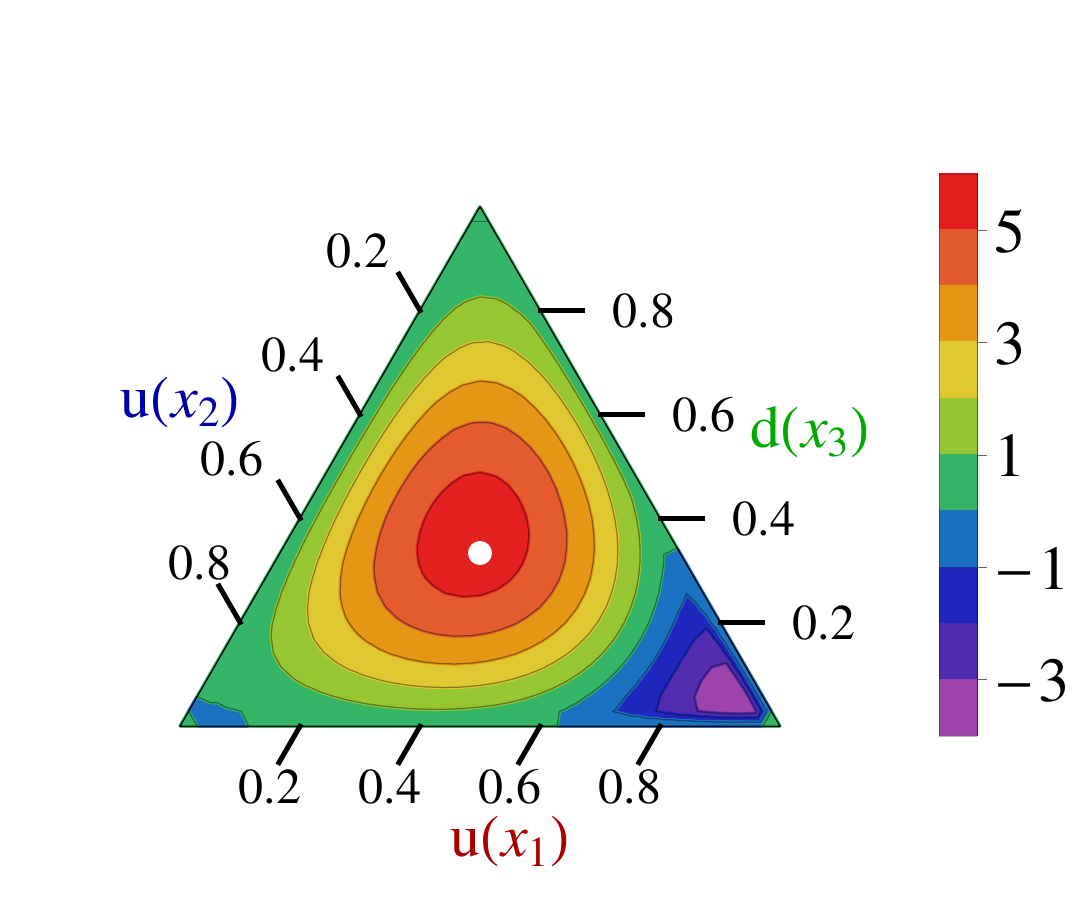}}
\end{tabular}\vspace*{-5ex}
\end{center}
\caption{\label{PlotPDAs} Barycentric plots from Ref.~\cite{Mezrag:2017znp}: \emph{left panel} -- conformal limit PDA, $\varphi_N^{\rm cl}([x])=120 x_1 x_2 x_3$; \emph{middle panel} -- computed proton PDA evolved to $\zeta=2\,$GeV, which peaks at $([x])=(0.55,0.23,0.22)$; and \emph{right panel} -- Roper resonance PDA at $\zeta=2\,$GeV.  The white circle in each panel serves only to mark the centre of mass for the conformal PDA, whose peak lies at $([x])=(1/3,1/3,1/3)$.
}
\end{figure*}

As a first step, Refs.\,\cite{Mezrag:2017znp, Mezrag:2018hkk}, developed algebraic models for the nucleon and Roper-resonance Faddeev amplitudes, informed by results obtained in the QCD-kindred framework \cite{Segovia:2014aza, Segovia:2015hra}.  Evolving the PDAs obtained therewith, from the hadronic scale to $\zeta=2\,$GeV, comparison with existing lQCD calculations became possible.

The evolved PDAs are depicted in Fig.\,\ref{PlotPDAs}.  These images are barycentric plots, in which the support of the DAs ($0\leq x_1,x_2,x_3\leq1$ with the additional constraint $x_1+x_2+x_3=1$) is mapped onto an equilateral triangle; and their structure reveals valuable insights.  For instance, the proton's PDA is a broadened, unimodal function, whose maximum is shifted relative to the peak in QCD's asymptotic profile, $\varphi_N^{\rm cl}([x])$.  This effect signals the presence of both scalar and pseudovector diquark correlations in the nucleon, with the $[ud]$ diquark generating approximately 65\% of the proton's normalisation.  The Roper-resonance has a similar diquark content, but the pointwise form of its PDA is negative on a material domain as a result of marked interferences between the contributions from both types of diquark.  Moreover, the associated, prominent locus of zeros in the lower-right corner of the Roper barycentric plot (rightmost figure) mirrors features seen in the wave function for the first radial excitation of all quantum mechanical system.  Similar behaviour is also found in the leading-twist PDAs of radially excited mesons \cite{Li:2016dzv, Li:2016mah}.

\subsection{Lattice-regularised QCD}
\label{sec:lattice}

Lattice QCD is a first-principles approach to investigate nonperturbative properties of QCD, and is expected to play an important role in the investigation of possible diquark correlations. However, the number of quarks and gluons contained within a colour-neutral hadron is not well-defined in the low energy regime. As such the colour-charged constituent quarks and diquarks are phenomenological objects and are difficult to be measured directly from the point of view of lQCD. Generally speaking, the calculable observables in lQCD are Schwinger functions $\langle O[\psi,\bar{\psi};U]\rangle $; namely, vacuum expectation values of (Euclidean) time-ordered operator products $O$ made up of quark ($\psi, \bar{\psi}$) and gluon (expressed by gauge links, $U$) fields, from which one can extract physical quantities, such as energies and matrix elements of hadronic states. LQCD studies of diquark correlations also follow this logic and the physical information is derived from different aspects of the related Schwinger functions.

The most straightforward approach is to extract the effective masses of diquarks from the temporal fall-off of diquark propagators \cite{Hess:1998sd, Babich:2007ah, Bi:2015ifa}, as is usually done to extract hadron masses. There are two conceptual issues in this approach. Firstly, a diquark operator by itself is not a colour singlet and should be treated within a specific gauge and thereby the conclusions may be gauge dependent. Secondly, if one interprets the temporal fall-off parameters as the effective masses of diquarks, one has to perform the intermediate state insertion using unphysical, colour-charged states. One has to keep these limitations in mind, if one takes such effective masses as the counterparts of phenomenological constituent quark masses.

A more rigorous treatment is to consider the possible diquark cluster within a hadron system, \emph{e.g}.\ a baryon \cite{Orginos:2005vr, Alexandrou:2006cq, DeGrand:2007vu, Green:2010vc}. The contribution of different types of diquarks to the hadron's mass was investigated by calculating the masses of baryons with a static heavy quark~\cite{Orginos:2005vr, Alexandrou:2006cq, Green:2010vc}. It is interesting to see that, for these kinds of baryons, the mass splitting between an axial-vector- and a scalar-diquark is compatible with the difference between effective-masses of related diquarks mentioned in the sections above and is commensurate with the nucleon-$\Delta(1232)$ mass splitting. The spatial correlation of the two light quarks has been studied in these baryons. On the other hand, one can also investigate spatial correlations amongst the quarks inside baryons through lQCD calculations of the baryon's wave function \cite{Babich:2007ah}, defined by beginning with a standard baryon correlator and displacing quarks at the sink.  The resulting function of spatial displacements is then evaluated in a fixed gauge.

Hard exclusive reactions involving large momentum transfer between the initial and final state hadron are most sensitive to the leading Fock states with a small number of partons and to the distribution of the longitudinal light-front momentum amongst these constituents.  This information is encoded in light-front PDAs \cite{Lepage:1979zb, Lepage:1980fj, Efremov:1979qk}.  They are universal functions that reveal features of hadron structure that are complementary to those obtained with parton distribution functions (PDFs) and form factors, which do not provide information on the individual Fock states. This section therefore also contains lQCD results for the wave function normalisation constants and the first moments of PDAs associated with the lowest-lying baryon octet \cite{Braun:2009jy, Bali:2015ykx, Bali:2019ecy}. These results can be used as a benchmark for models of hadron wave functions and might indicate signatures of diquark formation.


\begin{table}[t]
\centering
\caption{Parameters of configurations with 2+1 flavour dynamical domain wall fermions (RBC-UKQCD). $am_l$ and $am_s$ are the bare mass parameters of 	the degenerate $u,d$ sea quarks and strange sea quark, respectively. The residual masses are from Ref.\,\cite{Aoki:2010dy}. The lattice spacings are from Ref.\,\cite{Yang:2014sea}.}
\begin{ruledtabular}
\begin{tabular}{ccccc}
$a^{-1}$(GeV) & label & $am_l/am_s$ & volume &  $am_{res}$\\
\hline
$\sim 1.75(4)$ & c005  & 0.005/0.04 & $24^3\times64$ & 0.003152(43) \\
& c02 & 0.02/0.04 & $24^3\times64$ &  \\
\hline
$\sim 2.33(5)$ & f004 & 0.004/0.03 & $32^3\times64$ &  0.0006664(76) \\
\end{tabular}		
\end{ruledtabular}
\label{tab:confs}
\end{table}

\subsubsection{Effective masses of diquarks in Landau gauge}
This perspective focuses on the two $ud$ diquark configurations, scalar and axial-vector, which have longest been of interest in phenomenological studies. The related diquark operators are:
\begin{subequations}
\begin{align}
\mbox{scalar}, J^P=0^+\!: & \quad  J_c^5= \varepsilon_{abc}u^{a,T}C\gamma_5 d^b \,,
\quad J_c^{05} = \varepsilon_{abc} u^{a,T}C\gamma_5\gamma_4 d^b \,; \\
\mbox{axial-vector},
J^P=1^+\!: & \quad  J_c^i=\varepsilon_{abc} u^{a,T}C\gamma_i d^b \,, \quad
J_c^{0i}=\varepsilon_{abc}u^{a,T}C\gamma_i\gamma_4 d^b \,.
\end{align}
\end{subequations}
Since these operators are gauge dependent, their correlation functions, taking $J^{05}_c$ for example,
\begin{equation}\label{eqn:corr}
C(t)=\sum\limits_{\vec x}\langle 0|TJ^{05}_c(\vec{x},t)\bar{J}^{05}_c(0)|0\rangle \,,
\end{equation}
should be calculated from lQCD in a fixed gauge.

A recent full-QCD lattice study of diquarks was carried out with lattice chiral fermions \cite{Bi:2015ifa}. Lattice chiral fermions have well-defined chiral symmetry on the lattice and can access pion masses close to the physical value. The calculation was performed on the RBC/UKQCD configurations generated with $N_f=2+1$ domain wall fermions \cite{Aoki:2010dy}. The ensemble parameters are listed in Table~\ref{tab:confs}. Overlap fermions \cite{Neuberger:1997fp} were adopted for the valence quarks in the calculation of correlation functions in Eq.\,\eqref{eqn:corr} after the gauge configurations were fixed to Landau gauge. In this mixed-action lattice setup, quite a few valence quark masses, $am_q$, were available for use in extrapolating the results to the chiral limit.

\begin{figure}[!t]
\begin{center}
\includegraphics[width=0.5\textwidth]{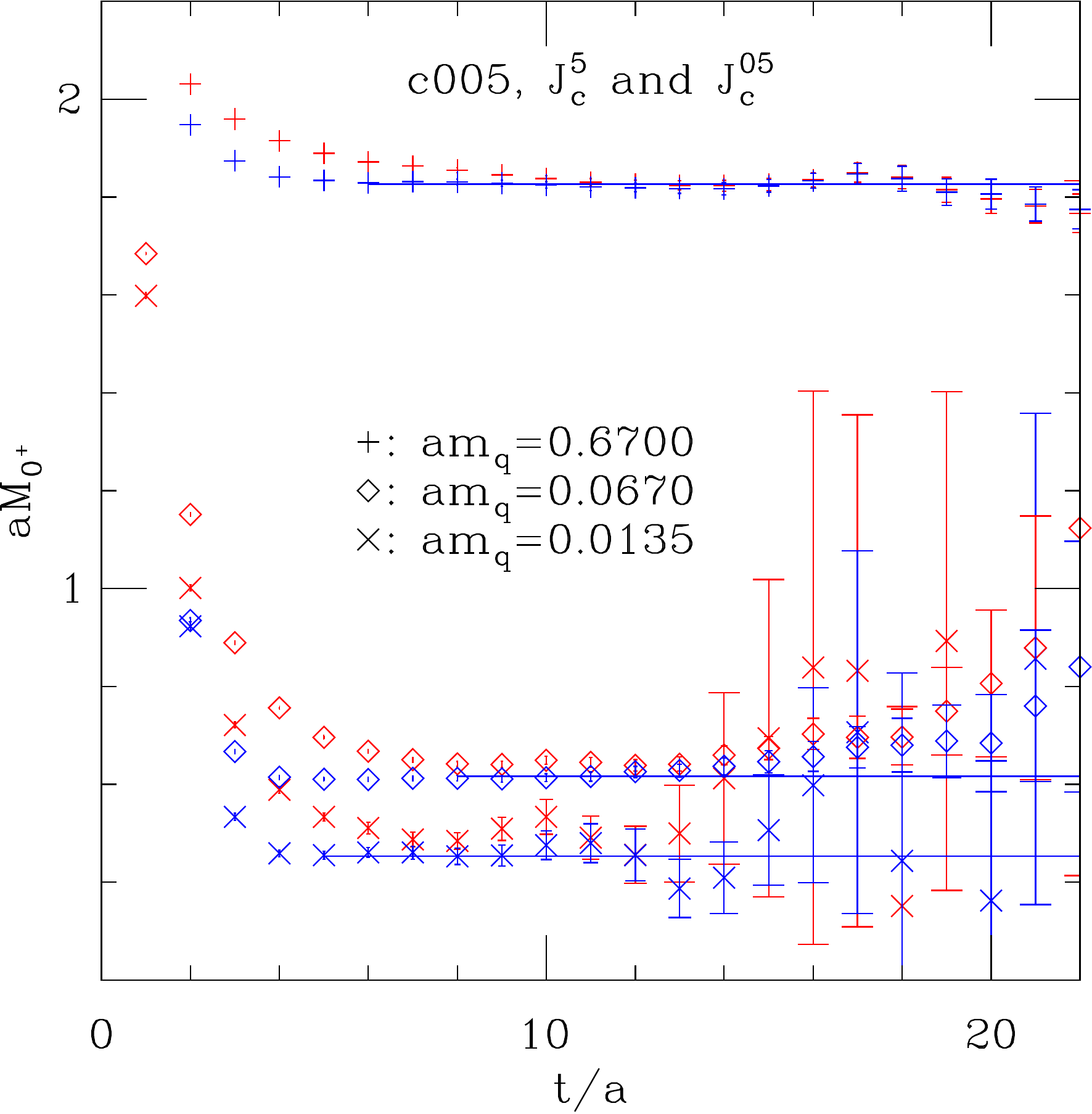}
\end{center}
\caption{\label{fig:scalar_c005} Effective scalar diquark masses at various valence-quark masses on ensemble c005 \cite{Bi:2015ifa}. The red and blue points are from the correlators $J^{5}_c$ and $J^{05}_c$, respectively. The straight lines illustrate the fit results obtained using single-exponential functions.}
\end{figure}

The temporal fall-off of $C(t)$ is usually monitored by introducing the effective-mass
function with respect to the time $t$,
\begin{equation}
M_{\rm eff}(t)a=\ln \frac{C(t)}{C(t+1)} \,.
\end{equation}
Fig.\,\ref{fig:scalar_c005} shows $M_{\rm eff}(t)$ for scalar diquark correlators at different valence-quark masses $am_q$ on gauge ensemble c005, where plateaux appear in the large-time range and those from operator $J^{5}_c$ and $J^{05}_c$ at the same valence-quark mass merge together. The case of the axial-vector diquark is similar. This implies that $C(t)$ decays exponentially at large $t$, \emph{i.e}.\ $C(t)\sim e^{-Mt}~~(t\gg 0)$. Through a single-exponential fit to $C(t)$, the parameter $M$ can be derived, as illustrated in Fig.~\ref{fig:scalar_c005} by a straight line for a specific valence quark mass $am_q$, which is usually interpreted as the effective mass of the corresponding diquark.

\begin{table}[!b]
\centering
\caption{\label{tab:mass}
Effective masses $M_q$ of $u,d$ quarks, $M_s$ of the strange quark, and those of diquarks ($m_{0^+}$ and $m_{1^+}$), computed in Landau gauge and extrapolated to the chiral limit.}
\begin{ruledtabular}
\begin{tabular}{cccccc}
& $M_q$    & $M_s$   & $m_{0^+}$ & $m_{1^+}$  & $m_{1^+}-m_{0^+}$ \\
& (MeV)    & (MeV)   & (MeV)     &  (MeV)     & (MeV) \\
\hline
c02  & 492(19) & 575(23) & 797(24) & 1127(28) & 330(35)\\
c005 & 427(25) & 586(16) & 725(20) & 1022(44) & 297(48)\\
f004 & 413(12) & 603(15) & 690(47) & 990(60)  & 300(76)	
\end{tabular}
\end{ruledtabular}
\end{table}

Similarly, an effective quark mass can be determined from the temporal fall-off of the quark propagators $S_q(t)=\sum_{\vec{x}}\text{Tr}\, S_q(\vec{x},t;\vec{0},0)\sim e^{-M t}$.  In this way, effective masses for the valence $u,d$ quarks (denoted by $M_q$) and the valence strange quark (denoted by $M_s$) are obtained. It is found that, on each gauge ensemble, the effective masses of quarks and diquarks depend linearly on the valence-quark mass $m_q$ or equivalently $m_\pi^2$ when $m_\pi <600$ MeV. Thus the chiral limit can be reached after linear extrapolations $M(m_q)=M(0)+c m_q$ or $M(m_\pi)=M(0)+c'm_\pi^2$.

Table~\ref{tab:mass} lists the computed effective masses $M_q$ of $u,d$ quarks, $M_s$ of the strange quark, and those of diquarks ($m_{0^+}$ and $m_{1^+}$) in the chiral limit. It is seen that while the results from ensemble c005 and f004 are consistent with each other, the values from ensemble c02 are larger. This is because the light sea quarks $u,d$ of the ensemble c02 have a larger mass than those of the other two ensembles. However, the mass difference $\delta_{{1^+0^+}} $ seems less sensitive to sea-quark masses.

\begin{figure}[!t]
\begin{center}
\begin{tabular}{ccc}
\includegraphics[clip, height=15em, width=0.45\textwidth]{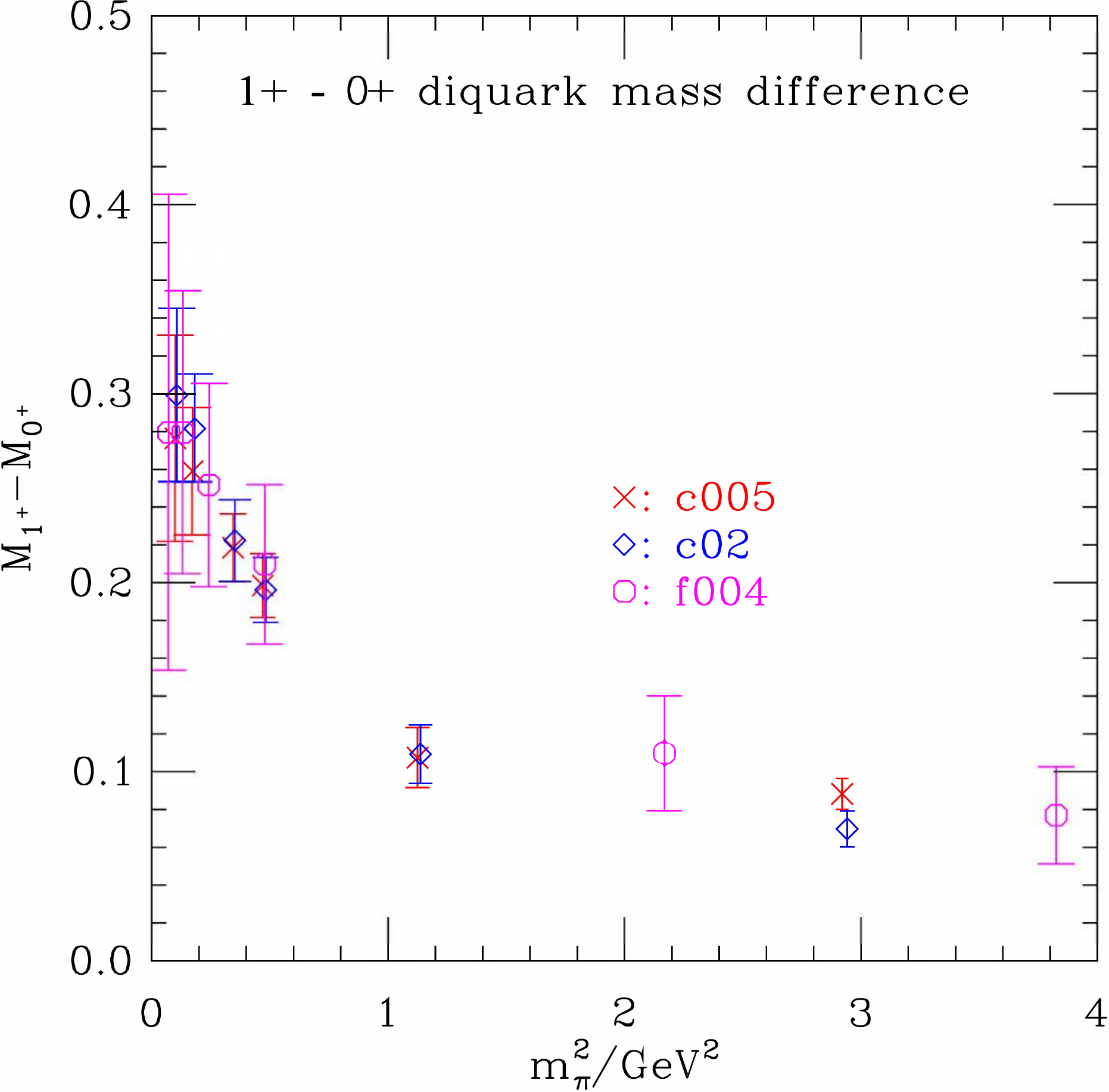} & \hspace*{2em} &
\includegraphics[clip, height=15em, width=0.45\textwidth]{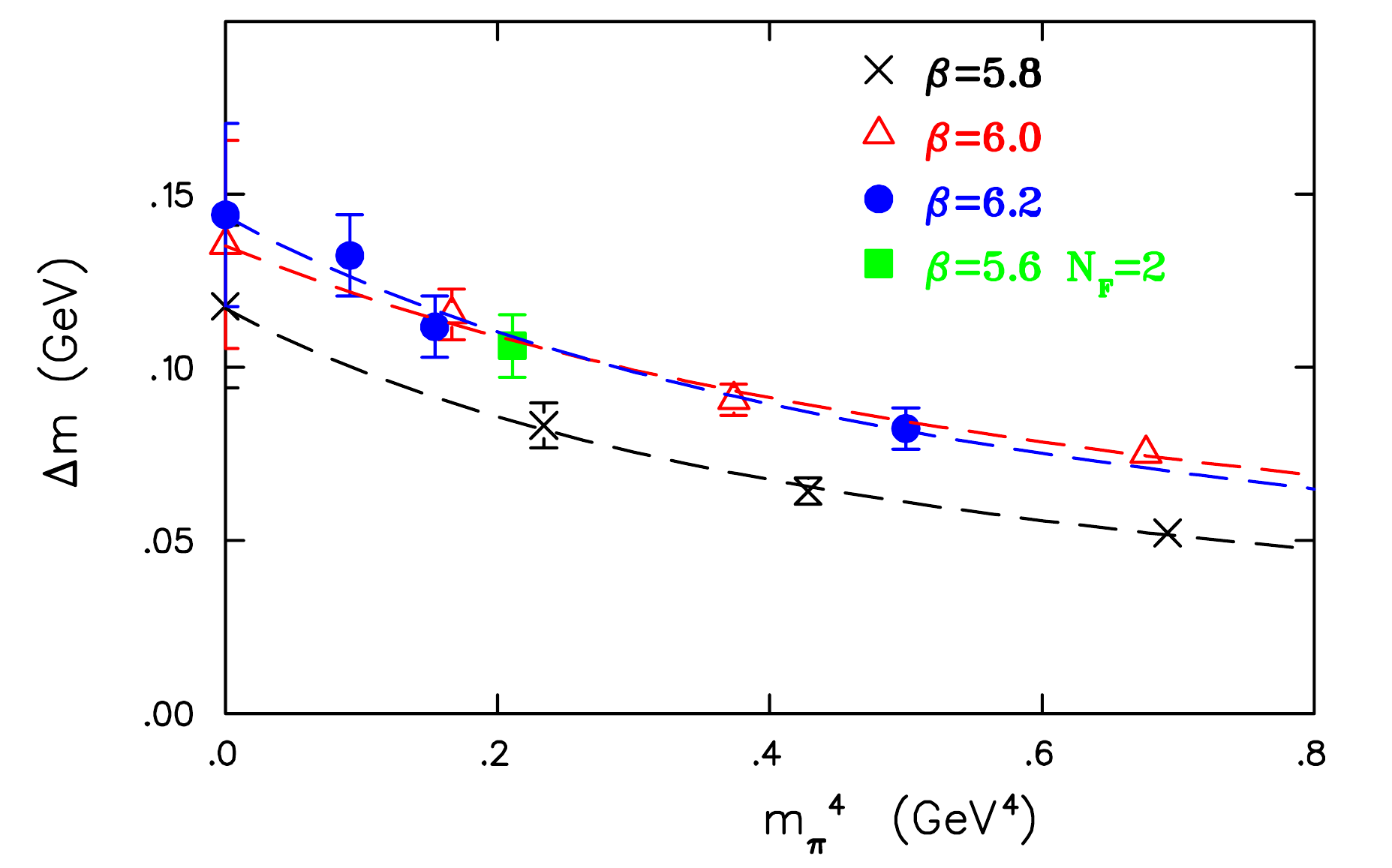}
\end{tabular}
\end{center}
\caption{\label{fig:diff_bad_good}
\emph{Left panel}.
$m_\pi^2$ dependence of $\delta_{{1^+0^+}}$ (in physical units) from all the three ensembles \cite{Bi:2015ifa}.
\emph{Right panel}.
Mass difference $\delta_{{1^+0^+}}$ between scalar and axial vector diquarks as a function of $m_\pi^4$ on different lattices (labeled by $\beta$) $\beta = 5.8$ (crosses), $\beta = 6.0$ (open triangles), $\beta = 6.2$ (filled circles) and unquenched result (filled square). The dashed lines are fits of the form $\Delta m :=\delta_{{1^+0^+}} =b_1/[1+b_2 m_\pi^4]$. (Adapted from Ref.\,\cite{Alexandrou:2006cq}.)
}
\end{figure}

Figure\,\ref{fig:diff_bad_good}--left panel shows the $m_\pi^2$ dependence of $\delta_{{1^+0^+}}$ (in physical units) from all the three ensembles. All data points seem to lie on a universal curve, which implies that the sea quark mass dependence and discretisation effects are less important in comparison with statistical errors. After chiral extrapolations, using both a  linear function in $m_\pi^2$ and the \emph{Ansatz} \cite{Alexandrou:2006cq} $\delta_{{1^+0^+}} =b_1/[1+b_2 m_\pi^4]$  in the range $m_\pi^2\le 1.2\,{\rm GeV}^2$, the final result for $m_{1^+}-m_{0^+}$ in the chiral limit is
\begin{equation}
\delta_{{1^+0^+}} =m_{1^+}-m_{0^+}=0.285(25)(45)\,{\rm GeV} \,,
\end{equation}
where the first error is statistical and the second owes to the different extrapolation functions and different fit range. This result can be compared with the $\Delta(1232)$-nucleon mass difference $\delta_{\Delta N}=0.272(56)$ GeV on ensemble c005 and $0.304(108)\,$GeV for ensemble c02, as well as the experimental value $\delta_{\Delta N}\approx 0.27\,$GeV. Evidently, as found using CSMs, Sec.\,\ref{sec:DSEs}, $\delta_{{1^+0^+}}$ is similar to $\delta_{\Delta N}$.  Furthermore, from Table~\ref{tab:mass}, one can see that the (effective) mass difference between the scalar diquark and the light quarks is roughly $M_{0^+}-M_q\approx 0.3\,$GeV. Such differences agree with estimates from hadron spectroscopy in, \emph{e.g}.\ Ref.\,\cite{Jaffe:2004ph}, and CSM calculations, \emph{e.g}.\ Ref.\,\cite[Tables\,1,\,3]{Yin:2019bxe}.


\subsubsection{Diquark correlations within baryons}
Since diquarks are not colour singlets and cannot exist as asymptotic states, one might argue that it is better to investigate their physical significance within hadron systems, for instance, baryons. There are lQCD studies on possible diquark correlations in the background of a static quark~\cite{Orginos:2005vr, Alexandrou:2006cq, Green:2010vc}. An objective diquark and the static quark form a baryon system, which can be produced by the operator
\begin{equation}
J_\Gamma(x)=\varepsilon_{abc}\left[u^{a,T}(x)C\Gamma d^b(x)\pm d^{a,T}(x)C\Gamma u(x)^b\right]Q^c(x),
\end{equation}
where $\Gamma=I,\gamma_\mu,\gamma_5,\gamma_5\gamma_\mu,\sigma_{\mu\nu}$, the $\pm$ sign corresponds to the flavour symmetric (antisymmetric) combination, and $Q^c(x)$ is the static quark field. The mass of the baryon can be extracted from the temporal fall-off of the correlation function $C_\Gamma (t)=\langle J_\Gamma(\vec{x},t)J_\Gamma^\dagger(\vec{x},0)\rangle$. The lattice calculation of this kind of correlator is similar to that of normal baryons with the propagator of the static quark being expressed as
\begin{align}
S_Q(\vec{x}_2,t_2;\vec{x}_1,t_1) &= e^{-m_Q(t_2-t_1)}\delta^3(\vec{x}_1-\vec{x}_2) \left(\frac{1+\gamma_4}{2}\right)\left[\prod\limits_{t=t_1}^{t=t_2-a} U_4(\vec{x}_1,t)\right]^\dagger \,,
\end{align}
for $t_2>t_1$, where $a$ is the lattice spacing and $U_4(x)$ is the temporal gauge link at $x$.


The scalar and pseudovector diquarks can be generated by $J_\Gamma$ with $\Gamma=\gamma_5 $ and $\gamma_i$, respectively. Thus, the effective mass difference $\delta_{{1^+0^+}}\!$ can be extracted from the ratio $C_{\gamma_i}(t)/C_{\gamma_5}\sim e^{-t \delta_{{1^+0^+}}}$ when $t\gg 0$, since the contribution of the static quark cancels out in the exponential prefactor. In Ref.\,\cite{Alexandrou:2006cq}, the above calculation was carried out on several quenched gauge ensembles and an ensemble generated with $N_f=2$ Wilson fermions. Figure~\ref{fig:diff_bad_good}--right panel shows the results for $\delta_{{1^+0^+}}$ as a function of $m_\pi^4$ \cite{Alexandrou:2006cq}. On quenched (labeled by $\beta=6.0$ and $\beta=6.2$) fine lattices, the data points fall almost on a universal line, as expected when close to the continuum limit.
On the unquenched (labeled by $\beta=5.6, N_f=2$) fine lattice, $\delta_{{1^+0^+}}$ also falls nicely on the same curve as the quenched results. This indicates quenching effects at these quark masses are small, whereas for the coarsest lattice, labeled by $\beta=5.8$, scaling violations are apparent. The dashed lines are the fit with the ansatz $\delta_{{1^+0^+}} =b_1/[1+b_2 m_\pi^4]$, which is suggested by a prediction from effective colour-spin Hamiltonian arguments \cite{Jaffe:2004ph}, \emph{viz}.\ $\delta_{{1^+0^+}}$ scales like $1/(M_{q_1}M_{q_2})$, where $M_{q_{1,2}}$ are the masses of the constituent quarks. Obviously, the \emph{Ansatz} describes the data well.
In this connection, it is worth remarking that on the pictured $m_\pi^2$-domain, CSM predictions are described by $\delta_{{1^+0^+}} = b_1/[1+b_2 m_\pi^2]$ (see, \emph{e.g}.\ Ref.\,\cite[Fig.\,5]{Roberts:2011cf}), suggesting one has not entered a domain wherein constituent-quark degrees-of-freedom are relevant.  This contradiction should be understood and resolved.

The $\Delta$-nucleon mass splitting, $\delta_{\Delta N}$, can be also calculated on these lattices. On the quenched fine lattice $\beta=6.0$, the ratio $\delta_{{1^+0^+}} /\delta_{\Delta N}$ is found to be $0.67(7)$, $0.73(8)$ and $0.67(8)$ at three different valence quark masses, respectively. This prediction is lower than that obtained when computing effective masses of diquarks in the Landau gauge. However, this value of $\delta_{\Delta N}$, which is $\sim 2/3$ the $\Delta$-nucleon mass difference, matches well with the CSM prediction, Eq.\,\eqref{CSMavsc}.

\begin{figure}[!t]
\begin{center}
\includegraphics[clip, width=0.55\textwidth]{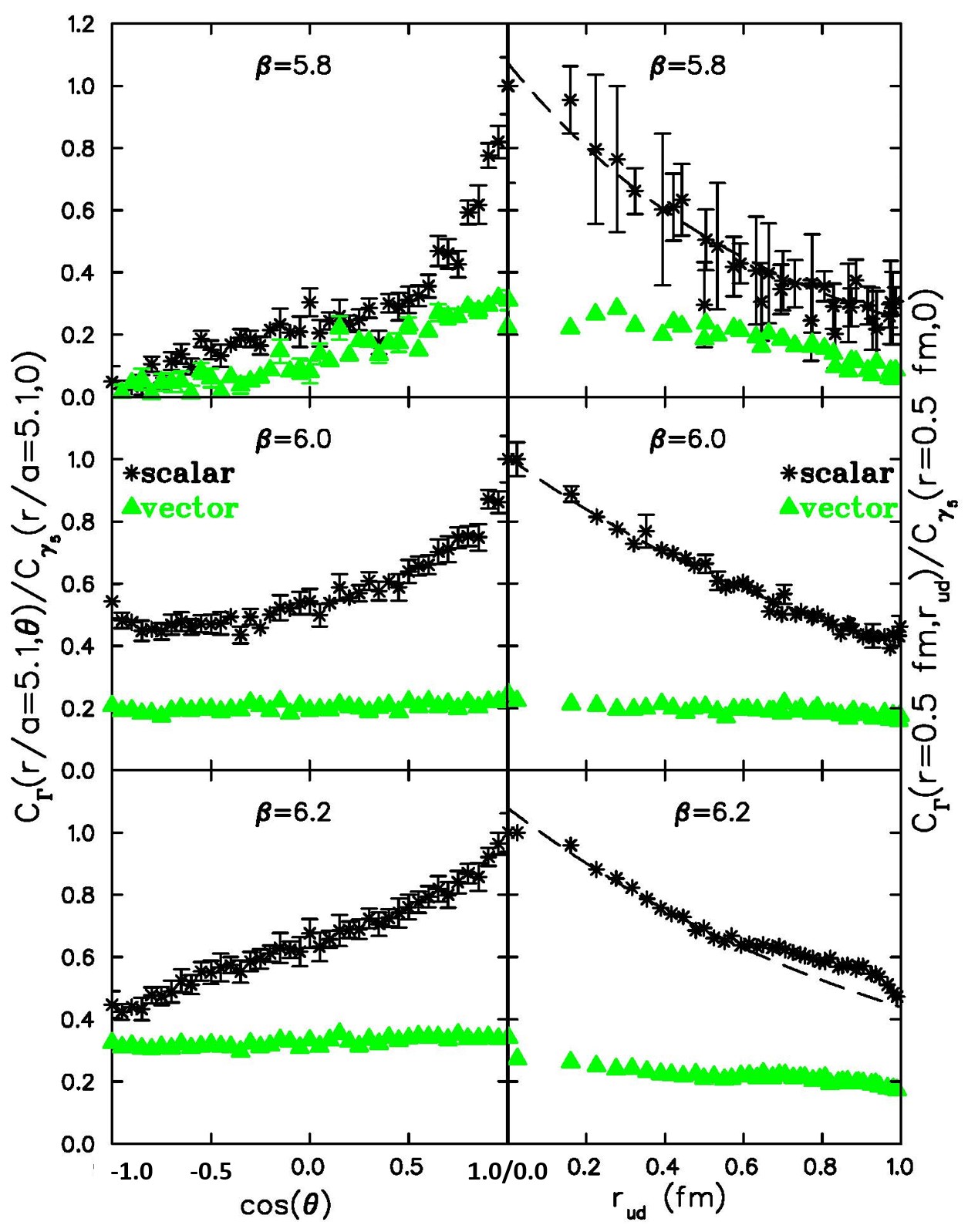}
\end{center}
\caption{\label{fig:correlation}
\emph{Left panels}: $C_\Gamma(r/a = 5.1,\theta)/C_{\gamma_5}(r/a = 5.1,0)$ versus $\cos(\theta)$.
\emph{Right panels}: $C_\Gamma(r = 0.5\,{\rm fm},r_{ud})/C_{\gamma_5}(r = 0.5\,{\rm fm},0)$ versus $r_{ud}$.
Both panels: asterisks -- scalar diquark and filled triangles -- axial-vector diquark; both obtained using the lightest pion on the three quenched lattices. (Figure adapted from Ref.~\cite{Alexandrou:2006cq}).}
\end{figure}

In addition to the masses, diquark correlations can be probed directly by investigating the spatial distribution of two quarks within a diquark via their density-density correlators within baryons.  The correlators are generated by an operator $J_\Gamma(x)$,
\begin{equation}
C_\Gamma(\vec{r}_u, \vec{r}_d,t) = \langle 0|J_\Gamma(\vec{0},2t) J^u_0(\vec{r}_u,t) J_0^d(\vec{r}_d,t) J_\Gamma^\dagger(\vec{0},0)|0\rangle \,,
\end{equation}
where $J_0^f(\vec{r},t)= :\!\bar{f}(\vec{r},t)\gamma_0f(\vec{r},t)\!:$ is the density operator of quark flavour $f=u,d$ and $\vec{r}_{u,d}$ are the distances of $u,d$ quarks from the static quark.  On the spherical shells $|\vec{r}_u|=|\vec{r}_d|=r$ with respect to the location of the static quark, the angle $\theta = \arccos(\hat{\vec{r}}_u \cdot \hat{\vec{r}}_d)$ can be a meaningful variable to show the diquark correlation in the sense that any attraction between the two quarks will be reflected by the enhancement of $C_\Gamma(r,\theta)$ at small angles, \emph{i.e.} near $\cos(\theta)=1$.

Fig.\,\ref{fig:correlation} shows the density correlators for the scalar and the axial-vector diquarks for the three quenched lattices at the lightest quark mass~\cite{Alexandrou:2006cq}. On the left hand side, the correlator of the scalar diquark (black points) as a function of $\cos(\theta)$ grows faster when $\cos(\theta)$ approaches 1 and thereby shows stronger spatial correlations relative to the axial-vector diquark (green points). Furthermore, if the $u-d$ separation $r_{ud}=2r\sin (\theta/2)$ is introduced at a fixed $r$, the spatial size of diquarks can be estimated from the fall-off of $C_\Gamma(r,r_{ud})$ versus $r_{ud}$. The correlator as a function of $r_{ud}$ is shown on the right side of Fig.~\ref{fig:correlation} for a fixed shell radius $r = 0.5$ fm, where the curves are obtained from fits with the form $C_{\gamma_5}(r,r_{ud})\propto \exp(-r_{ud}/r_0(r))$. The parameter $r_0(r)$ provides a gauge invariant definition of the scalar diquark size at a given $r$. It is found that $r_0(r)$ increases mildly and saturates around $r_0\approx 1.1\pm 0.2$ fm. By this measure, a scalar diquark is a large object, with a characteristic size of $O(1)$ fm.  This is also the size predicted by CSMs.

Ref.\,\cite{Green:2010vc} follows a similar strategy, but with a more general geometry such that the positions $\vec{r}_{u,d}$ of the $u,d$ quarks are not restricted on the same spherical shells centered at the static quark. The conclusion is that the correlation of the scalar diquark is stronger than that of the axial-vector diquark and the diquark size is comparable to the typical hadron size.

The analyses reviewed in this section add to the arguments against relying heavily on hadron models built using point-like (hard) diquarks.


\subsubsection{Bethe-Salpeter wave function approach}
\label{lQCDBSWF}
There are also lQCD efforts aimed at exploring the inner structure of hadrons by calculating their Bethe-Salpeter (BS) wave functions.  Taking the nucleon as an example, one starts by defining spatially extended lattice operators based on the conventional nucleon operator $\eta(x) = \varepsilon_{abc} [u^{a,T}(x) C\gamma_5 d^b(x)] u^c(x)$. A straightforward way is to shift the diquark component from the third quark field by a spatial separation $\vec{R}$ \cite{Chen:2007zzc},
\begin{equation}
\eta(\vec{x},t;\vec{R}) = \epsilon_{abc}[u^{a,T}(\vec{x}+\vec{R},t)C\gamma_5 d^b(\vec{x}+\vec{R},t)]u^c(\vec{x},t) \,.
\end{equation}
Obviously, $\eta(\vec{x},t;\vec{R})$ is not gauge invariant, so its correlator with a source operator $\eta_s$,
\begin{equation}
C(R,t)=\frac{1}{N_R}\sum\limits_{\vec{x},|\vec{R}|=R}Tr\left[(1+\gamma_4)\langle 0|\eta(\vec{x},t;\vec{R})\bar{\eta}_s(0)|0\rangle\right] \,,
\end{equation}
should be calculated in a fixed gauge.  Here, the summation over $\vec{x}$ is constrained to the same $|\vec{R}|=R$, in order to fix the correct quantum numbers, and $N_R$ is the degeneracy of $\vec{R}$. $C(R,t)$ can be parameterised as
\begin{equation}\label{wvf}
C(R,t)=\sum\limits_n \Phi_n(R) e^{-m_n t} \,,
\end{equation}
where $m_n$ is the mass of the $n$-state and the spectral weight $\Phi_n(R)$ is interpreted as the (gauge fixed) Bethe-Salpeter wave function of the $n$-th state, up to a normalisation constant.

\begin{figure}[!t]
\begin{center}
\includegraphics[width=0.5\textwidth]{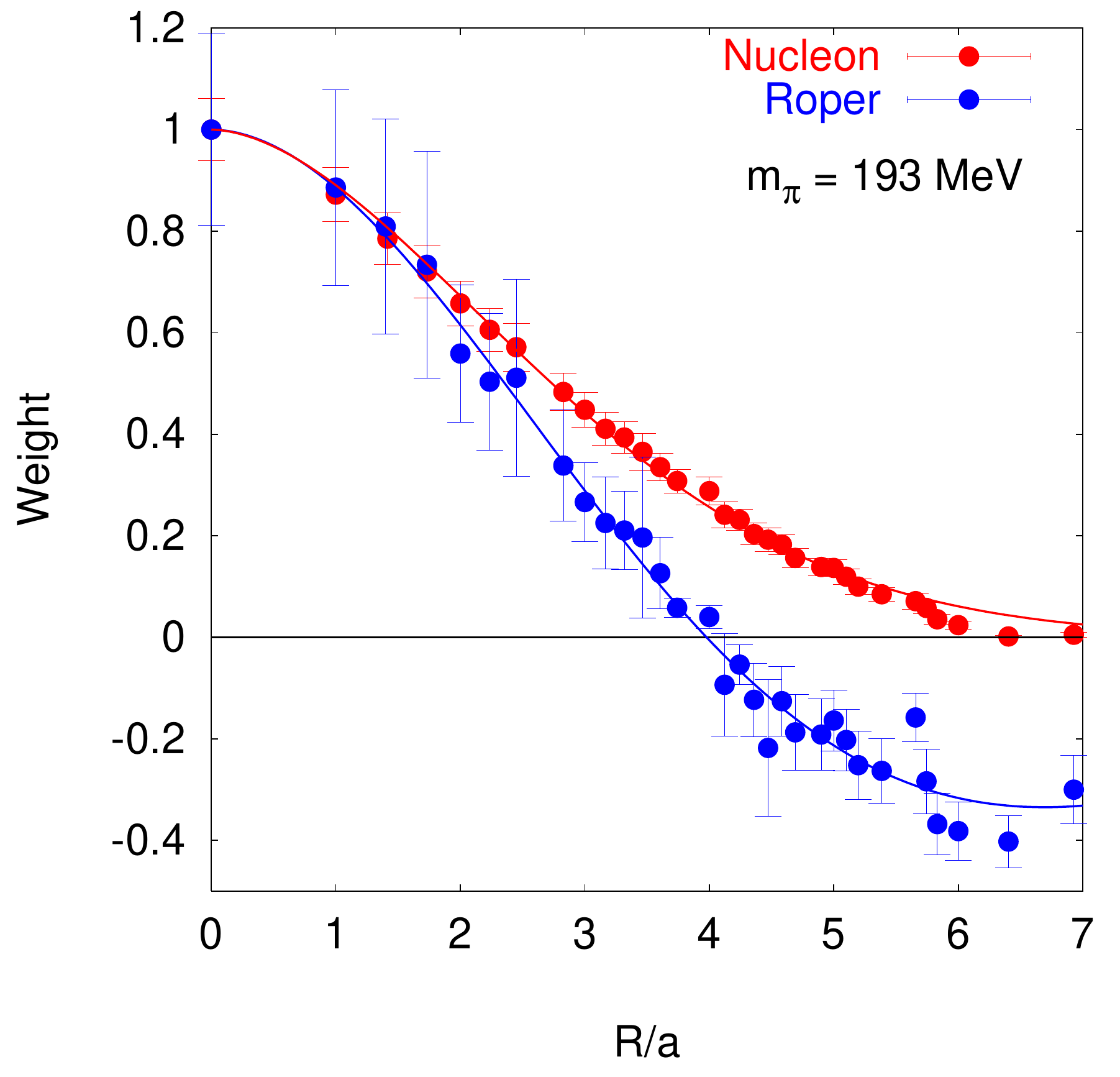}
\end{center}
\caption{\label{fig:nuc_wvf}
Lattice-QCD results for Bethe-Salpeter wave functions of the nucleon and its lightest like-parity excitation, computed with $m_\pi=193$ MeV and normalised such that $\Phi_n(0)=1$ .}
\end{figure}

In Refs.\,\cite{Chen:2007zzc, Sun:2019aem}, the wall-source correlation functions, $C(R,t)$, were calculated in Coulomb gauge, with quenched gauge configurations generated on $16^3\times 28$ lattices using lattice spacing
$a \sim 0.2\,$fm.
Overlap fermions were adopted as the valence quarks, with quark masses in a wide range.  The analysis yielded the masses and BS wave functions of the ground and the first excited states at different quark masses.  With $m_\pi\sim 193$ MeV, the masses of the ground state and first excited state are, respectively, $0.939(28)$ GeV and $1.40(18)$ GeV.  The first excited state was therefore tentatively identified with the Roper resonance.  The BS wave functions of the nucleon and the Roper state at $m_\pi=193$ MeV are plotted in Fig.\,\ref{fig:nuc_wvf}, where the wave function of Roper has a clear radial node.  Since $R$ is the separation of the (scalar) diquark component and the third quark field in the operator $\eta$, the BS wave functions seem compatible with the Roper as the first radially excited state of the nucleon, as a quark-diquark system.  (These conclusions match those of CSM analyses, see \emph{e.g}.\ Sec.\,\ref{QCDkindred} and Fig.\,\ref{PlotPDAs} above, and Ref.\,\cite{Burkert:2017djo}.)


\subsubsection{Light-cone distribution amplitudes}
Baryon distribution amplitudes (DAs) \cite{Lepage:1980fj, Efremov:1979qk, Chernyak:1983ej} are defined as matrix elements of renormalised three-quark operators at light-like separations (here the scheme proposed in Ref.\,\cite{Kraenkl:2011qb} is used):
\begin{align}
\label{eq_BDA}
&
\langle 0 | \big[ f_\alpha(a_1 n) g_\beta(a_2 n) h_\gamma(a_3 n) \big]^{\MSbar} | B_{p,\lambda} \rangle =  \frac14 \int \! [dx] \; e^{-ip \cdot n \sum_i a_i x_i}\nonumber \\
&
\quad \times  \Bigl( v^B_{\alpha\beta;\gamma} V^B(x_1,x_2,x_3) + a^B_{\alpha\beta;\gamma} A^B(x_1,x_2,x_3) + t^B_{\alpha\beta;\gamma} T^B(x_1,x_2,x_3)+\ldots \Bigr) \,. 
\end{align}
On the left-hand-side, the Wilson lines and the colour antisymmetrisation are not written explicitly but implied. $| B_{p,\lambda} \rangle$ is the baryon state with momentum $p$ and helicity $\lambda$, while $\alpha,\beta,\gamma$ are Dirac indices, $n$ is a light-like vector ($n^2=0$), the $a_i$ are real numbers, and $f,g,h$ are quark fields of the given flavour, chosen to match the valence quark content of the baryon $B$ (assuming exact isospin symmetry, one can choose a single representative for each isospin multiplet \cite{Franklin:1968pn}: $N := uud$; $\Sigma :=  dds$; $\Xi := ssu$; $\Lambda := uds$).  In the Lorentz decomposition on the right-hand-side of Eq.\,\eqref{eq_BDA},  only the three leading-twist DAs are shown: $V^B$, $A^B$, and $T^B$.  They appear in conjunction with particular Dirac structures.  The general decomposition consists of $24$ terms (see, \emph{e.g}.\ Ref.\,\cite{Braun:2000kw}). The exponential factor in combination with the integration measure for the light-front longitudinal momentum fractions,
\begin{align}
\int \! [dx] &= \int_0^1 \!\! dx_1 \int_0^1 \!\! dx_2 \int_0^1 \!\! dx_3 \; \delta (1-x_1-x_2-x_3)\,
\end{align}
ensure correct translational behaviour and momentum conservation in the light-front ``plus'' direction.

\begin{figure*}[tb]
\centering
\includegraphics[clip, width=.925\textwidth]{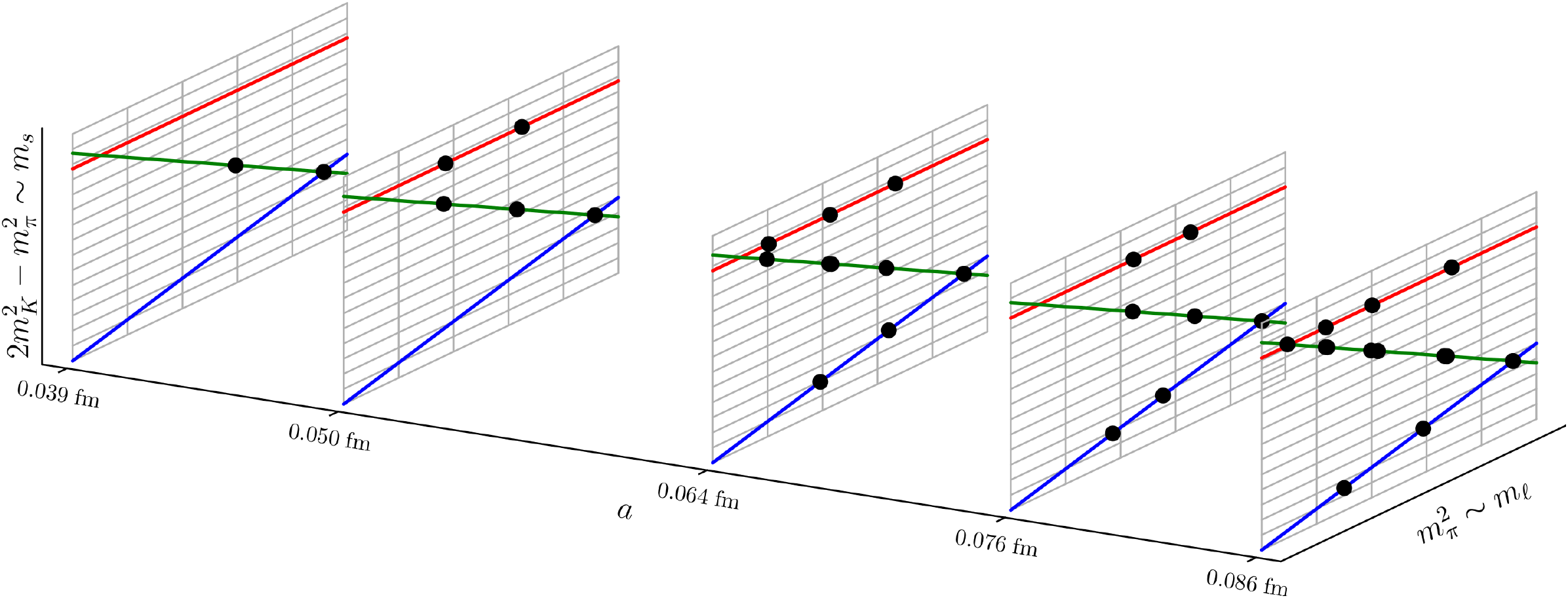}
\caption{\label{figure_ensembles}Schematic image showing the analysed CLS ensembles in the space spanned by the lattice spacing and quark masses. The different quark mass trajectories correspond to the limit of exact flavour symmetry (blue), the case of approximately physical mean quark mass (green), and to a nearly physical strange quark mass (red). Physical masses are reached at the intersections of green and red lines.}
\end{figure*}

To exploit the benefits of SU$_{\rm f}$(3) symmetry it is useful to define the following set of DAs:
\begin{align}
\Phi_{\pm}^{B\neq\Lambda}(x_{123}) & =\tfrac{1}{2} \Bigl([\VmA]^B(x_{123}) \pm [\VmA]^B(x_{321})\Bigr) \,, \notag \\
\Pi^{B\neq\Lambda}(x_{123}) & = T^B (x_{132}) \,, \notag \\
\Phi_{+}^{\Lambda}(x_{123}) & =\sqrt{\tfrac{1}{6}} \Bigl([\VmA]^\Lambda(x_{123}) + [\VmA]^\Lambda(x_{321}) \Bigr) \,, \notag \\
\Phi_{-}^{\Lambda}(x_{123}) & =-\sqrt{\tfrac{3}{2}} \Bigl([\VmA]^\Lambda(x_{123}) - [\VmA]^\Lambda(x_{321}) \Bigr) \,, \notag \\
\Pi^{\Lambda}(x_{123}) &= \sqrt{6} \; T^\Lambda (x_{132}) \,, \label{eq_convenient_DAs}
\end{align}
where $(x_{ijk})\equiv (x_i,x_j,x_k)$.  (For more details, see Refs.\,\cite{Wein:2015oqa, Bali:2015ykx}.)  In the limit of SU$_{\rm f}$(3) symmetry (subsequently indicated by a $\star$), where $m_u=m_d=m_s$, the following relations hold:
\begin{align}
\Phi_{+}^\star &\equiv \Phi_{+}^{N\star} = \Phi_{+}^{\Sigma\star} = \Phi_{+}^{\Xi\star} = \Phi_{+}^{\Lambda\star} = \Pi^{N\star} = \Pi^{\Sigma\star} = \Pi^{\Xi\star} \,, \notag
\\*
\Phi_{-}^\star &\equiv \Phi_{-}^{N\star} = \Phi_{-}^{\Sigma\star} = \Phi_{-}^{\Xi\star} = \Phi_{-}^{\Lambda\star} = \Pi^{\Lambda\star}\,.\label{eq_SU3_leadingtwist}
\end{align}
Therefore, the amplitudes $\Pi^B$ (or $T^B$) only need to be considered when SU$_{\rm f}$(3) symmetry is broken. In the case of SU(2) isospin symmetry, which is exact in a typical $N_f=2+1$ simulation ($m_u=m_d\equiv m_\ell$) and is only broken very mildly in the real world, the nucleon DA $\Pi^N$ is equal to $\Phi_+^N$ in the whole $m_\ell$-$m_s$-plane.

DAs can be expanded in terms of orthogonal polynomials $\mathcal{P}_{nk}$ in such a way that the coefficients have autonomous scale dependence at one loop (conformal partial wave expansion). Taking into account the corresponding symmetry of the DAs defined in Eqs.\,\eqref{eq_convenient_DAs}, this expansion reads
\begin{align}
\Phi_{+}^{B} &= 120 x_1 x_2 x_3 \bigl( \varphi^B_{00} \mathcal P_{00} + \varphi^B_{11} \mathcal P_{11} + \dots \bigr) \, , \quad
\Phi_{-}^{B} = 120 x_1 x_2 x_3 \bigl( \varphi^B_{10} \mathcal P_{10} + \dots \bigr) \,, \notag \\
\Pi^{B\neq\Lambda} &= 120 x_1 x_2 x_3 \bigl( \pi^B_{00} \mathcal P_{00} + \pi^B_{11} \mathcal P_{11} + \dots \bigr) \, , \quad
\Pi^{\Lambda} = 120 x_1 x_2 x_3 \bigl( \pi^{\Lambda}_{10} \mathcal P_{10} + \dots \bigr) \,.\label{eq_moments}
\end{align}
Here, all nonperturbative information is encoded in the set of scale-dependent coefficients $\varphi^B_{nk}$, $\pi^B_{nk}$ (often called shape parameters), which can be related to matrix elements of local operators that are calculable using lQCD.  All $\mathcal{P}_{nk}$ have definite symmetry (being symmetric or antisymmetric) under the exchange of $x_1$ and $x_3$ \cite{Anikin:2013aka} and in each DA (labeled with $+$ and $-$) only polynomials of one type, either symmetric or antisymmetric, appear (see, \emph{e.g.} Ref.\,\cite{Braun:2008ia}). The leading contributions in Eqs.\,\eqref{eq_moments} are $120 x_1 x_2 x_3 \varphi^B_{00}$ and $120 x_1 x_2 x_3 \pi^{B\neq\Lambda}_{00}$.  They are usually referred to as the asymptotic DAs. The corresponding normalisation coefficients $\varphi^B_{00}=: f^B$ and $\pi^{B\neq\Lambda}_{00}=: f^B_T$ can be thought of as the wave functions at the origin and are also called wave function normalisation constants.

\begin{figure*}[!t]
\centering
\includegraphics[width=\textwidth]{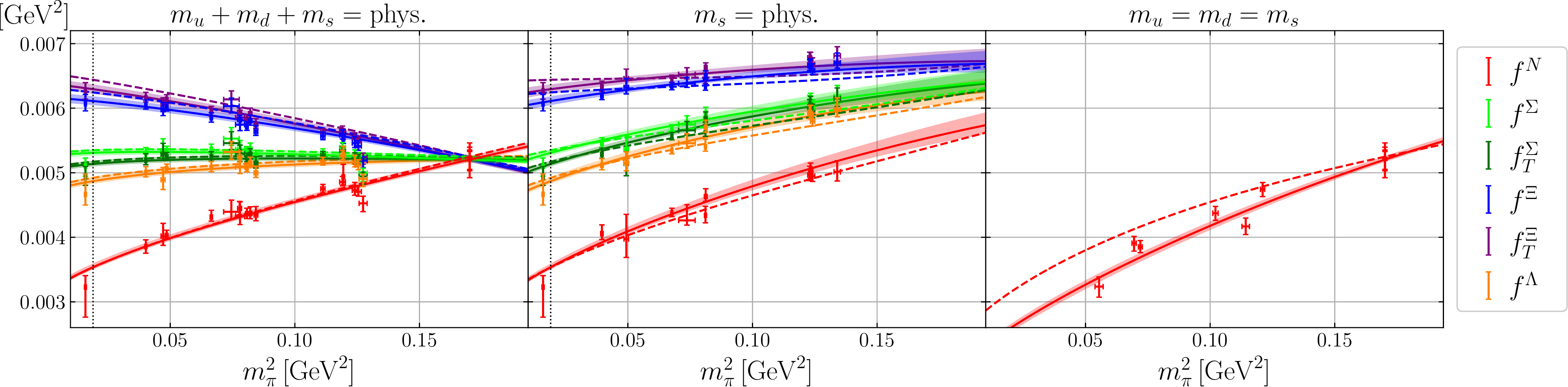}\\[\baselineskip]
\includegraphics[width=\textwidth]{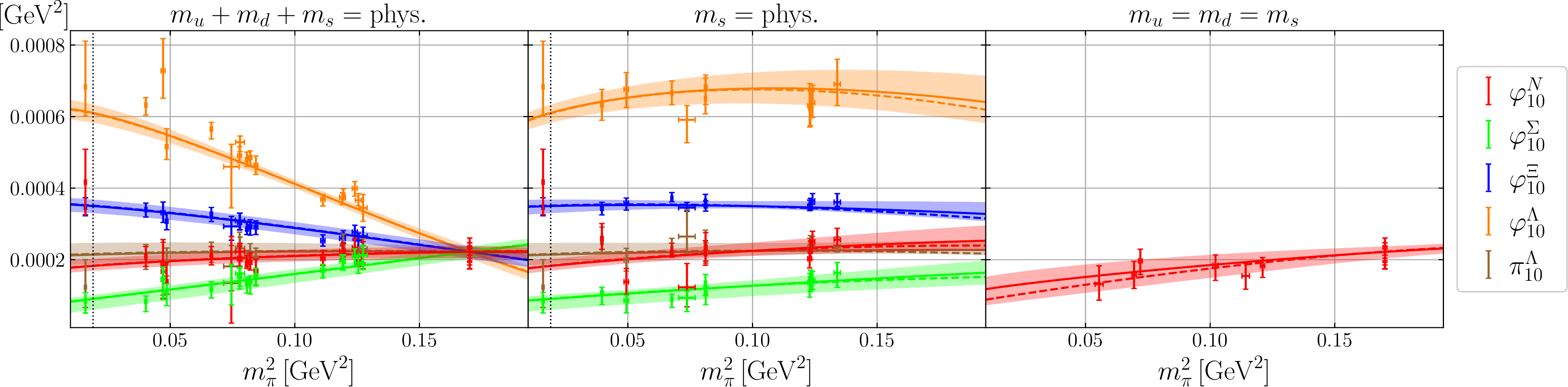}
\caption{\label{figure_mP2}Mass dependence of the normalisation constants $f$, $f_T$ (upper panels) and the first moments $\phi_{10}$, $\pi_{10}$ (lower panels), along the three quark mass trajectories shown in Fig.~\ref{figure_ensembles} after taking the continuum and infinite volume limits. The points are corrected for discretisation and volume effects.}
\end{figure*}

The two-point correlation functions that have to be evaluated on the lattice in order to obtain the normalisation constants and first moments of baryon octet DAs are given in Ref.\,\cite{Bali:2015ykx}. For the analysis reviewed here, a large set of lattice ensembles generated within the coordinated lattice simulations (CLS) effort was used. These $N_f=2+1$ simulations employed the nonperturbatively order-$a$ improved Wilson (clover) quark action and the tree-level Symanzik improved gauge action. A special feature of CLS configurations is the use of open boundary conditions in the time direction \cite{Luscher:2011kk, Luscher:2012av} for ensembles with small lattice spacings, which avoids topological freezing \cite[Appendix\,A]{Bietenholz:2016mgn}. The ensembles cover a wide range of volumes with $2.9\leq m_\pi L\leq6.5$, where most have $m_\pi L>4$.

The results were renormalised using a two-step procedure. First, the renormalisation factors were computed nonperturbatively on the lattice \cite{Martinelli:1994ty} within the RI$^\prime$\nobreakdash-SMOM scheme \cite{Sturm:2009kb}, which was adapted to three-quark operators in Refs.\,\cite{Kaltenbrunner:2008zz, Gockeler:2008we, Bali:2015ykx, Gruber:2017ozo}. These factors were then converted to the $\MSbar$\,scheme using one-loop (continuum) perturbation theory. The conversion factors can be found in Ref.\,\cite{Gruber:2017ozo}.

\begin{figure}[!t]
\centering
\begin{tabular}{lcr}
\includegraphics[width=0.45\textwidth]{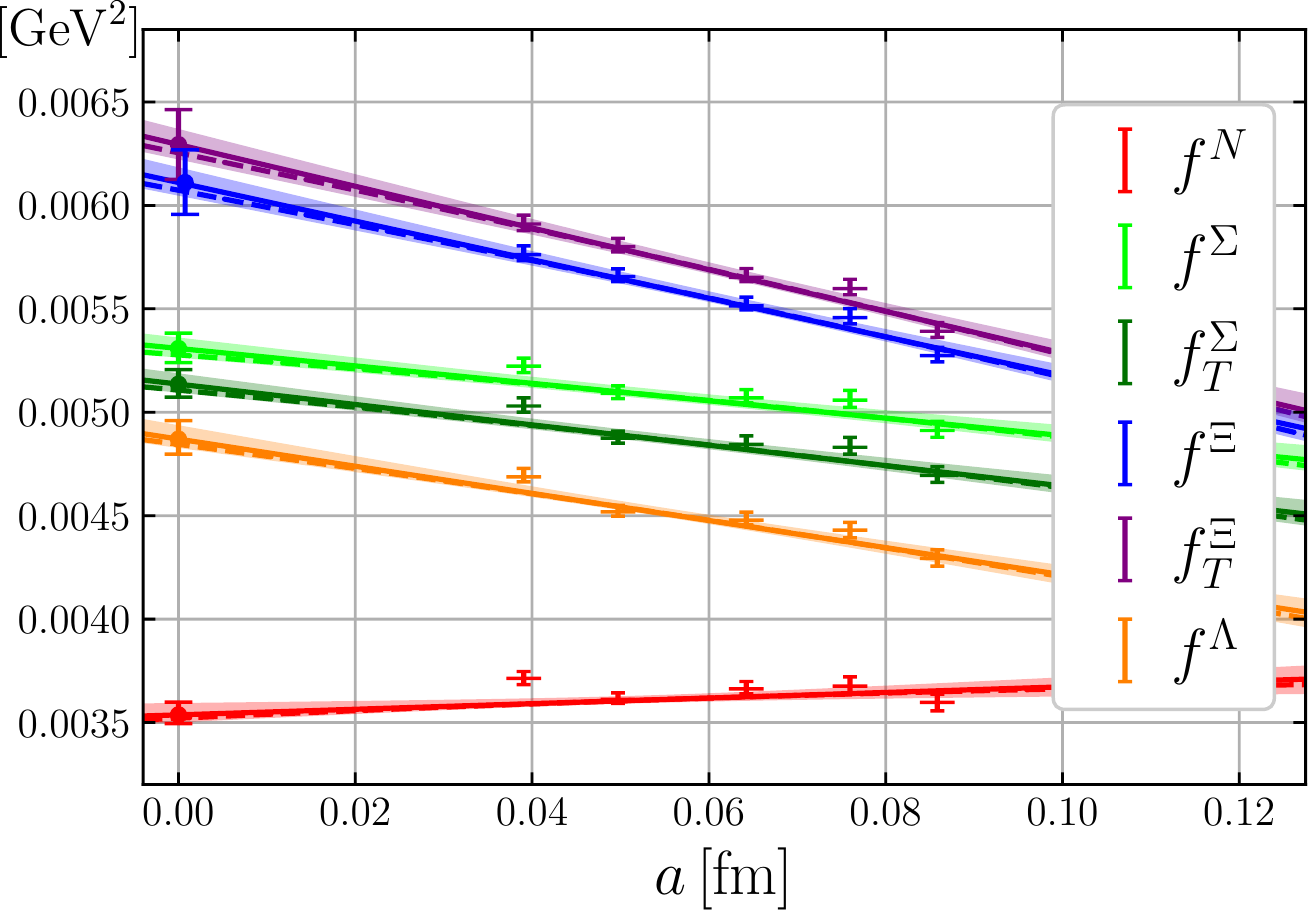} & \hspace*{1em} &
\includegraphics[width=0.45\textwidth]{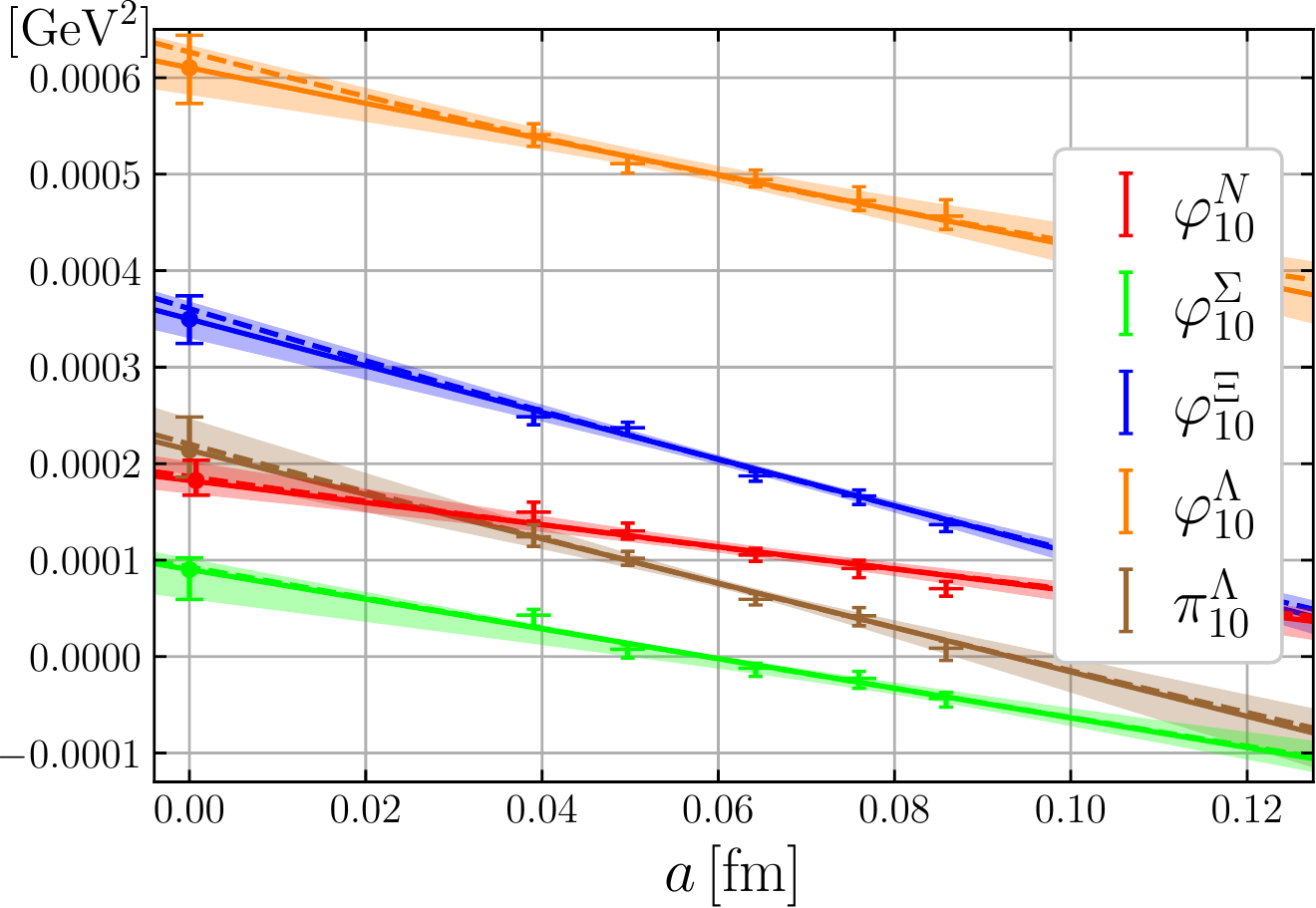}
\end{tabular}
\caption{\label{figure_a}Continuum extrapolation of the normalisation constants $f$, $f_T$ (upper panel) and the first moments $\phi_{10}$, $\pi_{10}$ (lower panel), after taking the limits to physical masses and to infinite volume. The shown data points are obtained by correcting for mass and volume effects, and, subsequently, taking the average of all ensembles with similar lattice spacing.}
\end{figure}

\begin{figure}[!t]
\centering
\begin{tabular}{c}
\includegraphics[width=0.45\textwidth]{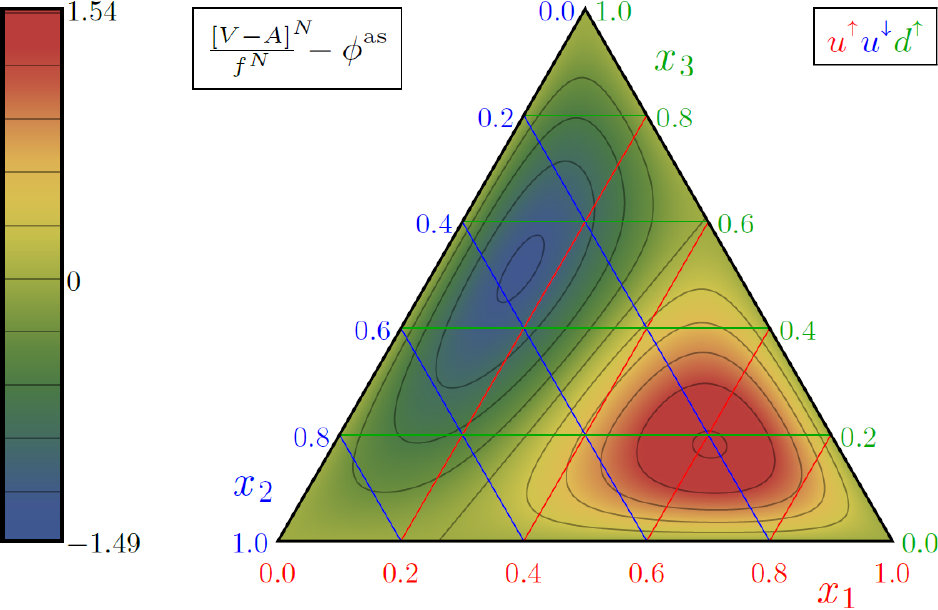} \\
\includegraphics[width=0.45\textwidth]{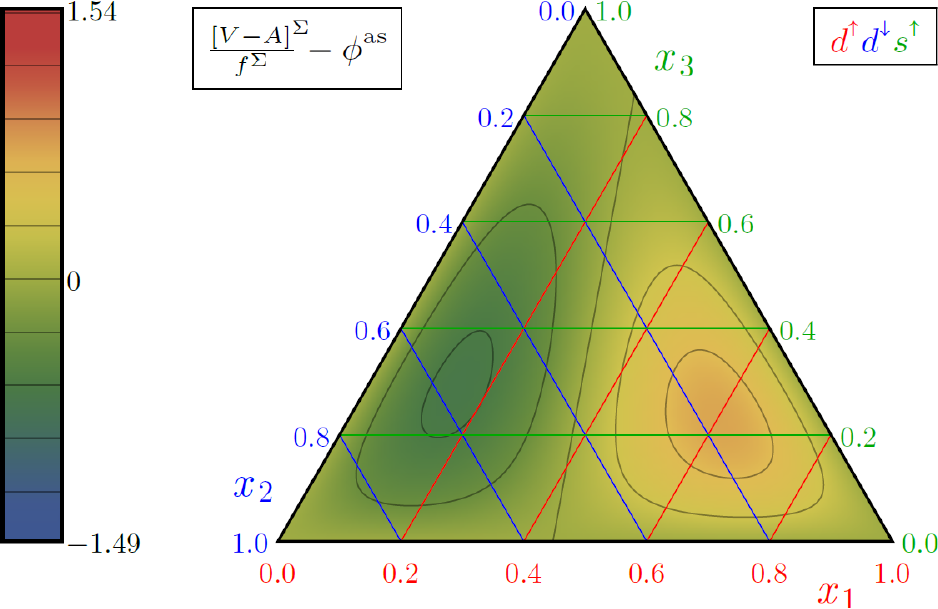} \\
\includegraphics[width=0.45\textwidth]{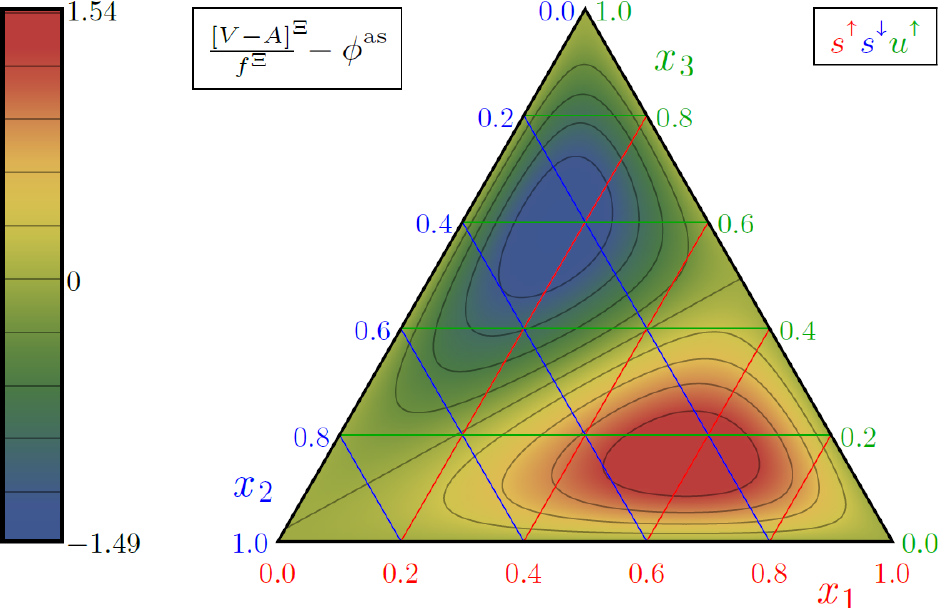} \\
\includegraphics[width=0.45\textwidth]{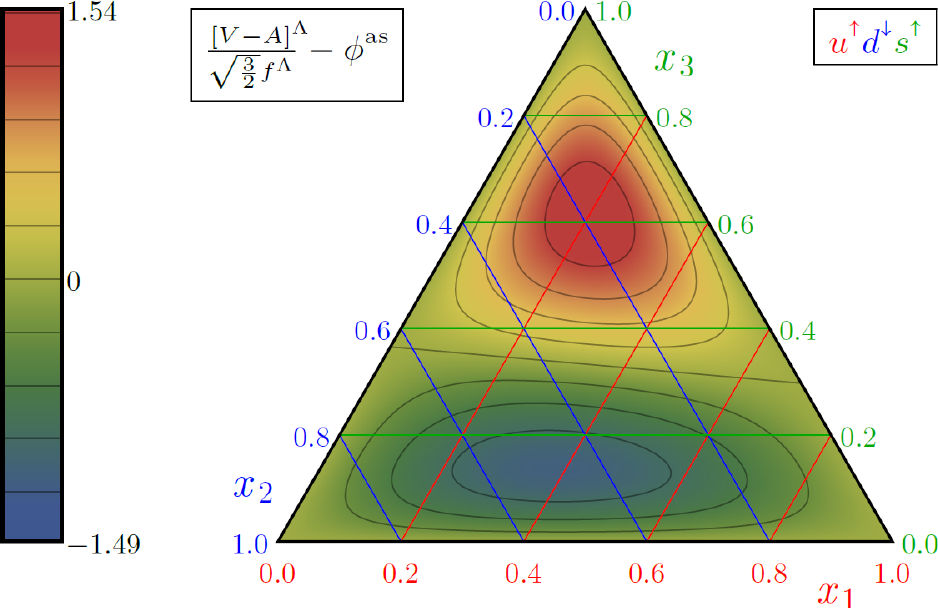}
\end{tabular}
\caption{\label{figure_barycentric} Barycentric plots of the deviations from the asymptotic shapes of the baryon DAs.}
\end{figure}

As schematically represented in Fig.\,\ref{figure_ensembles}, the available ensembles were generated along three different trajectories in the quark-mass plane. The combination of multiple quark-mass trajectories with a wide range of lattice spacings and volumes enabled a simultaneous extrapolation to physical masses, to infinite volume, and to the continuum by means of a global fit to all $40$\,ensembles.  To accomplish that, the following strategy was adopted.
In the continuum limit, the mass dependence calculated using one-loop BChPT in Ref.\,\cite{Wein:2015oqa} was used (amended by the leading finite-volume behaviour \cite{Bali:2019ecy}; see also Ref.\,\cite{Wein:2011ix}) including the correct flavour symmetry breaking patterns.
Discretisation effects were then parametrised\footnote{Since the three-quark operators used were not order-$a$ improved, the leading terms were treated as linear in $a$.}, allowing for mass-dependence.
With the quark-mass and continuum extrapolations treated thus, one can show that the expected flavour symmetry breaking patterns, which are broken at finite lattice spacing \cite{Bali:2015ykx}, are recovered in the continuum.
Considering the intricate interplay between quark-mass and discretisation effects, resolving both dependences simultaneously is pivotal.

Fig.\,\ref{figure_mP2} displays the quark mass dependence. In particular, in the left panels, one can observe the fan-like structure, which is typical of the transition from a flavour symmetric world (where the baryons form exact flavour multiplets) to the physical point. It is notable that the SU$_{\rm f}$(3) breaking in octet baryon DAs turns out to be very large. Some shape parameters even assume opposite signs for different baryons at the physical point. The effect of SU$_{\rm f}$(3) breaking on the leading-twist normalisation constants can be as large as $80$\%, for instance \mbox{$(f_T^\Xi - f^N)/f^N\approx 0.78$}, and is much stronger than estimated in QCD sum rule calculations \cite{Chernyak:1987nu} where $\lesssim 10\%$ SU$_{\rm f}$(3) breaking is found. For the shape parameters, such effects are even more pronounced.

The approach to the continuum limit is depicted in Fig.~\ref{figure_a}. While discretisation effects are important for the normalisation constants (up to $\sim 20\%$), they can have a dramatic impact on the moments: between $a=0.086$ fm (the coarsest lattice spacing) and $a=0$, there can be huge variations of the moments, which even affect the sign, \emph{e.g}.\ $\varphi_{10}^\Sigma$. This nicely demonstrates both the vital importance of the continuum limit for hadron structure observables and that a relatively wide range of lattice spacings is necessary to obtain trustworthy results.

The numerical results and estimates for systematic uncertainties of the normalisation constants and the first moments can be found in Ref.\,\cite[Table\,2]{Bali:2019ecy}. The shape of the DAs can be visualised using barycentric plots.  Fig.\,\ref{figure_barycentric} shows the deviation from the asymptotic shape for the standard combination $[\VmA]^B$, which directly corresponds to the Fock state $f^\uparrow g^\downarrow h^\uparrow$.

Considering the nucleon, in agreement with earlier lattice studies \cite{Braun:2014wpa, Bali:2015ykx} and with results from the Faddeev wave function model \cite{Mezrag:2017znp, Mezrag:2018hkk}, Sec.\,\ref{CedricMezrag}, one can see that the ``leading'' $u^\uparrow$ quark, which has the same helicity as the nucleon, carries a larger momentum fraction. Historically, this statement has also been the main finding of the QCD sum rule approach \cite{Chernyak:1984bm, Chernyak:1987nu}. Assuming exact isospin symmetry, the spin-flavour structure of the nucleon light-cone wave function can be represented, schematically, as $[\VmA]^N u^\uparrow (u^\downarrow d^\uparrow - d^\downarrow u^\uparrow)$. In this picture, the result for $[\VmA]^N$ corresponds to a shift of the momentum distribution towards the $u^\uparrow$ quark, which carries the nucleon helicity, and there is some deviation from the approximate symmetry under $x_2\leftrightarrow x_3$. This symmetry could be interpreted as a scalar diquark structure for the remaining valence quarks, which is assumed in many models.  As seen in Fig.\,\ref{PlotPDAs}, that symmetry is undermined by the presence of axial-vector diquark correlations.

Continuing with inspection of Fig.\,\ref{figure_barycentric}, one can identify two competing patterns.
First, strange quarks carry an increased fraction of the momentum.
Second, in the $\lvert\uparrow\downarrow\uparrow\rangle$ state, the first quark has a larger momentum fraction than the second.  (More information on the $\lvert\uparrow\uparrow\downarrow\rangle$ state can be found in Ref.\,\cite{Bali:2019ecy}.)  Also in the $u^\uparrow d^\downarrow s^\uparrow$ spin orientation of the $\Lambda$-baryon, the maximum of the distribution is shifted towards the $s$-quark.

\begin{table}[!t]
\centering
\caption{Continuum results for the normalised first moments of the DAs $[\VmA]^B$ and~$T^{\BnoL}$ in the $\MSbar$~scheme at a scale $\zeta=\unit{2}{\giga\electronvolt}$, see Eqs.~\eqref{eq_not_momentum_fractions}. All uncertainties in the calculation have been added in quadrature. \label{tab_not_momentum_fractions}}
\begin{ruledtabular}
\begin{tabular}{c r@{\hspace{.5em}}l r@{\hspace{.5em}}l r@{\hspace{.5em}}l r@{\hspace{.5em}}l}
$B$ & \multicolumn{2}{c}{$N$} & \multicolumn{2}{c}{$\Sigma$} & \multicolumn{2}{c}{$\Xi$} & \multicolumn{2}{c}{$\Lambda$}\\
\hline
\mystrut$\langle x_1 \rangle^B$   & $u^\gooduparrow$   & $0.396_{-6}^{+7}$ & $d^\gooduparrow$   & $0.363_{-7}^{+4}$ & $s^\gooduparrow$   & $0.390_{-4}^{+4}$ & $u^\gooduparrow$   & $0.308_{-3}^{+3}$\\
\mystrut$\langle x_2 \rangle^B$   & $u^\gooddownarrow$ & $0.311_{-5}^{+5}$ & $d^\gooddownarrow$ & $0.309_{-5}^{+5}$ & $s^\gooddownarrow$ & $0.335_{-2}^{+2}$ & $d^\gooddownarrow$ & $0.300_{-7}^{+7}$\\
\mystrut$\langle x_3 \rangle^B$   & $d^\gooduparrow$   & $0.293_{-6}^{+5}$ & $s^\gooduparrow$   & $0.329_{-3}^{+6}$ & $u^\gooduparrow$   & $0.275_{-5}^{+5}$ & $s^\gooduparrow$   & $0.392_{-5}^{+5}$\\
\hline
\mystrut$\langle x_1 \rangle^B_T$ & $u^\gooduparrow$   & $0.344_{-2}^{+2}$ & $d^\gooduparrow$   & $0.327_{-2}^{+2}$ & $s^\gooduparrow$   & $0.354_{-5}^{+5}$ & \multicolumn{2}{c}{---} \\
\mystrut$\langle x_2 \rangle^B_T$ & $u^\gooduparrow$   & $0.344_{-2}^{+2}$ & $d^\gooduparrow$   & $0.327_{-2}^{+2}$ & $s^\gooduparrow$   & $0.354_{-5}^{+5}$ & \multicolumn{2}{c}{---} \\
\mystrut$\langle x_3 \rangle^B_T$ & $d^\gooddownarrow$ & $0.311_{-5}^{+5}$ & $s^\gooddownarrow$ & $0.345_{-3}^{+3}$ & $u^\gooddownarrow$ & $0.291_{-9}^{+9}$ & \multicolumn{2}{c}{---} \\
\end{tabular}
\end{ruledtabular}
\end{table}

To make these statements quantitative, one can consider normalised first moments of $[\VmA]^B$ and~$T^{\BnoL}$,
\begin{align}
\langle x_i \rangle^B &= \frac{1}{f^B}\int \![dx]\, x_i\, [\VmA]^B\, \,, \quad \langle x_i \rangle^{\BnoL}_T = \frac{1}{f_T^B} \int \![dx]\, x_i \, T^B\,,
\label{eq_not_momentum_fractions}
\end{align}
see also Ref.\,\cite[Eqs.\,(6.3)]{Bali:2015ykx}.  These are sometimes referred to as momentum fractions and interpreted as the portions of the hadron's total light-front momentum carried by the individual valence quarks. This notion is somewhat imprecise since the averaging is done with a DA instead of a squared wave function; furthermore, the interpretation as momentum fractions breaks down completely in the case of $T^\Lambda$, which has no asymptotic part. Overlooking such caveats, these objects are nevertheless interesting because they provide a simple quantitative measure for the relative deviations of a DA from the asymptotic case $\langle x_1 \rangle^\text{as}=\langle x_2 \rangle^\text{as}=\langle x_3 \rangle^\text{as}=1/3$.  The numerical results are summarised in Table~\ref{tab_not_momentum_fractions} and they clearly agree with the qualitative picture suggested by the above discussion of Fig.\,\ref{figure_barycentric}.

\section{Diquarks in Experiment and Phenomenology}
\label{sec:Experiment}

\subsection{Space-like Nucleon Form Factors}
\label{SLNFFS}

Nucleon structure investigations using high energy electron scattering have been a successful field of discoveries since 1955, with the determination of the proton size \cite{Hofstadter:1955ae}. The status of the current knowledge of nucleon electromagnetic form factors is reviewed in Refs.\,\cite{Punjabi:2015bba, Pacetti:2015iqa}. To a large extent, this success owes to the dominance of the one-photon exchange mechanism in electron scattering as proposed in the original theory \cite{Rosenbluth:1950yq}.

The most decisive studies of the partonic structure of the nucleon (and its excitations) could be performed when the dominant part of the wave function is a three-quark Fock state. This requires large momentum transfers, $Q^2$ larger than several GeV$^2$, where the contribution of the so-called pion-cloud is suppressed. In the early 1990s, the elastic scattering cross-section data sets at large $Q^2$ for the proton and the neutron were in agreement with the dipole fit, $G_D=(1+Q^2/[0.843\,\text{GeV}]^2)^{-2}$, see Ref.\,\cite{Bosted:1994tm}. Moreover, the Stanford Linear Accelerator Center (SLAC) experimental data \cite{Arnold:1986nq, Sill:1992qw} on the proton Dirac form factor $F_1^p$ at $Q^2$ above 10 GeV$^2$ were in fair agreement with the scaling prediction \cite{Lepage:1979za} based on perturbative QCD (pQCD), $F_1^p \propto Q^{-4}$, where $Q^2$ is the spacelike four-momentum transfer squared.

\begin{figure}[!t]
\begin{tabular}{ccc}
\includegraphics[clip, width=0.45\textwidth] {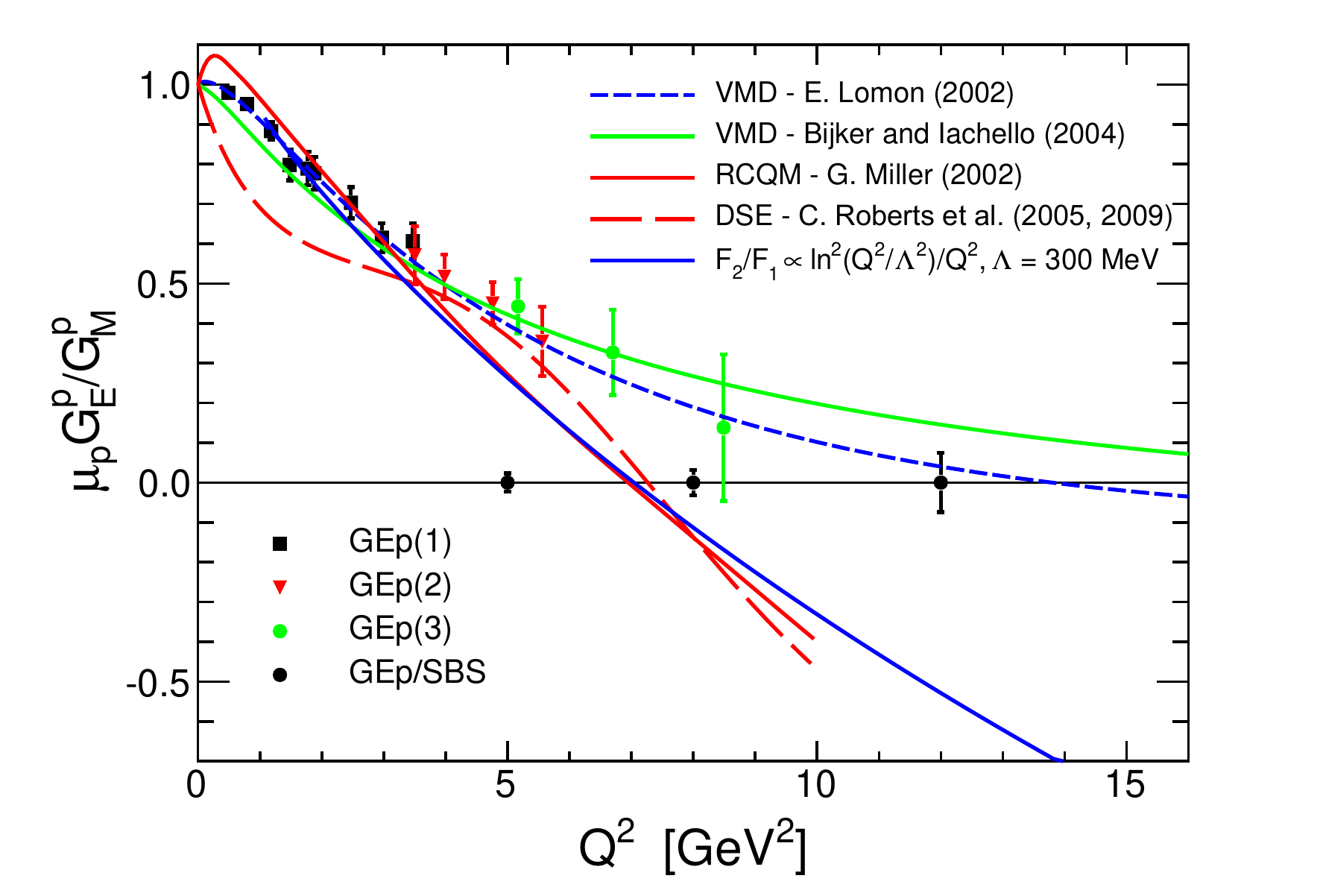} & &
\includegraphics[clip, width=0.45\textwidth, height=0.23\textheight] {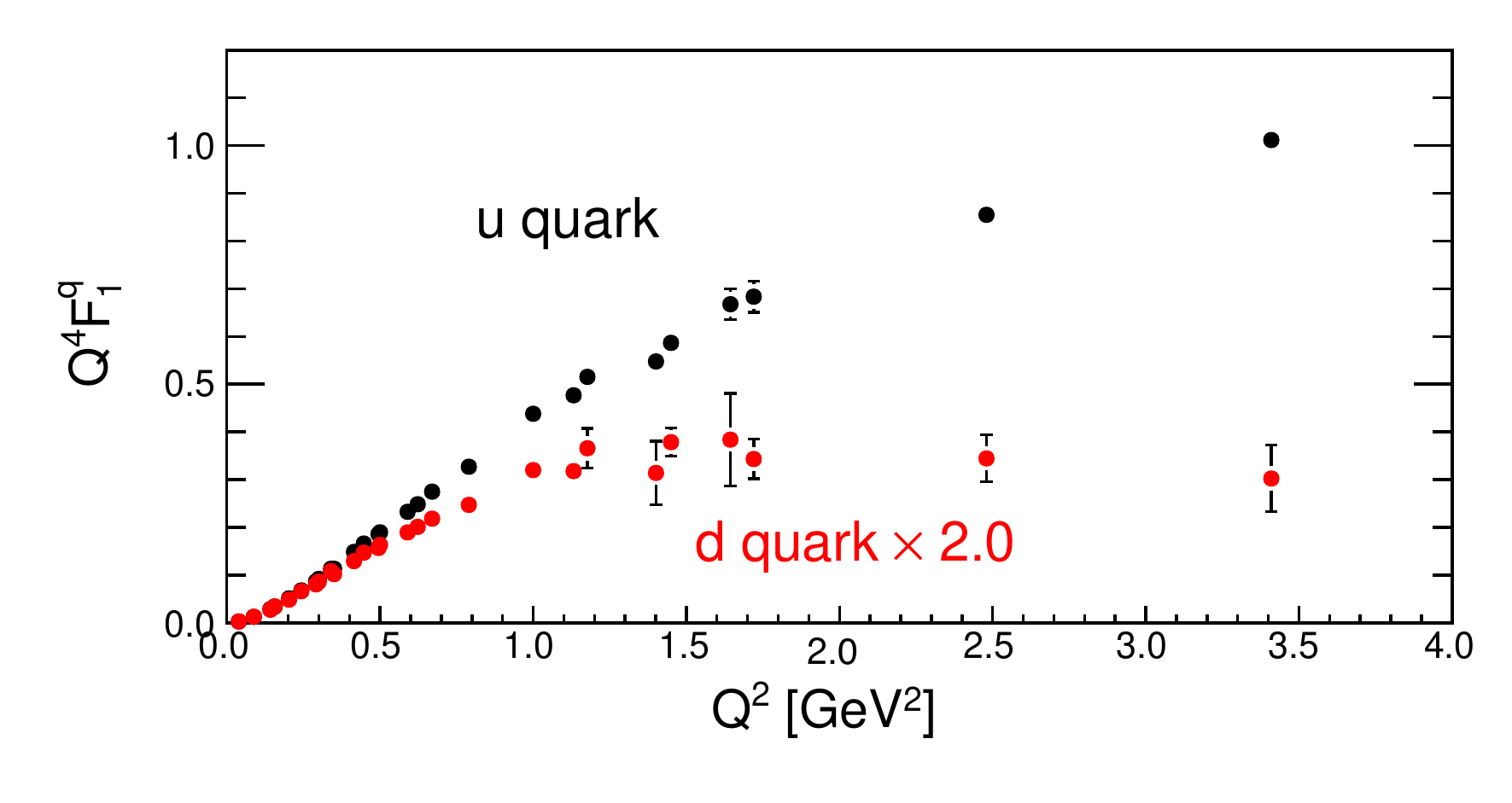}
\end{tabular}
\caption{\label{fig:GEp} {\it Left panel}. Existing data and projected data accuracy for the ratio $\mu _p\,G_E^p/G_M^p$. {\it Right panel}. Separated $d$- and $u$-quark contributions to the proton form factor $F_1^p$ from the measurement of $G_M^n/G_M^p$ (see more in the text).}
\end{figure}

A new era began with a precision measurement \cite{Perdrisat1989} that realised experimentally the double polarisation method suggested in Refs.\,\cite{Scofield:1959zz, Akhiezer:1968ek, Akhiezer:1974em, Arnold:1980zj}. This particular double polarisation method has large sensitivity to the typically small electric form factor owing to the interference nature of the corresponding double polarisation asymmetry. It is also less sensitive to two-photon exchange contributions, which are believed to complicate the Rosenbluth extraction of $[G_E^{p}]^2$. The experimental results from JLab \cite{Jones:1999rz, Gayou:2001qt, Gayou:2001qd, Punjabi:2005wq, Puckett:2010ac} are shown in Fig.\,\ref{fig:GEp} (left panel). The ratio of the proton's Pauli form factor, $F_2^p$, and the Dirac form factor, $F_1^p$, have been found to be in disagreement with the scaling law $F_2^p/F_1^p \propto 1/Q^2$ (which requires $G_E^p$ to be proportional to $G_M^p$ for large momentum transfer, \emph{i.e}.\ $G_E^p \approx G_M^p/\mu_p \approx G_D$, with $\mu_p=2.79$ the magnetic moment of the proton) suggested in Ref.\,\cite{Lepage:1979za}.

The experimental data on $\mu _p\,G_E^p/G_M^p$ revealed an unexpected, almost linear, decrease with $Q^2$, which also translates into a different $Q^2$ dependence of $F_1^p$ and $Q^2\,F_2^p$ for the proton. The violation of the scaling prediction has been attributed to quark orbital angular momentum inside the proton, leading to a logarithmic scaling.  This hypothesis provides a very efficient fit of the proton data over a wide range of the transferred momentum above $1\,\text{GeV}^2$ \cite{Belitsky:2002kj}.  Notably, however, it does not describe analogous neutron data \cite{Cloet:2008re}; hence, the proton success is likely accidental.

The measurement of the proton to neutron cross-section ratio in quasi-elastic nucleon knockout reactions off the deuteron was used in JLab's precision experiment to extract the neutron magnetic form factor for $Q^2$ up to $4\,\text{GeV}^2$ \cite{Lachniet:2008qf}.  Combined with the latest JLab experiment on the neutron electric form factor \cite{Riordan:2010id}, experimental data on all four nucleon electromagnetic (Sachs) form factors became available on a $Q^2$ domain anticipated to ensure three-quark dominance. Using this information and assuming SU(2)-isospin symmetry, the first flavour-decomposition analysis of the JLab data for the nucleon form factors was reported in Ref.\,\cite{Cates:2011pz}. The right panel of Fig.\,\ref{fig:GEp} reveals a large, unexpected reduction in the relative size of the $d$-quark contribution to the $F_1$ form factor. A similar result was found for the Pauli form factor but at larger photon momenta. This behaviour is predicted by a GPD-based analysis \cite{Diehl:2013xca} of the form factors and also in DSE studies \cite{Segovia:2014aza, Segovia:2015ufa, Cui:2020rmu}.

The flavour decomposition results of Dirac and Pauli form factors lead to two simple conclusions. The $u$-quark and $d$-quark contributions to the electric and magnetic form factors of the proton each have different $Q^2$ dependence; and the contribution of the $d$-quark to the $F_1^p$ form factor at $Q^2=3.4\,\text{GeV}^2$ is three times less than the contribution of the $u$-quarks, when already corrected for the number of quarks and their charge. The latter suggests that the probability for a proton to survive the absorption of a massive virtual photon is much higher when the photon interacts with a valence $u$-quark that occurs twice within a proton.
%
%
This may be an indication of a $u$-$u$ correlation -- correlations usually enhance high-momentum components and the interaction cross-section.  Similarly, the relatively weak $d$-quark contribution to $F_1^p$ might indicate a suppression of the $u$-$d$ correlation or a mutual cancellation between different types of $u$-$d$ correlations.  On the other hand, these features could simply express a preference in the proton wave function for $d$-quarks to be sequestered in a soft $[ud]$ correlation, as described in Ref.\,\cite{Segovia:2016zyc} and in the discussion of Fig.\,\ref{fig:DSE-Roper-F1F2}.

\begin{figure}[!t]
\includegraphics[width=0.55\textwidth] {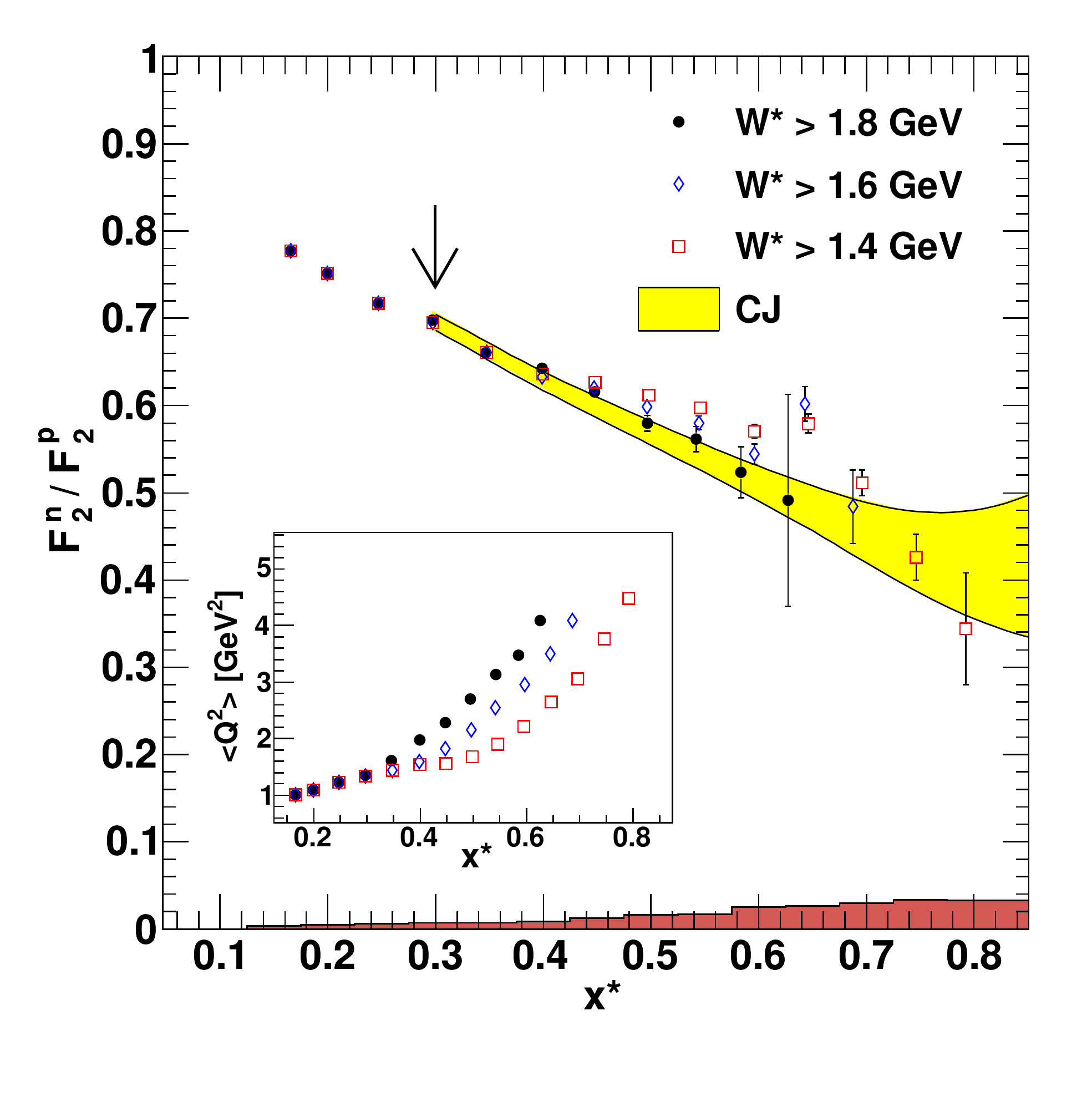}
\caption{\label{fig:Bonus}
The structure function ratio $F^n_2/F^p_2$ \cite{Baillie:2011za} versus $x^\ast$ for various lower limits on the invariant mass $W^\ast$, where $^\ast$ refers to their definition of the kinematic variables for bound nucleons. All data were collected at JLab with the BoNuS detector in Hall B at one beam energy, $5.262\,$GeV. The error bars are statistical, with the total (correlated and uncorrelated) systematic uncertainties indicated by the band along the abscissa. This band does not include the overall 3\% normalisation uncertainty or the 3\% spectator approximation uncertainty. The data are compared with a recent parametrisation \cite{Accardi:2011fa}.}
\end{figure}

An alternative approach to pin down the kinematics dependence of diquark correlations in nucleons is the flavour decomposition in the limit of large Bjorken $x$, \emph{i.e}.\ when one valence quark carries the full nucleon momentum. If the $F^n_2$ over $F^p_2$ structure function ratio shown in Fig.\,\ref{fig:Bonus} and reported in Ref.\,\cite{Baillie:2011za} was $\frac{1}{4}$ in the limit of $x\rightarrow$1, then it would indicate that only the $[ud]$-diquark survives in this limit.  For all other values, the ratio would reveal the nature and mixture of additional contributing diquark correlations \cite{Roberts:2013mja}, \emph{e.g}.\ an $x=1$ value of $\sim 0.4$ corresponds to a $\sim 30$\% contribution from axial-vector diquark correlations.  This connection is discussed further in Ref.\,\cite[Fig.\,8]{Chen:2020ijn}.

\subsection{Time-like Nucleon Form Factors}
\label{TLNFFS}

\begin{figure}[!t]
\includegraphics[width=0.55\textwidth] {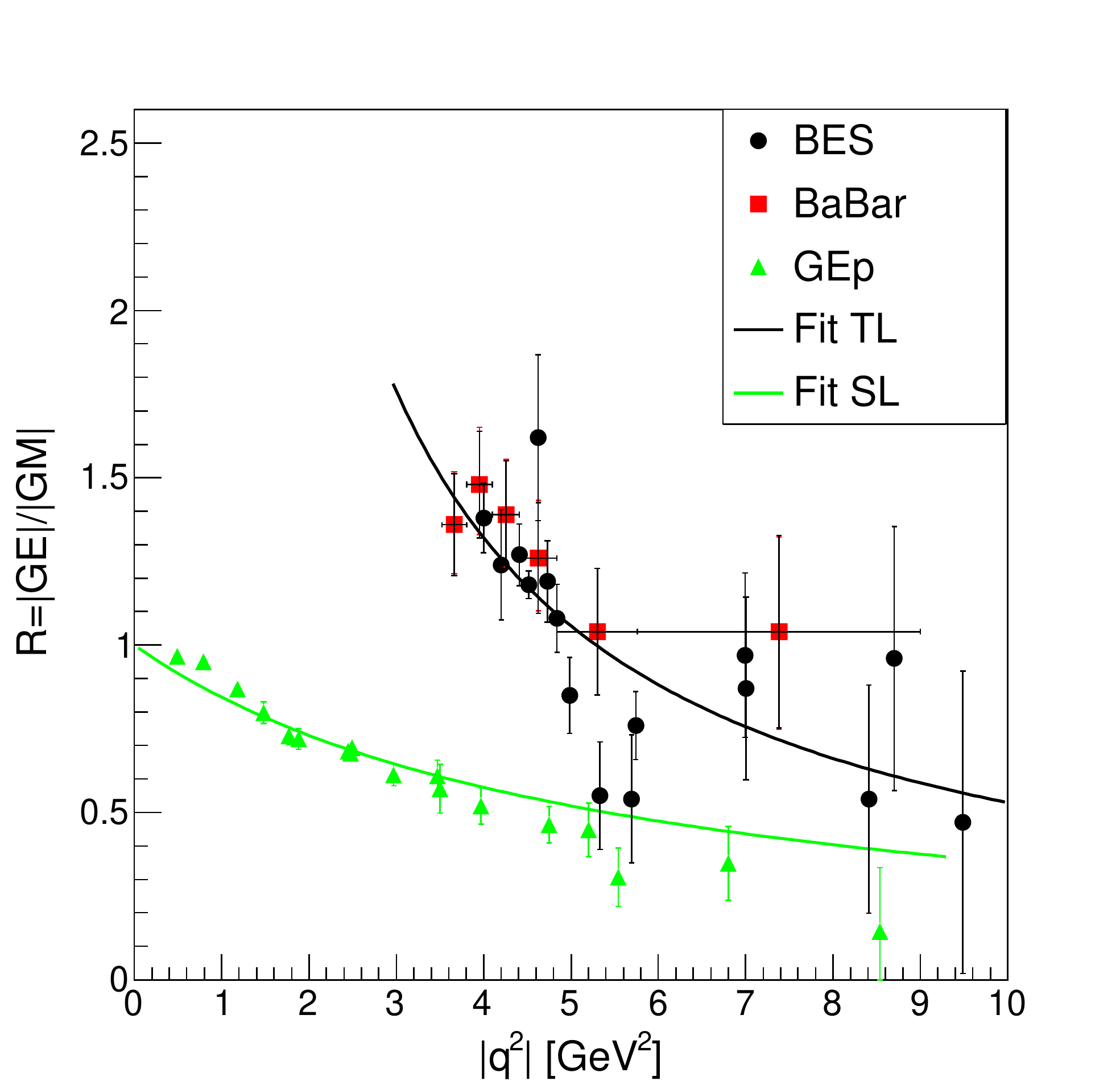}
\caption{\label{fig:RatioTLSL}
$|G_E|/|G_M|$ in the timelike region from BaBar (red squares), BESIII (black circles) and in the spacelike region from the GEp collaboration (green triangles). The solid lines are monopole-like fits.}
\end{figure}

Recent measurements in the time-like (TL) region from the BESIII Collaboration \cite{Ablikim:2019eau} at BEPCII led to the first individual determination of the electric and magnetic form factors (FFs) in the $q^2=-Q^2>0$ GeV$^2$ region. At the kinematic threshold, only one amplitude corresponding to the $S$-wave state characterises the reaction; therefore, $G_E=G_M$ or $R=G_E/G_M$=1 (when appropriately normalised to the magnetic dipole moment $\mu_p$). A measurement near threshold is very hard \cite{CMD-3:2018kql}. However, the existing data show that $|R|$ increases up to 1.4 at about 400\,MeV above threshold, see Fig.\,\ref{fig:RatioTLSL}, before it decreases, confirming findings from BABAR \cite{Lees:2013ebn, Lees:2013uta}.

A dip in the TL range $5-6$ GeV$^2$ could be the hint of a node in $|G_E|$. Note that a zero crossing of $|G_E|$ above $Q^2 = -q^2 = 9$ GeV$^2$ is not excluded by the space-like data, see left panel of Fig.\,\ref{fig:GEp} and the discussion of Fig.\,\ref{GEonGM}. The direct comparison of the measured TL and space-like (SL) results is illustrated in Fig.\,\ref{fig:RatioTLSL}, together with a simple fit according to the model of Ref.\,\cite{Kuraev:2011vq}. This model gives a prediction of FFs based on a coherent picture that connects $p\bar p$ annihilation into a lepton pair (or $\bar p p$ creation) in the TL region to $ep$ elastic scattering in the SL region and assumes that the system evolves through a diquark configuration.

The underlying assumption \cite{Kuraev:2011vq} is that the proton, usually described as an antisymmetric state of coloured quarks, is constituted from three valence quarks and a sea of gluons that are not held in the spatial center of the nucleon, which is electrically neutral. The strong gluonic field creates a gluonic condensate of clusters with a randomly oriented chromo-magnetic field.

In the most central region of the strong chromo-magnetic fields, the colour quantum number of quarks does not play a role, owing to stochastic averaging. When the colour quantum number of quarks with the same flavour vanishes, the $uu$ (or $dd$) quarks are repelled outwards due to the Pauli principle and hence away from the central region of the proton (or neutron). The third quark is attracted by one of the identical quarks and forms a compact diquark.

In the region of less intense gluonic fields, the colour state of quarks is restored; and the creation of a quark+diquark dipole system occurs when the attractive force exceeds the stochastic force of the gluon field. One can estimate, knowing the strength of the chromo-electric field, that the minimal distance where the quark+diquark picture appears is $r_0=0.22$ fm. The distribution in momentum space, as revealed by the Fourier transform, gives an additional monopole decrease for the electric form factor reflected by the form factor ratio.

A similar picture can be drawn in the annihilation region above the physical threshold, $q^2\ge 4 m_p^2$, where the vacuum state transfers all the energy released by the electron-positron annihilation to an S-wave state with total spin 1, composed of at least six massless valence current quarks, a set of gluons, and a sea of $q\bar q$ current quarks with total energy $q_0>2m_p$ and total orbital momentum unity. Such a state, created in a small spatial volume of the order $ 1/\sqrt{q^2}$, starts to expand and cool down.

In the first stages of cooling, the strong chromo-electric (chromo-magnetic) field leads to an effective loss of colour freedom of the quarks and antiquarks. As a result of Fermi statistics, the identical (colourless) quarks ($uu$ in the proton and $dd$ in the neutron) are repelled. The remaining quark (antiquark) of different flavour is attracted to one of the quarks at the surface, creating a compact diquark ($ud$ state). Then, the long range colour forces create a stable colourless state of proton and antiproton, using part of the initial energy to transform valence current-quarks and -antiquarks into constituent quarks/antiquarks. In analogy with charge screening in a plasma, the model leaves the quark counting (dipole-like) QCD-prediction for the magnetic form factor unchanged and suggests an additional suppression mechanism for the electric form factor, consistent with the data.

In closing this section it is worth noting that hadron induced reactions may also give evidence of diquark configurations, appearing as a deviation from pQCD scaling \cite{Kim:1988gg} that can be investigated in prospective programs at FAIR (Facility for Antiproton and Ion Research, Darmstadt) and NICA (Nuclotron-based Ion Collider fAcility, Dubna).


\subsection{Nucleon to Resonance Transition Form Factors}
\label{secNRTFFs}
Developing a unified description of electromagnetic elastic and transition form factors involving the nucleon and its resonances has become of great importance. On the theoretical side, it is via the $Q^2$-evolution of form factors that one gains access to the running of QCD's coupling and masses \cite{Cloet:2013gva, Chang:2013nia}. Moreover, QCD-kindred approaches that compute form factors at large photon virtualities are needed because the meson-cloud screens the dressed-quark core of all baryons at low momenta \cite{Tiator:2003uu, Kamano:2013iva, Mokeev:2015lda}.  Experimentally, substantial progress has been made in the extraction of transition electrocouplings, $g_{{\rm v}NN^\ast}$, from meson electroproduction data, obtained primarily with the CLAS detector at JLab \cite{Aznauryan:2011qj, Mokeev:2018zxt, Isupov:2017lnd, Mokeev:2012vsa, Markov:2019fjy}. The electrocouplings of all low-lying $N^\ast$ have been determined via independent analyses of $\pi^+ n$, $\pi^0p$ and $\pi^+ \pi^- p$ exclusive channels \cite{Aznauryan:2009mx, Mokeev:2012vsa, Mokeev:2015lda, Tanabashi:2018oca}; and preliminary results for the $g_{{\rm v}NN^\ast}$ of some high-lying $N^\ast$ states, with masses below $1.8\,{\rm GeV}$, have also been obtained from CLAS meson electroproduction data~\cite{Aznauryan:2012ba, Mokeev:2018zxt}. Complete, up-to-date information on the $Q^2$ evolution of $g_{{\rm v}NN^\ast}$ electro-couplings at $Q^2<6.0\,\text{GeV}^2$ for most resonances in the mass range up to $1.8\,\text{GeV}$ from analyses of exclusive meson electro-production with CLAS can be found in Ref.\,\cite{Blin:2019fre}.

During the next decade, CLAS\,12 will deliver resonance electroproduction data up to $Q^2 \approx 12\,$GeV$^2$~\cite{Mokeev:2018zxt, Burkert:2018nvj, Burkert:2019opk} and thereby empirical information which can address a wide range of issues that are critical to understanding strong interactions, \emph{e.g}.: is there an environment sensitivity of DCSB; and are quark+quark correlations an essential element in the structure of all baryons?  Existing experiment-theory feedback suggests that there is no environment sensitivity for the $N(940)$, $N(1440)$, $\Delta(1232)$ and $\Delta(1600)$ baryons: DCSB in these systems is expressed in ways that can readily be predicted once its manifestation is understood in the pion, and this includes the generation of diquark correlations with the same character in each of these baryons. Resonances in other channels, however, probably contain additional diquark correlations with different quantum numbers (Sec.\,\ref{sec:DSEs}), and can potentially be influenced in new ways by meson-baryon final state interactions (MB\,FSIs). Therefore, these channels, and higher excitations, open new windows on nonperturbative QCD and its emergent phenomena whose vistas must be explored and mapped if the most difficult part of the Standard Model is finally to be solved.

\begin{figure*}[!t]
\centering
\includegraphics[width=0.45\textwidth]{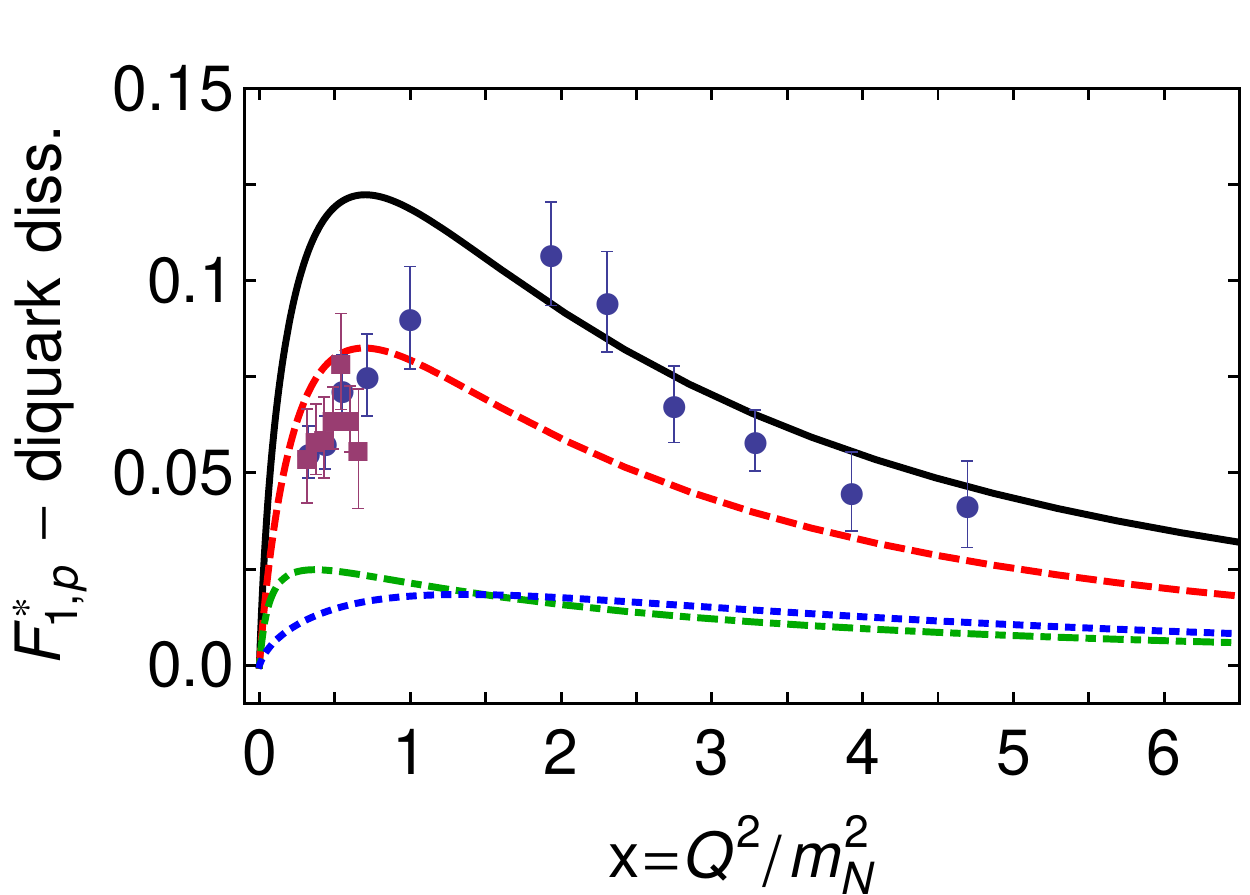}
\hspace*{0.40cm}
\includegraphics[width=0.45\textwidth]{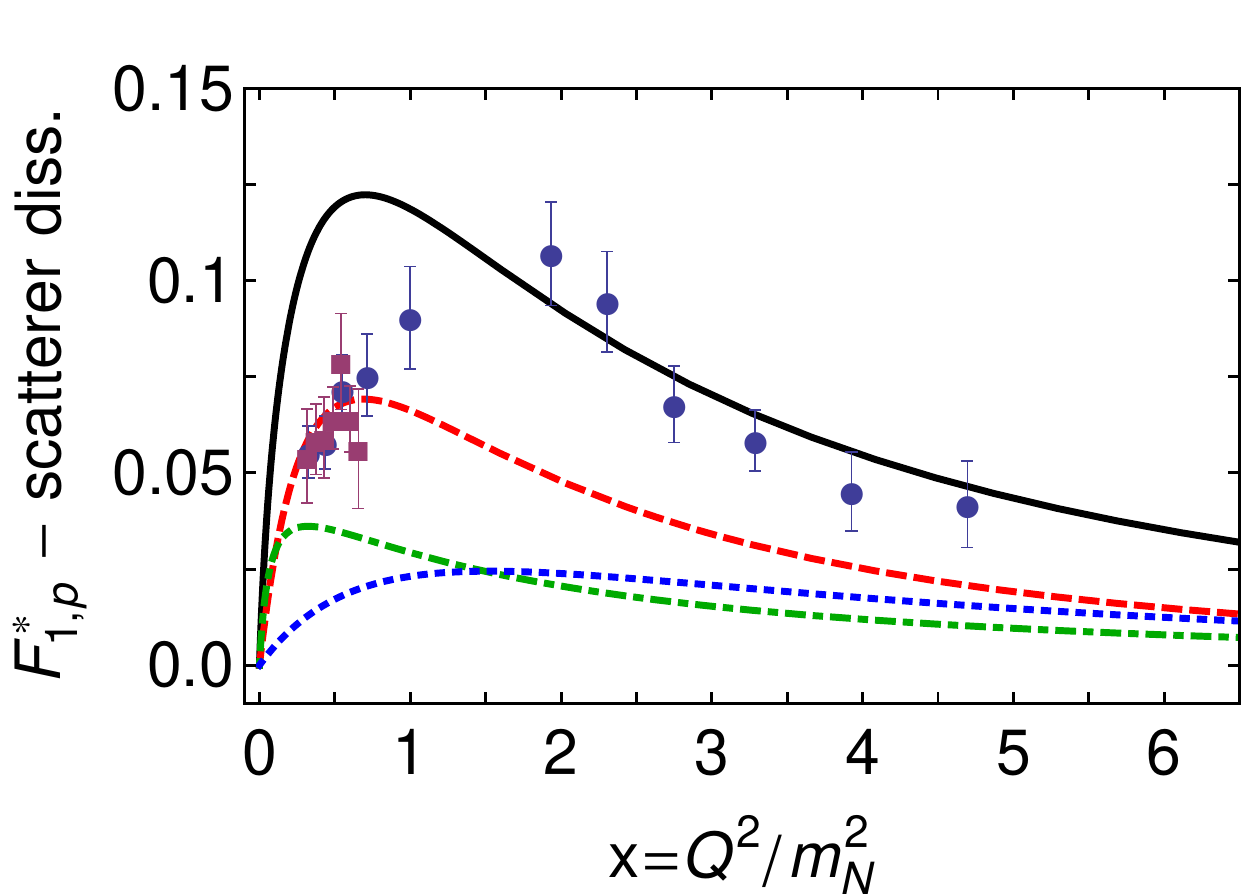} \\
\includegraphics[width=0.45\textwidth]{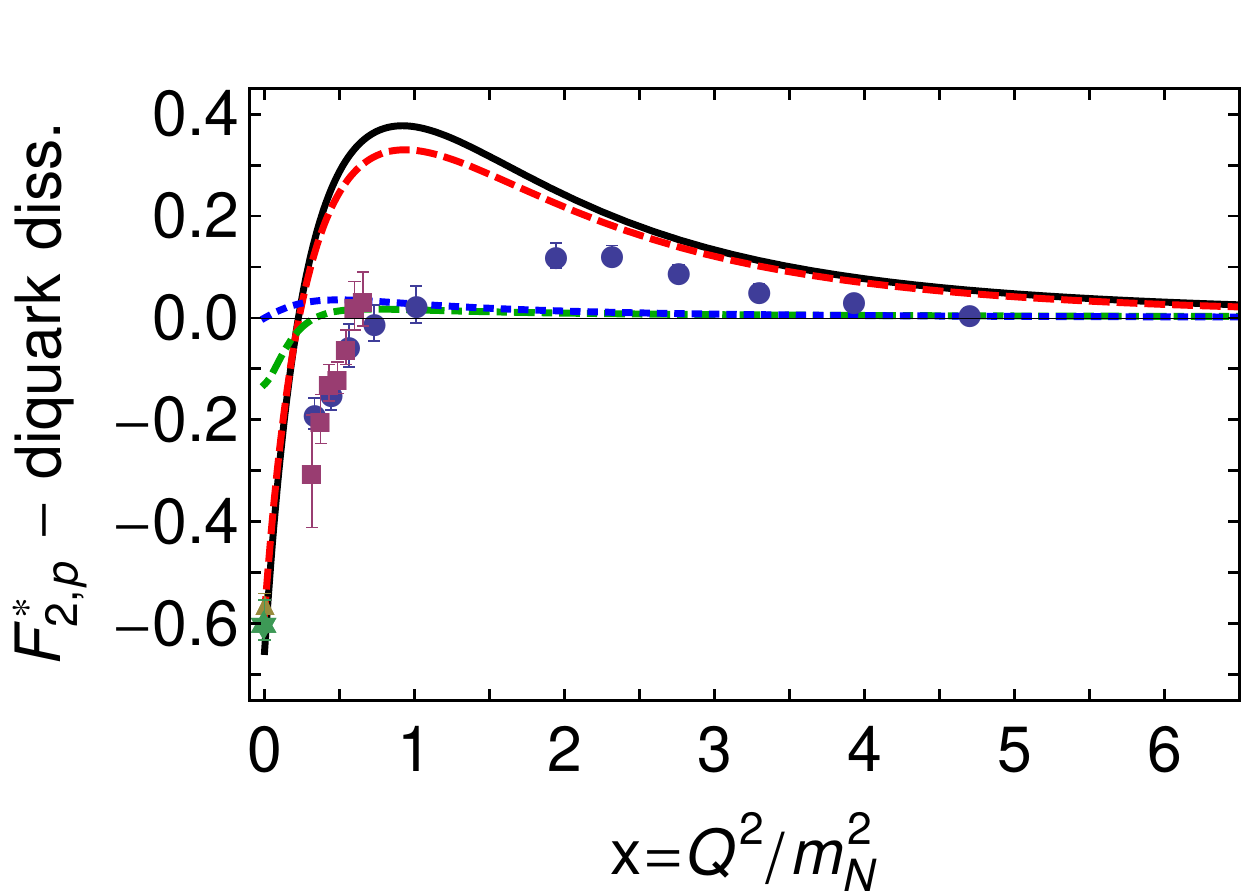}
\hspace*{0.40cm}
\includegraphics[width=0.45\textwidth]{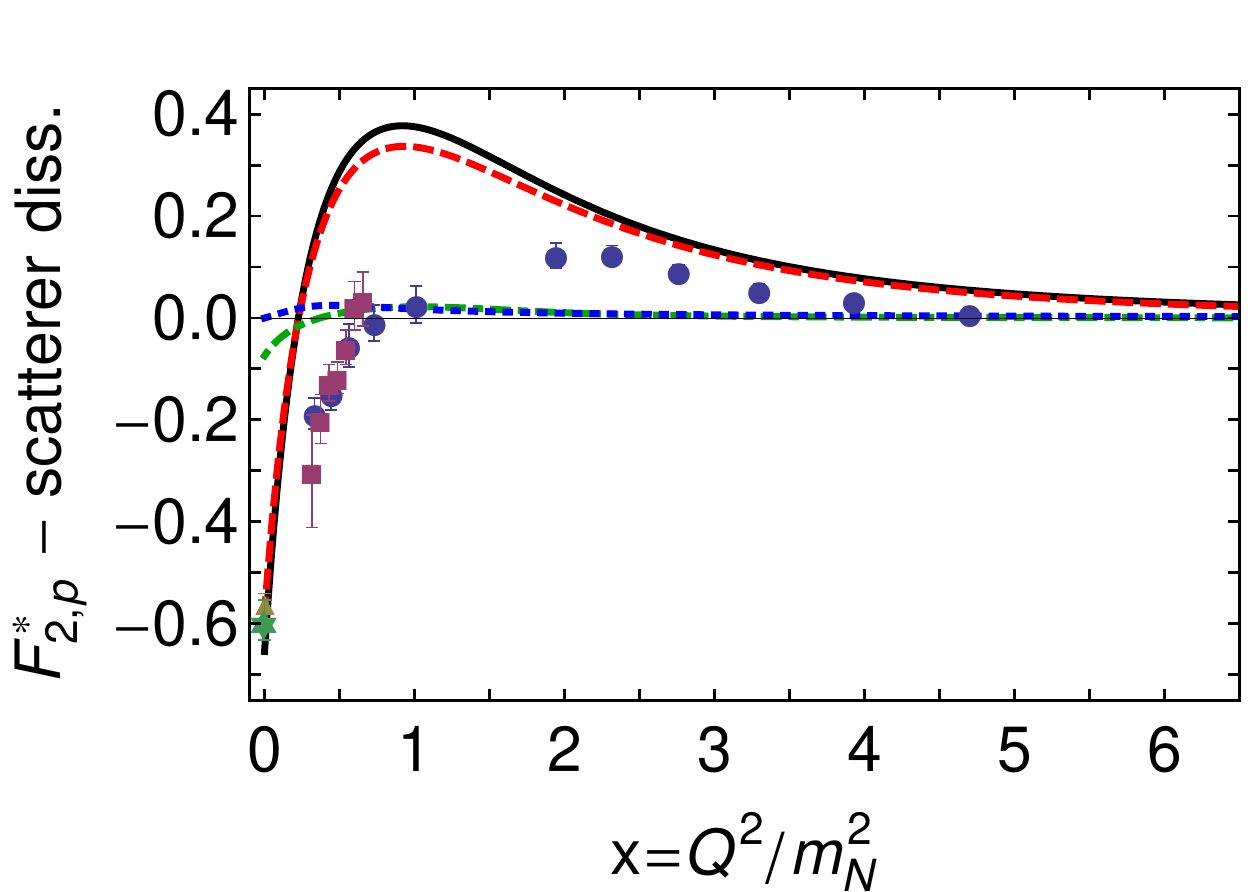}
\caption{Solid black curves are full DSE computations \cite{Chen:2018nsg} of the Dirac $F_1^\ast$ (top) and Pauli $F_2^\ast$ (bottom) proton to Roper transition form factors.  JLab data, circles (blue) \cite{Aznauryan:2009mx}, squares (purple) \cite{Mokeev:2012vsa, Mokeev:2015lda}, triangle (gold) \cite{Dugger:2009pn}, and particle data group (PDG) value star (green) \cite{Tanabashi:2018oca}. Left panels show the diquark breakdown: dashed red -- scalar diquark in both nucleon and Roper; dot-dashed green -- pseudovector diquark in both nucleon and Roper; and dotted blue -- scalar diquark in nucleon and pseudovector diquark in Roper.  Right panels show the scatterer breakdown: red dashed -- photon strikes an uncorrelated dressed quark; dot-dashed green -- photon strikes a diquark; and dotted blue -- diquark breakup contributions, including photon striking a dressed-quark in flight between diquarks.
\label{fig:DSE-Roper-F1F2}}
\end{figure*}

The Dirac and Pauli form factors of the $\gamma^{\ast}p\to R^+$ transition, where $R^+$ is the positively-charged Roper resonance, are displayed in Fig.\,\ref{fig:DSE-Roper-F1F2}. The results obtained using QCD-based propagators and vertices agree with the data on $x\gtrsim 2$ \cite{Segovia:2015hra, Segovia:2016zyc, Chen:2018nsg}. The disagreement between the QCD-kindred result and data on $x\lesssim 2$ owes to meson-cloud contributions, which are expected to be important on this domain \cite{Suzuki:2009nj, Segovia:2015hra, Roberts:2016dnb, Aznauryan:2016wwm, Burkert:2017djo}.

The anatomy of the $\gamma\,p\to R^+$ Dirac transition form factor is revealed in the upper panels of Fig.~\ref{fig:DSE-Roper-F1F2}. Plainly, this component of the transition proceeds primarily through a photon striking a bystander dressed quark that is partnered by a scalar-diquark: $[ud]$, with lesser but non-negligible contributions from all other processes. In exhibiting these features, $F_{1,p}^{\ast}$ shows marked qualitative similarities to the proton's elastic Dirac form factor~\cite{Cloet:2008re}. The $\gamma\,p\to R^+$ Pauli transition form factor is dissected in the lower panels of Fig.~\ref{fig:DSE-Roper-F1F2}. In this case, a single contribution is overwhelmingly important, \emph{viz}.\ photon strikes a bystander dressed-quark in association with $[ud]$ in the proton and $R^+$. No other diagram makes a significant contribution.

In hindsight, given that the diquark content of the proton and $R^+$ are almost identical, with the $\psi_0 \sim u+[ud]$ component contributing roughly 60\% of the charge of both systems, the qualitative similarity between the proton elastic and proton-Roper transition form factors is not surprising \cite{Segovia:2015hra, Segovia:2016zyc, Chen:2018nsg}.

\begin{figure*}[!t]
\begin{center}
\begin{tabular}{ll}
\includegraphics[clip, width=0.40\textwidth]{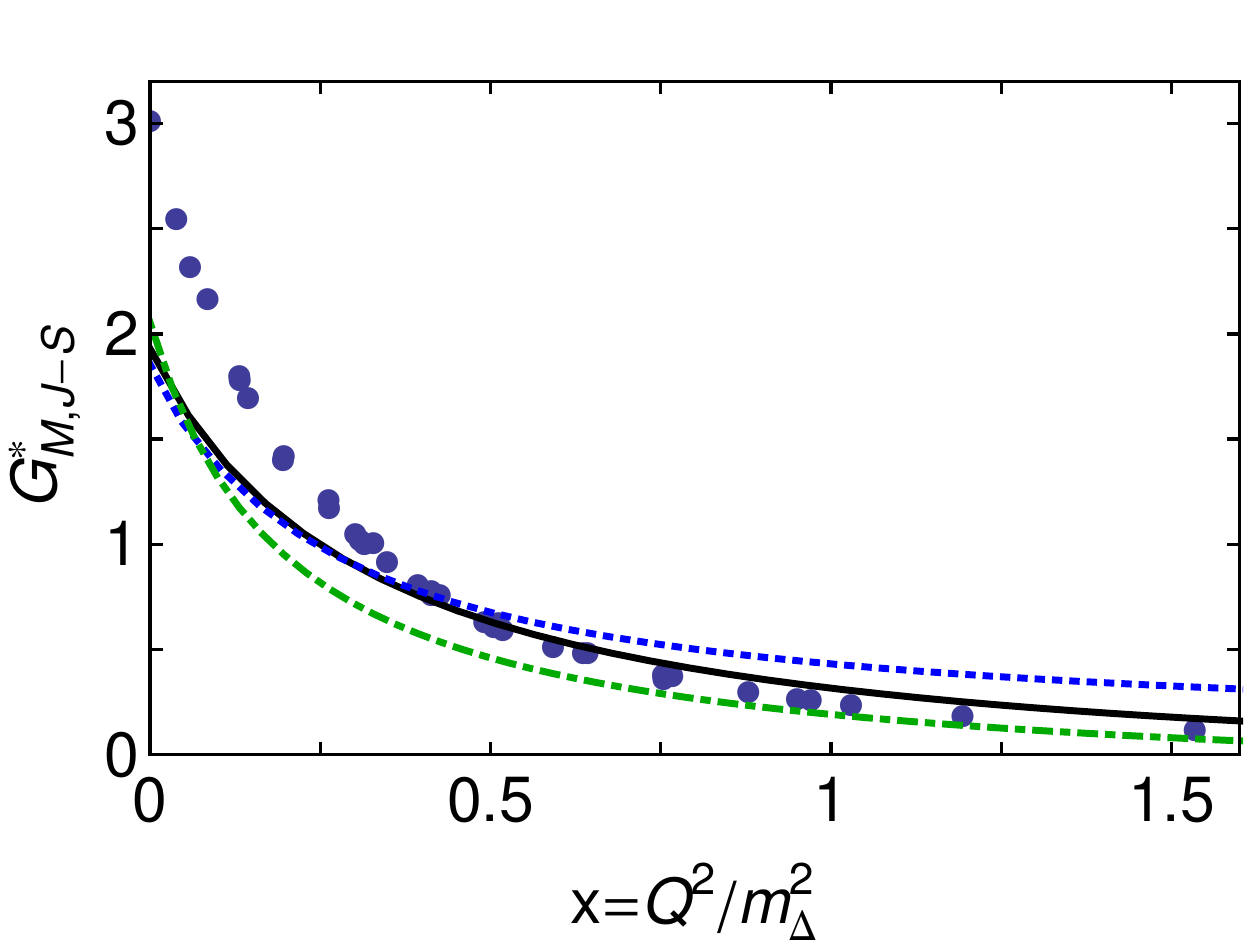}
&
\hspace*{-0.20cm}
\includegraphics[clip, height=0.22\textheight, width=0.45\textwidth]{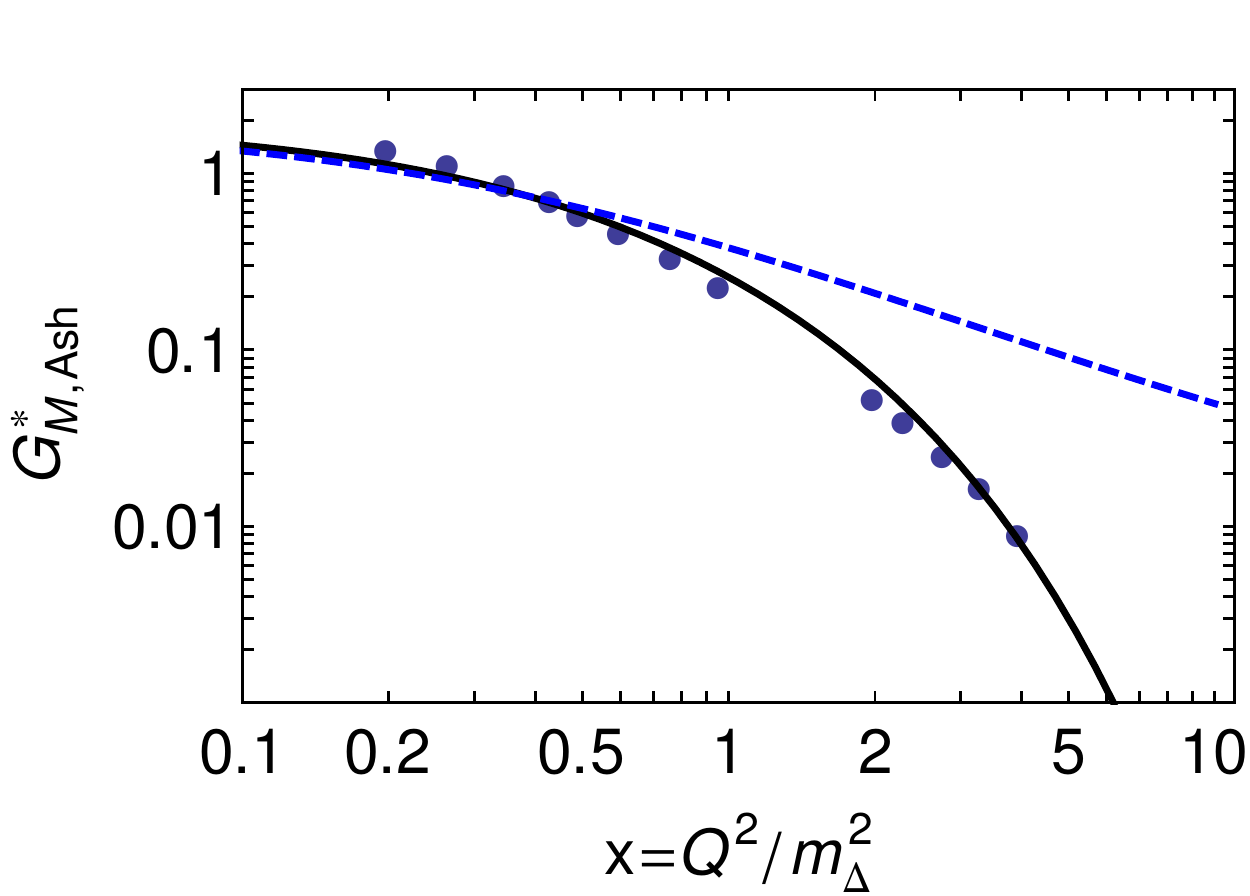} \\[-2ex]
\hspace*{-0.30cm}
\includegraphics[clip, width=0.42\textwidth]{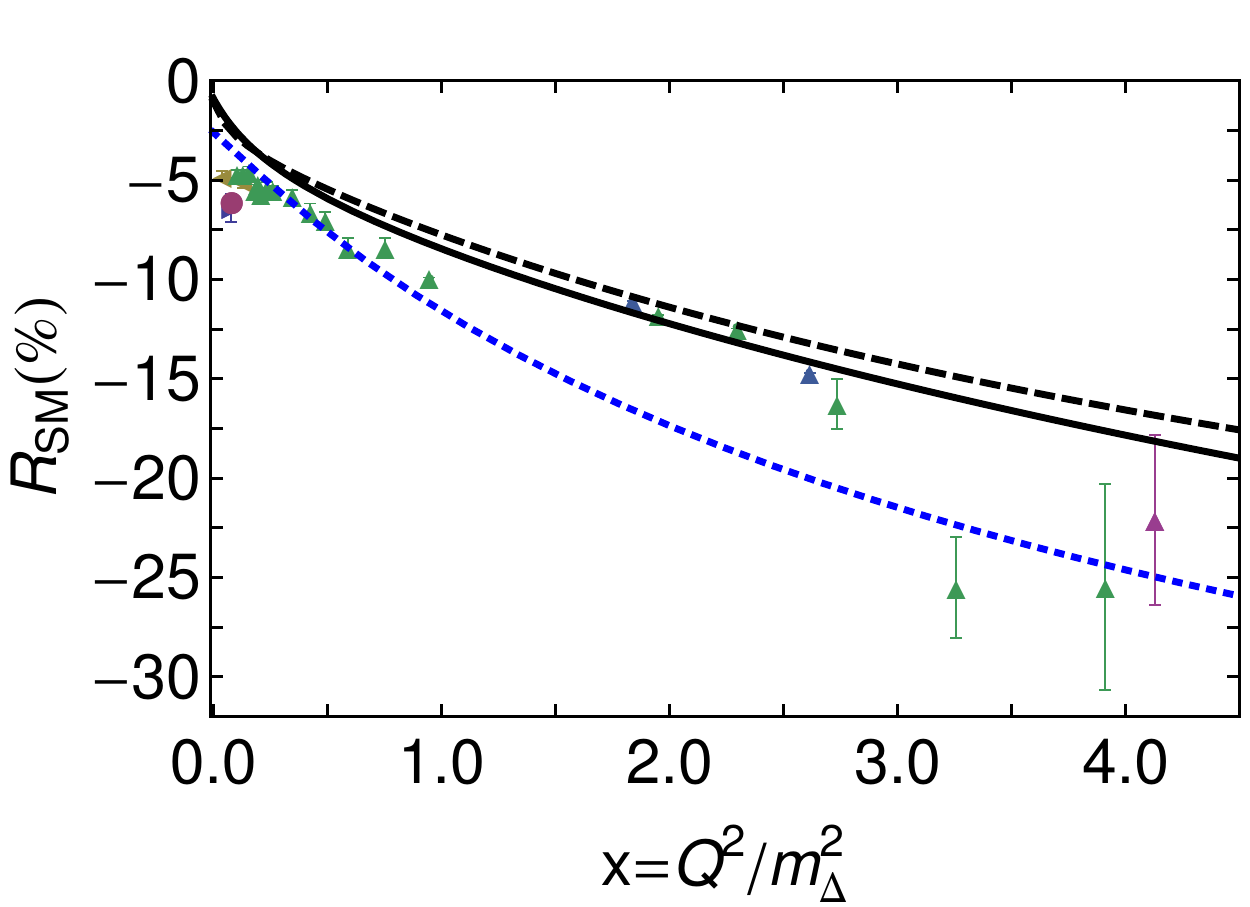}
&
\hspace*{+0.20cm}
\includegraphics[clip, width=0.42\textwidth]{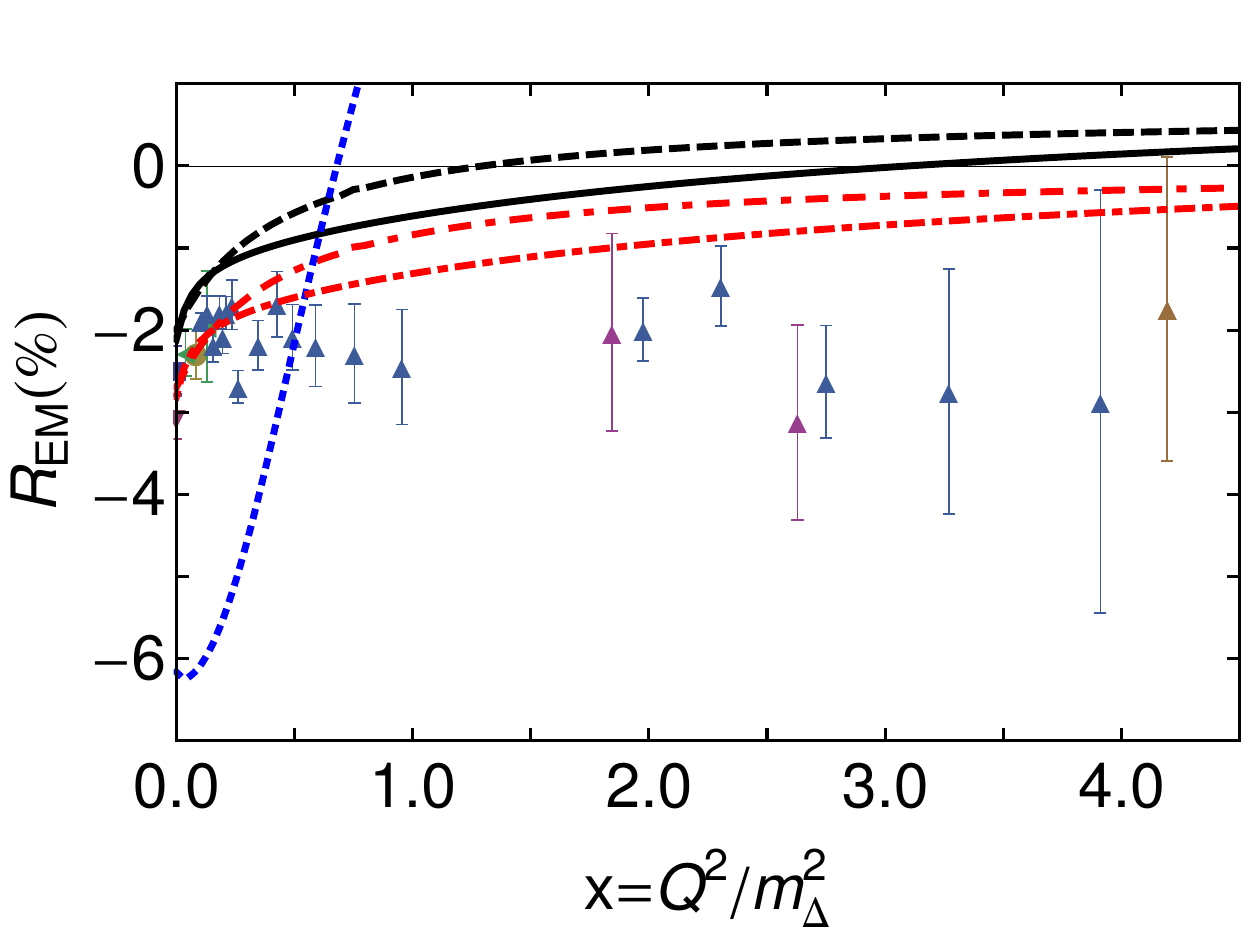}
\end{tabular}
\caption{\label{fig:NucDel}
\emph{Upper-left panel} -- $G_{M,J-S}^{\ast}$ result obtained with QCD-kindred interaction (solid, black) and with contact-interaction (SCI) (dotted, blue).  The green dot-dashed curve is the dressed-quark core contribution inferred using the dynamical meson-exchange model in Ref.\,\protect\cite{JuliaDiaz:2006xt}.
\emph{Upper-right panel} -- $G_{M,Ash}^{\ast}$ result obtained with QCD-kindred interaction (solid, black) and with SCI (dotted, blue).
\emph{Lower-left panel} -- $R_{SM}$ prediction of QCD-kindred kernel including dressed-quark anomalous magnetic moment (DqAMM) (black, solid), not including DqAMM (black, dashed), and SCI result (dotted, blue).
\emph{Lower-right panel} -- $R_{EM}$ prediction obtained with QCD-kindred framework (solid, black); same input but without DqAMM (dashed, black).
The following results are renormalised (by a factor of $1.34$) to agree with experiment at $x=0$: dot-dashed, red - zero at $x\approx 14$; and dot-dash-dashed, red - zero at $x\approx 6$).  The SCI result is the dotted, blue curve.
All data are from references listed in Ref.\,\protect\cite{Segovia:2014aza}.
}
\vspace*{-0.40cm}
\end{center}
\end{figure*}

Figure\,\ref{fig:NucDel} displays the transition form factors that characterise the $\gamma^\ast N(940)\to \Delta(1232)$ reaction \cite{Segovia:2013rca, Segovia:2014aza}. The upper-left panel shows the magnetic transition form factor in the Jones-Scadron convention \cite{Jones:1972ky}. DSE results within both SCI and QCD-kindred frameworks agree with the data on $x\gtrsim 0.4$. On the other hand, both curves disagree markedly with the data at infrared momenta. This mismatch owes to the fact that DSE computations ignore meson-cloud effects, an observation confirmed by the similarity between the DSE curves and the bare result determined using the dynamical meson-exchange model in Ref.\,\cite{JuliaDiaz:2006xt}.

The upper-right panel of Fig.\,\ref{fig:NucDel} shows the $\gamma^\ast N(940)\to \Delta(1232)$ magnetic transition form factor in the Ash convention \cite{Ash1967165}, which is traditionally adopted for the presentation of experimental results. One can see that the normalised QCD-kindred curve is in fair agreement with the data, indicating that the Ash form factor falls faster than a dipole for two main reasons: (\emph{i}) meson-cloud effects provide up-to $35\%$ of the form factor for $x \lesssim 2$; and (\emph{ii}) the additional kinematic factor $\sim 1/\sqrt{Q^2}$ that connects the Ash and Jones-Scadron conventions provides material damping for $x\gtrsim 2$ (see Ref.\,\cite{Segovia:2014aza} for additional details).

The lower-left panel of Fig.\,\ref{fig:NucDel} displays the $\gamma^\ast N(940)\to \Delta(1232)$ Coulomb quadrupole ratio, $R_{\rm SM}$. The results computed using either the QCD-kindred or the SCI formalism are broadly consistent with available data. This shows that even a contact-interaction, judiciously employed, can produce correlations between dressed-quarks within Faddeev wave-functions and related features in the current that are comparable in size with those observed empirically. Moreover, suppressing the dressed-quark anomalous magnetic moment (DqAMM) \cite{Chang:2010hb} in the transition current has little impact. These remarks highlight that $R_{SM}$ is not particularly sensitive to details of the Faddeev kernel and transition current.

In contrast, the lower-right panel in Fig.\,\ref{fig:NucDel} shows that $R_{\rm EM}$, the $\gamma^\ast N(940)\to \Delta(1232)$ electric quadrupole ratio, is a particularly sensitive measure of diquark and orbital angular momentum correlations. The SCI result is negative at low photon virtualities, it crosses zero at an experimentally accessible momentum transfer and then increases with $x$ in order to reach the helicity-conservation limit \cite{Carlson:1985mm}. On the other hand, four variants of the QCD-kindred result are presented.  They differ primarily in the location of the zero that is a feature of this ratio in all cases that have been considered. The inclusion of a DqAMM shifts the zero to a larger value of $x$. Given the uniformly small value of this ratio and its sensitivity to the DqAMM, it appears that MB\,FSIs must play a large role on the entire momentum domain that is currently accessible to experiment.

\begin{figure*}[!t]
\includegraphics[clip, width=0.32\textwidth]{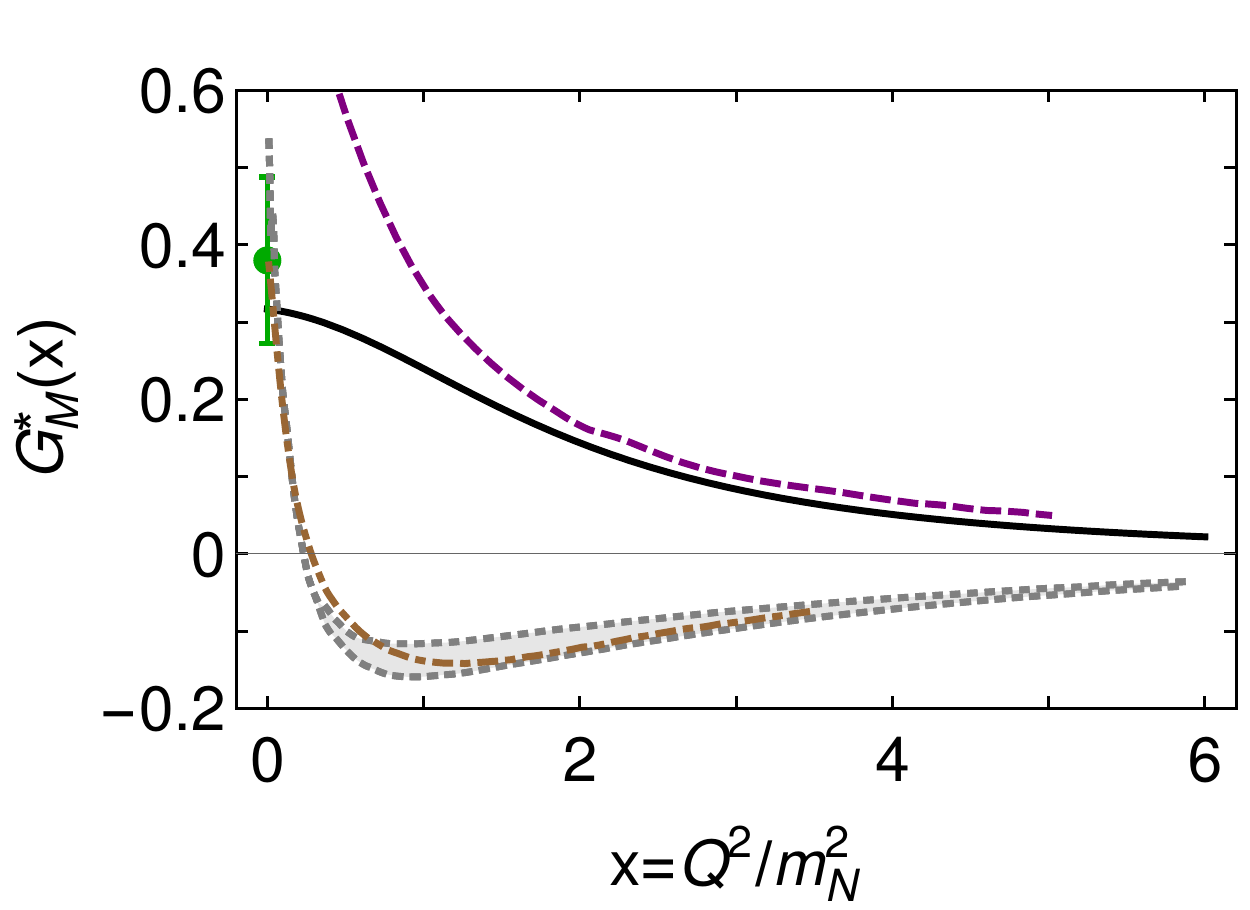}
\includegraphics[clip, width=0.32\textwidth]{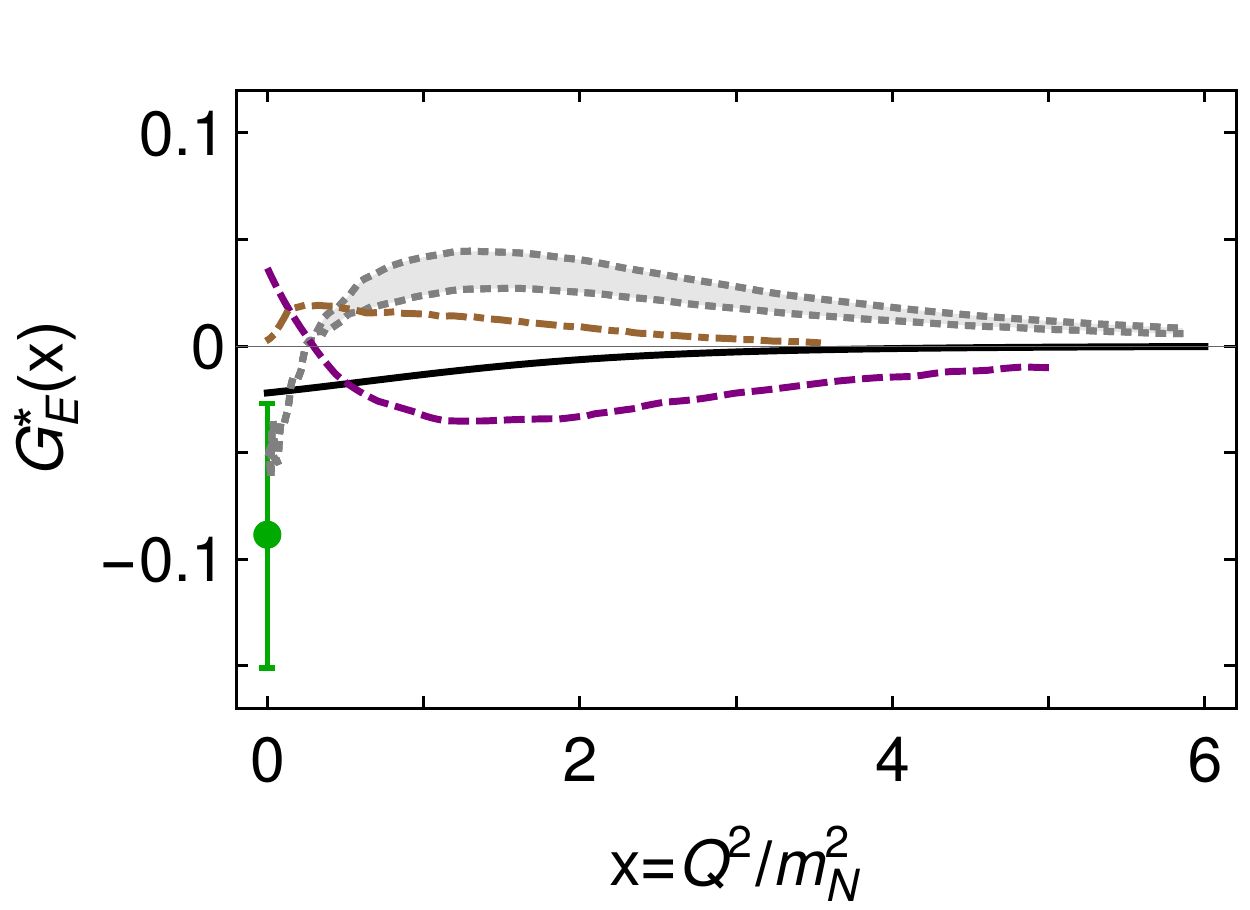}
\includegraphics[clip, width=0.32\textwidth]{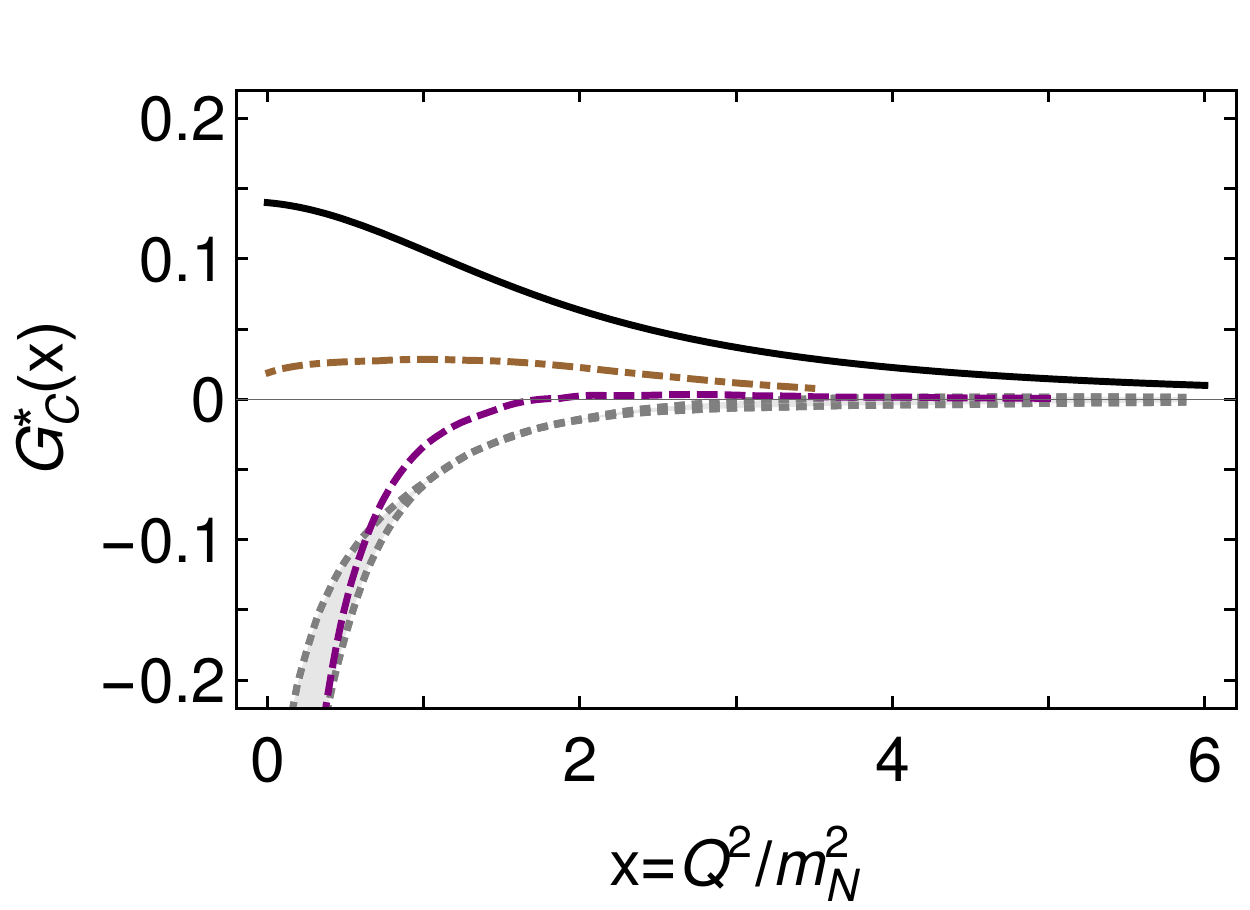}
\\[-2ex]
\includegraphics[clip, width=0.32\textwidth]{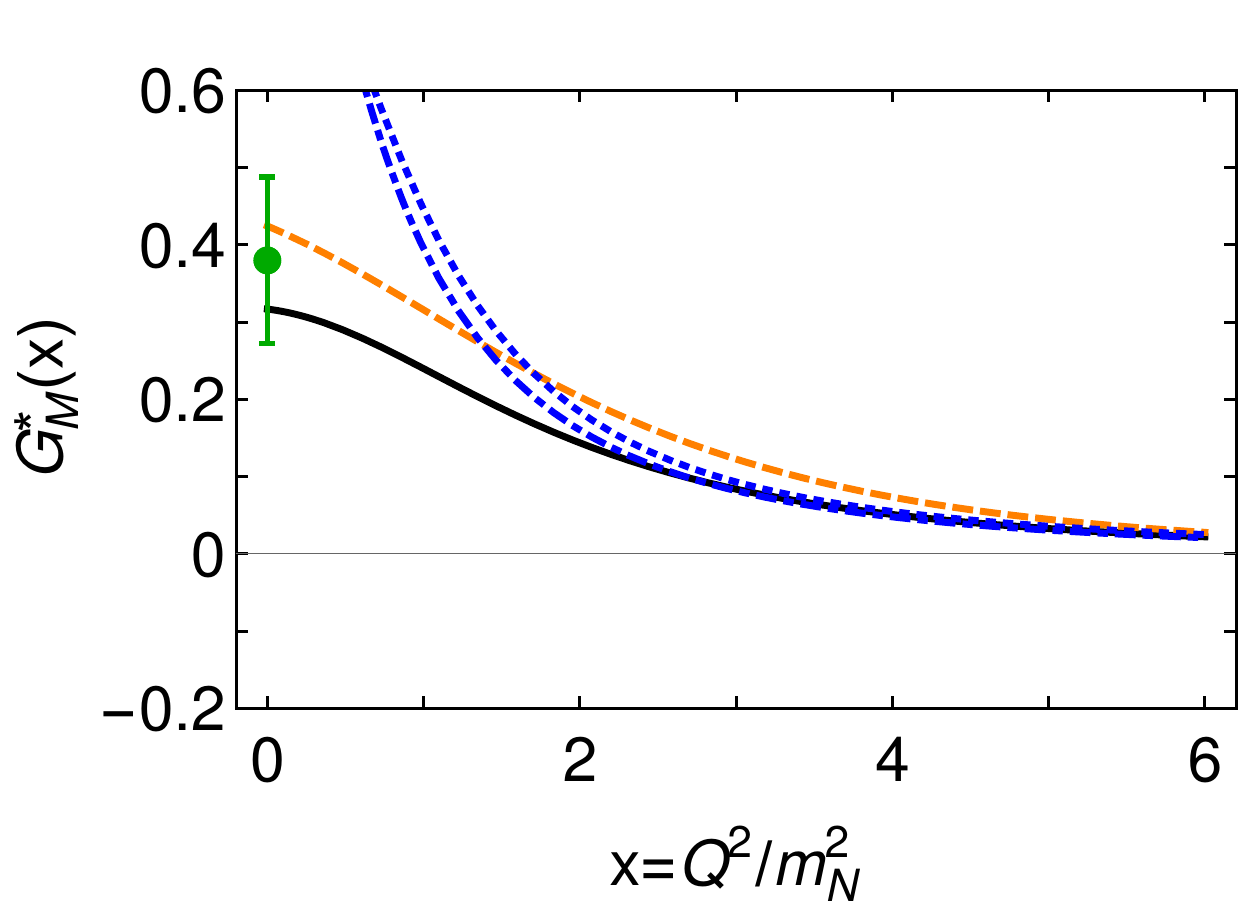}
\includegraphics[clip, width=0.32\textwidth]{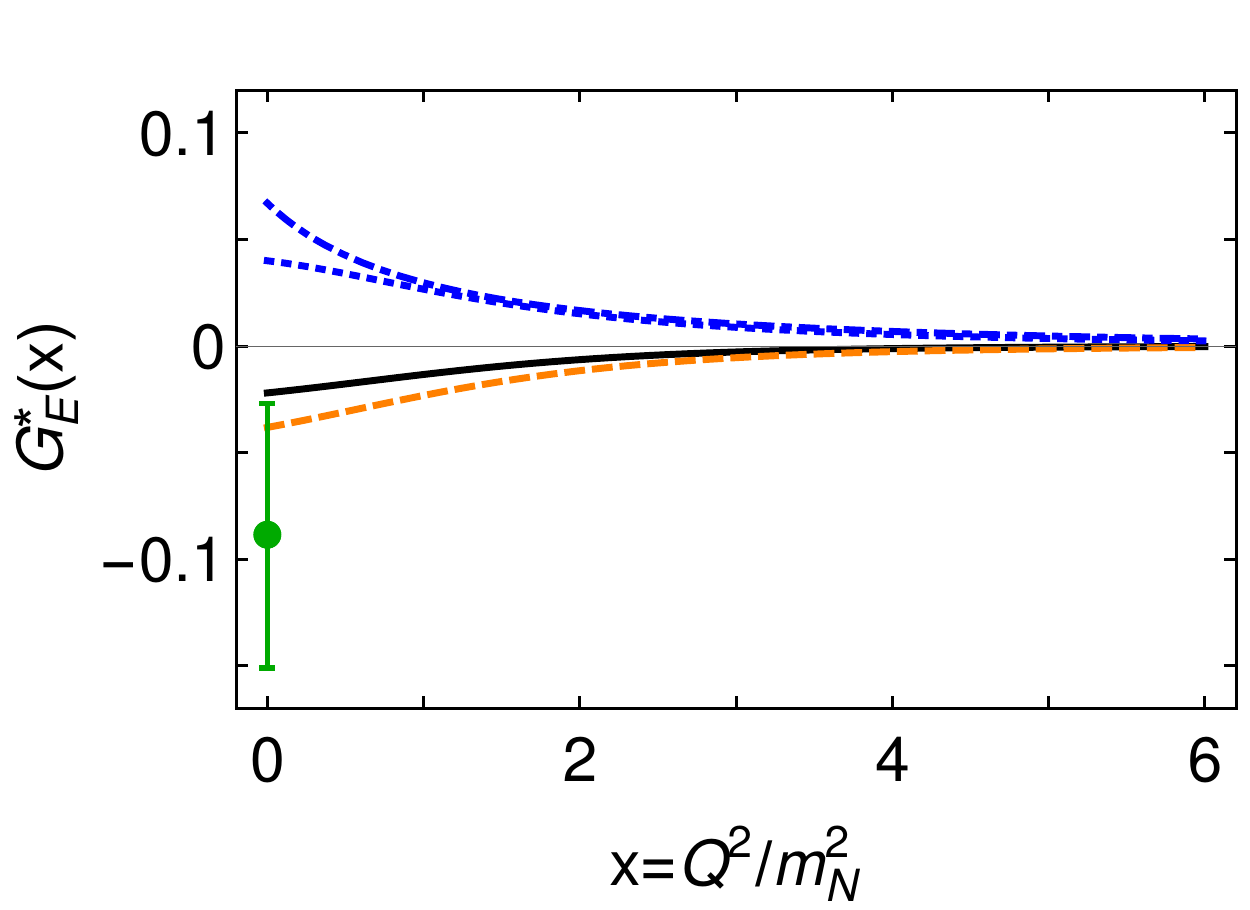}
\includegraphics[clip, width=0.32\textwidth]{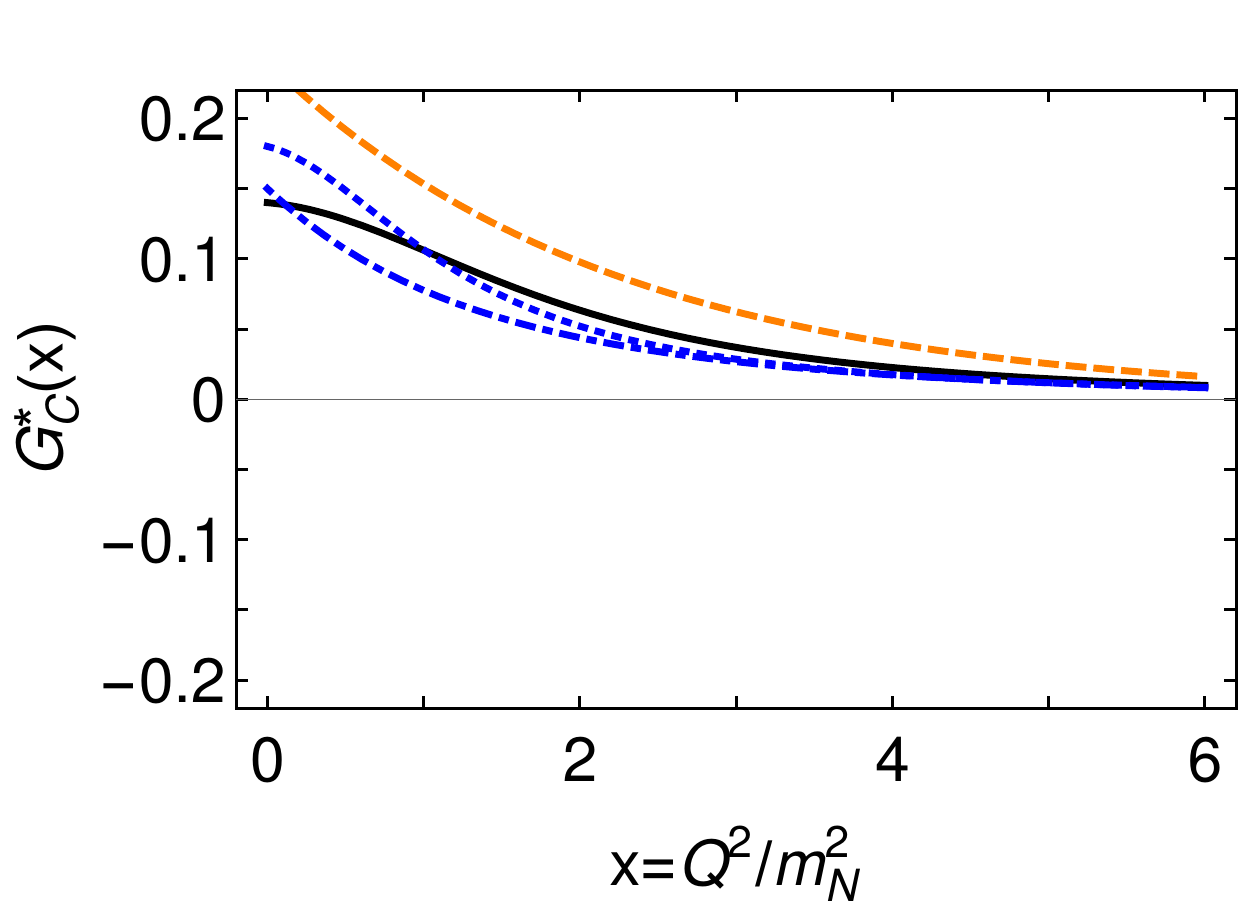}
\caption{\label{D1600TFFs}
\emph{Left panels} -- Magnetic dipole $\gamma^\ast p\to \Delta^+(1600)$ transition form factor; \emph{middle} -- electric quadrupole; and \emph{right}:  Coulomb quadrupole.
Data from Ref.~\cite{Tanabashi:2018oca}; and the conventions of Ref.~\cite{Jones:1972ky} are employed.
Panels on the top: solid (black) curve, complete result; shaded (grey) band, light-front relativistic Hamiltonian dynamics (LFRHD)~\cite{Capstick:1994ne};
dot-dashed (brown) curve, light-front relativistic quark model (LFRQM) with unmixed wave functions~\cite{Aznauryan:2015zta}; and dashed (purple) curve, LFRQM with configuration mixing~\cite{Aznauryan:2016wwm}.
Panels on the bottom: solid (black) curve, complete result; dotted (blue) curve, both the proton and $\Delta(1600)$ are reduced to $S$-wave states; Dot-dashed (blue) curve, result obtained when $\Delta(1600)$ is reduced to $S$-wave state; dashed (orange) curve, obtained by enhancing proton's axial-vector diquark content.
}
\end{figure*}

Predictions for the $\gamma^\ast p\to \Delta^+(1600)$ transition form factors are displayed in Fig.\,\ref{D1600TFFs}.  Empirical results are only available at the real-photon point for two of the three form factors: $G_M^\ast(Q^2=0)$, $G_E^\ast(Q^2=0)$.  Evidently, the quark model results (shaded grey band \cite{Capstick:1994ne}, dot-dashed brown curve \cite{Aznauryan:2015zta} and dashed purple curve\,\cite{Aznauryan:2016wwm}) are very sensitive to the wave functions employed for the initial and final states.  Furthermore, inclusion of relativistic effects has a sizeable impact on transitions to positive-parity excited states \cite{Capstick:1994ne}.

The DSE prediction within the QCD-kindred framework \cite{Lu:2019bjs} is the solid (black) curve in each panel of Fig.\,\ref{D1600TFFs}. In this instance, every transition form factor is of unique sign on the domain displayed. Notably, the mismatches with the empirical results for $G_M^\ast(Q^2=0)$, $G_E^\ast(Q^2=0)$ are commensurate in relative sizes with those in the $\Delta(1232)$ case, suggesting that MB\,FSIs are of similar importance in both channels.

One can mimic some effects of a meson cloud by modifying the axial-vector diquark content of the participating hadrons.  Accordingly, to illustrate the potential impact of MB\,FSIs, the transition form factors were computed using an enhanced axial-vector diquark content in the proton.  This was achieved by setting $m_{1^+} = m_{0^+} = 0.85\,$GeV, values with which the proton's mass is practically unchanged. The procedure produced the dashed (orange) curves in the bottom panels of Fig.~\ref{D1600TFFs}; better aligning the $x\simeq 0$ results with experiment and suggesting thereby that MB\,FSIs will improve the predictions.

The dotted (blue) curve in the bottom panels of Fig.~\ref{D1600TFFs} is the result obtained when only rest-frame $S$-wave components are retained in the wave functions of the proton and $\Delta(1600)$-baryon; and the dot-dashed (blue) curve is that computed with a complete proton wave function and a $S$-wave-projected $\Delta(1600)$. Once again, the higher partial-waves have a visible impact on all form factors, with $G_E^\ast$ being most affected.  These observations are clear pointers to intrinsic deformation of the nucleon and $\Delta$-baryons \cite{Eichmann:2011aa, Roberts:2019wov, Brodsky:2020vco, Carman:2020qmb}.

In the near future, the electro-excitation $N\to \Delta(1600)\frac{3}{2}^+$ amplitudes will become publicly available at photon virtualities $2.0\,\text{GeV}^2 < Q^2 < 5.0\,\text{GeV}^2$ from analysis of CLAS data on $\pi^+ \pi^- p$ electro-production off the proton \cite{Isupov:2017lnd, Trivedi:2018rgo}.  Preliminary indications are that the DSE predictions will be validated \cite{MokeevPrivate2020}.


\subsection{Multidimensional Structure of Baryons}
\label{MDSoBs}

\begin{figure}[!t]
\begin{tabular}{ccc}
\includegraphics[width=0.40\textwidth]{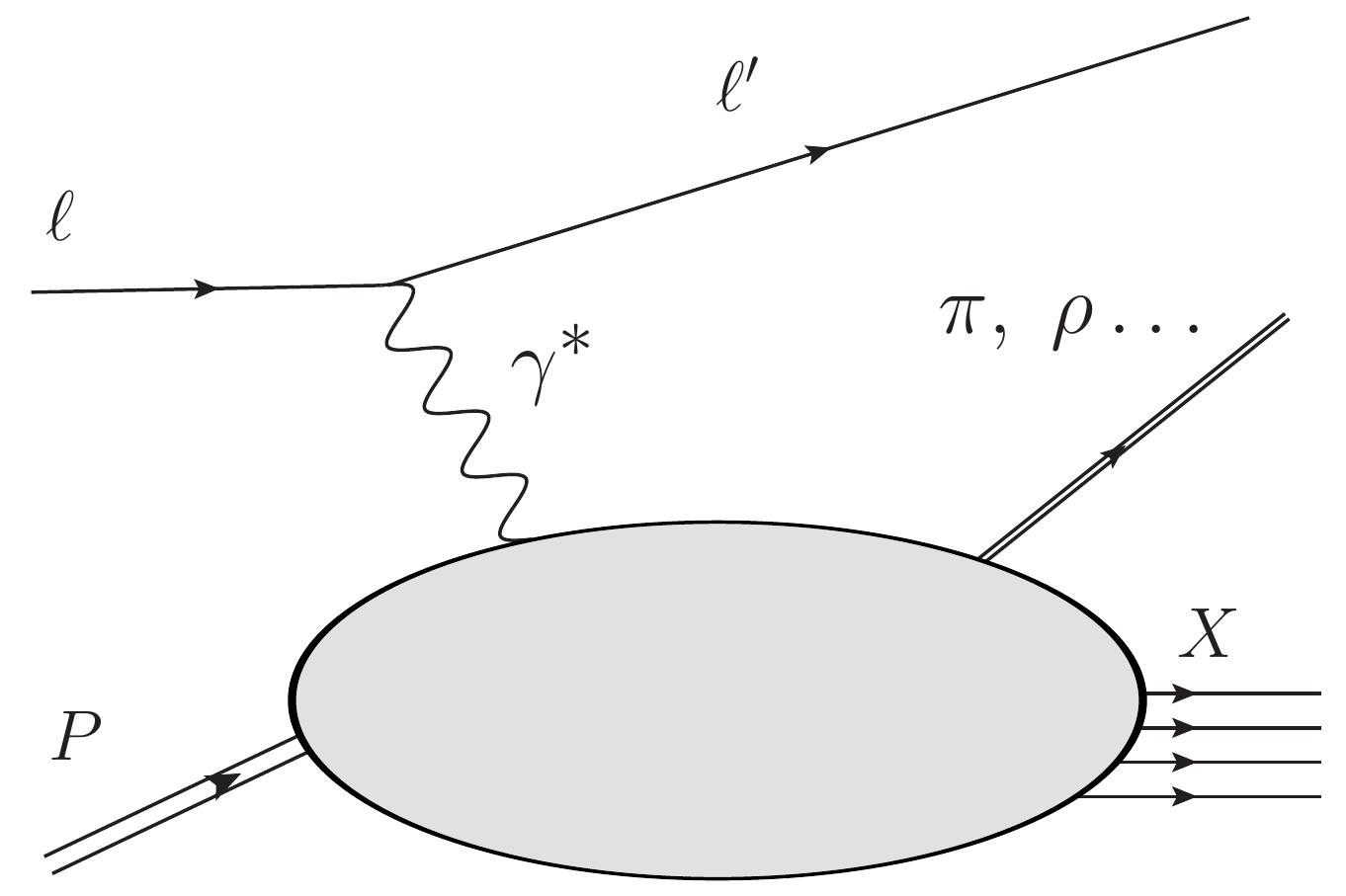} & \hspace*{2em}  &
\includegraphics[width=0.40\textwidth]{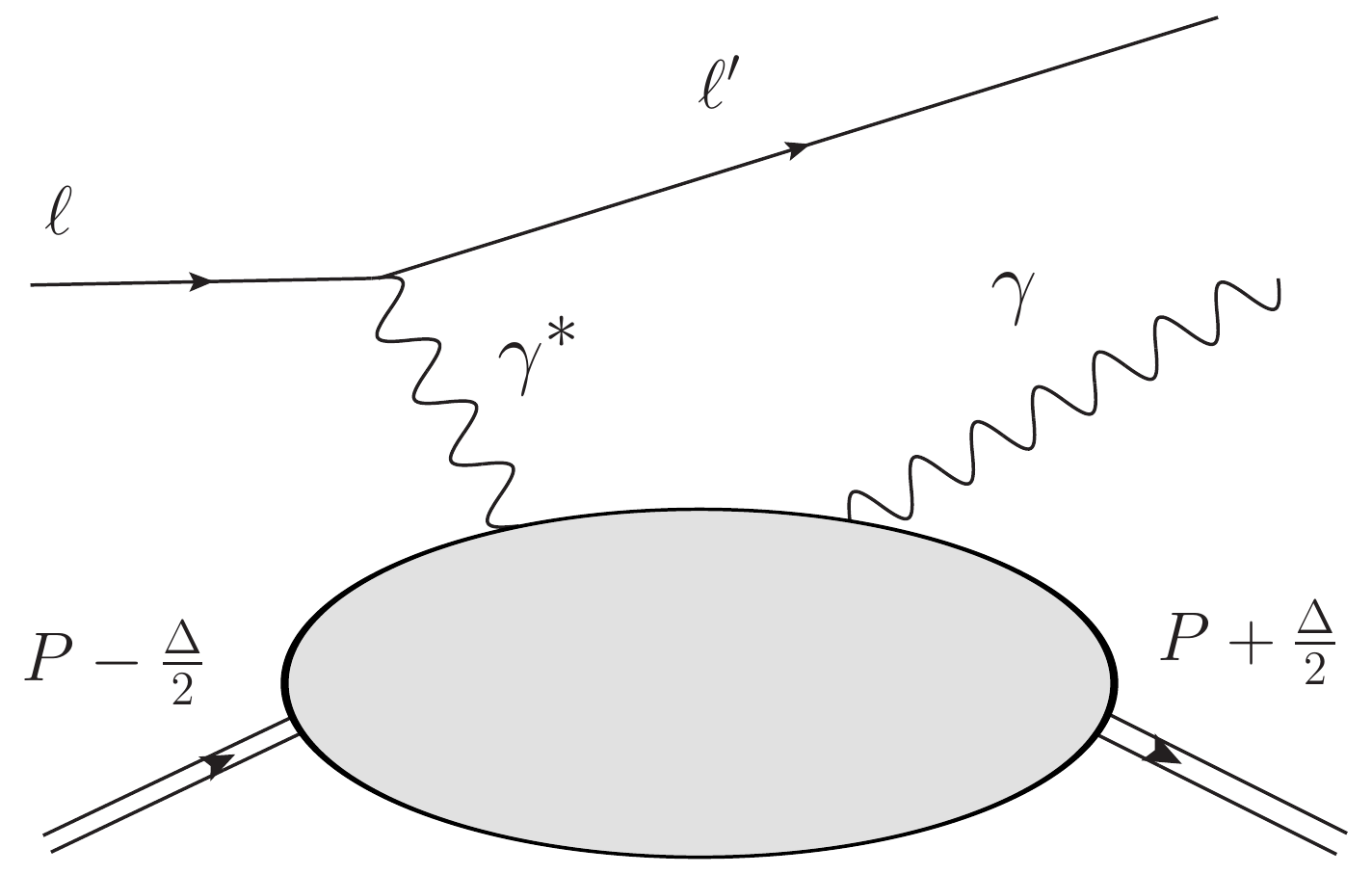}
\end{tabular}
\caption{\label{fig:SIDIS_DVCS} Examples of processes giving access to nucleon 3D structure. \emph{Left panel} -- semi-inclusive deep inelastic scattering (SIDIS); and \emph{right panel} -- deeply virtual Compton scattering (DVCS).
$\ell$, $\ell^\prime$ indicate leptons; $\gamma$ is a photon; $\gamma^\ast$ is a deeply-virtual photon; spaced double lines indicate incoming or outgoing nucleons; and near double lines are outgoing mesons.}
\end{figure}

Since the first deep inelastic scattering (DIS) experiments at SLAC in the late 1960s, the understanding of hadron structure has been enriched, with numerous matrix elements now identified as encoding the nonperturbative distribution of quarks and gluons within hadrons; in particular, inside the nucleon. Among them, it is usual to highlight transverse momentum dependent parton distributions (TMDs) \cite{Mulders:1995dh, Boer:1997nt} and generalised parton distributions (GPDs) \cite{Mueller:1998fv, Ji:1996nm, Radyushkin:1997ki}. Both yield a light-front 3D picture of the nucleon, the former in momentum space, whereas the latter can be related to a coordinate space probability. These functions are not simply Fourier transforms of each other; but rather different projections of so-called Wigner distributions \cite{Ji:2003ak}. GPDs and TMDs can be extracted from experimental data via a range of DIS processes: Fig.\,\ref{fig:SIDIS_DVCS} depicts a couple of examples. Their multidimensional character is appealing and one can wonder how material nonpointlike diquark correlations could affect the nucleon's 3D shape.

\subsubsection{Transverse Momentum Dependent parton distributions}
Compared to DIS, semi-inclusive deep inelastic scattering (SIDIS) experiments provide more complete information since, in the final state, a hadron is detected, together with the scattered electron, and its energy and transverse momentum, $\vec{P}_T$, are measured. SIDIS experiments (see the left panel of Fig.\,\ref{fig:SIDIS_DVCS}) enable access to TMDs, characterised by the parton longitudinal momentum fraction $x$ and its intrinsic transverse momentum $\vec{k}_T$ (which generates the hadron transverse momentum $\vec{P}_T$). These new parton distribution functions encode the motion of partons in the light-front transverse plane and complement the information given by the much studied PDFs. Therefore, SIDIS has emerged as a powerful means of probing strong interaction dynamics.  It provides access to TMDs through measurements of spin and azimuthal asymmetries. Studies of spin-azimuthal asymmetries in the semi-inclusive production of hadrons have been given high-priority at the JLab 12\,GeV facility \cite{Dudek:2012vr}; and they are one of the driving forces for the future electron ion collider (EIC) in the USA \cite{Accardi:2011mz, Anselmino:2011ay, Accardi:2012qut, Aschenauer:2014twa} and in developing a proposal for an electron ion collider in China (EicC) \cite{EicCWP, EicCWPEL, Chen:2020ijn}.

Assuming single photon exchange, the SIDIS cross-section can be decomposed, in a model-independent way, into a sum of various azimuthal modulations coupled to corresponding structure functions \cite{Bacchetta:2006tn}. Using tree-level factorisation, the structure functions can be calculated up to subleading order in $1/Q$ (twist three) using transverse-momentum-dependent quark-quark and quark-gluon-quark correlators. The eight leading-twist TMDs are probability densities for finding a (polarised) parton with a longitudinal momentum fraction $x$ and transverse momentum $\vec{k}_T$ in a (polarised) nucleon; and the sixteen twist-3 TMDs provide information on quark-gluon correlations (see Table~\ref{tab:TMDTable}). Interpretation of leading-twist structure functions in terms of convolutions of TMDs and TMD fragmentation functions are based on factorisation theorems \cite{Collins:2011zzd}.

Subleading structure functions require a proof of validity of TMD factorisation at higher twist; as yet, no proof exists.  However, studies of sub-leading twists are also important for two primary reasons: (\emph{i}) they are important to understanding long-range quark-gluon dynamics; and (\emph{ii})  they are not small in the kinematics of fixed target experiments, hence should be properly accounted for if TMDs are to be reliably extracted. Good examples of sizeable twist-3 TMDs are the $\cos\phi_h$ moment of the unpolarised cross-section $F_{UU}^{\cos\phi_h}$ and the $\sin\phi_h$ moment depending on the longitudinal polarisation of the beam $F_{LU}^{\sin\phi_h}$.  $F_{UU}^{\cos\phi_h}$ was measured at JLab with the $5.5\,$GeV electron beam and its contribution to the asymmetry appeared to be of the same order as the leading-twist moment $\cos 2\phi_h$ \cite{HarrisonThesis}. $F_{LU}^{\sin\phi_h}$ was first measured at JLab \cite{Avakian:2003pk, Aghasyan:2011ha, Gohn:2014zbz}, with later confirmation by different experiments at JLab and measurements at other facilities.
In this case, large spin-azimuthal asymmetries were also observed.

\begin{table}[t]
\caption{\label{tab:TMDTable}
\emph{Left panel} -- leading-twist TMDs; and \emph{right panel} -- twist-3 TMDs.}
\centering
\begin{tabular}{|c|c|c|c|}
\hline
& \makecell{Unpol.\\ quark} & \makecell{L-pol.\\ quark} & \makecell{T-pol. \\ quark} \\
\hline
\makecell{Unpol. \\ nucleon}& $f_1$ & $\emptyset$ & $h_1^\perp$ \\
\hline
\makecell{Unpol. \\ nucleon} & $ \emptyset $ & $g_1$ & $h^\perp_{1L}$ \\
\hline
\makecell{Unpol. \\ nucleon} & $f^\perp_{1T}$ & $g_{1T}$ & $h_1,~h_{1T}^\perp$\\
\hline
\end{tabular}
\qquad
\begin{tabular}{|c|c|c|c|}
\hline
& \makecell{Unpol.\\ quark} & \makecell{L-pol.\\ quark} & \makecell{T-pol. \\ quark} \\
\hline
\makecell{Unpol. \\ nucleon} & $f^\perp$ & $g^\perp$ & $h,~e$\\
\hline
\makecell{Unpol. \\ nucleon} & $f^\perp_L$ & $g^\perp_L $ & $h_L,~e_L$ \\
\hline
\makecell{Unpol. \\ nucleon} & $f_T,~ f^\perp_T$ & $g_T,~g^\perp_T$ & \makecell{$h_T,~e_T$\\$h^\perp_T,~e^\perp_T$}  \\
\hline
\end{tabular}
\end{table}

Being nonperturbative in nature, TMDs are very difficult to compute in QCD. Therefore, they have been studied in a variety of low-energy QCD-inspired models, \emph{e.g}.: a light-cone constituent quark model \cite{Pasquini:2008ax}; a light+front quark+diquark model \cite{Maji:2017bcz}; the spectator model \cite{Bacchetta:2008af}; and a bag model \cite{Avakian:2010br}.  (For a review, see Ref.\,\cite{Pasquini:2012jm}.) They all have in common a tendency to oversimplify the complexity of the QCD dynamics in hadrons; but studies in different models, based on different assumptions, may help to unravel  nonperturbative aspects of TMDs. Models might also play a useful role as a first step in the description of experimental observations, potentially providing an intuitive way to connect the physical observables to the dynamics of partons.

Model calculations can shed light on the important question of whether twist-3 functions should be different from zero or not. In this respect, Ref.\,\cite{Mao:2013waa} investigated the beam spin asymmetries (SSAs) $F_{LU}^{\sin\phi_h}$ of $\pi^+$, $\pi^-$ and $\pi^0$ production in the SIDIS process using a diquark bystander model.  Two different contributions to the beam SSAs were considered; namely, $e H_1^{\perp}$ and $g^{\perp} D_1$ (where $e$ and $g^{\perp}$ are twist-3 TMDs). By using two different choices for the propagator of the axial-vector diquark, together with different relations between the quark flavours and the diquark types, Ref.\,\cite{Mao:2013waa} obtained two different sets of $e$ and $g_{\perp}$. Comparing these predictions with the CLAS and HERMES data, they concluded that even though their model can describe the asymmetries for certain pion production in some kinematic regions, it was difficult to explain the asymmetries for all three pions in a consistent way. The applicability of quark models to TMDs beyond the leading twist approximation remains debatable. However, the additional information on higher twist TMDs from models may become very important for both phenomenology and experimental event generators.

There have been many model calculations of the leading twist TMDs. A review of the different models and their comparisons with the experimental data is beyond the scope of this document. A single example is given by Ref.\,\cite{Maji:2016yqo}, which studies the T-even TMDs in a light front quark+diquark model. The model contains both scalar and axial-vector diquark bystanders, with light-front wave functions modelled after AdS/QCD phenomenology. For the worm-gear $h_{1L}^{\perp}$ TMD, which describes transversely polarised quarks in a longitudinally polarised proton, Ref.\,\cite{Maji:2016yqo} predicts negative distributions for both $u$- and $d$-quarks, in disagreement with the predictions of a light-front constituent quark model \cite{Pasquini:2008ax}, wherein the distribution is negative for the $u$-quark but positive for the $d$-quark. This discrepancy needs to be resolved.  In any event, it will be essential in future model studies to account for the interaction of the probe with the diquark, which is known to be crucial in describing nucleon elastic and transition form factors -- Secs.\,\ref{sec:DSEs}, \ref{secNRTFFs}.

Another approach to accessing TMDs is via phenomenological extractions. The assumption involved in modern extractions of TMDs from available data relies on a Gaussian \emph{Ansatz} for the transverse momentum dependence of distribution and fragmentation functions \cite{Anselmino:2005nn, Anselmino:2008sga}. If this phenomenological approach is taken, it is then crucial to develop an analysis framework that will allow testing different extraction procedures and estimating systematic uncertainties related to different models and assumptions.  Assessing the sensitivity to kinematic limitations and radiative corrections, and validating the extracted functions are also key points. This is the main goal of the Extraction and VAlidation framework (EVA), which is being developed at JLab \cite{Avakian:2015vha} and will serve to help both the experimental and phenomenological communities to test results and ensure model-independence of inferred data.

\subsubsection{Generalised Parton Distributions}

GPDs can be accessed experimentally through exclusive processes such as DVCS, in which the nucleon remains intact, Fig.\,\ref{fig:SIDIS_DVCS}\,--\,right panel. A factorisation theorem ensures that the DVCS amplitude can be split into a hard part, calculable in perturbation theory, and a nonperturbative part, encoded in GPDs. Past, current (such as JLab 12\,GeV and COMPASS) and future (EIC) experimental programs allocate a significant part of their beam time to GPDs studies, highlighting the attraction of GPDs in the hadron physics community (for a review of recent phenomenology see Ref.\,\cite{Kumericki:2016ehc}). As noted above, GPDs provide access to quark  and gluon  spatial density distributions, $\rho^{q,g}(x,b_\perp)$, where $b_\perp$ is the light-front transverse spatial coordinate.

Being the Fourier transform of a matrix element of a non-local operator depending on a light-like distance, GPDs are simpler than TMDs in some aspects, \emph{e.g}.\ evolution equations, Wilson lines, \emph{etc}.  Still, they remain nonperturbative objects that cannot yet be directly computed in full QCD.  Various models have been developed, for instance Refs.\,\cite{Vanderhaeghen:1999xj, Mukherjee:2002gb, Goloskokov:2005sd, Kumericki:2009uq, Mezrag:2013mya, Mezrag:2016hnp, Moutarde:2018kwr}, taking advantages of theoretical features of GPDs.  Note, too, that neural networks were also used to extract GPDs \cite{Kumericki:2011rz, Moutarde:2019tqa}. However, none of these approaches was able to fulfill \emph{a priori} all the theoretical constraints which apply to GPDs. A promising framework to do so has recently been developed~\cite{Chouika:2017dhe, Chouika:2017rzs} but still needs to be confronted with experimental data.

Diquark bystander models of GPDs already exist, in both covariant and light-front formulations (see, \emph{e.g}.\ Ref.\,\cite{Mezrag:2016hnp} for a discussion).  One may expect that a quark+diquark picture could be valid within the valence region; and if questions about connections with QCD can be raised for some models, at least one has been fitted to data, yielding the so-called GPD hybrid model \cite{Ahmad:2006gn, Goldstein:2010gu, GonzalezHernandez:2012jv}. Here, the proton-quark+diquark vertex coupling exhibits a modified monopole behaviour as a function of the struck, off-shell quark's four-momentum squared, $k^2$, which is regulated by two mass parameters describing,  respectively, the quark, $m_q$, and monopole, $M_\Lambda^q$, masses.  While $m_q$ and $M_\lambda^q$ are parameters to be fitted to data, the diquark propagator is expressed through a spectral representation (\`a la K\"all\'en-Lehmann) such that the spectral function introduces Regge behavior of the GPDs at low $x$. In particular, by imposing the sum rule constraints on the quark GPDs, $H^{u,d}$ and $E^{u,d}$, relating them to the Dirac ($F_1$) and Pauli ($F_2$) form factors:
\begin{equation}
\int dx H^{u,d}(x,\xi,t) = F_1^{u,d}(t) \,, \quad \int dx E^{u,d}(x,\xi,t) = F_2^{u,d}(t) \,,
\end{equation}
and using flavour separated data \cite{Cates:2011pz}, the diquarks' radii can be estimated: $\sqrt{\langle r^2 \rangle}_{[ud]}  = 0.48\,$fm for the $[ud]$ diquark, and $\sqrt{\langle r^2 \rangle}_{\{uu\}}  = 0.56\,$fm for the $\{uu\}$ diquark. (DSE predictions for diquark electromagnetic radii have the same ordering \cite{Maris:2004bp}: $r_{[ud]} \gtrsim r_\pi$, $r_{\{uu\}} \gtrsim r_\rho$.)  These results may be compared with the proton radius ($\approx 0.84\,$fm) and most importantly with the separate $u$- and $d$-quark radii in the two configurations: the study of Ref.\,\cite{GonzalezHernandez:2012jv} found $\sqrt{\langle r^2 \rangle}_{d} \approx \sqrt{\langle r^2 \rangle}_{u}$ for $F_1$, while the $d$-quark radius exceeds that of the $u$ quark in $F_2$.  (These conclusions are also consistent with those reached via DSE analyses \cite[Sec.\,V.B]{Cui:2020rmu}.)

As reported above (Fig.\,\ref{PlotPDAs}), DSE predictions are available for the leading-twist PDAs of the nucleon, and its first radial excitation, obtained using dynamical diquark correlations \cite{Mezrag:2017znp, Mezrag:2018hkk}.  This could open the door to a dynamical computation of nucleon GPDs, following previous work on the pion \cite{Mezrag:2014jka, Chouika:2017dhe, Chouika:2017rzs}.  In this way, one could connect basic QCD considerations, such as DCSB and the formation of diquark correlations, to the 3D structure of hadrons; and from there proceed to experimental data on exclusive processes using, \emph{e.g}.\ PARTONS \cite{Berthou:2015oaw}, phenomenology software that has recently become publicly available.

It is anticipated that forthcoming experiments will add greatly to the empirical store of information about GPDs. Indeed, new data are expected from at least two main sources: COMPASS and JLab, in various kinematical regions; and in the case of JLab, also through various exclusive processes.  This might bring into sharper relief those questions which relate to the role of higher-twist and higher-order $\alpha$-strong corrections \cite{Moutarde:2013qs, Defurne:2017paw}, which are known to be important in TMD processes.  Experimental access is provided by a careful choice of kinematics, minimising the Bethe-Heitler contributions and exposing the modulations present in DVCS amplitudes.  The corrections could then be revealed via harmonic decomposition.  Such studies should be possible at JLab.


\subsection{Meson Structure as a Window onto Diquark Structure}
\label{MSaaWoDS}
It is worth highlighting that the character of diquark correlations depends on $N_c$, the number of colours. For instance, the Lagrangian of two-colour QCD, $N_c=2$, respects Pauli-G\"ursey (PG) symmetry \cite{Pauli:1957, Gursey:1958}.  In this case, DCSB in the $N_f$-flavour theory yields $N_f(2N_f-1)-1$ degenerate NG modes: $N_f^2-1$ meson- and $N_f(N_f-1)$ diquark+anti-diquark modes. The NG mesons are the usual pion-like states, and the lowest-mass diquarks are the scalar diquark and its antiparticle. The degeneracy of these five states indicates that, for two-colour QCD, the mechanisms responsible for meson structure are also those which determine the character of diquarks.  Hence, by improving our understanding of pion structure and interactions, we move closer to a sound description of the static and dynamic features of scalar diquark correlations.

In the $N_c=3$ case, the fundamental and conjugate representations of the Lie group are inequivalent so PG symmetry is broken.  Consequently, the diquark is not a colour-singlet and the degeneracy between pions and scalar diquarks is lifted. However, as discussed in Sec.\,\ref{sec:diquarks-dse}, using the DSE RL truncation, the diquark Bethe-Salpeter equation differs from that of its meson partner by only a multiplicative factor of $1/2$.  Thus, one may reasonably expect that understanding meson structure will shed light on that of diquarks.

Pion properties are largely defined by DCSB, which is forcefully expressed in the chiral-limit dressed-quark propagator, $S(k)$, that can be obtained by solving QCD's gap equation in the absence of Higgs couplings.  Writing $S(k) = 1/[-i \gamma\cdot k \, A(k^2) + B(k^2)]$, then owing to DCSB, one obtains a solution for $B(k^2)$ that is large at infrared momenta: $B(0)\simeq 0.5\,$GeV, and vanishes logarithmically faster than $1/k^2$ in the ultraviolet \cite{Lane:1974he, Politzer:1976tv}.  Moreover, in the chiral limit, DCSB is also necessary and sufficient to ensure \cite{Maris:1997hd, Qin:2014vya, Binosi:2016rxz}:
\begin{equation}
\label{gtrE}
f_\pi^0 E_\pi(k;0) = B(k^2)\,,
\end{equation}
where $f_\pi^0$ is the chiral-limit value of the pion's leptonic decay constant and $E_\pi$ is the leading piece of the pion's Bethe-Salpeter amplitude.

Eq.\,\eqref{gtrE} is remarkable.  It is true in any covariant gauge and independent of the renormalisation scheme; and it means that the two-body problem in the flavour-nonsinglet pseudoscalar meson channel is solved, nearly completely, once the solution to the one body problem is known.  Recalling now the parallel between mesons and diquarks, it follows immediately that DCSB must play a major role in forming strong diquark correlations and determining their structure.  This is true for all channels.  Hereafter, however, the pion--scalar-diquark connection will be the focus in a discussion of the $x$-dependence of their parton distribution functions.

In the valence region, most of our knowledge of pion structure functions comes from pionic Drell-Yan scattering \cite{Badier:1983mj, Conway:1989fs}; and in the sea region, from hard diffractive processes measured in $ep$ collisions at HERA \cite{Adloff:1998yg}.  Still, data remain sparse; and none are available for the kaon.  This situation has triggered a longstanding controversy concerning the large-$x$ behaviour of the pion valence-quark DF ${\mathpzc q}^\pi(x;\zeta)$.  (See,  \emph{e.g}.\ Refs.\,\cite{Holt:2010vj, Wijesooriya:2005ir, Aicher:2010cb}.)

Briefly, QCD predicts \cite{Ezawa:1974wm, Farrar:1975yb, Berger:1979du}:
\begin{equation}
\label{eq:LargexBehaviour}
{\mathpzc q}^{\pi}(x;\zeta =\zeta_H) \underset{x\to 1}{\sim} (1-x)^{2}\,,
\end{equation}
where $\zeta_H$ is the hadronic scale at which the dressed quasiparticles emerging from the valence-quark and -antiquark degrees of freedom express all properties of the pion \cite{Ding:2019qlr, Ding:2019lwe, Cui:2020dlm, Cui:2020piK}; in particular, they carry all the pion's light-front momentum.  Moreover, the exponent evolves as $\zeta$ increases beyond $\zeta_H$, becoming $2+\gamma$, where $\gamma\gtrsim 0$ is an anomalous dimension that increases logarithmically with $\zeta$.  Yet, a leading-order (LO) pQCD analysis of Drell-Yan data, Ref.\,\cite{Conway:1989fs} -- the E615 experiment, yields ($\zeta_5 = 5.2\,$GeV) a marked contradiction of Eq.\,\eqref{eq:LargexBehaviour}, \emph{viz}.
\begin{align}
{\mathpzc q}_{\rm E615}^{\pi}(x; \zeta_5) &\underset{x\simeq 1}{\approx} (1-x)^{1}\,.
\end{align}

Subsequent calculations \cite{Hecht:2000xa} confirmed Eq.\,\eqref{eq:LargexBehaviour}, eventually prompting reconsideration of the E615 analysis, with the result that, at next-to-leading order (NLO) and including soft-gluon resummation \cite{Wijesooriya:2005ir, Aicher:2010cb}, the E615 data can be viewed as being consistent with Eq.\,\eqref{eq:LargexBehaviour}. Notwithstanding these advances, uncertainty over Eq.\,\eqref{eq:LargexBehaviour} will remain until other analyses of the E615 data incorporate threshold resummation effects and, crucially, new data are obtained.  One can be optimistic for two reasons: firstly, relevant tagged deep-inelastic scattering (DIS) experiments have been approved at JLab \cite{JlabTDIS1, JlabTDIS2}; and secondly, experimental access to the pion PDF should be possible at future facilities  \cite{Peng:2016ebs, Peng:2017ddf, Horn:2018fqr, Denisov:2018unj, Chang:2020rdy, Chen:2020ijn}.

\begin{figure*}[!t]
\centering
\includegraphics[width=0.8\textwidth]{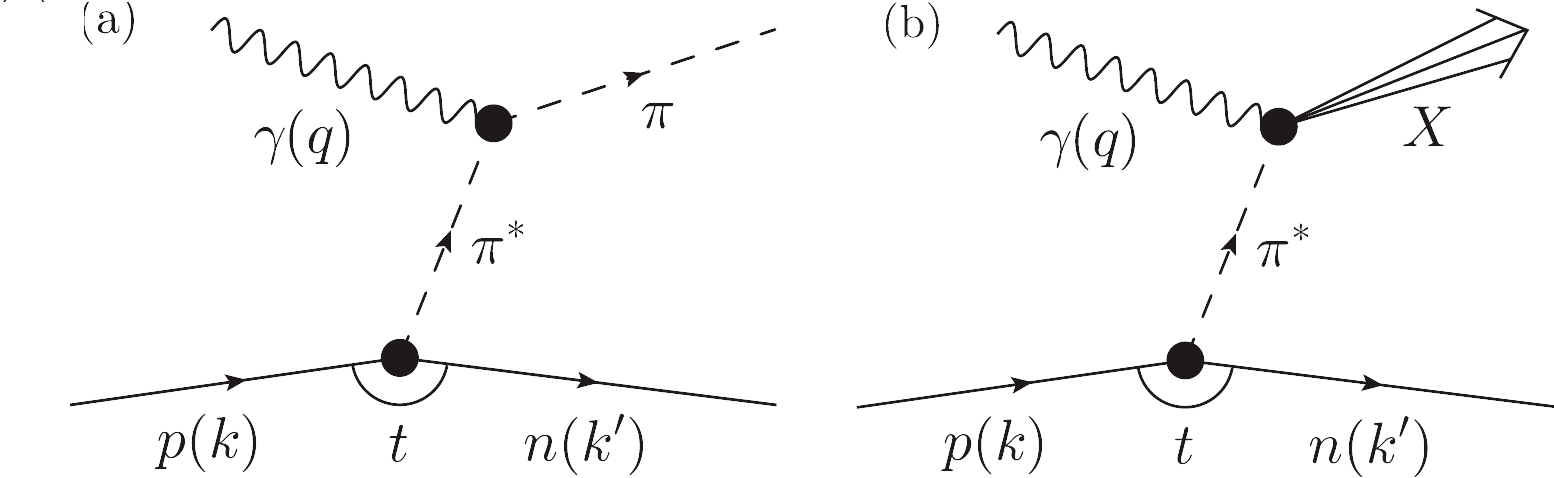}
\caption{\label{fig:Sullivan}
Sullivan processes \cite{Sullivan:1971kd}.  In these examples, a proton's pion cloud is used to provide access to the pion's (a) elastic form factor and (b) parton distribution functions.  $t = –(k-k^\prime)^2$ is a Mandelstam variable and the intermediate pion, $\pi^\ast(P=k-k^\prime)$, $P^2= –t$, is off-shell.}
\end{figure*}

\begin{figure}[!t]
\centering
\includegraphics[width = 0.75\textwidth]{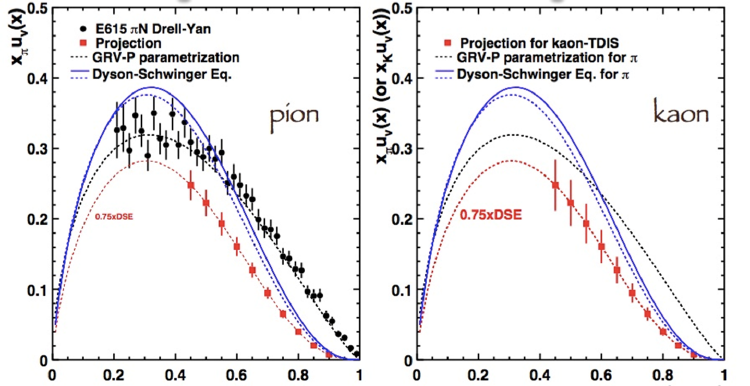}
\caption{Anticipated errors for pion and kaon PDF extractions.  The projected data are drawn on the DSE result from Ref.\,\cite{Hecht:2000xa}, which is multiplied by $0.75$ in order to increase clarity in the comparison.  Additional details are provided elsewhere \cite{JlabTDIS1, JlabTDIS2}.}
\label{fig:PionPDF}
\end{figure}

JLab's tagged DIS experiments will exploit a Sullivan process \cite{Sullivan:1971kd}, in which one draws a pion target from the proton's pion cloud, Fig.\,\ref{fig:Sullivan}(b).  Contemporary theory indicates \cite{Qin:2017lcd} that this is a valid approach on $-t < 0.6 \,$GeV$^2$ for the pion (and on $-t < 0.9 \,$GeV$^2$ for the kaon).  Studies during the past decade, based on JLab 6\,GeV measurements, have instilled confidence in the reliability of pion electroproduction as a tool for extracting the pion form factor.  In turn, this supports the study of the pion structure function using a similar approach.  Measurements at JLab\,12 will allow access to the kinematic domain: $-t < 0.2\, \textrm{GeV}^2$ , $Q^2 \ge 1\,\textrm{GeV}^2$, $M_X ^2> 1.0\, (\textrm{GeV/c})^2$, which will probe regions of intermediate and high $x$ within the pion.  An overlap with the domain of Drell-Yan measurements will enable cross-checking.  Projected uncertainties in pion (and kaon) structure measurements are depicted in Fig.\,\ref{fig:PionPDF}.

As new measurements are awaited, theory progress continues.  Novel algorithms within lQCD are beginning to yield results for the pointwise behaviour of the pion's valence-quark distribution \cite{Xu:2018eii, Chen:2018fwa, Karthik:2018wmj, Sufian:2019bol}.  In addition, extensions of the continuum analysis in Ref.\,\cite{Hecht:2000xa} are yielding new insights.

\begin{figure}[!t]
\begin{tabular}{ccc}
\includegraphics[clip, width=0.46\textwidth]{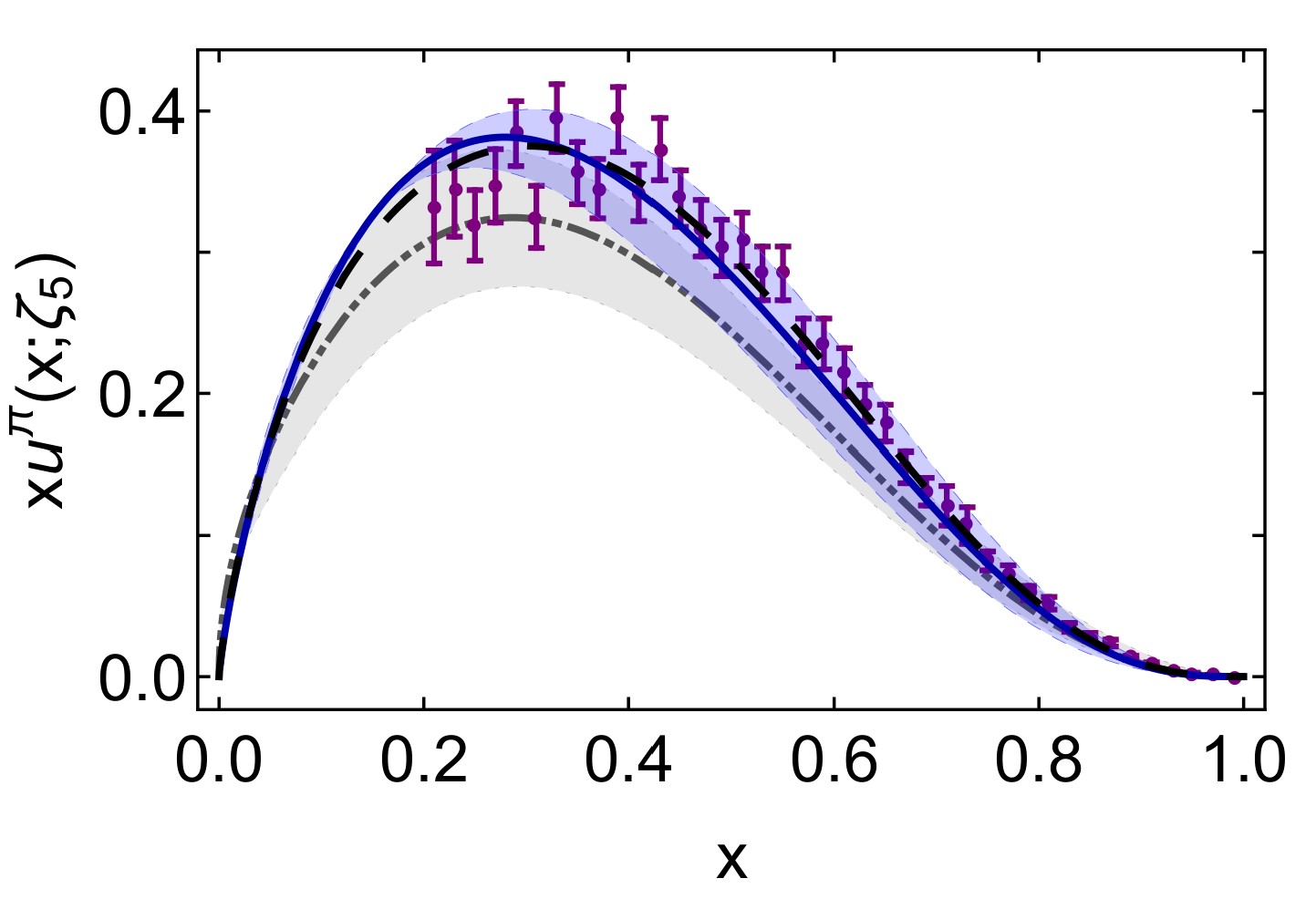} & \hspace*{1em} & \\[-32ex]
& & \includegraphics[clip, width=0.46\textwidth]{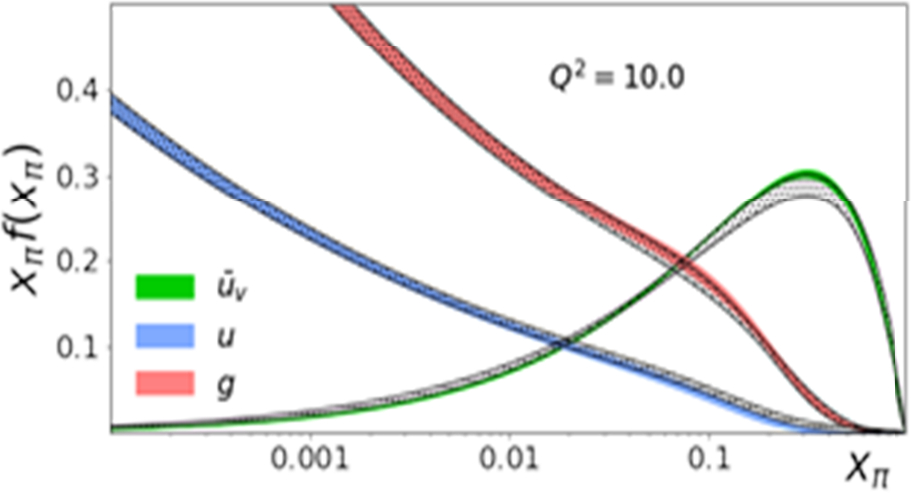}
\end{tabular}\vspace*{2ex}
\caption{\label{qpizeta5}
\emph{Left panel}. Pion valence-quark momentum distribution function, $x {\mathpzc q}^\pi(x;\zeta_5)$:
solid (blue) curve embedded in shaded band -- modern, continuum calculation \cite{Ding:2019qlr, Ding:2019lwe, Cui:2020dlm, Cui:2020piK};
long-dashed (black) curve -- early continuum analysis \cite{Hecht:2000xa}; and
dot-dot-dashed (grey) curve within shaded band -- lQCD result \cite{Sufian:2019bol}.
Data (purple) from Ref.~\cite{Conway:1989fs}, rescaled according to the analysis in Ref.~\cite{Aicher:2010cb}.
(In all cases, the shaded blue bands indicate the size of calculation-specific uncertainties, as described elsewhere \cite{Ding:2019qlr, Ding:2019lwe, Cui:2020dlm, Cui:2020piK}.)
\emph{Right panel}.
Sample EIC extraction of valence quark, sea quark and gluon DFs in the pion, at a scale $Q^2 = 10\,\textrm{GeV}^2$. The extraction is done with the following assumptions: the $u$ DF equals the $\bar{d}$ DF in the pion and the $\bar{u}$ DF is the same as the other sea quark DFs ($d$, $s$ and $\bar{s}$). The extraction at $x_\pi<10^{-2}$, at this $Q^2$ scale, is constrained by existing HERA data.
}
\end{figure}

Capitalising on these new developments, recent parameter-free continuum analyses have delivered predictions for the valence, glue and sea distributions within the pion \cite{Ding:2019qlr, Ding:2019lwe, Cui:2020dlm, Cui:2020piK}, unifying them with, \emph{inter alia}, electromagnetic pion elastic and transition form factors. Their predictions for the pion parton distributions at a scale relevant to the E615 experiment \cite{Conway:1989fs, Wijesooriya:2005ir} are depicted in Fig.\,\ref{qpizeta5}\,--\,left panel. The large-$x$ behaviour is \cite{Ding:2019qlr, Ding:2019lwe, Cui:2020dlm, Cui:2020piK}:
\begin{align}
\label{DSEbeta}
{\mathpzc q}^{\pi}(x; \zeta_5) &\underset{x\to 1}{\sim} (1-x)^{2.74(12)}\,.
\end{align}
Refs.\,\cite{Ding:2019qlr, Ding:2019lwe, Cui:2020dlm, Cui:2020piK} also produce the following apportioning of momentum at the scale $\zeta=\zeta_5$:
\begin{equation}
\langle x\rangle^\pi_{\mathrm{valence}} = 0.41(4)\,, \quad
\langle x\rangle^\pi_{\mathrm{glue}} = 0.45(1)\,, \quad
\langle x\rangle^\pi_{\mathrm{sea}} = 0.14(2)\,.
\end{equation}

Fig.\,\ref{qpizeta5}\,--\,left panel compares the DSE result for the pion's valence-quark DF with that obtained in an exploratory lQCD analysis \cite{Sufian:2019bol}: the pointwise form of the lQCD prediction agrees with the DSE result, as highlighted by the fact that one finds ${\mathpzc q}_{\rm LQCD}^{\pi}(x; \zeta_5) \sim (1-x)^{2.45(58)}$, in agreement with Eq.\,\eqref{DSEbeta}.  This agreement is significant.  Now, two distinct treatments of the pion bound-state problem have delivered the same prediction for the pion's valence-quark DF; and their behaviour on the valence-quark domain, $x\gtrsim 0.2$, agrees with the most complete analysis of extant data \cite{Aicher:2010cb}.   Evidently, Eq.\,\eqref{eq:LargexBehaviour} is stronger then ever before; and real progress is being made toward understanding pion structure and its relation to the emergence of mass.

It is straightforward to extend the continuum studies of the pion's DFs to the scalar diquark and thereby obtain predictions for all its DFs, too.  Comparison of their $x$-dependence with that of their analogues in the pion partner could be instructive.

It is worth mentioning that the anticipated US electron ion collider (EIC) \cite{PhysicsToday73} offers numerous access paths to pion and kaon structure functions on a large kinematic domain \cite{Aguilar:2019teb} -- see Fig.\,\ref{qpizeta5}\,--\,right panel.  It could deliver the critical results that are needed to (\emph{i}) test and improve the phenomenology tools used to connect experiment and theory; and (\emph{ii}) validate existing pictures of NG mode structure.  An electron ion collider is also being proposed in China (EicC) \cite{EicCWP}; and with current design specifications, the EicC could both \cite{Chen:2020ijn} neatly fill a gap between JLab at 12\,GeV and the EIC and develop a powerful synergy with new initiatives at CERN \cite{Denisov:2018unj}.


\subsection{Exotic Hadrons and their Connection to Diquarks}
\label{EHatCtDs}
As mentioned in Sec.\,\ref{sec:quark-models}, the past two decades have seen a rejuvenation of hadron spectroscopy by the discovery of many states that do not fit the typical QM pattern. The new systems are collectively called $XYZ$ states to highlight their still poorly understood character \cite{Chen:2016qju, Esposito:2016noz, Ali:2017jda, Olsen:2017bmm, Guo:2017jvc, Liu:2019zoy, Guo:2019twa}. Several are good candidates for tetra- and penta-quark systems; hence, understanding their properties could assist in confirming a role for diquark correlations in the spectrum of QCD.

%


\subsubsection{Experimental status at a glance}

\noindent \underline{Mesons at the $\mathbf{\bar D^{(*)} D^{*}}$, $\mathbf{\bar B^{(*)} B^{*}}$ thresholds.}\\
The first and best known exotic state is the $X(3872)$.  Its discovery as an unexpected charm-anticharm bound-state in 2003 gave birth to the long saga of charmonium- and bottomonium-like states. It was observed as an extremely narrow peak in the $B^+ \to K^+ (J/\psi\,\pi^+\pi^-)$ channel, exactly at the $\bar D^0 D^{*0}$ threshold \cite{Choi:2003ue, Aubert:2004ns}. It was later confirmed in the open charm channel \cite{Gokhroo:2006bt, Aubert:2007rva, Adachi:2008su, Ablikim:2020xpq}, and in other reactions.  Most notably, $X(3872)$ is produced promptly in  $p\bar{p}$ \cite{Acosta:2003zx, Abazov:2004kp}, $pp$ \cite{Aaij:2011sn, Chatrchyan:2013cld} and Pb\,Pb collisions \cite{CMS:2019vma}, at rates commensurate with that of the $\psi(2S)$ charmonium. The upper limit on the width is $\Gamma < 1.2\,\text{MeV}$ at 90\% confidence-limit \cite{Choi:2011fc} and its quantum numbers have been established to be $1^{++}$ \cite{Aaij:2013zoa, Aaij:2015eva}. It follows that the pion pair in $X(3872)\to J/\psi\,\pi^+\pi^-$ must have relative orbital momentum $L=1$ and isospin one.  (The di-pion distribution is indeed dominated by the $\rho$ meson.)  Hence, for this decay mode to be significant, large isospin violation is required, much larger than expected for an ordinary charmonium; a result confirmed by comparison with the isospin conserving mode $X(3872)\to J/\psi \,\omega$ \cite{delAmoSanchez:2010jr, Ablikim:2019zio}.  COMPASS claimed an excess of events in $\mu^+ N \to \mu^+ N (J/\psi \pi^+ \pi^-) \pi^+$ at the $X(3872)$'s mass \cite{Aghasyan:2017utv}. However, the di-pion distribution better matches a $1^{+-}$ assignment, which points to a degenerate $\tilde X(3872)$.

Ten years later, two axial-vector states were seen in this region.  BESIII and Belle observed a peak in the $J/\psi\,\pi^+$ invariant mass of the $e^+e^- \to J/\psi\,\pi^+\pi^-$ reaction, close to the $(\bar D^{*} D)^{+}$ threshold~\cite{Ablikim:2013mio, Liu:2013dau}. The state is called $Z_c(3900)$, with mass and width $M = 3887.2 \pm 2.3\,\text{MeV}$ and $\Gamma = 28.2 \pm 2.6 \,\text{MeV}$, respectively. The minimal quark content for such a state is $c\bar c u \bar d$; hence, it is manifestly exotic. Its quantum numbers are $1^{+-}$ \cite{Collaboration:2017njt}. The state is seen as a threshold enhancement in the open charm channel \cite{Ablikim:2013xfr, Ablikim:2015swa}, and the neutral partner is also observed \cite{Ablikim:2015tbp, Ablikim:2015gda}. A second $1^{+-}$ state, called $Z_c^\prime(4020)$, has been found in $e^+e^- \to (\bar D^{*0} D^{*+})\,\pi^-$ \cite{Ablikim:2013emm} and $e^+e^- \to (h_c\, \pi^+)\, \pi^-$ \cite{Ablikim:2013wzq}, with mass $M = (4023.9 \pm 2.4) \,\text{MeV}$, slightly above the $D^*D^*$ threshold, and width $\Gamma= (10\pm 6)\,\text{MeV}$. A neutral partner has also been reported \cite{Ablikim:2014dxl, Ablikim:2015vvn}.

The two $Z_c^{(\prime)}$ have heavier replicas in the bottomonium sector. Two charged states appear at the $\bar B B^*$ and $\bar B^* B^*$ thresholds, named $Z_b(10610)$ and $Z_b^\prime(10650)$. They have been seen in several hidden-bottom final states, $\Upsilon(5S) \to (X_{b\bar b} \pi^+) \pi^-$, with $X_{b \bar b} = \Upsilon(1S),\Upsilon(2S),\Upsilon(3S),h_b(1P),h_b(2P)$. The $b \bar b$ pair in $\Upsilon$ and $h_b$ have spin $1$ and $0$, respectively. The spin flip transition is forbidden in the static limit, thus the decay $\Upsilon(5S)\to h_b\, \pi^+\pi^-$ should be heavily suppressed. On the contrary, the rate is sizeable and dominated by the intermediate $Z_b^{(\prime)}$, which appears to be superpositions of heavy-quark spin singlet and triplet. The averaged masses and widths are $M=(10607.2 \pm 2.0) \,\text{MeV}$,  $\Gamma = (18.4 \pm 2.4) \,\text{MeV}$, and $M^\prime=(10652.2 \pm 1.5) \,\text{MeV}$, $\Gamma^\prime = (11.5 \pm 2.2) \,\text{MeV}$, respectively~\cite{Belle:2011aa, Krokovny:2013mgx, Garmash:2014dhx}. They both decay into the closest open bottom pair \cite{Garmash:2015rfd}.  Notably, the analogous $X_b$ with $1^{++}$ has not yet been seen.

\smallskip

\noindent\underline{The $Y$ vector states.}\\
Electron+positron colliders can directly produce $J^{PC}=1^{--}$ states. This process occurs if the center-of-mass energy coincides with the mass of a resonance, or if an energetic photon $\gamma_\text{ISR}$ is emitted by the initial state, effectively reducing the center-of-mass energy to the resonance mass.  Consequently, $B$-factories were able to discover many unexpected charmonium-like $J^{PC}=1^{--}$ systems, usually called $Y$ states.  Their identification as exotics owed mainly to the overpopulation of the sector: all quark model slots for $\psi$ and $\Upsilon$ systems had already been filled.

The $Y(4260)$ was found in the reaction $e^+e^- \to J/\psi\,\pi^+\pi^-$ \cite{Aubert:2005rm, He:2006kg, Yuan:2007sj, Lees:2012pv, Liu:2013dau}:  mass $M= 4251 \pm9 \,\text{MeV}$ and width $\Gamma = 120 \pm 12 \,\text{MeV}$. The higher statistics analysis by BESIII suggests that this peak actually results from the interference between two resonances, called $Y(4230)$ and $Y(4390)$ \cite{Ablikim:2016qzw}. These two are loosely compatible with peaks seen in $h_c\, \pi^+\pi^-$, $\psi'\pi^+\pi^-$ \cite{BESIII:2016adj, Ablikim:2017oaf}, and the lighter system also seems to appear in $\chi_{c0}\, \omega$ and $\pi^+ D^0 D^{*-}$ \cite{Ablikim:2014qwy, Ablikim:2018vxx} (see the summary in Table~\ref{ytab}). The radiative decay $Y(4260)\to \gamma X(3872)$ has also been reported \cite{Ablikim:2013dyn}.

A heavier $Y(4630)$ has been seen near the $\Lambda_c^+\Lambda_c^-$ threshold, with $M = (4634^{+8}_{-7}{}_{-8}^{+5})\,\text{MeV}$ and $\Gamma = (92^{+40}_{-24} {} ^{+10}_{-21})\,\text{MeV}$.  The $Y(4630)$ state decays into $\psi^{\prime}\,\pi^+\pi^-$ and $\Lambda_c^+\Lambda_c^-$ \cite{Pakhlova:2008vn, Lees:2012pv}.  The baryonic decay mode is dominant, with the following branching ratio ${\cal B}(Y(4660)\to \Lambda_c^+\Lambda_c^-) / {\cal B}(Y(4660)\to\psi^{\prime}\,\pi^+\pi^-)= 25\pm7$ \cite{Cotugno:2009ys}.

In the hidden bottom sector, a $Y(10750)$ system has recently been reported: mass $M = 10752.7 \pm 5.9^{+0.7}_{-1.1}\,\text{MeV}$ and width $\Gamma = 35.5^{+17.6}_{-11.3}{}^{+3.9}_{-3.3}\,\text{MeV}$ \cite{Abdesselam:2019gth}.

\begin{table}[!t]
\caption{\label{ytab} Mass and width determinations of the $Y(4230)$ and $Y(4390)$ states, in MeV.}
\begin{ruledtabular}
\begin{tabular}{ccccc}
\tstrut
& Mass (MeV) & Width (MeV) & Process & Source \\
\hline
\tstrut
\multirow{5}{*}{$Y(4230)$}
& $4218.4_{-4.5}^{+5.5} \pm 0.9$ & $66.0_{-8.3}^{+12.3}\pm 0.4$ & $h_c \,\pi^+\pi^-$ & \cite{BESIII:2016adj} \\
& $4230\pm 8\pm 6$ & $38\pm 12\pm 2$ & $\chi_{c0}\,\omega$ & \cite{Ablikim:2014qwy} \\
&$4209.5\pm 7.4\pm 1.4$ & $80.1\pm 24.6\pm 2.9$ & $\psi' \pi^+\pi^-$ & \cite{Ablikim:2017oaf} \\
& $4222.0\pm 3.1\pm 1.4$ & $44.1\pm 4.3\pm 2.0$ & $J/\psi \, \pi^+\pi^-$ & \cite{Ablikim:2016qzw} \\
& $4228.6\pm 4.1\pm 5.9$ & $77.1\pm 6.8\pm 6.9$ & $\pi^+D^{0}D^{*-}$ & \cite{Ablikim:2018vxx} \\
\hline
\tstrut
\multirow{3}{*}{$Y(4390)$} & $4320.0\pm 10.4\pm 7.0$ & $101.4_{-19.7}^{+25.3}\pm10.2$ & $J/\psi \, \pi^+\pi^- $ & \cite{Ablikim:2016qzw}\\
& $4391.5_{-6.8}^{+6.3}\pm1.0$ & $139.5_{-20.6}^{+16.2}\pm 0.6$ & $h_c \,\pi^+\pi^-$ & \cite{BESIII:2016adj} \\
& $4383.8\pm 4.2\pm 0.8$ & $84.2\pm 12.5\pm 2.1$ & $\psi' \pi^+\pi^-$ & \cite{Ablikim:2017oaf} \\
\end{tabular}
\end{ruledtabular}
\end{table}

\smallskip

\noindent\underline{Mesons seen in B decays.}\\
The large sample of $B$ mesons collected at LHCb and with Belle enables refined amplitude analyses to probe for the presence of exotic resonances. These typically appear with small fit fractions in Dalitz plots dominated by ordinary resonances in the crossed channels. Notably, they appear to be substantially broader than the other exotic candidates mentioned so far. While the $J/\psi\, \phi$ neutral resonances also admit an ordinary charmonium explanation \cite{Ortega:2016hde}, the charged systems are manifestly exotic. The $Z(4430)$ and the $Z(4200)$ appear in both $J/\psi\, \pi^+$ and $\psi' \pi^+$ final states, with different significance. The list of Z systems is summarised in Table~\ref{XZs}.

\begin{table}[!t]
\caption{\label{XZs} Potentially exotic states seen in $B$ meson decays.  The quantum numbers reported are favoured in the fits, but need confirmation.  \emph{Upper panel}. -- charged $Z$ states; and \emph{lower panel} neutral $X$ systems seen in $J/\psi \,\phi$ resonances. The $Z(4200)$ seen in $\psi'\pi^+$ at LHCb has $0^{--}$ as the most favoured assignment, although $1^{+-}$ is compatible within $1\sigma$ \cite{Aaij:2014jqa}. Identification with the state seen by Belle in $J/\psi\, \pi^+$ \cite{Chilikin:2014bkk} supports the latter assignment.}
\centering
\begin{ruledtabular}
\begin{tabular}{cccccccccc}
\tstrut
& $J^{PC}$ & Mass (MeV) & Width (MeV) & Process & Source \\
\hline
\tstrut
$Z(4050)$ & $?^{?+}$ & $4051^{+24}_{-40}$ & $82^{+50}_{-28}$ & \multirow{2}{*}{$\bar B^0\to K^- (\chi_{c1}\,\pi^+)$} & \multirow{2}{*}{\cite{Mizuk:2008me}} \\
$Z(4250)$ & $?^{?+}$ & $4248^{+190}_{-50}$ & $177^{+320}_{-70}$ \\
$Z(4100)$ & $1^{-+}$ & $4096\pm20^{+18}_{-22}$ & $152\pm58^{+60}_{-35}$ & $\bar B^0\to K^- (\eta_c\,\pi^+)$ & \cite{Aaij:2018bla} \\
\multirow{2}{*}{$Z(4200)$} & $1^{+-}$ & $4196^{+35}_{-32}$ & $370^{+100}_{-150}$&$\bar B^0\to K^- (J/\psi \,\pi^+)$ & \cite{Chilikin:2014bkk} \\
& $1^{+-}$ & $4239^{+50}_{-21}$ & $220^{+120}_{-90}$ & $\bar B^0\to K^- (\psi' \pi^+)$ & \cite{Aaij:2014jqa} \\
$Z(4430)$ & $1^{+-}$ & $4478^{+15}_{-18}$ & $181\pm31$ & $\bar B^0\to K^- (\psi' \pi^+)$ & \cite{Aaij:2014jqa} \\
\hline
\tstrut
$X(4140)$ & $1^{++}$& $4146.5 \pm 4.5^{+4.6}_{-2.8}$ & $83 \pm 21^{+21}_{-14} $ & \multirow{4}{*}{$B^-\to K^- (J/\psi\, \phi)$} & \multirow{4}{*}{\cite{Aaij:2016nsc}} \\
$X(4274)$ & $1^{++}$ & $4273.3 \pm 8.3^{+17.2}_{-\ 3.6}$ & $56 \pm 11^{+\ 8}_{-11}$ \\
$X(4500)$ & $0^{++}$ & $4506 \pm 11^{+12}_{-15}$ & $92 \pm 21^{+21}_{-20}$ \\
$X(4700)$ & $0^{++}$ & $4704 \pm 10^{+14}_{-24}$ & $120 \pm 31^{+42}_{-33}$ \\
\end{tabular}
\end{ruledtabular}
\end{table}

\smallskip

\noindent\underline{Pentaquarks.}\\
The LHCb Collaboration has observed several pentaquark candidates in the $\Lambda_b^0 \to (J/\psi p) K^-$ decay. The first amplitude analysis found two pentaquarks $P_c(4380)$ and $P_c(4450)$, with masses and widths, respectively, $M_1= 4380\pm 8\pm 29 \,\text{MeV}$, $\Gamma_1=205\pm 18\pm 86\,\text{MeV}$, and $M_2=4449.8\pm 1.7\pm 2.5\,\text{MeV}$, $\Gamma_2 =39\pm 5\pm 19\,\text{MeV}$ \cite{Aaij:2015tga}. The quantum number assignment is not conclusive, but the interference pattern suggests that the systems have opposite parities, with $J^{PC}_1 = {\frac{3}{2}}^-$ and $J^{PC}_2 = {\frac{5}{2}}^+$ favoured. A more recent, higher statistics one-dimensional analysis \cite{Aaij:2019vzc} was able to separate the latter into two states, $P_c(4440)$ and $P_c(4457)$, with mass and width, respectively, $M_{2a} = 4440.3 \pm 1.3^{+4.1}_{-4.7} \,\text{MeV}$, $\Gamma_{2a} = 20.6\pm4.9_{-10.1}^{+8.7}\,\text{MeV}$ and $M_{2b} = 4457.3 \pm 0.6^{+4.1}_{-1.7} \,\text{MeV}$, $\Gamma_{2b} = 6.4 \pm 2.0_{-1.9}^{+5.7} \,\text{MeV}$. Furthermore, a new $P_c(4312)$ is also indicated, $M_3 = 4457.3 \pm 0.6^{+4.1}_{-1.7} \,\text{MeV}$, $\Gamma_3 = 6.4\pm2.0_{-1.9}^{+5.7}$. No amplitude analyses are available to determine the quantum numbers of these new states, so the quantum number assignments must be read with caution.


\subsubsection{Theoretical tools for analyses of exotics}
\noindent\underline{Amplitude analysis.}\\
Some of the experimental analyses suffer from poor amplitude models. This can lead to misleading statements about the existence of exotic resonances and artificially inflate the number of states, as might have occurred in the $Y$ sector.  (Ref.\,\cite{Rodas:2018owy} describes an example in a light-quark system.)  Unfortunately, implementing more refined amplitudes in data analysis is not easy and requires an interplay between the work of theorists and experimentalists. On the basis of published data, some conclusions can be drawn. The properties of the amplitude can hint toward the nature of potentially exotic states; in particular, whether they are more likely to be a consequence of short-range QCD physics (as diquarks), or be driven by long-range exchange forces (as molecules). It can also happen that hadron interactions produce peaks that are not indications of a bound-state or resonance, \emph{e.g}.\ triangle re-scattering mechanisms \cite{Szczepaniak:2015eza} or virtual (unbound) states \cite{Pilloni:2016obd}.

A global analysis of available data on the $Z_c(3900)$ challenges several hypotheses, each of which was found to be consistent with the present quality of data \cite{Albaladejo:2015lob, Pilloni:2016obd}. Conversely, a local analysis of data in the neighbourhood of the $P_c(4312)$ peak points to a virtual state interpretation, \emph{i.e}.\ an enhancement owing to hadron-hadron interactions not strong enough to bind a new state \cite{Fernandez-Ramirez:2019koa}. It is worth mentioning, however, that there is an analysis based on chiral perturbation theory that suggests the existence of several bound states in the $J/\psi \,p$ invariant mass \cite{Du:2019pij}.

Another reaction where pentaquarks are expected to be seen is direct photoproduction \cite{Karliner:2015voa, Kubarovsky:2015aaa, Blin:2016dlf, Winney:2019edt}. Such observations would help in excluding rescattering mechanisms, which feature in multibody final states.  The available GlueX analysis does not see any substantial peaks and places upper limits on the branching ratios for $P_c \to J/\psi \,p$ by assuming a vector meson dominance (VMD) model \cite{Ali:2019lzf}.  (N.B.\ VMD is a poor tool for such analyses and likely leads to overestimates \cite{Wu:2019adv, Xu:2019ilh}; hence, it is difficult to judge the significance of the quoted bounds.)

\smallskip

\noindent\underline{Diquarks as building blocks of exotics.}\\
When considering tetra- and penta-quark states, one approach is to use the model Hamiltonian of Ref.\,\cite{Godfrey:1985xj} and solve the associated few-body problem, \emph{e.g}.\ Ref.\,\cite{Richard:2018yrm}).
On the other hand, as noted above, since one gluon exchange in the $\bar {\bm 3}_c$ colour channel is attractive and repulsive for ${\bm 6}_c$, it is common to suppose that $\bar {\bm 3}_c$ diquark correlations act as dominant collective degrees-of-freedom in such systems.  Evidence that the two quarks in a tetraquark system arrange their colour in a diquark configuration before interacting with the antiquarks has also been found in static-limit lQCD simulations \cite{Cardoso:2011fq}.  This simulation also indicates that the four constituents arrange themselves into a H-shaped configuration.  This picture can explain large isospin mixing amongst neutral states and a preference to decay into baryons \cite{Rossi:2016szw}.

The crudest approximation is to work with pointlike diquarks by effectively absorbing all spatial dependence of the quark-quark interaction in Ref.\,\cite{Godfrey:1985xj} into a renormalised diquark mass that is fitted to data.  As canvassed above, taken literally, this approximation conflicts with many more rigorous analyses; nevertheless, the scheme may be useful in developing insights.  Adopting this perspective, the colour-spin Hamiltonian can be reduced to \cite{Jaffe:2004ph}
\begin{equation}
V_{ij}=-2 \kappa_{ij} \, \bm S_i\cdot \bm S_j\,\frac{\lambda^a_i}{2}\cdot \frac{\lambda^a_j}{2},
\label{colorspin}
\end{equation}
where $\kappa_{ij}$ are unknown effective couplings.

\begin{figure}[!t]
\includegraphics[width=0.65\textwidth]{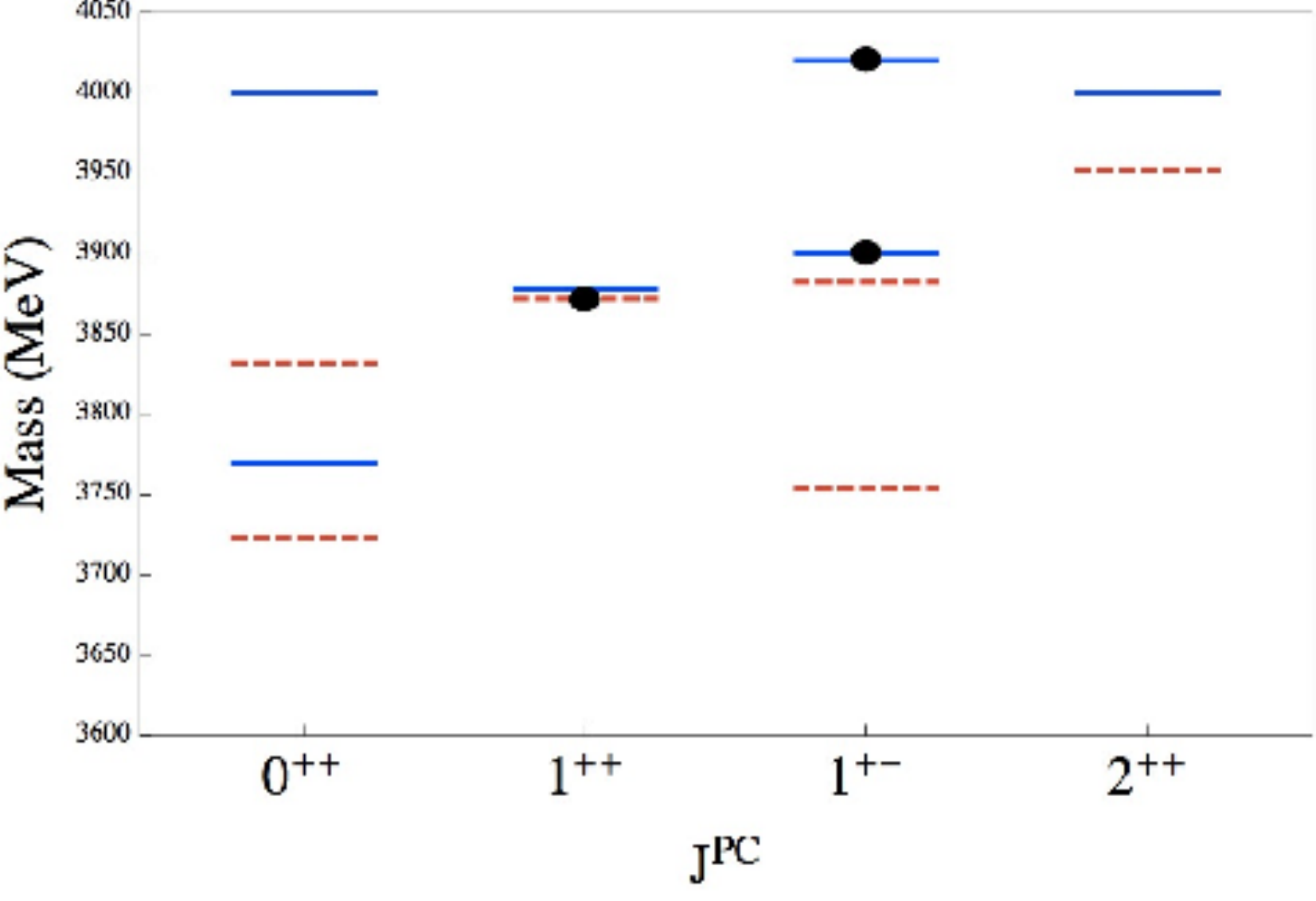}
\caption{\label{spettroXf}
Spectrum of $[cq][\bar c \bar q]$ tetraquarks as described in the diquark+antidiquark model of Ref.\,\cite{Maiani:2014aja}, which is a modification of an earlier model that produced the spectrum marked by dashed red lines \cite{Maiani:2004vq} and depicted in Fig.\,\ref{fig:X(3872)} above.}
\end{figure}

As discussed in connection with Fig.\,\ref{fig:X(3872)}, this approach has been used \cite{Maiani:2004vq} to analyse the $X(3872)$.  A $1^{+-}$ state appears, almost degenerate with the $X(3872)$, and can be identified with the $Z_c(3900)$. The existence of a lighter $Z'_c(3750)$ state with same quantum numbers as the $Z_c(3900)$ can be also justified \cite{Faccini:2013lda}.  However, the discovery of a heavier $Z'_c(4020)$ challenged the picture.  Subsequently, Ref.\,\cite{Maiani:2014aja} revised Eq.\,\eqref{MaianiTI}, positing that spin interactions within a compact diquark dominate over all other possible two-body pairings.  This implies that the Hamiltonian in Eq.\,\eqref{colorspin} is diagonal in the diquark basis, and that the spectrum can be reproduced by tuning $\kappa_{cq} = 67\,\text{MeV}$ \cite{Maiani:2014aja}, as shown in Fig.\,\ref{spettroXf}. By including a spin-orbit term, the spectrum of the vector $Y$ state can also be described, as the $P$-wave excitations of the ground-state multiplet \cite{Ali:2017wsf}. The $Z(4430)$ can also be accommodated as the radial excitation of the $Z_c(3900)$.  This picture is summarised in Table\,\ref{tab:states}.

Following a similar track, compact-diquark interpretations of the $J/\psi \,\phi$ states \cite{Maiani:2016wlq}, the $Z_b^{(\prime)}$ and $Y_b$ \cite{Ali:2019okl} and the pentaquarks \cite{Maiani:2015vwa, Ali:2019clg} have been presented. Other calculations of hidden charm tetraquarks based on pointlike diquarks are found in Ref.\,\cite{Anwar:2018sol}. Doubly heavy tetraquarks have not been seen, although evidence for a $bb \bar q \bar q'$ state is available from lQCD \cite{Francis:2016hui}. For these states, a diquark+antidiquark description is potentially favourable, because the heavy+heavy diquark could be much smaller than the size of the state \cite{Karliner:2017qjm, Eichten:2017ffp}. Conversely, the existence of a double-charm state is still unclear \cite{Carames:2011zz, Esposito:2013fma, Guerrieri:2014nxa, Cheung:2017tnt}. The main limitation of these models is the proliferation of states, much more than are observed. For each level predicted, an isovector and an isoscalar degenerate state appears, which is inconsistent with extent data. Also, the proximity of several states to open charm thresholds is not a natural consequence within compact diquark+antidiquark models.  Proposals to ameliorate this feature have been presented \cite{Blitz:2015nra, Esposito:2016itg}.

\begin{table}[!t]
\caption{\label{tab:states} $J^{PC}=1^{--}$ tetraquarks involving a diquark+antidiquark $[cq][\bar c \bar q']$ pair in $S$- and $P$-wave. The $Y(4220)$ and $Y(4330)$ are the two components of the $Y(4260)$ peak,  disentangled by BESIII in $J/\psi\,\pi^+\pi^-$. The $Y(4390)$ is identified as a separate state decaying into $h_c \pi^+\pi^-$.}
\begin{ruledtabular}
\begin{tabular}{lcc|lcc}
 &  $\left| S_{cq}, S_{\bar c\bar q'}; S, L \right\rangle_J$
 & experiment
 &
 &
$\left| S_{cq}, S_{\bar c\bar q'}; S, L \right\rangle_J$ & experiment\\
\colrule
\hline
$X_0$ & $\left| 0, 0; 0, 0 \right\rangle_0$ & & $Y_1$ & $\left| 0, 0; 0, 1 \right\rangle_1$ & $Y(4220)$
\\
$X_{1^{++}}$ & $\tfrac{1}{\surd 2}\left( \left| 1, 0; 1, 0 \right\rangle_1 + \left| 0, 1; 1, 0 \right\rangle_1 \right)$ & $X(3872)$ & $Y_2$ & $\tfrac{1}{\surd 2}\left( \left| 1, 0; 1, 1 \right\rangle_1 + \left| 0, 1; 1, 1 \right\rangle_1 \right)$ & $Y(4330)$
\\
$Z_{1^{+-}}$ & $\tfrac{1}{\surd 2}\left( \left| 1, 0; 1, 0 \right\rangle_1 - \left| 0, 1; 1, 0 \right\rangle_1 \right)$ & $Z_c(3900)$ & & &\\
$Z'_{1^{+-}}$ & $\left| 1, 1; 1, 0 \right\rangle_1$ & $Z_c'(4020)$ & $Y_3$ &
$\left| 1, 1; 0, 1 \right\rangle_1$ & $Y(4390)$
\\
$X_0'$ & $\left| 1, 1; 0, 0 \right\rangle_0$ & & & \\
$X_2'$ & $\left| 1, 1; 2, 0 \right\rangle_2$ & & $Y_4$ &
$\left| 1, 1; 2, 1 \right\rangle_1$ & $Y(4660)$
\\
\end{tabular}
\end{ruledtabular}
\end{table}

Some longstanding questions concerning diquark+antidiquark models of exotic resonances can be answered by supposing that compact diquarks and antidiquarks in tetraquark systems are separated by a potential barrier \cite{Maiani:2017kyi, Esposito:2018cwh}.  This picture also explains the larger branching ratio into open charm mesons with respect to the hidden charm systems.  Tuning the parameters of the barrier, the widths of several states are well reproduced. Another dynamical picture of diquarks (and triquarks) being produced in bottom meson (and baryon) decays at finite distance is presented in Ref.\,\cite{Brodsky:2014xia}.

The material in Secs.\,\ref{sec:DSEs}, \ref{sec:lattice} provides ample evidence that QCD does not support pointlike diquarks and reveals how nonpointlike diquark degrees-of-freedom can be exploited to describe a wide variety of observable hadronic phenomena.  Within QMs, too, the pointlike-diquark restriction can be lifted. For instance, diquark+antidiquark dynamics can be disentangled from the dynamics within the correlation by using a Born-Oppenheimer approximation \cite{Giron:2019bcs, Giron:2019cfc}.  Alternatively, one can separate the dynamics of the $c\bar c$ pair: if the latter is found in colour octet, the seed for a repulsive barrier is given, and a double well potential may be justified \cite{Maiani:2019cwl, Maiani:2019lpu}. 


\subsubsection{Production of exotic states in $pp$ and heavy-ion collisions}

The large prompt-production cross-section of the $X(3872)$ in high-energy collisions has triggered many debates on whether or not it is compatible with a pure molecular nature.  At issue is whether the $D^0$ and $\bar D^{*0}$ pair (constituting the $X(3872)$) can be produced with a relative momentum, $k_\text{rel}$, that is small enough for binding to occur when the initial collision happens at TeV energies. The distribution of pairs with momentum smaller than a given $k_\text{rel}$ can be estimated using Monte Carlo generators. However, this relies on estimating a $k_\text{max}$ that makes the binding possible. Several choices have been made, leading to mutually conflicting conclusions \cite{Artoisenet:2010uu, Esposito:2013ada, Guerrieri:2014gfa, Barabanov:2016jjv, Albaladejo:2017blx, Esposito:2017qef, Wang:2017gay, Braaten:2018eov}. Alternatively, in a scenario where the molecule mixes with the unobserved $\chi_{c1}(2P)$ \cite{Ortega:2009hj, Ferretti:2013faa, Ferretti:2014xqa, Ferretti:2018tco}, the production proceeds through the charmonium component.

\begin{figure}[t]
\begin{tabular}{ccc}
\includegraphics[width=.45\textwidth]{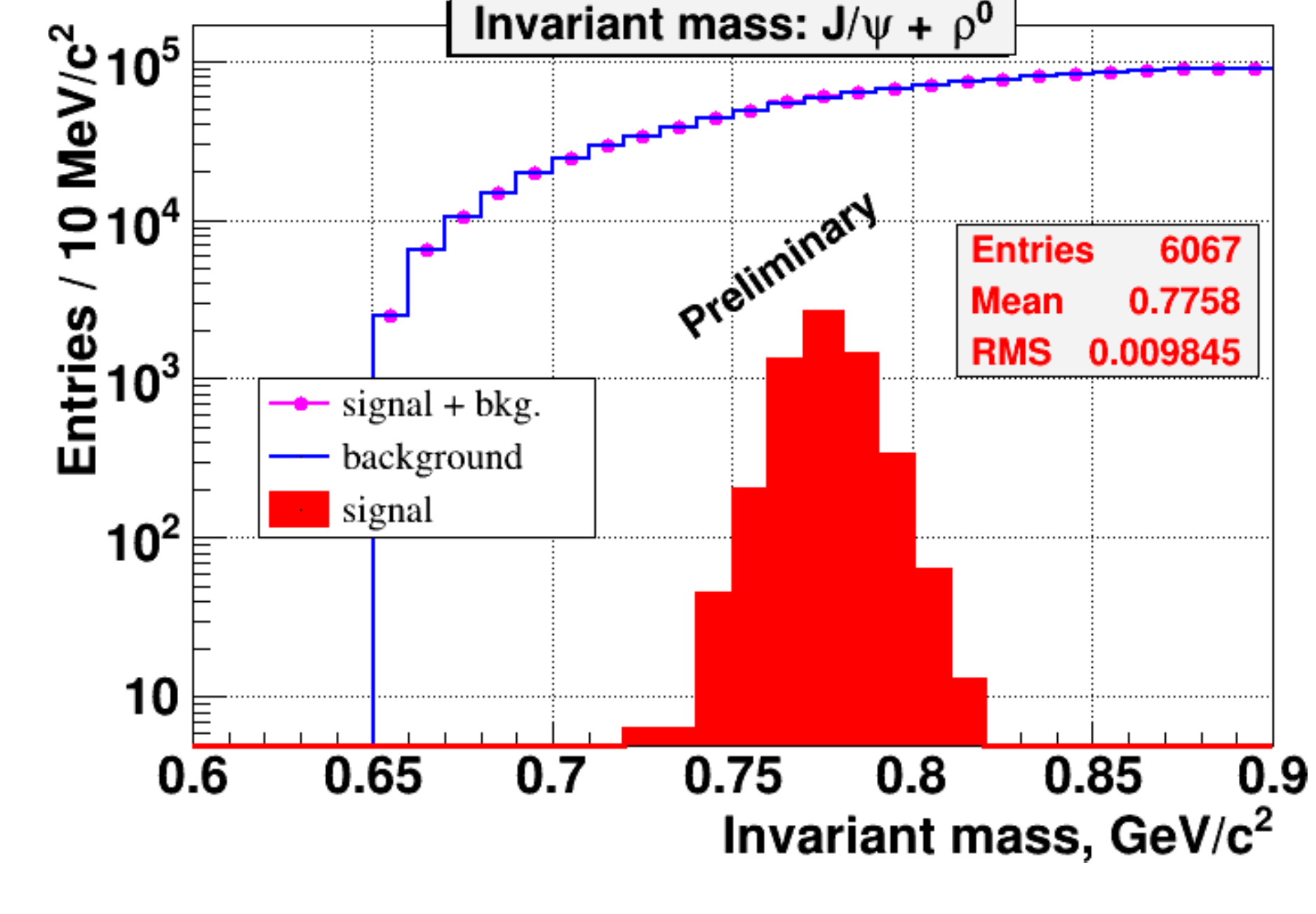} & \hspace*{2em} &
\includegraphics[width=.45\textwidth]{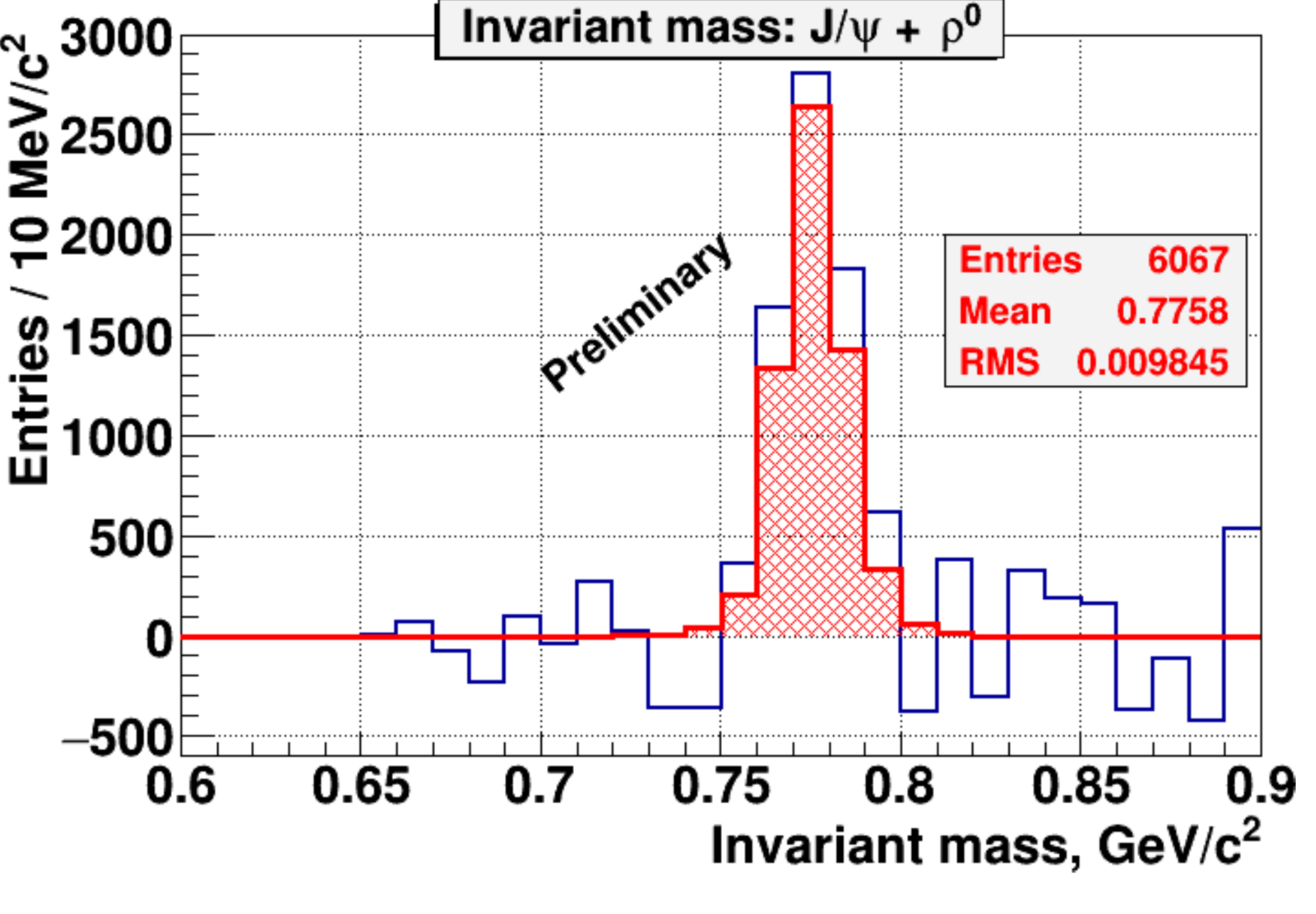}
\end{tabular}
\caption{\label{bara1} Invariant mass combination $M_{e^+ e^- \pi^+\pi^-} - M_{e^+ e^-}$ reconstructed from PYTHIA simulations. \emph{Left panel} -- signal and background are shown separately; and \emph{right panel} -- signal is reconstructed, \emph{viz}. the blue line is the background-subtracted histogram, while the red line shows the true $X(3872)$ events.}
\end{figure}

In Ref.\,\cite{Esposito:2015fsa}, the $pp$ production cross-section for the $X(3872)$ and deuteron states were compared. The $X(3872)$ exceeds the latter by a few orders of magnitude, suggesting that the two do not share the same nature. The behaviour in heavy ion collisions is also sensitive to the character of the $X(3872)$ \cite{Cho:2013rpa}.
The experiments planned at NICA may be well suited for testing these hypotheses.  NICA will provide colliding heavy-ion beams with luminosity up to $10^{32}\,\text{cm}^{-2}\text{s}^{-1}$ at centre-of-mass energy $\sqrt{s_\text{NN}} = 26\,\text{GeV}$ \cite{Sissakian:2009zza}.

NICA's good tracking and particle identification performance over a significant fraction of the phase space can provide a good opportunity to extend its ambitious physics program to study heavy mesons via their decays to electrons, hadrons or photons \cite{Barabanov:2019izj}. The $X(3872)$ was simulated using PYTHIA8 \cite{Sjostrand:2014zea}, assuming it is a $1^{++}$ charmonium state, with branching ratio to $J/\psi\, \rho$ of 5\% \cite{Li:2019kpj}: 1000 $X(3872)\to J/\psi\rho\to e^+e^-\pi^+\pi^-$ events are predicted for a 10 day run. The difference $M_{e^+ e^- \pi^+\pi^-} - M_{e^+ e^-}$ is shown in Fig.~\ref{bara1}. Background events are also simulated. The plots correspond to statistics collected in 10 months at the nominal luminosity. The sidebands are fitted to a polynomial function. Subtracting that function from the original distribution, one observes a clear peak of the $X(3872)$ decay.

As an extension of this topic, one can consider looking at other decay modes of the $X(3872)$. Since the branching ratio of $X(3872)$ to open charm is dominant, one should consider the possibility of reconstructing this state from the hadronic decays of charm mesons. For such a study, it will be important to use the silicon microvertex detector to tag the $D$-meson flight length. Evidently, this physics topic can develop synergistically with the heavy ion charm program at NICA.

\section{Future prospects for diquarks}
\label{sec:Future}

\subsection{Super BigBite Spectrometer Programme on High-$Q^2$ Space-like Nucleon Form Factors}
\label{SBSSLNFSS}
The electromagnetic form factors (EMFFs) measured in elastic lepton-nucleon scattering are key elements in resolving the role of diquark correlations in nucleon structure. At large momentum transfers, the unpolarised differential scattering cross-section is dominated by the magnetic form factor, $G_M$. On the other hand, the ratio of the electric and magnetic form factors at high-$Q^2$ is best determined using polarisation observables. To measure the EMFFs at large values of $Q^2$ is very challenging owing to the rapid decrease of the elastic scattering cross-section $d\sigma/d\Omega_e$ as roughly $Q^{-12}$. Presently, the Continuous Electron Beam Accelerator Facility (CEBAF) at JLab is the only electron-beam facility in the world with the luminosity, duty cycle, energy and polarisation capabilities to measure nucleon form factors at large $Q^2$.  (\emph{N.B}.\ Herein, ``large'' is defined relative to the $Q^2$ coverage and precision of existing data.)

The most recent polarisation transfer measurements of the proton form factor ratio $G_E^p/G_M^p$ from JLab's experimental Halls A and C (see, for instance, Refs.~\cite{Puckett:2017flj, Puckett:2010ac}) reached $Q^2 = 8.5\,$GeV$^2$, at which value the ratio $G_E^p/G_M^p$ was found to be consistent with zero, albeit with relatively large statistical uncertainty. To extend these measurements to larger values of $Q^2$ using small-acceptance spectrometers, such as those used in previous experiments of this type \cite{Jones:1999rz, Gayou:2001qd, Punjabi:2005wq, Puckett:2010ac, Meziane:2010xc, Puckett:2011xg, Puckett:2017flj}, would require prohibitive beam time.

The most promising path to enlarge the statistical figure-of-merit for measurements of polarisation observables in high-$Q^2$ elastic and quasi-elastic electron-nucleon scattering is to increase the solid angle and $Q^2$ acceptance of both the electron and proton arms of the experiments.  Experiment E12-07-109 \cite{GEP5, GEP5_PAC47} was approved by the Jefferson Lab Program Advisory Committee to measure $G_E^p/G_M^p$ to $Q^2 = 12$ GeV$^2$ using the polarisation transfer method.
With this experiment as the original motivation, the collection of apparatus known as the Super BigBite Spectrometer (SBS) was designed and constructed to carry out a comprehensive programme of high-$Q^2$ nucleon EMFF measurements, including: E12-09-019 \cite{GMN} to measure the neutron magnetic form factor $G_M^n$ to 13.5\,GeV$^2$ using the ``ratio method'' on a liquid deuterium target; E12-09-016 \cite{GEN2} to measure $\mu_n G_E^n/G_M^n$ to 10.2\,GeV$^2$ using a high-luminosity polarised $^3$He target, enabled by convection-driven circulation of polarised gas \cite{Wojtsekhowski:2017kti, Helium3article, Helium3_target2015}; and E12-17-004 \cite{GENRP} to measure $\mu_n G_E^n/G_M^n$ at $Q^2 = 4.5$\,GeV$^2$ using the technique of charge-exchange recoil polarimetry \cite{Basilev:2019sno} for the first time in the context of a nucleon form factor measurement.
Several additional hadron structure measurements using SBS have also been approved, including single-spin asymmetries in semi-inclusive deep inelastic scattering \cite{SBS_SIDIS}, and tagged deep inelastic scattering from the nucleon's pion cloud \cite{JlabTDIS1}.

\begin{figure*}
\begin{center}
\includegraphics[width=0.95\textwidth,height=0.40\textheight]{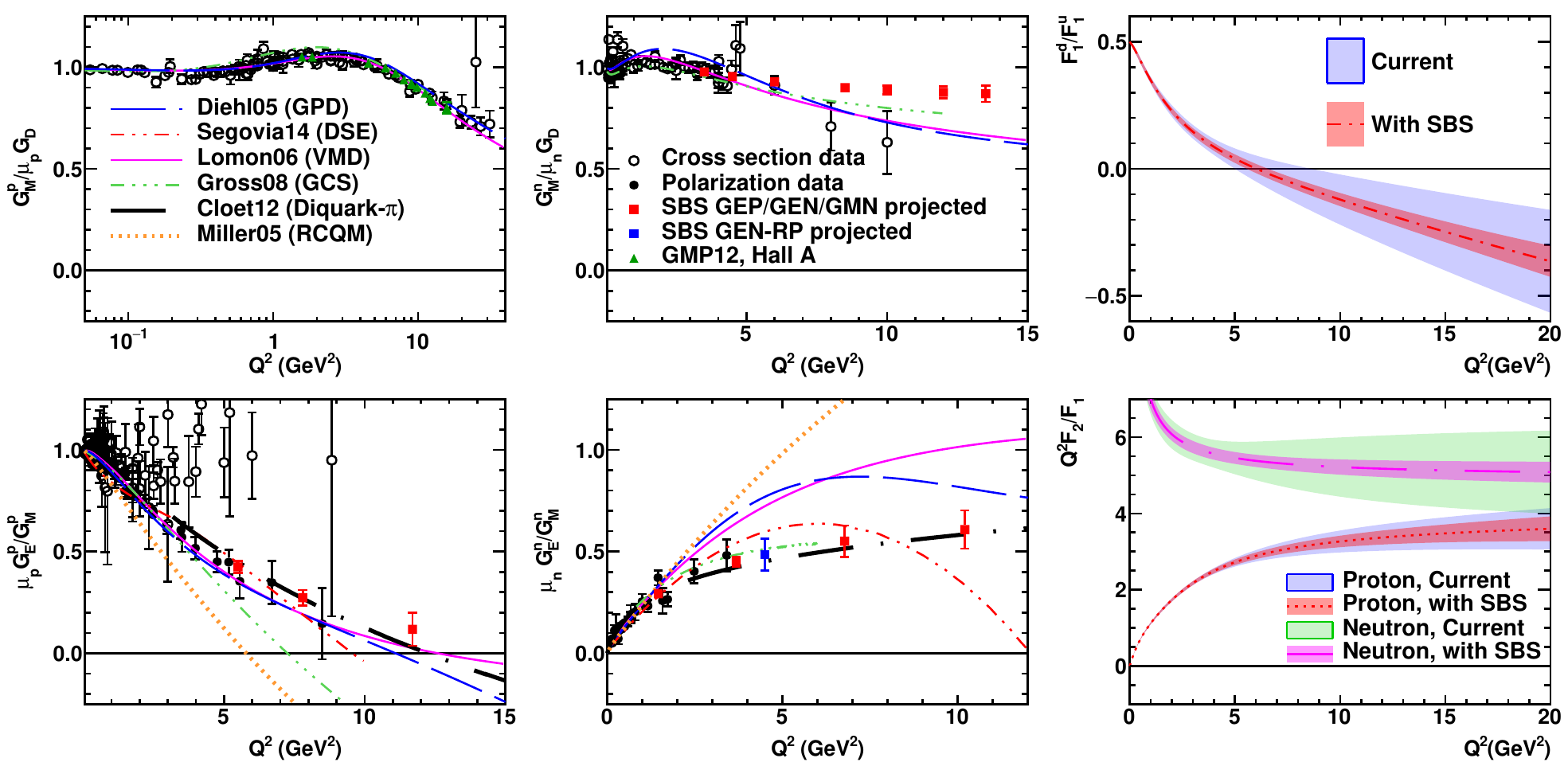}
\end{center}
\vspace*{-0.20cm}
\caption{\label{fig:SBSGEP} \label{fig:SBSFF_summaryfig} Projected results from the SBS form factor programme, compared to existing data, including preliminary results for $G_M^p$ extracted from the recent high-$Q^2$ elastic $ep$ cross-section measurements in Hall A \cite{E12-07-108}, and selected theoretical predictions.  Projected SBS results are plotted at values extrapolated from the global fit described in the appendix of Ref.\,\cite{Puckett:2017flj} for the proton, the Ref.\,\cite{Kelly:2004hm} fit (FFF2004) for $G_M^n$, and Ref.\,\cite{Riordan:2010id} for $G_E^n$. Theoretical curves are from the GPD-based model of Ref.\,\cite{Diehl:2004cx} (Diehl05), the DSE calculation of Ref.\,\cite{Segovia:2014aza} (Segovia14), the VMD model of Ref.\,\cite{Lomon:2006xb, Lomon:2012pn} (Lomon06), the covariant spectator model from Ref.\,\cite{Gross:2006fg} (Gross08), the constituent quark+diquark model calculation from Ref.\,\cite{Cloet:2012cy} (Cloet12), and a relativistic constituent quark model calculation in Ref.\,\cite{Miller:2002ig} (Miller05). Top right: projected impact of SBS programme on the flavor separated form factor ratio $F_1^d/F_1^u$, with central value and uncertainty band evaluated using the fit from Ref.\,\cite{Kelly:2004hm}. The improvement in this ratio owes mainly to the new $G_M^n$ data. Bottom right: projected impact of SBS programme on the ratio $Q^2F_2/F_1$ of Pauli and Dirac form factors for the proton and neutron, also evaluated using the FFF2004 parametrisation. The improvement in these ratios is driven by the $G_E^p$ and $G_E^n$ data.}
\end{figure*}

Figure\,\ref{fig:SBSFF_summaryfig} shows the $Q^2$ coverage and projected precision of the data expected from the SBS EMFF programme, compared to existing data and selected theoretical predictions. It also shows the projected impact of the SBS programme on the flavor separated ratio of Dirac form factors $F_1^d/F_1^u$ and the Pauli-Dirac ratios $F_2^{p,n}/F_1^{p,n}$, in an analysis based on FFF2004 \cite{Kelly:2004hm}. The analysis suggests that $F_1^d$, which is mainly constrained by $G_M^n$, could become negative around $Q^2 = 7$ GeV$^2$.
This possibility poses challenges to many theory models, particularly the GPD framework; but according to DSE analyses -- Refs.\,\cite{Segovia:2014aza, Segovia:2015ufa, Cui:2020rmu} and Sec.\,\ref{QCDkindred}, it follows naturally from the presence of both scalar and axial-vector diquark correlations in the proton.

Note that the extrapolation of fit-based uncertainties beyond the $Q^2$ range of existing data understates the uncertainty in the true form factor behaviour in the unexplored $Q^2$ regime. This is underscored by the dispersion of theoretical models when extrapolating beyond the $Q^2$ range of existing data. Such an exercise is nonetheless useful in visualising the impact of new data. In approximately 125 days of approved beam time, the SBS programme will extend the $Q^2$ reach and precision of spacelike form factor data far into currently uncharted territory. Data of such kinematic coverage and precision will severely challenge the most sophisticated theoretical descriptions of nucleon structure in the transition region between nonperturbative and perturbative QCD, and the flavour-separation enabled by combined proton and neutron measurements is amongst the most sensitive experimental signatures of diquark degrees of freedom. The first SBS experiment is slated for installation later in 2020, coinciding with a planned shutdown of CEBAF for machine improvements. Currently, the SBS programme is projected to begin in the summer of 2021.

The SBS programme is enabled by several key conceptual innovations. First, a large dipole magnet with $\int B dL \approx 2.5\ T\cdot m$ and a cut in the yoke for passage of the beam pipe is used to realise moderately large solid-angle acceptance and large momentum bite at the forward scattering angles of elastically (and quasi-elastically) scattered nucleons in $eN \rightarrow eN$ scattering at large $Q^2$ values. The SBS dipole field provides momentum analysis and also precesses the longitudinal component of the scattered nucleon's spin into a transverse component that can be measured in a subsequent analysing reaction in the recoil polarisation experiments. Specially designed active and passive magnetic shielding is used to suppress the transverse components of the SBS fringe field along the downstream beam pipe to avoid and/or correct the steering of the primary beam downstream of the target. Gas Electron Multiplier (GEM) technology \cite{Sauli:1997qp, Gnanvo:2014hpa} is used extensively for charged-particle tracking in all SBS experiments. Unlike traditional multiwire drift chambers or proportional chambers, GEMs can operate with stable gain and good tracking performance in the high-luminosity environment of Hall A (up to $6\times 10^{38}$ cm$^{-2}$ s$^{-1}$), even with direct line-of-sight to the target. While the SBS dipole field blocks low-energy charged particles from reaching the GEMs, the large flux of low-energy photons leads to a significant background counting rate. An iron-scintillator hadronic calorimeter (HCAL) located behind the SBS magnet is used in all the SBS experiments, playing several important roles. HCAL provides an efficient trigger for high-energy nucleons, preferentially selects forward elastic $\vec{p} + \text{CH}_2$ scattering events with high analysing power in the $G_E^p$ measurement, aids track reconstruction in the back GEMs by constraining the coordinates of the proton tracks, and also detects forward neutrons (protons) from charge-exchange $\vec{p}n \rightarrow np$ or $\vec{n}p\rightarrow pn$ scattering.

Clean selection of elastic or quasi-elastic $eN$ events in the presence of dominant inelastic backgrounds in large-$Q^2$ $eN$ scattering and clean reconstruction of the final-state particle kinematics in the presence of the aforementioned soft photon backgrounds in the GEMs requires coincident detection of both electron and nucleon in the final state. To match the kinematic acceptance of the electron arm to that of the proton arm generally requires an even larger solid angle acceptance for the electron than for the nucleon at large $Q^2$. In the neutron form factor measurements, the existing BigBite spectrometer \cite{Riordan:2010id, RiordanThesis, ObrechtThesis} is used for full momentum, vertex, and scattering angle reconstruction for scattered electrons. The SBS programme, with its higher luminosities and higher beam energies, requires several upgrades to the BigBite detector package, including GEM-based tracking, a highly segmented gas Cherenkov counter for pion rejection, and a more finely segmented scintillator plane for precise timing measurements. In the $G_E^p$ measurement, an even larger solid angle for electron detection is required; and as such, a novel high-temperature lead-glass electromagnetic calorimeter (ECAL) provides the required energy, spatial, and timing resolution at moderate cost \cite{ECAL_NIM}. The continuous thermal annealing of radiation damage to the glass in a 225$^\circ$C oven is required to maintain stable transparency of the glass, and therefore stable energy resolution of the calorimeter, in the high-radiation environment in Hall A. Unlike in previous experiments of this type, the calorimeter energy resolution is a critical performance parameter, because the ECAL needs to be triggered at a high threshold of approximately 80-90\% of the elastically scattered electron energy to achieve a manageable data rate while maintaining a high efficiency for the events of interest.


\subsection{Colour Propagation and Hadron Formation}
\label{CPHF}
As one of the last frontiers of nonperturbative QCD, the ubiquitous processes of colour propagation and hadron formation in quark fragmentation have not yet been understood on the basis of the fundamental theory. The very successful Lund String Model \cite{Andersson:1983ia} (LSM) embedded in PYTHIA \cite{Sjostrand:2006za}, which incorporates a QCD-inspired foundation, and other widely-used Monte Carlo models such as HERWIG \cite{Bellm_2016}, allow a good description of high-energy scattering data without addressing the unsolved theoretical problem of formulating a description of colour propagation based on the QCD Lagrangian.

An important step forward was made in the 1990's by the introduction of new data from the HERMES experiment at the \emph{Deutsches Elektronen-Synchrotron} laboratory. These data were the first from DIS on nuclear targets (nDIS) that involved identified final state hadrons \cite{Airapetian:2000ks, Airapetian:2003mi, Airapetian:2005yh, Airapetian:2007vu, Airapetian:2009jy, Airapetian:2011jp}. Previous nDIS experiments by the European muon collaboration (EMC) \cite{Ashman:1991cx} laid the foundation for these studies (see Ref.\,\cite{Accardi:2009qv}) but the introduction of fully identified hadrons at higher energies was a crucial step that allowed the experimental study of hadron-specific formation mechanisms in the clean environment of lepton DIS. The use of the nuclear medium allows colour propagation mechanisms to be probed at the fm scale; specifically, the use of a variety of nuclear targets and virtual photon energies allows studies spanning the spatial range from $0-10\,$fm using stable atomic nuclei. These distance scales, which greatly exceed any pQCD factorisation limits, allow study of how the colour propagation and hadron formation processes unfold.

As an heuristic example, to form a hadronic system of mass $M = 1\,$GeV and radius $R = 0.5\,$fm in an interaction at higher $x$-Bjorken, $x_{Bj}$, with four-momentum transfer $Q^2=4 \, {\rm GeV}^{\rm2}$ and photon energy $\nu=10\,$GeV, the recoiling hadronic mass $W$ can be estimated as 4\,GeV \cite{DelDuca:1992ru}. Two distinct stages of the process can be identified: the colour propagation stage and the hadron formation stage. The LSM string constant $\kappa$, equal to $1\,$GeV/fm, can be used to estimate the colour propagation stage of the process; namely, the passage of the struck coloured quark through space. Dividing the mass involved by the string constant gives a crude estimate of the colour lifetime of the system: $W/\kappa = 4 \,$fm/c is the distance over which the struck quark emits gluons, thus 4\,fm is the colour lifetime of the struck quark in the rest frame of the string.

More detailed LSM formulations \cite{BL1, Brooks_2019, Ferreres-Sole:2018vgo} add the dependence on the relative energy, $z_h$, as defined below, which can increase and decrease the colour lifetime, $\tau_c$, by approximately a factor of two, and the transverse momentum $p_T$, which increases it by a small amount for the example kinematics discussed here. As a lifetime,  $\tau_c$ increases via time dilation, $\gamma \tau_c$, when boosted to other reference frames, where $\gamma$ is the relativistic boost factor. In the hadron formation stage of this process, the colour-neutral systems formed in the colour propagation stage, such as $q \bar{q}$, $qqq$ and $\bar{q}\bar{q}\bar{q}$ clusters, begin as colour singlet systems without a definite mass and size, and they evolve into the known hadrons and baryons over a finite time interval. These are referred to as ``prehadrons" or hadrons in formation. Since the final-state hadrons have a finite size, $R$, one can estimate the formation time in the rest frame of the hadron as R/$c$ \cite{Dok_1991} which gets larger when boosted to other reference frames via time dilation. The relativistic gamma factor for this process can be approximated as ranging from $\nu z$/M to $\nu z$/Q, \emph{i.e.} $\gamma$ is approximately in the range of 5-10 for this example, to boost the leading hadron into the laboratory frame.

Therefore, nDIS in the range of atomic nuclei spanning $\sim 0-10\,$fm in diameter offers a new opportunity to probe hadron-dependent details of colour propagation and hadron formation and to extrapolate that information to the processes occurring in the absence of the nuclear medium.  Further, these studies are providing new information on hadron structure, in particular, on diquark correlations in baryons, as explained below. Strong evidence for the existence of diquarks can be obtained by comparison of meson production in nDIS with baryon production in nDIS, and quantitative information on their size can be inferred from the degree of interaction with the nuclear medium that they exhibit.

\subsubsection{Probes of meson production using nDIS}
The contact between the picture described above and recent experimental observations comes from two observables: the hadronic multiplicity ratio, $R^h_M$, and transverse momentum broadening, $\Delta p_T^2$. The hadronic multiplicity ratio is defined as:
\begin{equation}
R_M^h (Q^2,\nu,z_h,p_T;h) := {{{1}\over{N_e(Q^2,\nu)}} \cdot N_h (Q^2,\nu,z_h,p_T) \Big|_A \over {{{1}\over{N_e}(Q^2,\nu)} \cdot N_h (Q^2,\nu,z_h,p_T) \Big|_D} } \,,
\label{eq:MR}
\end{equation}
where the labels $D$ and $A$ refer to a lighter nucleus $D$ such as deuterium, and a heavier nucleus $A$ such as xenon; the label $h$ identifies which type of hadron is being measured; $Q^2$ and $\nu$ are the four-momentum transfer and energy transfer, respectively; $p_T$ is the momentum component of the hadron transverse to the direction of the three-momentum transfer; $z_h$ is the relative energy, defined either as the hadron total energy divided by $\nu$ or as the ratio of the light-cone variables of hadron energy-momentum $p_h^+$ to overall energy-momentum $p^+$ \cite{Collins:1997nm}; and $N_h$ and $N_e$ are the total number of DIS hadrons and electrons, respectively, measured from $D$ and $A$. The transverse momentum broadening is defined as:
\begin{equation}
\langle \Delta p_T^2 \rangle (Q^2,\nu,z_h;h) = \langle p_T^2 \rangle \Big|_A - \langle p_T^2 \rangle \Big|_D \,,
\label{eq:PT}
\end{equation}
where the labels are the same as in Eq.~\eqref{eq:MR} and the angle brackets denote an average over events containing one or more hadrons $h$.

To set the stage for what follows, it is necessary to briefly review the status of interpretation of meson production in nDIS. In all cases with DIS kinematics, a large amount of energy and momentum transfer is absorbed by a relatively small structure or subsystem within a nucleon inside the nucleus, most easily visualised as a valence quark. If a valence quark absorbs all the momentum and energy transferred by the scattered lepton, it propagates out of the nucleus accompanied by other quarks and gluons generated in the interaction, typically producing a spray of multiple hadrons in the final state.  one of the hadrons must contain the struck quark, while the others are produced in the interaction or are pre-existing protons or neutrons ejected from the nucleus, normally at lower energies. The hadron containing the struck quark can often be identified with some degree of certainty using kinematic variables such as rapidity, Feynman $x$, or the relative energy $z_h$ defined above.

A recent approach to the description of HERMES pion production data based on this picture can be found in Refs.\,\cite{BL1, Brooks_2019} (BL19).  Therein, two experimental observables are fitted simultaneously in $z_h$ bins. The first observable is the multiplicity ratio defined in Eq.\,\eqref{eq:MR}. This observable provides a measure of the degree of interaction between the medium and the propagating particles produced in the hard interaction. In the BL19 approach, it primarily signifies the existence of an hadronic interaction between a forming hadron, or a fully formed hadron, and the nuclear medium. Secondarily, it can also be influenced by the loss of energy of quarks and gluons in the nuclear medium via gluon bremsstrahlung.

The second observable is the broadening of the transverse momentum distribution of the hadron defined in Eq.\,\eqref{eq:PT}. The transverse momentum is defined with respect to the direction of the three-momentum transfer, often referred to as the direction of the ``virtual photon" in the single-photon-exchange approximation. Within the BL19 formalism, it is the result of medium-stimulated emission of soft gluons in the partonic phase of colour propagation through the medium.

In the BL19 approach, the two observables mentioned above are simultaneously fitted to determine two to four parameters. The modelling categorises the instantaneous state of the struck quark as being either in the partonic (coloured) state or in an hadronic state (colour singlet, bound inside a prehadron or a full hadron). In the latter case, the hadron may still be forming and thus may not have its full mass and size. The three parameter fit includes (\emph{i}) the ``colour lifetime", defined as the time in which the struck quark travels without being incorporated into a colour singlet state; (\emph{ii}) an effective hadronic interaction cross-section, which only pertains to the colour singlet hadronic state; and (iii) a transport parameter related to the $\hat{q}$ theoretical quantity that describes the transverse momentum acquired by a parton owing to in-medium scattering, which only pertains to the partonic state. In the four-parameter version of the model, the magnitude of quark energy loss is extracted in addition.

This approach is successful in describing the production of positive pions from nuclear targets as heavy as xenon in the HERMES data in a one-dimensional analysis in $z_h$. The results obtained produce a good simultaneous fit to the two observables and, as a byproduct, the fit independently reproduces the LSM string constant $\kappa$ to an accuracy of better than 20\%, using the assumption that the data are dominated by the struck quark. Thus, in the BL19 approach, meson production from nuclear targets in nDIS kinematics can be described successfully. While this has only been demonstrated for the $z_h$-dependence of positive pions thus far, the systematic behavior of the observables for other mesons in the HERMES data is quite similar to that of the positive pions, suggesting that the same modelling approach should also be capable of describing those.

\begin{figure*}[!t]
\begin{center}
\includegraphics[width=0.9\textwidth]{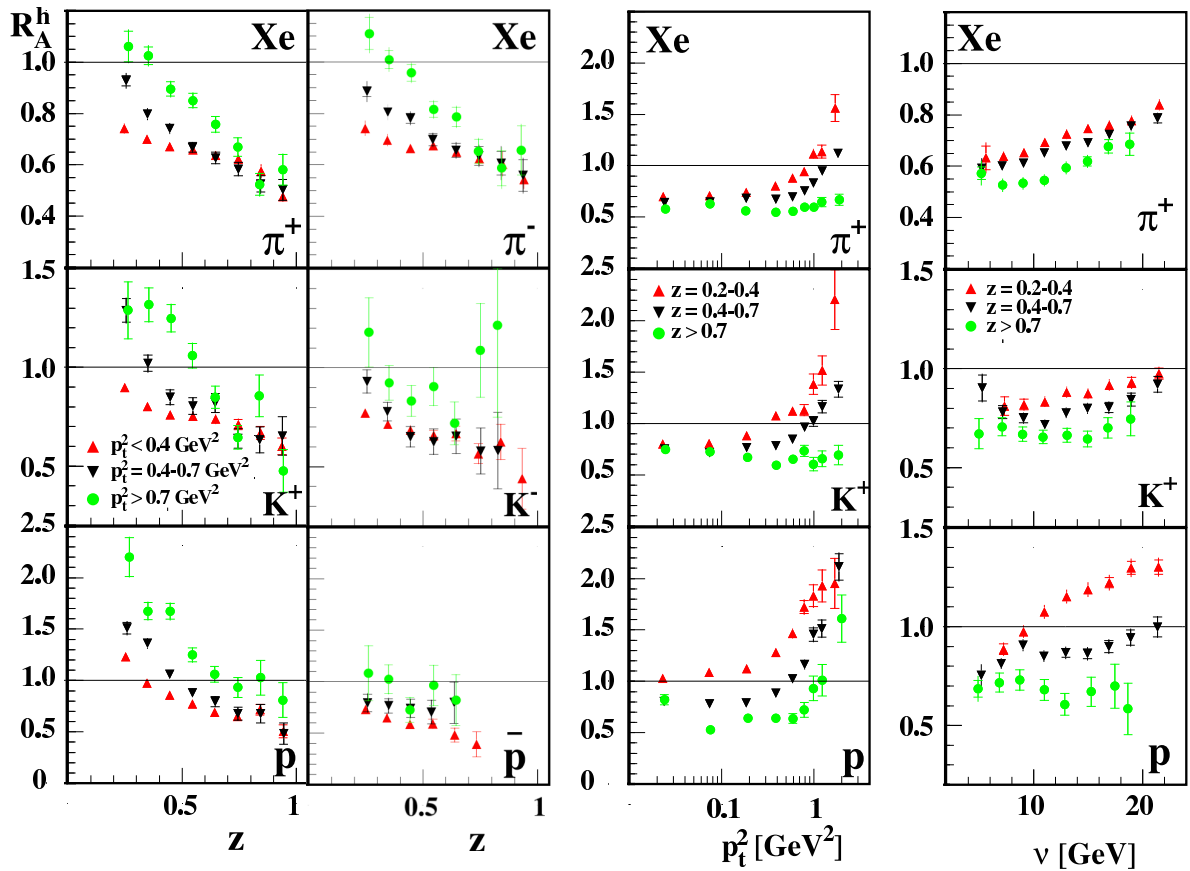}
\caption{\label{fig:hermes} Data from the HERMES Collaboration showing two-dimensional multiplicity ratios for positively charged hadrons as a function of $z_h$, $p_T^2$, and $\nu$, for three bins in a second variable \cite{Airapetian:2011jp}. \emph{Column 1a and 1b}: $R_M^h(z_h; p_T^2)$. \emph{Column 2}: $R_M^h(p_T^2; z_h)$. \emph{Column 3}: $R_M^h(\nu; z_h)$. Each of the top panels corresponds to positive or negative pions, the middle panels correspond to positive or negative kaons, and the bottom panels correspond to protons or antiprotons, as labelled. The multiplicity ratios shown are for the xenon target data compared to the deuteron target data, which show the most pronounced nuclear effects of the various targets discussed in Ref.\,\cite{Airapetian:2011jp}.}
\end{center}
\end{figure*}

However, the situation is different for proton production from nuclei. The systematic behaviour of proton observables in the HERMES data is markedly different from that for the mesons and antiprotons, as elaborated in the following section and illustrated in Fig.\,\ref{fig:hermes}. It can be argued that the systematic differences in the behaviour of the multiplicity ratio for protons owes to the importance of diquark degrees of freedom in the proton.  This theme is the topic of the next section.

\subsubsection{Probes of baryon production and connection to diquarks}
As mentioned above, the systematic features of proton production observables in the HERMES nDIS data are qualitatively different from those of meson and antiproton observables. HERMES published multiplicity ratios for protons and anti-protons ($\bar p$) in addition to five of the lightest mesons; and for all except the neutral pion, HERMES provides data on the two-dimensional behaviour of the multiplicity ratios, \emph{e.g}.\ $R_M^h(z_h,p_T^2)$ and $R_M^h(z_h,\nu)$. The $\bar p$ observables are qualitatively similar to those of the mesons, but the two-dimensional proton production data are qualitatively different from the data for the other five particles. The most visible differences are seen in the xenon target data, which provides the longest in-medium path length. Prominent differences are:
\begin{enumerate}
\item \label{B1} $R_M^h(z_h;p_T^2)$ for protons shows an unexpectedly large enhancement for low $z_h$ and high $p_T^2$, exceeding $1.0$ for all values of $p_T^2$ and exceeding $2.0$ for the highest $p_T^2$. All five other particles have smaller values and most are consistent with being less than or equal to unity, including $\bar p$. See Fig.\,\ref{fig:hermes}, left two columns.
\item \label{B2} $R_M^h(p_T^2;z_h)$ shows a strict ordering for positive pions and kaons, where the highest enhancement is for low $z_h$ and no enhancement is seen for high $z_h$; but for the proton, this ordering disappears at high $p_T^2$. See Fig.\,\ref{fig:hermes}, third column from the left.
\item \label{B3} $R_M^h(\nu;z_h)$ for low $z_h$ exceeds $1.0$, rising to $1.3$ at high $\nu$. All other measured particles remain well below $1.0$. This has very significant implications for the interpretation of the data because, in the BL19 picture, the behaviour with $\nu$ arises from a Lorentz boost of the colour lifetime proportional to $\nu$.  This causes the quark or hadron interaction to disappear at infinite $\nu$, implying that the multiplicity ratio must approach $1.0$ in the high $\nu$ limit, and also implying that it will never exceed $1.0$. This expected behavior is consistent with the HERMES data for all the measured particles, except the proton, for which it is strongly violated. See Fig.\,\ref{fig:hermes}, right-most column.
\end{enumerate}

These observations may be interpreted as follows. Concerning Bullet-\ref{B1}, the degree of the enhancement at high $p_T^2$ is a signal of the strength of the interaction with the medium, whether of the parton or of the forming hadron. This is a sensitive measurement because interactions with the medium in general \emph{add} to the transverse momentum of the final particle and the high $p_T^2$ part of the spectrum is naturally not populated with many events, so in that region the effect of additional transverse momentum is most visible. Thus, the proton production mechanism involves a stronger interaction with the medium than that for the mesons and antiproton.

Concerning Bullet-\ref{B2}, the strict ordering of the mesons in the BL19 picture is understood by asserting that at high $z_h$ the travelling particle is dominantly a parton, which has a gentle interaction with the medium via partonic multiple scattering; while at low $z_h$, the travelling particle typically comes from a prehadron or fully-formed hadron, initially at higher $z_h$, which had a violent interaction with the medium via inelastic hadronic reactions, producing more new hadrons at lower energies in an intranuclear cascade and resulting in an enhancement that can exceed unity.  The HERMES data for mesons are consistent with this picture; but for the proton data, this correlation in $R_M^h(p_T^2;z_h)$ between $z_h$ and $p_T^2$ vanishes at the highest $p_T^2$ values, suggesting that the physical picture is quite different.

In Bullet-\ref{B3}, the disappearance of the expected approach of $R_M^h(\nu;z_h)$ to $1.0$ at high $\nu$ signals that the picture of BL19 is not correct for the proton. Instead of a single struck quark in the initial state that is Lorentz boosted by a factor proportional to $\nu$, a different mechanism must be at play.

A possible explanation for the HERMES proton data can be found by appealing to a quark+di\-quark picture of the proton. A description of proton hadronisation rates that relied on a quark+diquark model was used in the 1980s \cite{Breakstone:1985ef} to analyse pp scattering at $\sqrt{s}$ = 62 GeV. Proton-K$^+$ production rate ratios were seen to be up to several times larger than antiproton-K$^-$ production rate ratios for transverse momenta of $3-6\,\text{GeV}$. Conventional model calculations were able to describe the antiproton-K$^-$ ratio, but not the proton-K$^+$ ratio. Including a quark+diquark component of the proton, Ref.\,\cite{Breakstone:1985ef} was able to describe the data.

The analysis in Ref.\,\cite{Breakstone:1985ef} suggests that the HERMES proton data anomalies listed above may well have an explanation rooted in diquarks. Further, the HERMES data, and present and future data from JLab-CLAS $5$ and $11\,$GeV, provide a very exciting opportunity to probe much more deeply into the diquark nature of the proton, using the nuclear medium as a spatial analyser. As explained below, this is particularly true for the transverse momentum observable defined in Eq.\,\eqref{eq:PT}, and for a particular set of baryons as follows.  In the theoretical work of Ref.\,\cite{Yin:2019bxe}, an analysis of the dominant diquark correlations contained in various hadrons is presented. According to that work, the proton, neutron, and lambda baryon are all dominated by the $[ud]$ diquark, while the $\Sigma$ and $\Xi$ baryons are dominated by the $[us]$ diquark. If the $[ud]$ diquark is a key to explaining the HERMES proton data, comparison of multiplicity ratios and transverse momentum broadening of accessible baryons will, in the simplest case, show the same patterns for $p$, $n$, and $\Lambda$, and a different pattern for $\Sigma$ and $\Xi$. The production of all these baryons from nuclear targets will be measured in CLAS at $11\,$GeV \cite{E12-06-117}.

A more provocative and intriguing statement can be made by revisiting the three bullets above, to connect those observations to the diquark concept. First, at large momentum transfer, $Q^2$, the hard interaction involves a very small volume of the proton. In the BL19 picture it involves only one quark. However, in the case that diquarks are an important component of the proton, when they are close together it is not excluded that both of the quarks in the diquark can be involved in the scattering. In that case the struck object would be a two-quark system, which would either remain intact or fall apart. In the case that it remains intact, it can be the foundation for a new proton to be formed that can emerge over the full range in $z_h$, including at high $z_h$, where more conventional mechanisms like single-quark scattering are much more challenged; and proton knockout from the nucleus at high $z_h$ is also strongly suppressed. If the moving system is an intact coloured diquark, it would clearly interact much more strongly with the nuclear medium than a single quark. In this picture, nDIS leading to final-state protons would show a much stronger interaction with the medium, consistent with the Bullet-\ref{B1} above, and the direct scattering of a diquark is a very different production mechanism than single quark scattering.  This explanation is consistent with the observations of the Bullet~\ref{B2}, \ref{B3}.

Even more interesting, if the travelling object is a coloured diquark, it will lead to a significant increase in the magnitude of transverse momentum broadening $\langle \Delta p_T^2 \rangle (Q^2,\nu,z_h;h)$, defined in Eq.\,\eqref{eq:PT}, because it will have a QCD colour field that is more extended in size than the single quark, as well as having substantial mass. This observation leads to a prediction: {\em the $p_{\rm T}$ broadening of a proton should be approximately equal to the $p_{\rm T}$ broadening of a neutron and of a $\Lambda$-baryon in nDIS, and all three should be much larger than the $p_{\rm T}$ broadening of any system produced by single-quark scattering, such as mesons and $\Sigma$ and $\Xi$ baryons.} Similarly, the patterns for the multiplicity ratios noted in the three bullets above should be very similar for the proton, neutron, and $\Lambda$-baryon, aside from small mass effects that will modify the accessible $z_h$ range, if the traditional definition of $z_h=E_h/\nu$ is used. One can also expect that the $\Sigma$ and $\Xi$ will be similar to each other in the two observables because they are both dominated by the $[us]$ diquark.

No published data are currently available for $p_T$ broadening on the proton or the $\Lambda$-baryon, and no published data are available for the $\Lambda$ multiplicity ratio. However, preliminary data from CLAS with $5\,$GeV beam for the $\Lambda$-baryon have been released and shown in conferences \cite{Chetry_2019}. Similarly, analyses of proton multiplicity ratios and $p_T$ broadening are fully feasible with the same data set. The preliminary $\Lambda$-baryon results are qualitatively similar to the published HERMES proton data discussed above, and the $p_T$ broadening is an order of magnitude greater than that seen for the HERMES meson studies. Although these results are not yet final, they strongly support the idea that direct scattering of diquarks can be measured experimentally. This result, if confirmed by the final data, opens a remarkable new era of studies of the structure of the nucleon and the light strange baryons.


%

\subsection{Production Cross-sections of Baryons at Belle}
\label{PCsBB}
In the context of exotic hadrons, it was shown \cite{Jaffe:2005zz} that the production rates of $\Lambda(1116)$ and $\Lambda(1520)$ in $e^+ e^- \to  \mathrm{hadron}$ at $\sqrt{s} = 92$ GeV were 2-3 times bigger than the estimated values from those of $p$, $\Sigma$, $\Delta$, $\Xi$, $\Omega$ baryons.  The enhancement was attributed to $\Lambda(1116)$ and $\Lambda(1520)$ having only $[ud]$ scalar diquark configurations.  However, this assumption conflicts with predictions from structure studies that employ dynamical diquark correlations \cite{Yin:2019bxe}: $[us]$, $[ds]$ and $\{us\}$, $\{ds\}$ are also significant, so the conclusion is dubious.  Notwithstanding that, the heavier mass of the $c$-quark can affect the structural properties of charmed baryons \cite{Yin:2019bxe} and this makes it worthwhile to study their production cross-sections.

Ref.\,\cite{Niiyama:2017wpp} reported production cross-sections for hyperons ($\Lambda$, $\Lambda (1520)$, $\Sigma^0$, $\Sigma (1385)^+$,  $\Xi^-$, $\Xi(1520)^0$, $\Omega^-$) and charmed baryons ($\Lambda_c^+$, $\Lambda_c (2595)^+$, $\Lambda_c (2625)^+$, $\Sigma_c (2455)^0$, $\Sigma_c (2520)^0$) obtained using Belle data \cite{Brodzicka:2012jm}.  In order to avoid contamination from $\Upsilon(4S)$ decay, Ref.\,\cite{Niiyama:2017wpp} used $89.4\,$fb$^{-1}$ of off-resonance data taken at $\sqrt{s} = 10.52$ GeV, which is 60\,MeV below the mass of the $\Upsilon(4S)$.

Since charmed baryon production rates are small, it is advantageous to use both off- and on-resonance data.  The latter has been recorded at the $\Upsilon(4S)$ energy ($\sqrt{s} = 10.58$ GeV) with a luminosity of $711\,$fb$^{-1}$. To eliminate $B$-meson decay contributions in the on-resonance data, one requires the charmed-baryon candidates to have the hadron-scaled momenta
$x_p = p /\sqrt{s/4 - M^2 } > 0.44$,
where $p$ and $M$ are the momentum and mass of the charmed baryon.

The Belle detector \cite{Brodzicka:2012jm} is a large-solid-angle magnetic spectrometer that comprises a silicon vertex detector, a central drift chamber (CDC), an array of aerogel threshold Cherenkov counters (ACC), time-of-flight scintillation counters (TOF), and an ECAL composed of CsI(Tl) crystals located inside a superconducting solenoid coil that provides a $1.5\,$T magnetic field.

Charged particles produced from the $e^+ e^-$ interaction point (IP) are selected by requiring small impact parameters with respect to the IP along the beam ($z$) direction and in the transverse plane ($r - \phi$) of $dz < 2$ cm and $dr < 0.1$ cm, respectively. For long-lived hyperons ($\Lambda$, $\Xi$, $\Omega$), the trajectories must be reconstructed carefully.

Particle identification is performed by utilising $dE/dx$ information from the CDC, time-of-flight measurements in the TOF, and Cherenkov light yield in the ACC. The likelihood ratios for selecting $\pi$, $K$ and $p$ are required to be greater than $0.6$ over the other particle hypotheses. This selection has an efficiency of $90 - 95$\% and a fake rate of $5 - 9$\%. The use of charge-conjugate decay modes is implied, and the cross-sections of the sum of the baryon and antibaryon production are recorded. In order to estimate the efficiencies, Monte-Carlo (MC) events are generated using PYTHIA\,6.2, and the detector response is simulated using GEANT3.

One first obtains the inclusive differential cross sections, $d\sigma/dx_p$, as a function of hadron-scaled momenta, $x_p$.  Integrating the differential cross-sections in the $0 \le x_p \le 1$ region, one obtains the cross-section without radiative corrections (visible cross-sections).
For $S = -1$ hyperons, the integrations can be completed using a third order Hermite interpolation describing the behaviour in the measured $x_p$ range, assuming the cross-section is zero at $x_p = 0,1$. Estimates of the contributions from the unmeasured regions can be accomplished using PYTHIA, with the differences between two estimates included as part of the total systematic error.  For $S=-2$ and $-3$ hyperons, the cross-sections can be measured in the entire $x_p$ region.
For charmed baryons, the contributions from the unmeasured $x_p$ regions may be estimated using MC simulations with various fragmentation models.
Radiative corrections are subsequently applied in each $x_p$ bin of the $d\sigma/dx_p$ distribution. The correction for initial-state radiation (ISR) and vacuum polarisation can again be estimated using PYTHIA, by enabling or disabling these processes; and the final-state radiation (FSR) from charged hadrons can be analysed using PHOTOS \cite{Golonka:2005pn}.
Finally, the feed-down contributions from the heavier particles are subtracted from the radiative-corrected total cross-sections to obtain the direct cross-sections, which may directly reflect the internal structures of baryons.

\begin{figure}[!t]
\begin{tabular}{ccc}
\includegraphics[width=0.45\linewidth]{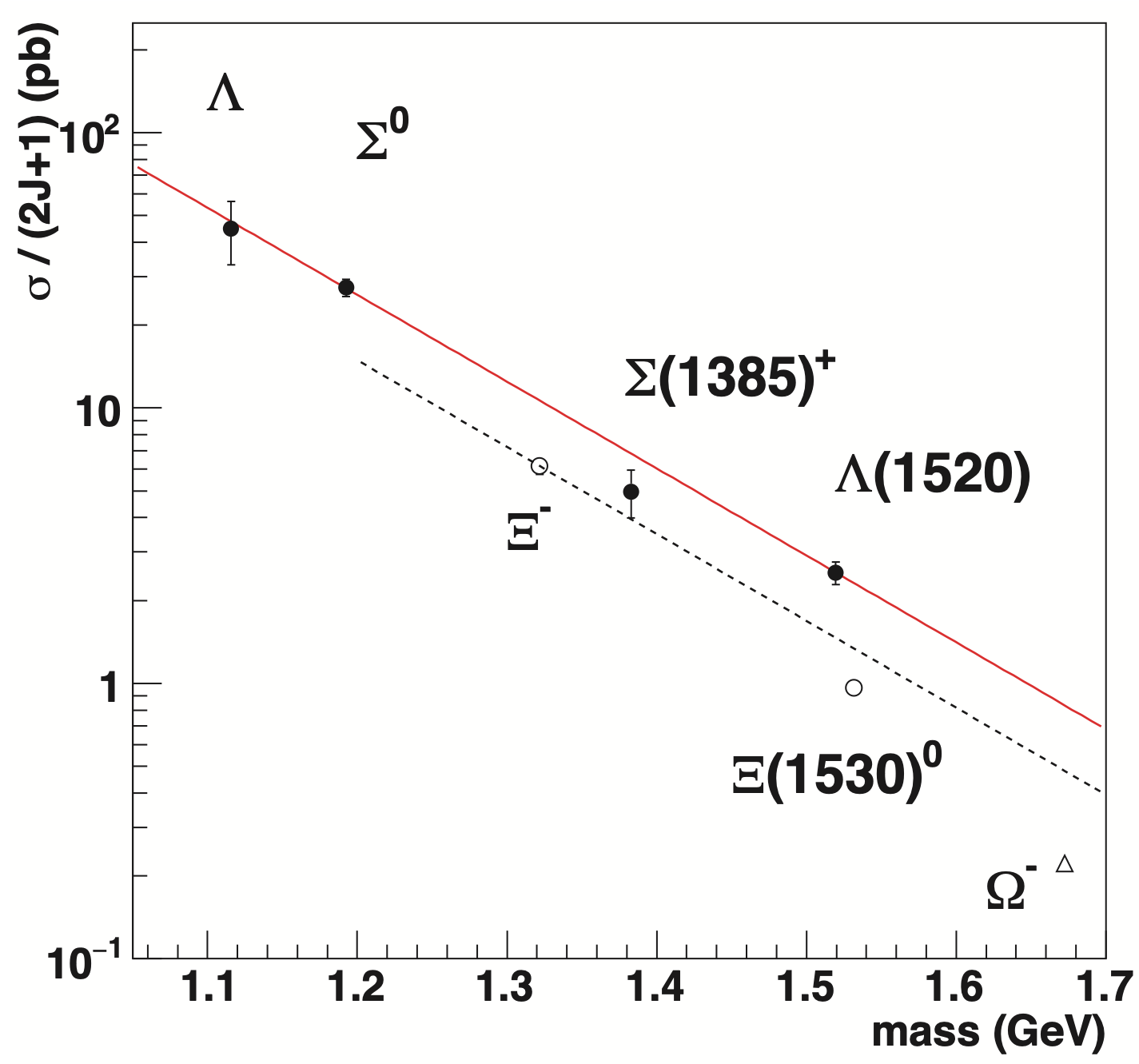} & \hspace*{2em} &
\includegraphics[width=0.45\linewidth]{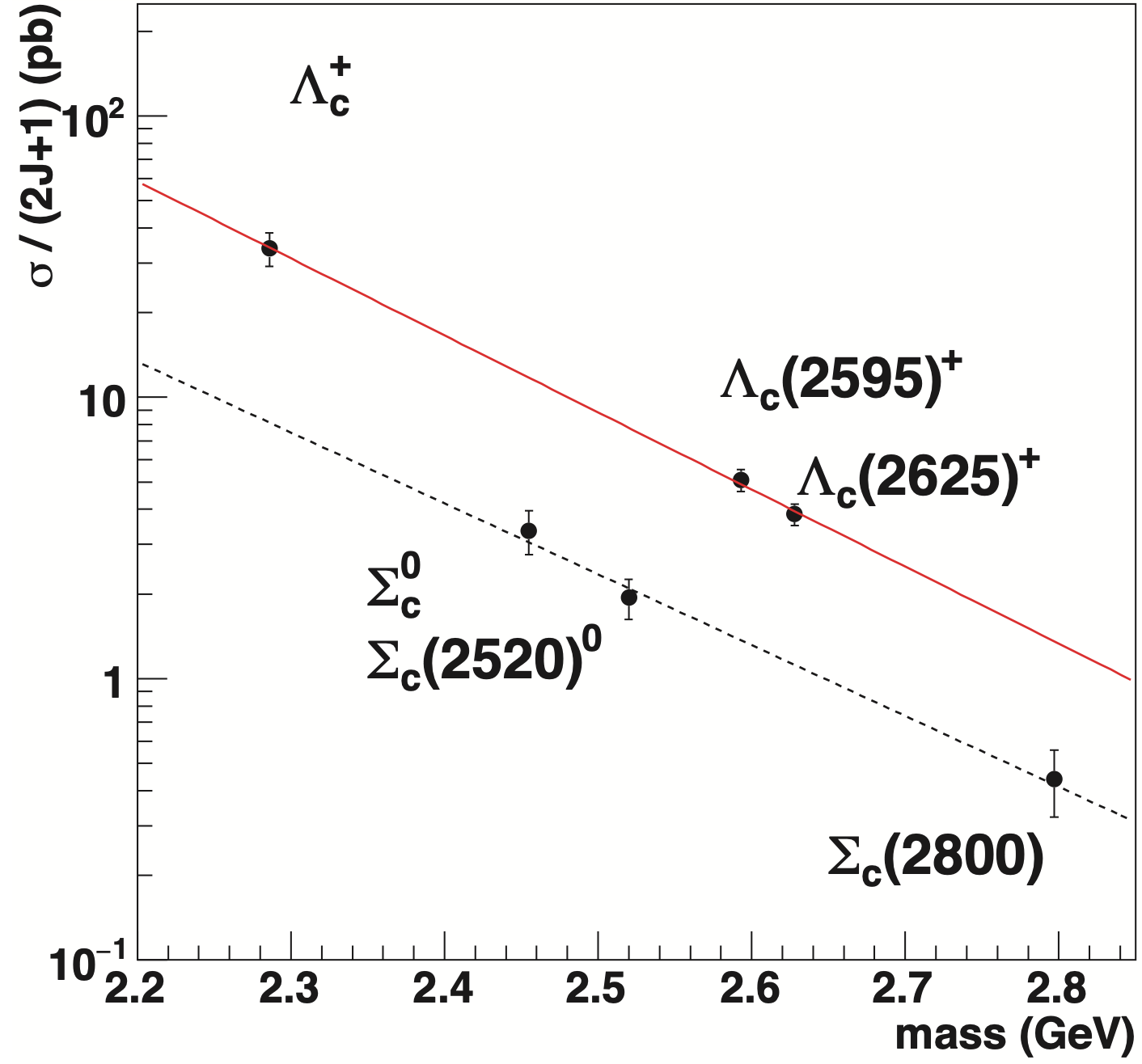}
\end{tabular}
\caption{\label{fig:hyperon}
\emph{Left panel}.
Direct production cross-sections divided by the number of spin states $(2J +1)$ as a function of hyperon masses. $S = -1, -2, -3$ hyperons are shown with filled circles, open circles, and a triangle, respectively. The solid line shows the fit result using an exponential function for $S = -1$ hyperons, except for $\Sigma (1385)^+$. The dashed line shows an exponential curve with the same slope parameter as $S = -1$ hyperons, which is normalised to the production cross section of $\Xi^-$.
%
\emph{Right panel}.
As left panel for charmed-baryon masses. The solid and dashed lines show the fit results using exponential functions for $\Lambda_c$ baryons and $\Sigma_c$ baryons, respectively.
}
\end{figure}

The procedure just described yields the results depicted in Fig.\,\ref{fig:hyperon}.  These are scaled direct production cross-sections, \emph{i.e}.\ direct production cross-sections divided by the number of spin states $(2J +1)$ as a function of baryon masses.
Fig.\,\ref{fig:hyperon}\,--\,left panel reveals exponential dependence of the scaled direct production cross-sections for $S = -1$ hyperons except for the $\Sigma(1385)^+$.  Since $\Sigma(1385)^+$ includes a $u$-quark component and the $e^+ e^- \to u \bar u$ cross-section is larger than those for $e^+ e^- \to d \bar d$ or $e^+ e^- \to s \bar s$, a low-lying $\Sigma(1385)^+$ runs contrary to na\"{\i}ve expectations.
Enhancements of the scaled direct production cross-sections for $\Lambda$ and $\Lambda (1520)$ are not evident, contradicting Ref.~\cite{Jaffe:2005zz}.  Notably, the feed-down contributions were not subtracted therein; and since the feed-down contribution is large for $\Lambda$, this may explain the discrepancy.
Fig.\,\ref{fig:hyperon} clearly indicates that the scaled production cross-sections for $S=-1$ hyperons are larger than those for $S=-2$ hyperons and that for the $S=-3$ $\Omega$ is the smallest amongst the strange baryons.

The results for charmed baryons are shown in Fig.\,\ref{fig:hyperon}\,--\,right panel. The scaled production cross-section for the $\Sigma_c(2800)$ measured by Belle \cite{Mizuk:2004yu} is shown in the same figure, computed using the weighted average of cross-sections for the three charged states and assuming that the $\Lambda_c^+ \pi$ decay mode dominates over the others. In Ref.\,\cite{Mizuk:2004yu}, the spin parity is tentatively assigned as $J^P = 3/2^-$, so a spin of $3/2$ was used for this state.

In the case of charmed baryon direct production in $e^+ e^-$ collisions, the first process should be $e^+ e^- \to c \bar c$, since the $c$-quark mass is well above the QCD energy scale, $\Lambda_{\mathrm{QCD}}$; hence, $c \bar c$ pair production should be relatively rare in hadronisation processes.
Fig.\,\ref{fig:hyperon}\,--\,right panel clearly shows that the scaled production cross-sections for the isoscalar charmed baryons are larger than those for the isovector charmed baryons.
Some may argue that this outcome is consistent with diquark+antidiquark pair production being easier in the lighter isoscalar-scalar channels than in the heavier isovector-pseudovector channels.  However, this assumes very simple spin-flavour wave functions for the systems involved, in conflict with calculations based on dynamical diquark degrees-of-freedom \cite{Yin:2019bxe}.
Evidently, therefore, further experimental and theoretical studies are necessary to unveil the information contained in Fig.\,\ref{fig:hyperon}.

\subsection{Next Steps for Continuum Schwinger Function Methods}
\label{sec:DSEsFuture}
Capitalising on the foundation provided by existing studies, sketched in Sec.\,\ref{sec:DSEs}, many new paths are open to the use of CSMs in exploring the origin and impact of diquark correlations in hadron physics.  Simplest amongst them is exploitation of the SCI for the computation of those nucleon-to-resonance transition form factors that have not yet been calculated using more sophisticated frameworks.  This can be useful because the inclusion of all possible diquark correlations is fairly straightforward when using the SCI; hence, such studies can provide guidance concerning those correlations and inter-diquark transitions that are most important when calculating the associated electroproduction reactions.

Informed by SCI studies, the QCD-kindred framework could profitably be employed to make predictions for the same array of transition form factors, expecting that the results should compare favourably with available experiments and provide guidance for conducting and planning others.  The advantage here is that the QCD-kindred framework can deliver realistic predictions on a $Q^2$ domain that extends far beyond that upon which MB\,FSIs play an important role.  This means that comparison with experiments can provide unclouded insights into the active degrees-of-freedom within the participating baryons and the correlations between them.  Given their anticipated diquark content and potential to reveal information about orbital angular momentum and intrinsic deformation, the following transitions are of most immediate importance:
\begin{equation}
\begin{array}{ccc}
N(940)\,\tfrac{1}{2}^+ \to N(1535)\,\tfrac{1}{2}^-; &
N(940)\,\tfrac{1}{2}^+ \to N(1520)\,\tfrac{3}{2}^-; &
N(940)\,\tfrac{1}{2}^+ \to \Delta(1700)\,\tfrac{3}{2}^- ; \\
N(940)\,\tfrac{1}{2}^+ \to N(1710)\,\tfrac{1}{2}^+; &
N(940)\,\tfrac{1}{2}^+ \to N(1700)\,\tfrac{3}{2}^-; &
N(940)\,\tfrac{1}{2}^+ \to \Delta(1620)\,\tfrac{1}{2}^- .
\end{array}
\end{equation}
This confidence is based on existing successful studies of transitions involving the $N(1440)\,\tfrac{1}{2}^+$, $\Delta(1232)\,\tfrac{3}{2}^+$, $\Delta(1600)\,\tfrac{3}{2}^+$ final states.

Information derived from the above analyses could be used as the foundation for computation of baryon DAs.  Defined on the light-front, these DAs provide the closest link in quantum field theory to a Schr\"odinger-like wave function with its probability interpretation.  In all likelihood, as discussed in connection with Figs.\,\ref{PlotPDAs}, \ref{figure_barycentric},  the pictures obtained therewith may readily be interpreted to reveal the importance, impact and size of diquark correlations within each baryon under study.

Owing to weaknesses in existing algorithms, \emph{ab initio} CSM calculations of nucleon elastic and transition form factors are currently limited in the range of $Q^2$ that is accessible: $Q^2 \lesssim 7\,$GeV$^2$.  JLab\,12 will probe far deeper than can be accessed using these methods.  Improvements are therefore necessary; at least so that comparisons can be made between \emph{ab initio} results and predictions delivered, \emph{e.g}.\ by the QCD-kindred quark+diquark framework.  Such contrasts at large-$Q^2$, beyond the range of MB\,FSIs, could be crucial in identifying unambiguous signals for the presence of diquark correlations.

A most pressing need is to improve upon leading-order RL truncation in \emph{ab initio} analyses of the baryon bound-state problem.  Whilst RL truncation does provide for SU$(6)$ spin-flavour symmetry breaking in baryon wave functions -- most notably because it enforces a coupling between wave function components in momentum-, spin- and flavour-spaces; and does contain the seeds for the formation of diquark correlations; it does not, \emph{e.g}.\ produce the mass-splittings amongst isospin partners that are achieved in quark+diquark truncations of the Faddeev equation.  It is necessary to identify a beyond-RL truncation of the Faddeev equation that can produce diquark correlations and explore its fidelity, features and flaws.


\subsection{Selected Advances (needed) in lattice-QCD}
\label{SAnilQCD}
Studies of nucleon elastic form factors using DSEs have demonstrated the significance of diquark correlations for nucleon structure at high transferred momentum \cite{Segovia:2014aza, Segovia:2015ufa, Cui:2020rmu}. In particular, the zero crossing in the electric Sachs form factor is sensitive to the presence of quark+quark correlations in the nucleon Faddeev amplitude, thus data from experiment or nonperturbative lQCD calculations can be used to determine their magnitude. The experimental programme to determine nucleon form factors up to $Q^2\approx 18\,\mathrm{GeV}^2$ is well underway \cite{GEP5, GEP5_PAC47, GMN, Wojtsekhowski:2017kti, Helium3article, Helium3_target2015, GENRP}.  A first-principles theoretical calculation of nucleon form factors with rigorous control of systematic effects is possible using modern lQCD methods.

\begin{figure}[!t]
\centering
\includegraphics[width=.48\textwidth]{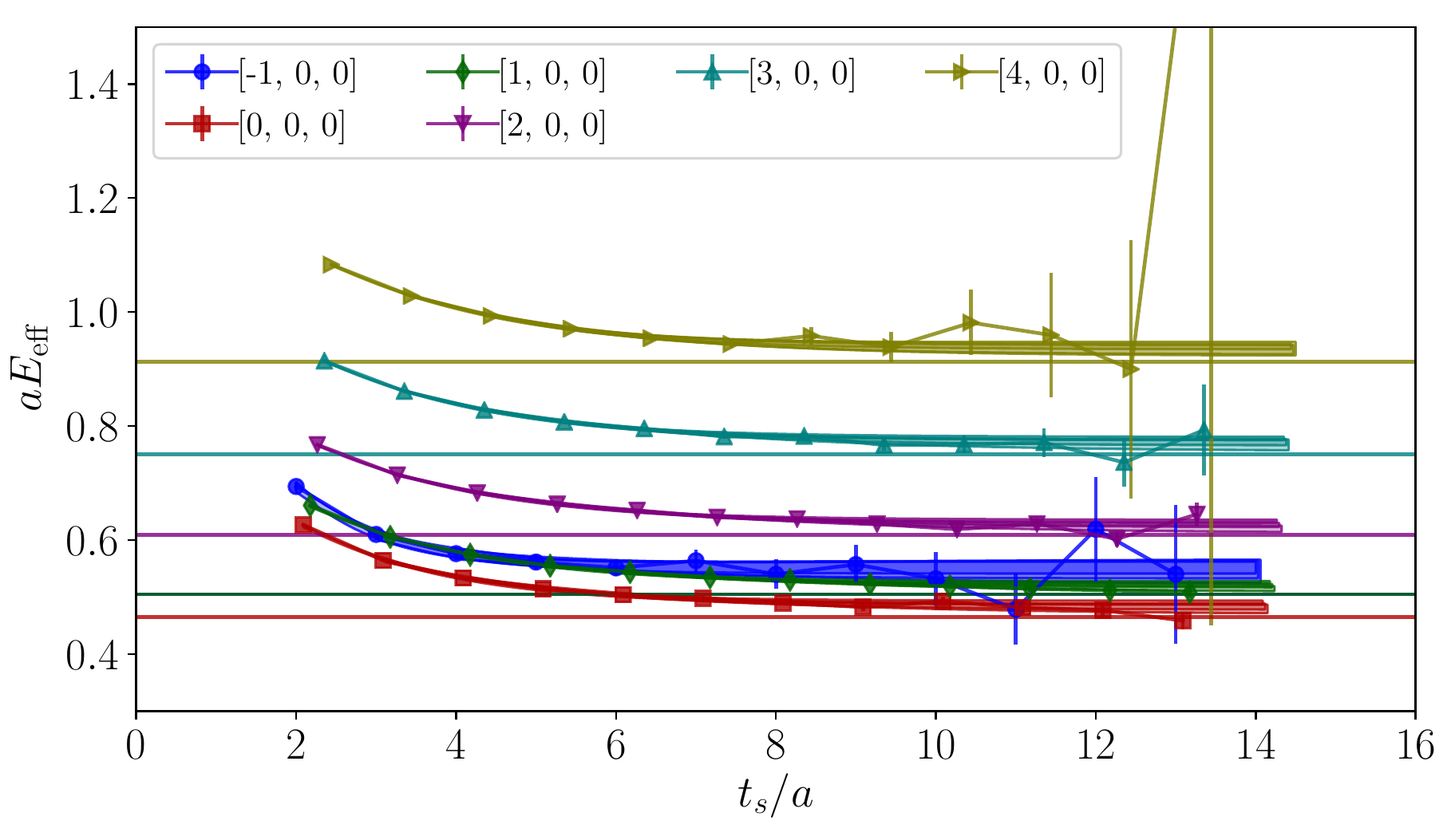}
\hspace{.02\textwidth}
\includegraphics[width=.48\textwidth]{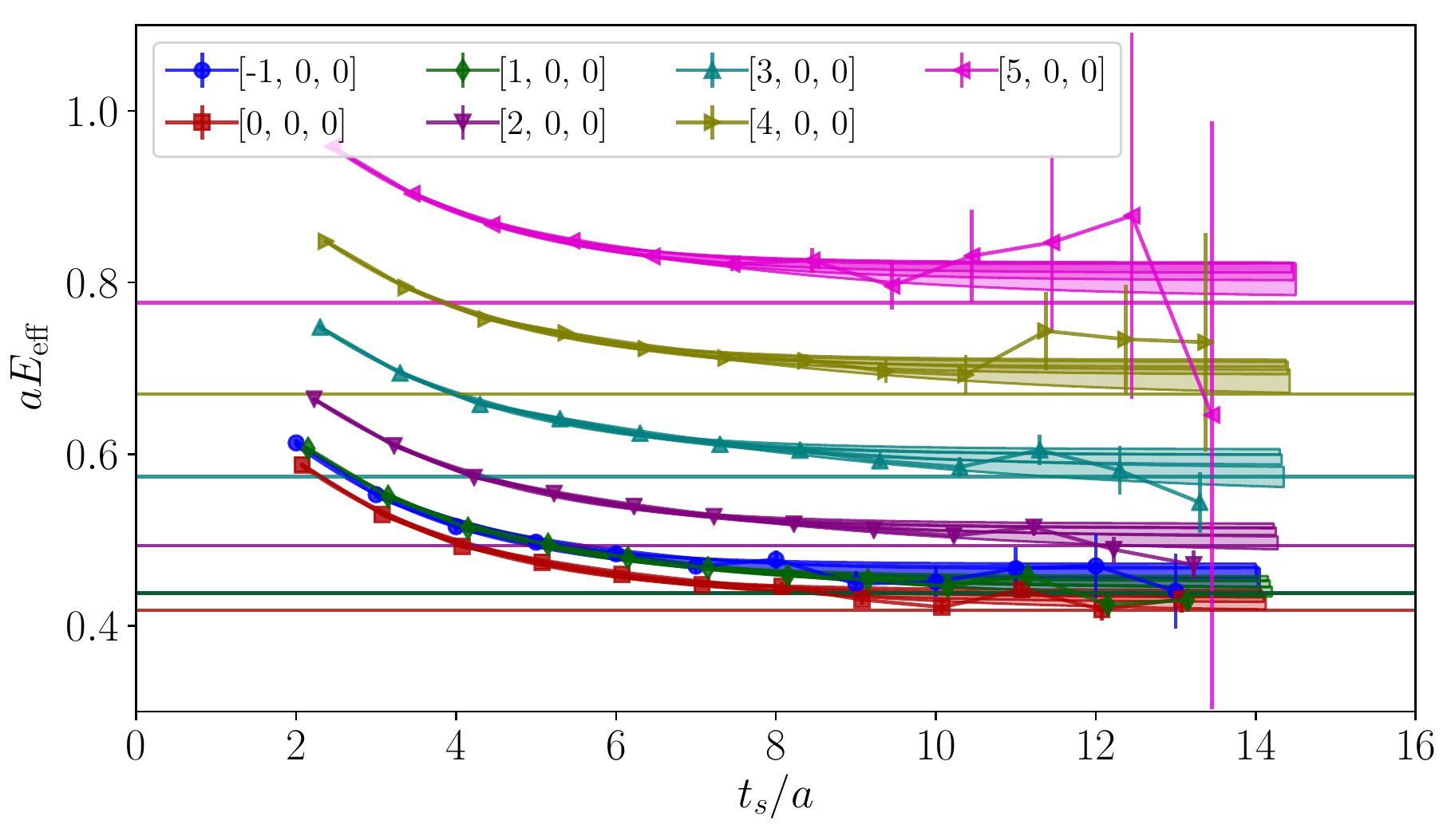}
\caption{Nucleon effective energies computed on a lattice with spacing $a\approx0.09\,\mathrm{fm}$ and pion masses $m_\pi\approx280\,\mathrm{MeV}$ (``D5'', left) and $170\,\mathrm{MeV}$ (``D6'', right). \label{fig:ff_largeQ2:Eeff}}
\end{figure}

Until recently, studies of nucleon form factors on a lattice were limited to $Q^2 \lesssim 1 - 2\,\mathrm{GeV}^2$.  One notable exception is the calculation of $G_{Ep}/G_{Mp}$ using a Feynman-Hellman method \cite{Chambers:2017tuf}.  Lattice calculations involving hadrons with large momentum $|\vec p\,|\gtrsim m_N$ are challenging for several reasons. First, MC fluctuations of lattice hadron correlators are governed by the energy of the state \cite{Lepage:1989hd}.  The signal-to-noise ratio for the nucleon is expected to decrease $\propto\exp\big[-(E_N(\vec p\,) - \frac32 m_\pi)\tau\big]$ with Euclidean time $\tau$, making high-momentum calculations especially ``noisy''.  At the same time, excited states of the nucleon, which are expected to introduce large systematic uncertainties, are less suppressed by Euclidean time evolution $\propto\exp\big[-\Delta E_N(\vec p\,)\tau\big]$ as the energy gap $\Delta E(\vec p\,) = E_{N,\text{exc}}(\vec p\,) - E_N(\vec p\,)$ shrinks with increasing relativistic nucleon momentum $|\vec p\,|$.  Both these challenges are best addressed by choosing the Breit frame on a lattice, so that the initial and final momenta of the nucleon are equal to $|\vec p^{\,(\prime)}|=\frac12\sqrt{Q^2}$. For example, momentum transfer $Q_1^2\approx10\,\mathrm{GeV}^2$ requires nucleon momentum  $p_1\gtrsim1.6\,\mathrm{GeV}$, which reduces the energy gap $\Delta E_N(0)\approx0.5\,\mathrm{GeV}$ to $\approx0.3\,\mathrm{GeV}$. Therefore, very large MC statistics combined with rigorous analysis of excited state contaminations become absolutely necessary to obtain credible results.


The calculations in Refs.\,\cite{Syritsyn:2017jrc, Kallidonis:2018cas} were performed with $N_f=2+1$ (light and strange) dynamical quarks using the clover-improved Wilson fermion action and lattice spacing $a\approx0.09\,\mathrm{fm}$.  Two values of pion mass ($m_\pi\approx280$ and $170\,\mathrm{MeV}$) were used in the calculations, enabling a check on the light-quark mass dependence of the results. Nucleon interpolating operators $N=\epsilon^{abc}[\tilde{u}^{aT} C\gamma_5 \tilde{d}^b] \tilde{u}^c$ were constructed on a lattice with ``momentum-smeared'' quark fields, $\tilde{q}$, to improve their overlap with the ground state of the boosted nucleon \cite{Bali:2016lva}.

\begin{figure}[!t]
\begin{tabular}{ccc}
\includegraphics[width=0.32\textwidth]{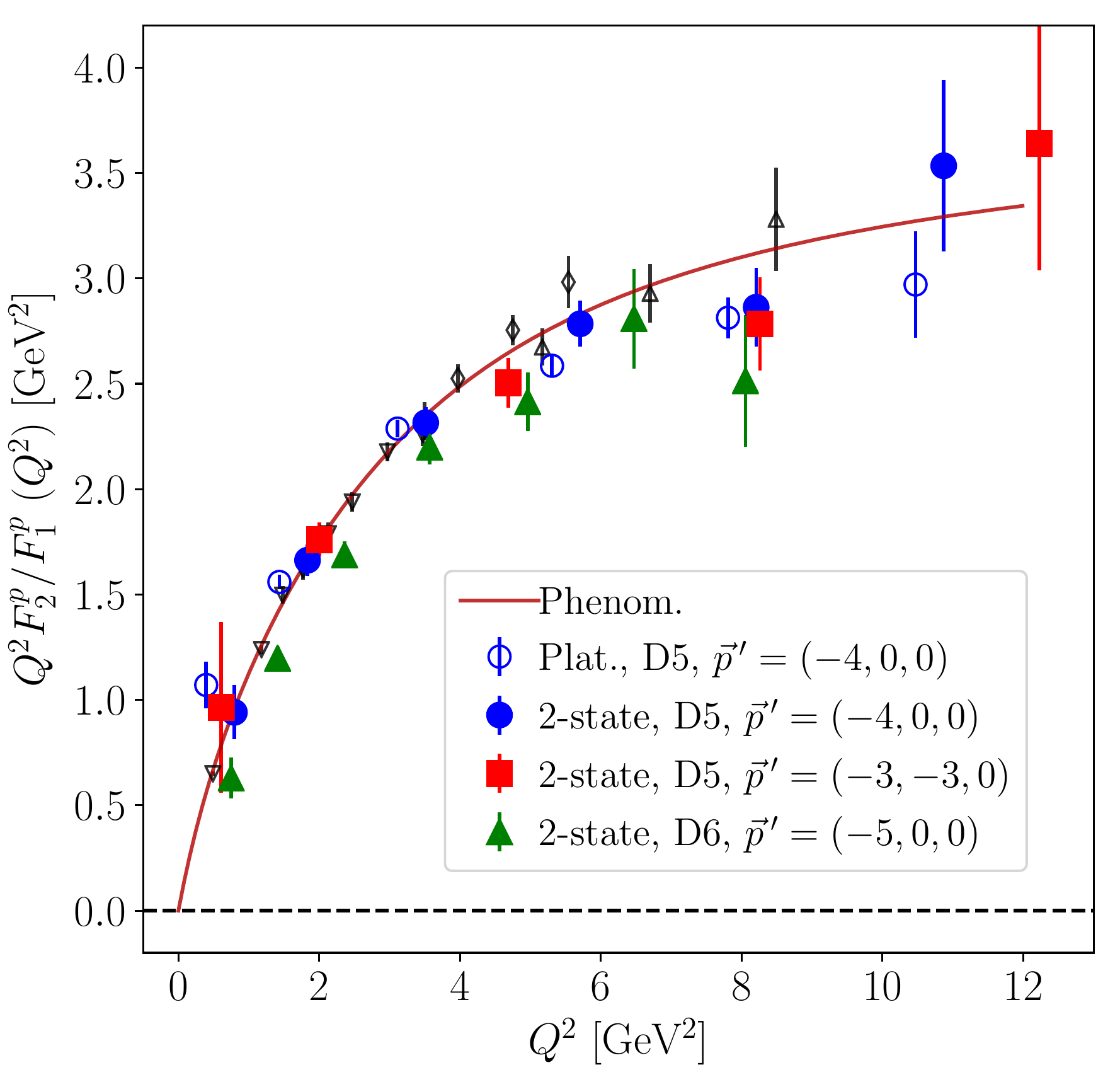} &
\hspace{.005\textwidth}
\includegraphics[width=0.32\textwidth]{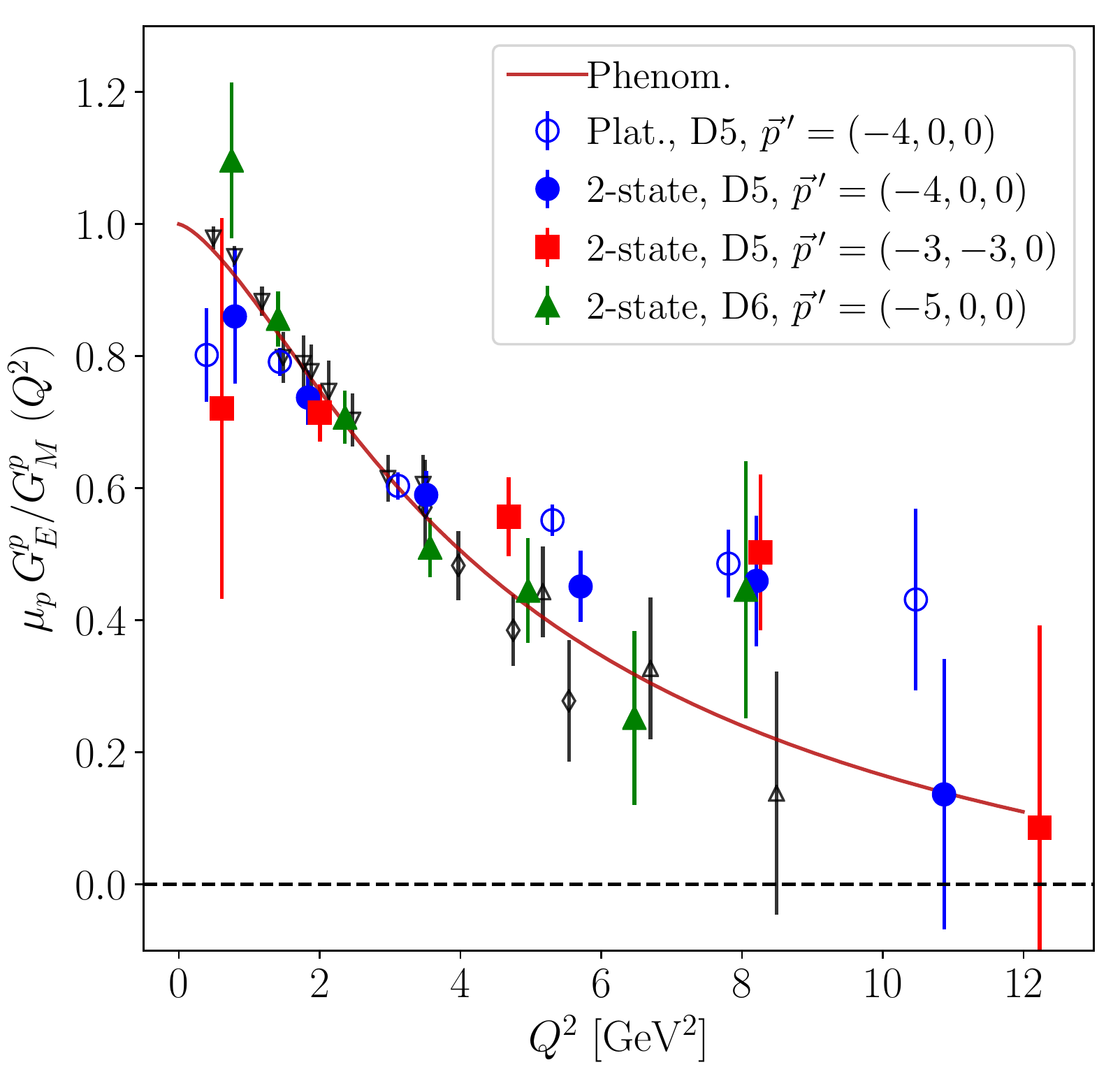} &
\hspace{.005\textwidth}
\includegraphics[width=0.32\textwidth]{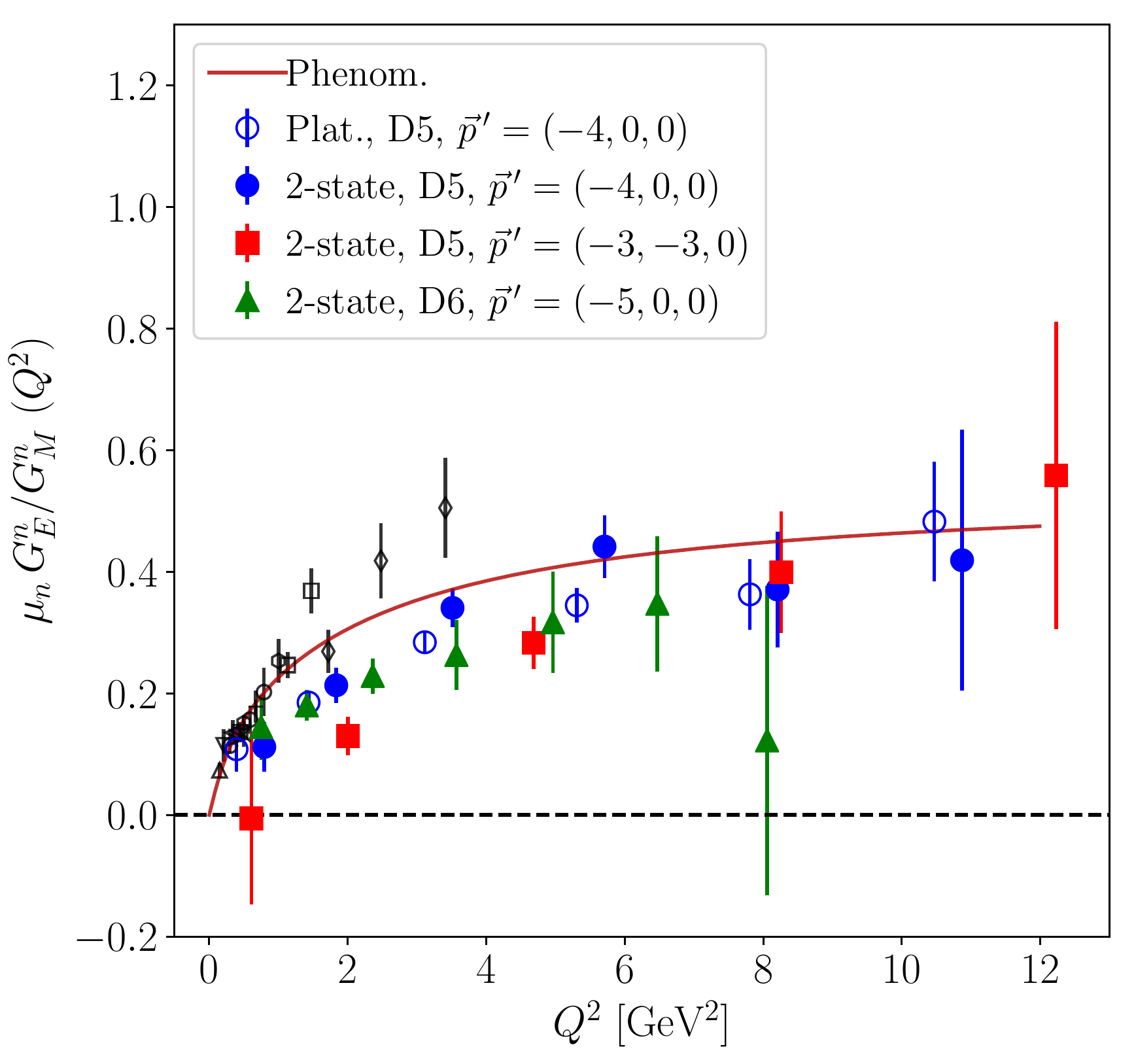}
\end{tabular}
\caption{Results from Ref.\,\cite{Kallidonis:2018cas}.
Ratio of proton form factors $Q^2 F_{2p}(Q^2) / F_{1p}(Q^2)$ (left);
ratio of proton form factors $\mu_p G_{Ep}(Q^2)/G_{Mp}(Q^2)$ (middle);
and ratio of neutron form factors $\mu_p G_{En}(Q^2)/G_{Mn}(Q^2)$ (right).
Disconnected quark contractions are neglected.
Results from calculations with pion masses $m_\pi\approx270\,\mathrm{MeV}$ (``D5'') and $180\,\mathrm{MeV}$ (``D6'') are compared to phenomenological fits of experimental data (black symbols) \cite{Alberico:2008sz}.
Additional comparisons with data and also DSE predictions are provided in Fig.\,\ref{GEonGM}.
\label{fig:ff_largeQ2:ff_ratios}}
\end{figure}

The nucleon correlators become dominated by the ground state $C(t) = \langle N(t) \ldots \bar{N}(0) \rangle \propto e^{-E_N t}$ as the Euclidean time $\tau$ is increased.  The approach to the ground state can be revealed by observing the plateau of the ``effective energy'' $E_N^\text{eff}(t) = \frac1a \log\big[C(t)/C(t+a)\big]$ as $t\to\infty$.  These plateaux are shown in Fig.\,\ref{fig:ff_largeQ2:Eeff} for both pion masses and momenta up to $p_N\approx1.5\,\mathrm{GeV}^2$.  As expected, there are substantial contributions from nucleon excited states.  Although more than one excited state is expected to contribute, the data are not precise enough to constrain  more than one, especially at large momenta.  Therefore, a simple two-state model was imposed on the results
\begin{align}
\langle N(\vec p, t) \bar{N}(0)\rangle &\sim C_0^2 e^{-E_{N0} t} + C_1^2 e^{-E_{N1} t}\,, \nonumber \\
\langle N(\vec p^{\,\prime}, t) J(\vec q, \tau) \bar{N}(0) \rangle
& \sim {\mathcal A}_{0^\prime 0} C_{0^\prime} C_0 e^{-E_{N0}^\prime(t-\tau) - E_{N0}\tau}
+ {\mathcal A}_{1^\prime 0} C_{1^\prime} C_0 e^{-E_{N1}^\prime(t-\tau) - E_{N0}\tau} \nonumber \\
& \quad + {\mathcal A}_{0^\prime 1} C_{0^\prime} C_1 e^{-E_{N0}^\prime(t-\tau) - E_{N1}\tau}
+ {\mathcal A}_{1^\prime 1} C_{1^\prime} C_1 e^{-E_{N1}^\prime(t-\tau) - E_{N1}\tau} \,,
\end{align}
to extract ground-state nucleon energies $E_{N0}^{(\prime)}$ and momentum-dependent matrix elements of nucleon operators $C_{0^{(\prime)}}=\langle\mathrm{vac}|N|N(\vec p^{\,(\prime)})\rangle$  and the vector current density ${\mathcal A}_{0^\prime 0}=\langle N(\vec p^{\,\prime})|J|N(\vec p\,)\rangle$. The latter are decomposed into form factors $F^q_{1,2}$ separately for each flavor $q$.

Wick contractions of lattice quark fields generate two types of diagrams: quark-connected and quark-disconnected. The latter have lattice quark ``loops'' that are connected to the valence quark lines only by gluons and are more difficult to compute.  One study \cite{Green:2015wqa} found their contribution to nucleon form factors at $Q^2\lesssim1.2\,\mathrm{GeV^2}$ to be small ($\lesssim1\%$).  Their effects at higher momenta remain to be explored.  Refs.\,\cite{Syritsyn:2017jrc, Kallidonis:2018cas} omitted these contributions.

The left panel of Fig.\,\ref{fig:ff_largeQ2:ff_ratios} displays the ratio of proton Pauli and Dirac form factors computed in Ref.\,\cite{Kallidonis:2018cas} along with a parametrisation of existing data \cite{Alberico:2008sz} that was constrained by proton experimental results on $Q^2\lesssim8.5\,\mathrm{GeV}^2$.  One pQCD analysis of this ratio has argued that it should scale as $F_{2p} / F_{1p} \sim \ln^2(Q^2/\Lambda^2)/Q^2$ \cite{Belitsky:2002kj}.  Evidently, although the general trend of the lQCD results is compatible with logarithmic growth, the current precision is insufficient to validate it.  Moreover, as remarked in Sec.\,\ref{SLNFFS}, existing empirical data on the analogous neutron ratio are inconsistent with such scaling \cite{Cloet:2008re}; hence, any success for the proton ratio is likely more apparent than real.

\begin{figure*}[!t]
\centering
\includegraphics[width=.44\textwidth]{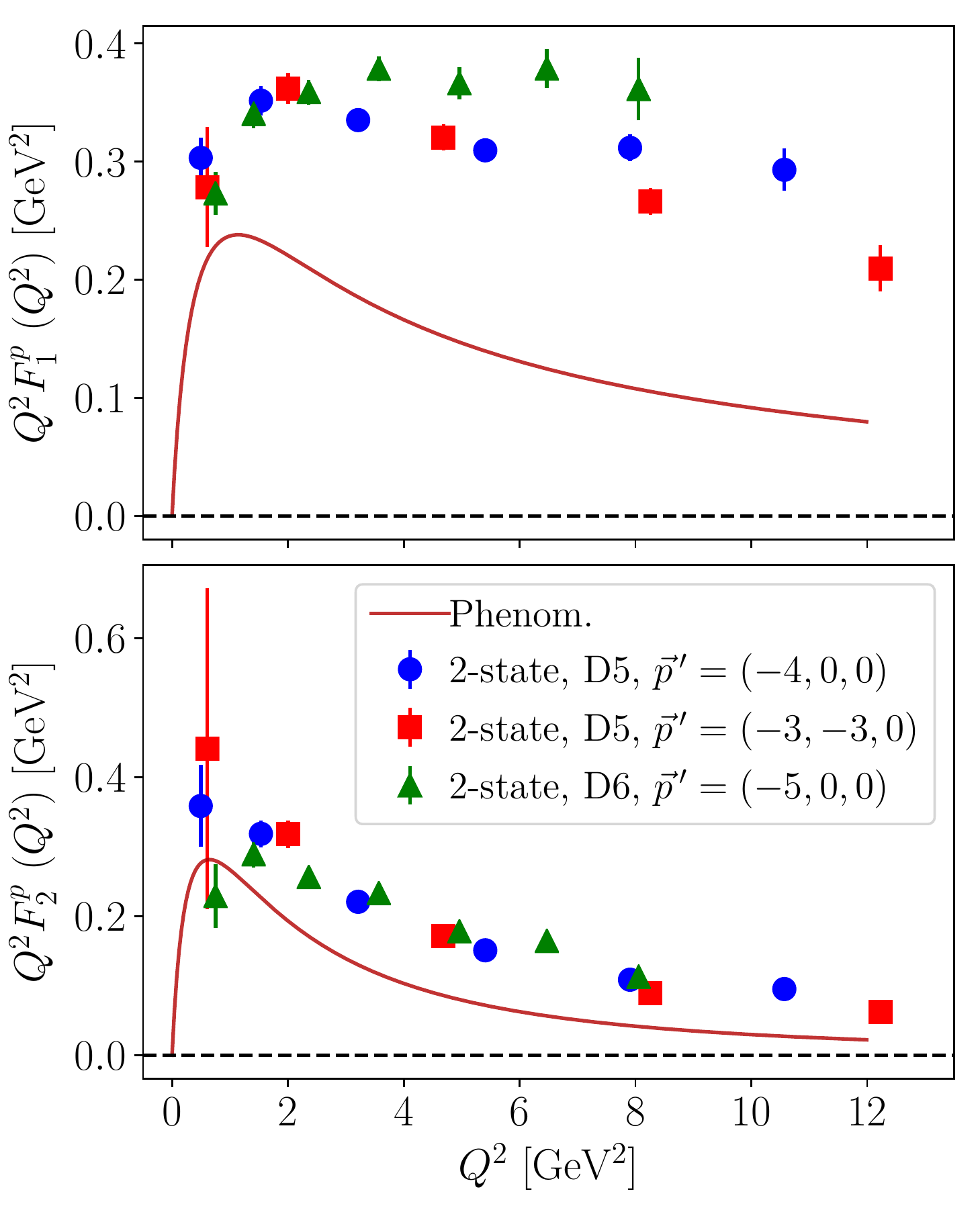}
\hspace{.02\textwidth}
\includegraphics[width=.46\textwidth]{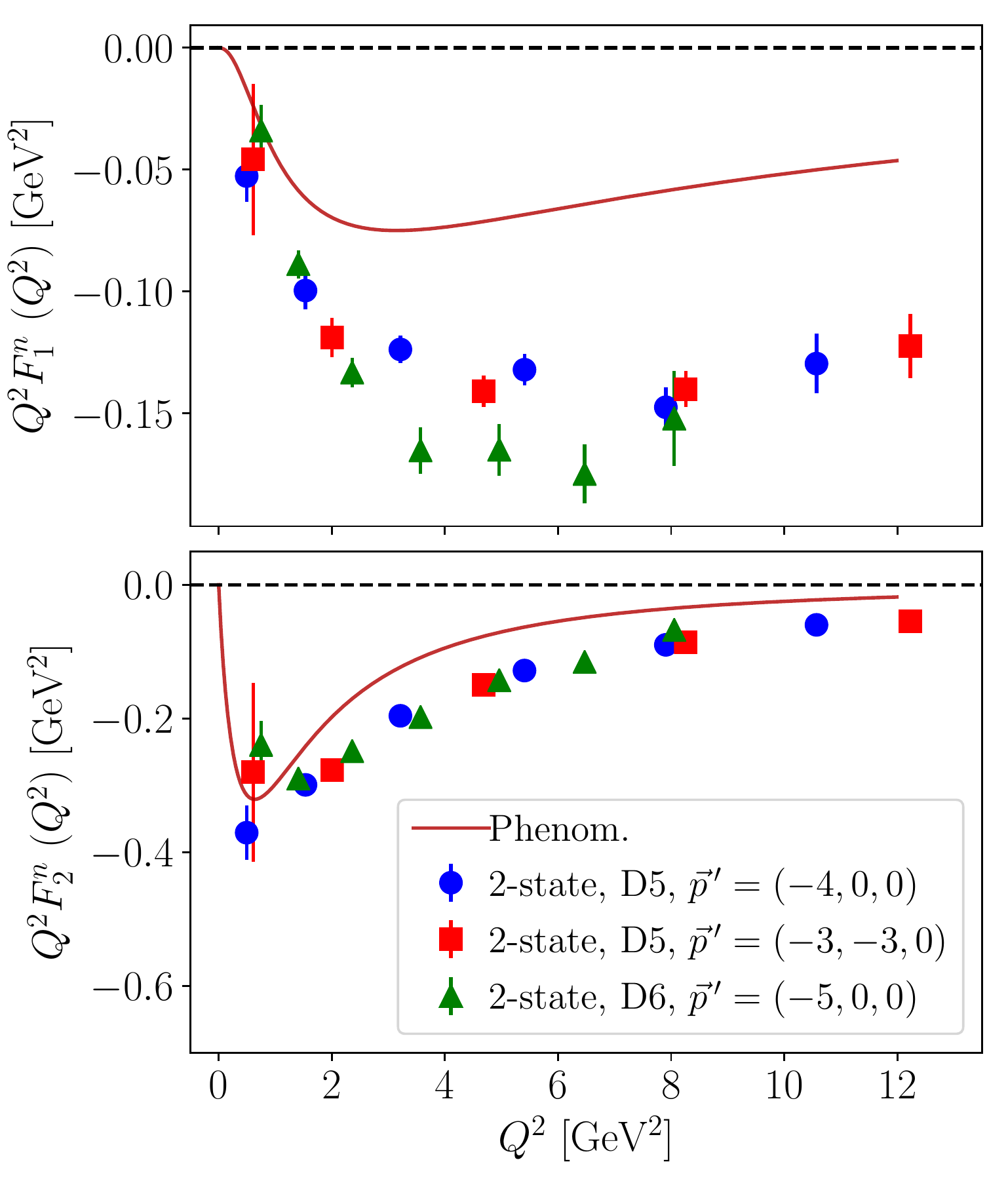}
\caption{\label{fig:ff_largeQ2:ff_comparison} Comparison of lattice results for form factors of the proton (left) and the neutron (right) \cite{Kallidonis:2018cas} with phenomenological fits of experimental data \cite{Alberico:2008sz}. Disconnected quark contractions are neglected.}
\end{figure*}

The lQCD results \cite{Kallidonis:2018cas} for the ratios of Sachs electric and magnetic form factors for the proton and neutron are, respectively, shown in the middle and right panels of Fig.\,\ref{fig:ff_largeQ2:ff_ratios}; again along with the phenomenological fits and some experimental data.  There is fair agreement between the lattice results and experiment (phenomenology) for the proton ratio, although better precision is required in light of forthcoming experiments at JLab.

In the case of the neutron, the lQCD prediction for $\mu_n\,G_{E}^n/G_{M}^n$ lies below the experimental data, as made clearer by the comparison in Fig.\,\ref{GEonGM}.
It may be that since the neutron is neutral, its electric form factor is more sensitive to systematic effects in the lQCD calculation \cite{Kallidonis:2018cas}; in particular, the omission of disconnected quark contractions and unphysically heavy pion masses.
It is worth concentrating effort here because this ratio is particularly interesting owing to the wide divergence between phenomenology and theory predictions on the domain beyond that for which empirical data is currently available -- see Fig.\,\ref{fig:SBSGEP}, lower-middle panel.
%

\begin{figure}[!t]
\centering
\includegraphics[width=.48\textwidth]{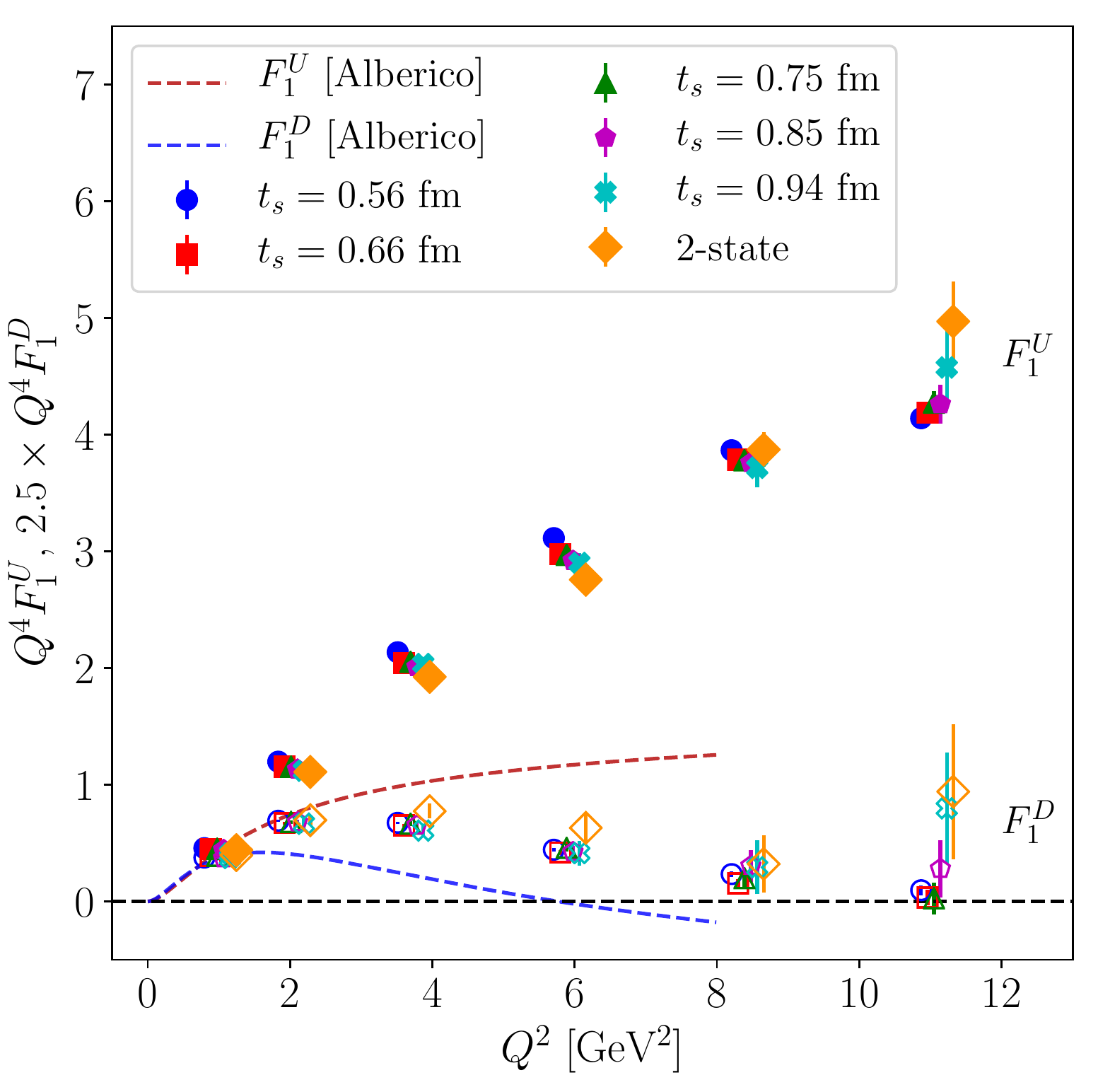}
\hspace{.02\textwidth}
\includegraphics[width=.48\textwidth]{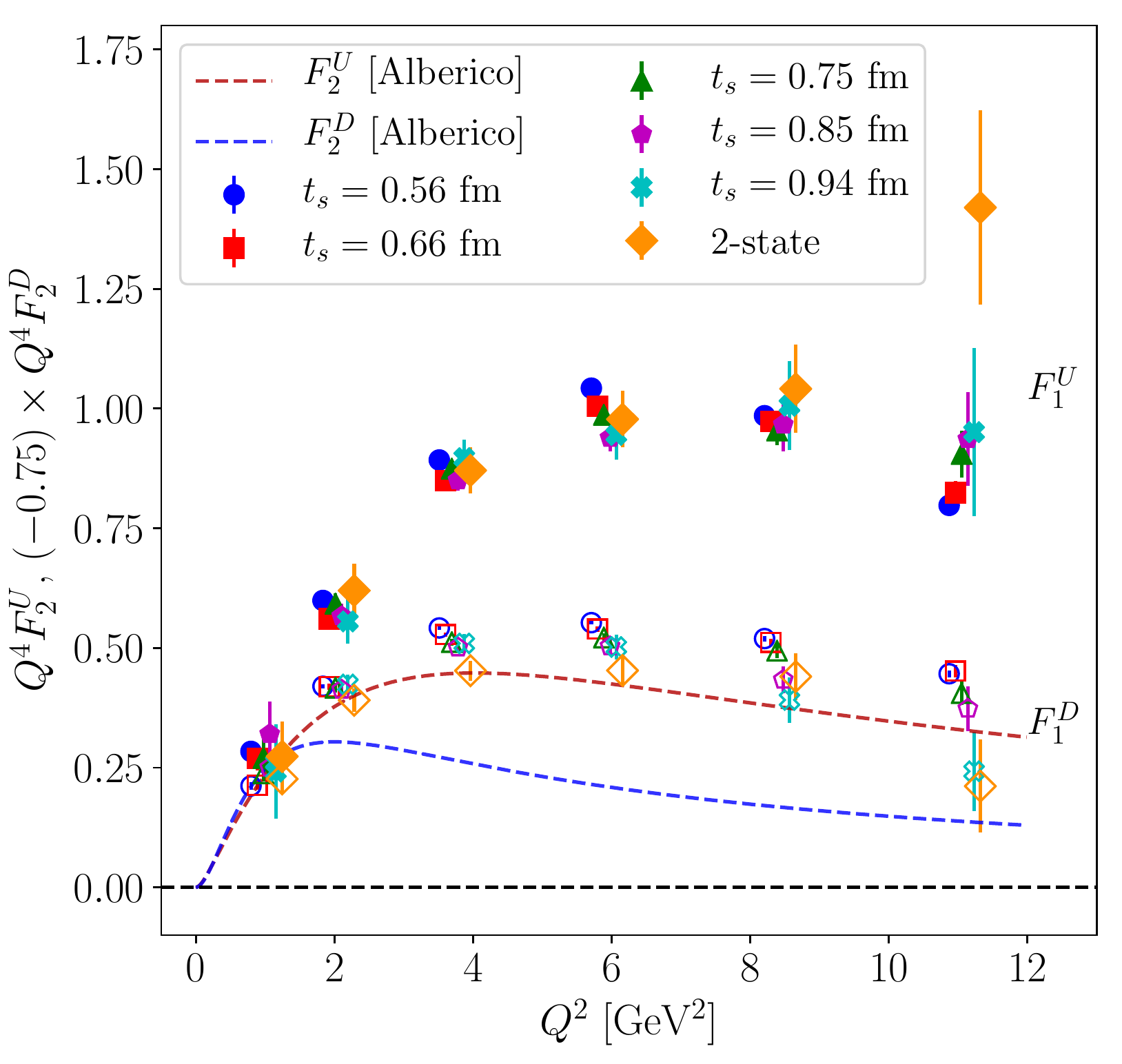}
\caption{Contributions of $u$ and $d$ quarks to Dirac $F_1$ (top) and Pauli $F_2$ (bottom) nucleon form factors. The scales are adjusted for comparison to figures in Ref.~\cite{Cates:2011pz}. Disconnected quark contractions are neglected. The phenomenological fits to experimental data are limited to $Q^2\le3.4\,\mathrm{GeV}^2$ in the neutron case~\cite{Alberico:2008sz}.
\label{fig:ff_largeQ2:ff_sepUD}}
\end{figure}

Individual proton and neutron form factors are shown in Fig.\,\ref{fig:ff_largeQ2:ff_comparison}, again compared to phenomenological fits. Although the lattice results have qualitatively similar $Q^2$ behavior, they overshoot the phenomenological fits by a factor of $2-2.5$. This substantial difference may owe to discretisation effects. Without a calculation on a smaller lattice spacing, these effects are difficult to assess. However, it may be expected that as momentum becomes large on a lattice, \emph{i.e}.\ comparable to the Brillouin zone size, rotational symmetry is violated. To test this, a calculation was repeated on the smaller lattice (``D5'') using nucleon state momenta along a 2D diagonal of the cubic lattice (shown with red squares in Fig.\,\ref{fig:ff_largeQ2:ff_comparison}). In general, lattice calculations with such momenta are expected to have smaller discretisation errors. Results from these two calculations largely agree. The largest difference is found in the case of $F_{1}^p$, where it is still much smaller than the deviation from experiment. A detailed study of $O(a)$-improved current operators and calculations at different lattice spacings are required to control this source of systematic effects.

Fig.\,\ref{fig:ff_largeQ2:ff_sepUD} shows flavour-separated form factors determined from the lQCD results in Ref.\,\cite{Kallidonis:2018cas}.  For comparison, these $u$ and $d$ contributions are shown rescaled in a fashion similar to that in Ref.\,\cite{Cates:2011pz}.  In experiment, the separated form factors are obtained by combining proton and neutron data and relying on SU$_{\rm f}(2)$ symmetry, which is exact in this lattice QCD calculation.  Since both the neutron and the proton data are required, the fit can only be relied upon for $Q^2\lesssim3.4\,\mathrm{GeV}^2$.  As seen with the nucleon form factors themselves, the flavour-separated lattice results overshoot experiment by a large factor.  Nevertheless, their is qualitative agreement on their $Q^2$ dependence and the relative size of $u$- and $d$- quark contributions.

At this juncture, the initial lQCD calculations of nucleon form factors overestimate empirical data by a large factor.  On the other hand, form factor ratios are in better agreement.  Calculations with smaller lattice spacings are underway and will lead to better understanding of current disagreements; validation of lQCD methods for high-momentum nucleon states on a lattice; and shed light on nucleon structure in the important region of transition from nonperturbative to perturbative quark-gluon dynamics.

\section{Epilogue}
\label{sec:Epilogue}

Modern facilities will probe hadronic interiors as never before, \emph{e.g}.\ JLab at 12 GeV will push form factor measurements to unprecedented values of momentum transfer and use different charge states, enabling flavour separations; an EIC and EicC would measure valence-quark distribution functions with previously unattainable precision; and elsewhere, collaborations like BaBar, Belle, BESIII, LHCb, are discovering new hadrons whose structure does not fit once viable paradigms. The wealth of new and anticipated information demands that the issue of correlations within hadrons be settled.

Fifty years ago, it was argued that pointlike diquarks might simplify treatment of the baryon bound-state problem and, subsequently, that they could explain the so-called missing resonance problem.  Today, analyses of the three valence-quark bound-state problem in quantum field theory predict that the nucleon, more generally a baryon, can be understood as a Borromean bound-state, in which non-Abelian features of QCD generate confined, non-pointlike yet strongly-correlated colour-antitriplet diquark clusters within. This diquark clustering is an emergent phenomenon, driven by the same mechanism: emergent hadronic mass (EHM), which is responsible for approximately 98\% of the visible mass in the Universe. There is evidence for such clusters in simulations of lQCD; and their presence within baryons is predicted to have numerous observable consequences, some of which already have strong experimental support. The idea of diquark clustering is also prominent amongst competing explanations of the existence and structure of tetra- and penta-quark bound-states; and there is extensive use of the diquark notion in nuclear and high-energy physics phenomenology.

Herein, our goal has been to provide a critical review of existing information, consolidate the facts, and therefrom develop a coherent, unified picture of soft quark+quark (diquark) correlations inside hadrons that answers the following key questions:
\begin{itemize}
\item[(i)] How firmly founded are continuum theoretical predictions of diquark correlations in hadrons? \\[1ex]
Within a widely-used leading-order truncation of the DSEs, the diquark Bethe-Salpeter equation differs from the meson analogue by only a $1/2$ colour factor; thus, one can find free diquark propagators whose poles are the diquark masses.  However, as noted in Sec.\,\ref{sec:diquarks-dse}, higher-order corrections remedy that defect, eliminating asymptotic diquark states from the spectrum.  Moreover, the continuum approach to the three valence-quark bound-state problem in relativistic quantum field theory has the potential to generate diquark correlations within the kernel of the Faddeev equation, \emph{i.e}.\ despite the kernel being blind to diquarks, it self-arranges in blocks that reflect a spin-flavour structure corresponding to diquark objects.
\item[(ii)] What does lattice QCD have to say about the existence and character of diquark correlations in baryons and multiquark systems? \\[1ex]
Lattice QCD has provided semi-quantitative information on diquark correlations.  The effective masses of quarks and diquarks can be derived from the related propagators in a fixed gauge.  In Landau gauge, the effective mass of $u$, $d$-quarks is roughly $0.4\,$GeV, which is close to the constituent mass value used in phenomenological models. The effective masses of the scalar and pseudovector diquarks are determined to be around $0.7\,$GeV and $1.0\,$GeV, respectively.  Although the mathematical meaning of these effective masses can be debated because diquarks are not asymptotic states, the values match those used in phenomenology and predicted by continuum Schwinger function methods.
\item[(iii)] Are there strategies for combining continuum and lattice methods in pursuit of an insightful understanding of hadron structure? \\[1ex]
Yes, there are many.  Some of them have been detailed herein, \emph{e.g}.\ DSE and lattice studies of baryon PDAs, where the way in which the total longitudinal momentum is shared by the three valence quarks can indicate quark+quark correlations. Another strategy sees the same physical observable studied as a function of the pion mass, \emph{i.e}.\ current quark mass, using both continuum and discrete formulations of QCD.  This could elucidate the role played by virtual quark+antiquark pairs, different gluonic environments, effects distinguishing light verses heavy quarks, \emph{etc}.
\item[(iv)] Can theory identify experimental observables that would constitute unambiguous measurable signals for the presence of diquark correlations? \\[1ex]
Yes. Some of them are related with spectroscopy of exotic hadrons, such as the fact that the diquark+antidiquark picture of tetraquark states inescapably implies the existence of charged or doubly charged partners of the XYZ particles.  More generally, for each level predicted, a degenerate isovector and isoscalar state should appear. On the other hand, most of the experimental observables identified herein are connected with structural properties of conventional hadrons, \emph{e.g}.\ the existence of zeros in the $d$-quark contribution to the Dirac and Pauli form factors highlights that any appearance of scaling in available data on nucleon electromagnetic form factors is incidental because the zero expresses a continuing role for correlations which distinguish between quark flavours and impose different features upon their scattering patterns.
\item[(v)] Is there a traceable connection between the so-called diquarks used to build phenomenological models of high-energy processes and the correlations predicted by contemporary theory; and if so, how can such models be improved therefrom? \\[1ex]
If such a connection is to be drawn, then the concept of diquarks as effective degrees of freedom must evolve to more closely resemble the contemporary view derived from continuum and discrete functional methods, \emph{viz}.\ modern diquarks are confined, with mass-scales that express the strength and range of the correlation inside the hadron; they are fully dynamical, with no quark holding a special place because each one participates in all correlations to the fullest extent allowed by its quantum numbers; they have electromagnetic sizes, which enforce certain distinct interaction patterns; and there are different species, amongst which  isoscalar-scalar and isovector-pseudovector correlations are the strongest but others play a key role in nucleon excited states.
\item[(vi)] Are diquarks the only type of two-body correlations that play a role in hadron structure? \\[1ex]
Inspired by the probable existence of quark+quark correlations inside baryons, the continuum quantum field theory approach has recently been applied to hybrid mesons (systems constituted from quark+antiquark+gluon systems) and glueballs (a system constituted only by gluons) exploiting the existence of strong two-body correlations in the gluon-quark and gluon-gluon channels. These studies have demonstrated that pursuing agreement with lQCD can provide insights into fundamental quantities of QCD such as the infrared-dressing of the three-gluon vertex and potential subtle effects of EHM in the quark-gluon and antiquark-gluon vertices that enter in the kernel of the Bethe-Salpeter equations for a hybrid.
\item[(vii)] Which new experiments, facilities and analysis tools are best suited to test the emerging picture of two-body correlations in hadrons? \\[1ex]
This point has been canvassed throughout, but there is merit in reiterating one example: deep inelastic scattering on nuclear targets (nDIS) with identified final state hadrons probes the mechanisms of colour propagation and hadron formation on fm distance scales.  A good description of the production of light mesons at HERMES has been achieved.  However, the production of protons in nDIS shows very different patterns, which imply a different production mechanism from that of light mesons.  A potential explanation involves direct knockout of diquarks, which subsequently form into new protons.  A powerful experimental test of this explanation would be the confirmation of transverse momentum broadening with a similar magnitude for proton, neutron, and $\Lambda$ baryons, and much greater magnitude than that seen for light mesons. This is motivated by theoretical analyses of the different kinds of diquark correlations inside baryons.
\end{itemize}

A great deal has changed since the introduction of the diquark concept more than fifty years ago.  The questions surrounding diquarks have evolved, as have the experimental and theoretical tools that can be deployed to answer them.  JLab at 12\,GeV will deliver a vast array of data.  It will challenge theory.  In answering that challenge, much will be learnt.  Continuing revelations in spectroscopy, and experiments at upgraded and new facilities will pose questions in different areas.  Within twenty years, as experiment moves into the realm where modern diquarks are supposed to live, answers will be found: Diquarks; if so, what and why?



\vspace*{0.10cm}
\section*{Acknowledgments}
We are particularly grateful to Daniele Binosi and Susan Driessen for their hospitality, support and help during the ECT$^\ast$ Workshop ``Diquark Correlations in Hadron Physics: Origin, Impact and Evidence", held in Trento, Italy, September 23-27, 2019, from which emerged the perspective contained herein.
Work supported by:
%
%
Consejo Nacional de Ciencia y Tecnolog{\'{\i}}a (CONACyT), under the Estancias posdoctorales en el extranjero (EPE-2019) program;
%
Chilean grants CONICYT PIA ACT-1413 and ACT-1409, BASAL FB-0821 and  BASAL AFB 180002; FONDECYT 1080564, 1120953, and 1161642, and ECOS-CONICYT C12E04;
%
%
the Helmholtz International Center for FAIR, within the LOEWE program of the State of Hesse;
DFG grant FI 970/11-1;
National Natural Science Foundation of China, under grant No. 11935017;
%
%
%
FCT Investigator Grant IF/00898/2015;
%
%
Academy of Finland, Project No. 320062;
%
%
US National Science Foundation under grants PHY-1714133 and PHY-1812382;
%
%
%
%
%
Jiangsu Province \emph{Hundred Talents Plan for Professionals};
%
U.S. Department of Energy, Office of Science, Office of Nuclear Physics, under contracts DE-AC05-06OR23177 and DE-SC0014230;
%
%
Ministerio Espa\~nol de Ciencia e Innovaci\'on, under grant No. PID2019-107844GB-C22;
Junta de Andaluc\'ia, under contract No. Operativo FEDER Andaluc\'ia 2014-2020 UHU-1264517;
European Union Horizon 2020 research and innovation programme, under grant agreement No. 824093;
%
%
%
%
Deutsche Forschungsgemeinschaft (collaborative research centre SFB/TRR-55);
%
and Jefferson Science Associates (JSA) Initiatives Fund Program grant No. 100-50-15 (FY2019).
%
SS also acknowledges support by the RHIC Physics Fellow Program of the RIKEN BNL Research Center and by the National Science Foundation under CAREER Award PHY-1847893.
PW expresses his gratitude to the Mainz Institute for Theoretical Physics (MITP) of the Cluster of Excellence PRISMA+ (Project ID 39083149) for its hospitality and support.
MYB is grateful to Prof. Dr. Vladimir Kekelidze, Prof. Dr. Alexander Vodopyanov and Prof. Dr. Adam Kisiel for fruitful discussions on this topic.
CM acknowledges the warm hospitality of INFN Sezione di Roma and the support of the NINPHA project.


\section*{Abbreviations}
\unskip
\label{abbreviations}
The following abbreviations are used in this manuscript:\\[-1ex]

\begin{longtable}{ll}
ACC & aerogel threshold Cherenkov counters \\
BABAR & detector at SLAC \\
Belle (Belle-II) & detector at Japan's high energy accelerator research complex in Tsukuba \\
BEPC (BEPCII) & Beijing Electron Positron Collider \\
BESIII & detector at BEPC \\
BL19 & Analysis framework in Refs.\,\cite{BL1, Brooks_2019} \\
BoNuS & detector and associated collaboration at JLab \\
BS (BSE) & Bethe-Salpeter (equation)\\
CDC & central drift chamber \\
CEBAF & Continuous Electron Beam Accelerator Facility at JLab \\
CERN & European Laboratory for Particle Physics \\
CL & confidence limit \\
CLAS & detector in Hall-B at JLab\\
CLAS\,12 & upgraded CLAS detector \\
CLS & coordinated lattice simulations \\
COMPASS & detector at CERN \\
CSM & continuum Schwinger-function method \\
DA & distribution amplitude \\
DCSB & dynamical chiral symmetry breaking \\
DF & distribution function \\
DIS & deep inelastic scattering \\
DSE & Dyson-Schwinger equation\\
DVCS & deeply virtual Compton scattering \\
ECAL & electromagnetic calorimeter \\
EIC & electron ion collider in the USA\\
EicC & electron ion collider in China\\
EMC & European muon collaboration \\
EMFF(s) & electromagnetic form factor(s) \\
EHM & emergent hadronic mass\\
FAIR & Facility for Antiproton and Ion Research, Darmstadt \\
FF(s) & form factor(s) \\
FFF2004 & nucleon form factor fit from 2004, Ref.\,\cite{Kelly:2004hm}\\
FSR & final-state radiation \\
GEM & gas electron multiplier \\
GPD & generalised parton distribution \\
HCAL & hadronic calorimeter \\
HERA & particle accelerator in Hamburg \\
HERMES & detector and associated collaboration at HERA \\
IP & interaction point \\
ISR & initial-state radiation \\
JLab & Thomas Jefferson National Accelerator Facility\\
JLab\,12 & Thomas Jefferson National Accelerator Facility with 12\,GeV $e^-$ beams\\
LFRHD & light-front relativistic Hamiltonian dynamics \\
LFRQM & light-front relativistic quark model \\
LHC & large hadron collider \\
LHCb & LHC beauty experiment \\
LO & leading-order \\
lQCD (LQCD)  & lattice-regularised quantum chromodynamics\\
LSM & Lund string model \\
MB\,FSIs & meson-baryon final state interactions \\
MC & Monte-Carlo \\
$\MSbar$ & modified minimal subtraction renormalisation scheme \\
nDIS & DIS on nuclear targets \\
NG (boson/mode) & Nambu-Goldstone (boson/mode) \\
NICA  & Nuclotron-based Ion Collider fAcility, Dubna \\
NLO & next-to-leading-order \\
PDA & parton distribution amplitude \\
PDF & parton distribution function \\
PDG & Particle Data Group \\
PG & Pauli-G\"ursey (symmetry) \\
pQCD & perturbative quantum chromodynamics\\
QCD & quantum chromodynamics \\
QM & quark model \\
RBC & Riken-Brookhaven-Columbia lattice-QCD collaboration \\
RI$^\prime$\nobreakdash-SMOM & modified Rome-Southampton lQCD regularisation and renormalisation scheme \\
RL & rainbow-ladder (truncation) \\
SBS & Super BigBite Spectrometer at JLab \\
SCI & symmetry-preserving treatment of the vector$\,\otimes\,$vector contact interaction \\
SIDIS & semi-inclusive deep inelastic scattering \\
SPM & Schlessinger point method \\
SM & Standard Model of Particle Physics \\
SSA & single spin asymmetry \\
TOF & time-of-flight scintillation counters \\
TMD &  transverse momentum dependent parton distribution \\
UKQCD & United Kingdom lattice-QCD collaboration\\
VMD & vector meson dominance
\end{longtable}

\medskip

\section*{References}
\label{references}

\bibliographystyle{../../../zProc/z10/z10KITPC/h-physrev4}
\bibliography{../../../CollectedBiB}

\begin{thebibliography}{100}

\bibitem{Pennington:1996dy}
M.~Pennington,
\newblock {Calculating hadronic properties in strong QCD -- hep-ph/9611242},
\newblock in {\em {Proceedings of the 2nd ELFE Workshop}}, 1996.

\bibitem{Politzer:2005kc}
H.~D. Politzer,
\newblock Proc. Nat. Acad. Sci. {\bf 102}, 7789 (2005).

\bibitem{Wilczek:2005az}
F.~Wilczek,
\newblock Proc. Nat. Acad. Sci. {\bf 102}, 8403 (2005).

\bibitem{Gross:2005kv}
D.~J. Gross,
\newblock Proc. Nat. Acad. Sci. {\bf 102}, 9099 (2005).

\bibitem{Binosi:2014aea}
D.~Binosi, L.~Chang, J.~Papavassiliou and C.~D. Roberts,
\newblock Phys. Lett. B {\bf 742}, 183 (2015).

\bibitem{Binosi:2016nme}
D.~Binosi, C.~Mezrag, J.~Papavassiliou, C.~D. Roberts and
  J.~Rodr{\'i}guez-Quintero,
\newblock Phys. Rev. D {\bf 96}, 054026 (2017).

\bibitem{Cui:2019dwv}
Z.-F. Cui {\em et~al.},
\newblock Chin. Phys. C {\bf 44}, 083102 (2020).

\bibitem{Bissey:2006bz}
F.~Bissey {\em et~al.},
\newblock Phys. Rev. D {\bf 76}, 114512 (2007).

\bibitem{Bissey:2009gw}
F.~Bissey, A.~Signal and D.~Leinweber,
\newblock Phys. Rev. D {\bf 80}, 114506 (2009).

\bibitem{Gribov:1999ui}
V.~N. Gribov,
\newblock Eur. Phys. J. C {\bf 10}, 91 (1999).

\bibitem{Munczek:1983dx}
H.~J. Munczek and A.~M. Nemirovsky,
\newblock Phys. Rev. D {\bf 28}, 181 (1983).

\bibitem{Stingl:1983pt}
M.~Stingl,
\newblock Phys. Rev. D {\bf 29}, 2105 (1984).

\bibitem{Cahill:1985mh}
R.~T. Cahill and C.~D. Roberts,
\newblock Phys. Rev. D {\bf 32}, 2419 (1985).

\bibitem{Stingl:1985hx}
M.~Stingl,
\newblock Phys. Rev. D {\bf 34}, 3863 (1986),
\newblock [Erratum: Phys. Rev. D {\bf 36}, 651 (1987)].

\bibitem{Cahill:1988zi}
R.~T. Cahill,
\newblock Austral. J. Phys. {\bf 42}, 171 (1989).

\bibitem{Krein:1990sf}
C.~D. Roberts, A.~G. Williams and G.~Krein,
\newblock Int. J. Mod. Phys. A {\bf 7}, 5607 (1992).

\bibitem{Burden:1991gd}
C.~J. Burden, C.~D. Roberts and A.~G. Williams,
\newblock Phys. Lett. B {\bf 285}, 347 (1992).

\bibitem{Hawes:1993ef}
F.~T. Hawes, C.~D. Roberts and A.~G. Williams,
\newblock Phys. Rev. D {\bf 49}, 4683 (1994).

\bibitem{Roberts:1994dr}
C.~D. Roberts and A.~G. Williams,
\newblock Prog. Part. Nucl. Phys. {\bf 33}, 477 (1994).

\bibitem{Roberts:2000aa}
C.~D. Roberts and S.~M. Schmidt,
\newblock Prog. Part. Nucl. Phys. {\bf 45}, S1 (2000).

\bibitem{Alkofer:2000wg}
R.~Alkofer and L.~von Smekal,
\newblock Phys. Rept. {\bf 353}, 281 (2001).

\bibitem{Roberts:2007ji}
C.~D. Roberts,
\newblock Prog. Part. Nucl. Phys. {\bf 61}, 50 (2008).

\bibitem{Brodsky:2012ku}
S.~J. Brodsky, C.~D. Roberts, R.~Shrock and P.~C. Tandy,
\newblock Phys. Rev. C {\bf 85}, 065202 (2012).

\bibitem{Strauss:2012dg}
S.~Strauss, C.~S. Fischer and C.~Kellermann,
\newblock Phys. Rev. Lett. {\bf 109}, 252001 (2012).

\bibitem{Gao:2015kea}
F.~Gao {\em et~al.},
\newblock Phys. Rev. D {\bf 93}, 094019 (2016).

\bibitem{Papavassiliou:2015aga}
J.~Papavassiliou,
\newblock J. Phys. Conf. Ser. {\bf 631}, 012006 (2015).

\bibitem{Binosi:2019ecz}
D.~Binosi and R.-A. Tripolt,
\newblock Phys. Lett. B {\bf 801}, 135171 (2020).

\bibitem{Segovia:2008zza}
J.~Segovia, D.~R. Entem and F.~Fernandez,
\newblock Phys. Lett. B {\bf 662}, 33 (2008).

\bibitem{Yang:2017qan}
G.~Yang, J.~Ping and J.~Segovia,
\newblock Few Body Syst. {\bf 59}, 113 (2018).

\bibitem{Nambu:1961tp}
Y.~Nambu and G.~Jona-Lasinio,
\newblock Phys. Rev. {\bf 122}, 345 (1961).

\bibitem{JonaLasinio:1964cw}
G.~Jona-Lasinio,
\newblock Nuovo Cim. {\bf 34}, 1790 (1964).

\bibitem{Nambu:2009zza}
Y.~Nambu,
\newblock Int. J. Mod. Phys. A {\bf 24}, 2371 (2009).

\bibitem{Maris:1997hd}
P.~Maris, C.~D. Roberts and P.~C. Tandy,
\newblock Phys. Lett. B {\bf 420}, 267 (1998).

\bibitem{Bali:2005fu}
G.~S. Bali, H.~Neff, T.~Duessel, T.~Lippert and K.~Schilling,
\newblock Phys. Rev. D {\bf 71}, 114513 (2005).

\bibitem{Prkacin:2005dc}
Z.~Prkacin {\em et~al.},
\newblock PoS {\bf LAT2005}, 308 (2006).

\bibitem{Cahill:1987qr}
R.~T. Cahill, C.~D. Roberts and J.~Praschifka,
\newblock Phys. Rev. D {\bf 36}, 2804 (1987).

\bibitem{Cahill:1988dx}
R.~T. Cahill, C.~D. Roberts and J.~Praschifka,
\newblock Austral. J. Phys. {\bf 42}, 129 (1989).

\bibitem{Oettel:1998bk}
M.~Oettel, G.~Hellstern, R.~Alkofer and H.~Reinhardt,
\newblock Phys. Rev. C {\bf 58}, 2459 (1998).

\bibitem{Bloch:1999vk}
J.~C.~R. Bloch, C.~D. Roberts and S.~M. Schmidt,
\newblock Phys. Rev. C {\bf 60}, 065208 (1999).

\bibitem{Bender:2002as}
A.~Bender, W.~Detmold, C.~D. Roberts and A.~W. Thomas,
\newblock Phys. Rev. C {\bf 65}, 065203 (2002).

\bibitem{Eichmann:2008ef}
G.~Eichmann, I.~C. Cloet, R.~Alkofer, A.~Krassnigg and C.~D. Roberts,
\newblock Phys. Rev. C {\bf 79}, 012202(R) (2009).

\bibitem{Roberts:2011cf}
H.~L.~L. Roberts, L.~Chang, I.~C. Cloet and C.~D. Roberts,
\newblock Few Body Syst. {\bf 51}, 1 (2011).

\bibitem{Segovia:2015ufa}
J.~Segovia, C.~D. Roberts and S.~M. Schmidt,
\newblock Phys. Lett. B {\bf 750}, 100 (2015).

\bibitem{Eichmann:2016yit}
G.~Eichmann, H.~Sanchis-Alepuz, R.~Williams, R.~Alkofer and C.~S. Fischer,
\newblock Prog. Part. Nucl. Phys. {\bf 91}, 1 (2016).

\bibitem{Hess:1998sd}
M.~Hess, F.~Karsch, E.~Laermann and I.~Wetzorke,
\newblock Phys. Rev. D {\bf 58}, 111502 (1998).

\bibitem{Orginos:2005vr}
K.~Orginos,
\newblock PoS {\bf LAT2005}, 054 (2006).

\bibitem{Alexandrou:2006cq}
C.~Alexandrou, {\mbox{Ph}}.~de~Forcrand and B.~Lucini,
\newblock Phys. Rev. Lett. {\bf 97}, 222002 (2006).

\bibitem{DeGrand:2007vu}
T.~DeGrand, Z.~Liu and S.~Schaefer,
\newblock Phys. Rev. D {\bf 77}, 034505 (2008).

\bibitem{Babich:2007ah}
R.~Babich {\em et~al.},
\newblock Phys. Rev. D {\bf 76}, 074021 (2007).

\bibitem{Green:2010vc}
J.~Green, J.~Negele, M.~Engelhardt and P.~Varilly,
\newblock PoS {\bf LATTICE\,2010}, 140 (2010).

\bibitem{Bi:2015ifa}
Y.~Bi {\em et~al.},
\newblock Chin. Phys. C {\bf 40}, 073106 (2016).

\bibitem{Pauli:1957}
W.~Pauli,
\newblock Nuovo Cim. {\bf 6}, 204 (1957).

\bibitem{Gursey:1958}
F.~G{\"u}rsey,
\newblock Nuovo Cim. {\bf 7}, 411 (1958).

\bibitem{Maris:2002yu}
P.~Maris,
\newblock Few Body Syst. {\bf 32}, 41 (2002).

\bibitem{Maris:2004bp}
P.~Maris,
\newblock Few Body Syst. {\bf 35}, 117 (2004).

\bibitem{Aznauryan:2012ba}
I.~Aznauryan {\em et~al.},
\newblock Int. J. Mod. Phys. E {\bf 22}, 1330015 (2013).

\bibitem{Segovia:2015hra}
J.~Segovia {\em et~al.},
\newblock Phys. Rev. Lett. {\bf 115}, 171801 (2015).

\bibitem{Roberts:2015lja}
C.~D. Roberts,
\newblock J. Phys. Conf. Ser. {\bf 706}, 022003 (2016).

\bibitem{Eichmann:2016hgl}
G.~Eichmann, C.~S. Fischer and H.~Sanchis-Alepuz,
\newblock Phys. Rev. D {\bf 94}, 094033 (2016).

\bibitem{Burkert:2017djo}
V.~D. Burkert and C.~D. Roberts,
\newblock Rev. Mod. Phys. {\bf 91}, 011003 (2019).

\bibitem{Lu:2017cln}
Y.~Lu {\em et~al.},
\newblock Phys. Rev. C {\bf 96}, 015208 (2017).

\bibitem{Chen:2017pse}
C.~Chen {\em et~al.},
\newblock Phys. Rev. D {\bf 97}, 034016 (2018).

\bibitem{Chen:2019fzn}
C.~Chen, G.~I. Krein, C.~D. Roberts, S.~M. Schmidt and J.~Segovia,
\newblock Phys. Rev. D {\bf 100}, 054009 (2019).

\bibitem{Yin:2019bxe}
P.-L. Yin {\em et~al.},
\newblock Phys. Rev. D {\bf 100}, 034008 (2019).

\bibitem{Lu:2019bjs}
Y.~Lu {\em et~al.},
\newblock Phys. Rev. D {\bf 100}, 034001 (2019).

\bibitem{Anselmino:1992vg}
M.~Anselmino, E.~Predazzi, S.~Ekelin, S.~Fredriksson and D.~B. Lichtenberg,
\newblock Rev. Mod. Phys. {\bf 65}, 1199 (1993).

\bibitem{Ripani:2002ss}
M.~Ripani {\em et~al.},
\newblock Phys. Rev. Lett. {\bf 91}, 022002 (2003).

\bibitem{Burkert:2012ee}
V.~D. Burkert,
\newblock EPJ Web Conf. {\bf 37}, 01017 (2012).

\bibitem{Kamano:2013iva}
H.~Kamano, S.~X. Nakamura, T.~S.~H. Lee and T.~Sato,
\newblock Phys. Rev. C {\bf 88}, 035209 (2013).

\bibitem{Edwards:2011jj}
R.~G. Edwards, J.~J. Dudek, D.~G. Richards and S.~J. Wallace,
\newblock Phys. Rev. D {\bf 84}, 074508 (2011).

\bibitem{Cates:2011pz}
G.~Cates, C.~de~Jager, S.~Riordan and B.~Wojtsekhowski,
\newblock Phys. Rev. Lett. {\bf 106}, 252003 (2011).

\bibitem{Wojtsekhowski:2020tlo}
B.~Wojtsekhowski,
\newblock {Flavor Decomposition of Nucleon Form Factors -- arXiv:2001.02190
  [nucl-ex]},
\newblock 2020.

\bibitem{Chen:2016qju}
H.-X. Chen, W.~Chen, X.~Liu and S.-L. Zhu,
\newblock Phys. Rept. {\bf 639}, 1 (2016).

\bibitem{Esposito:2016noz}
A.~Esposito, A.~Pilloni and A.~D. Polosa,
\newblock Phys. Rept. {\bf 668}, 1 (2017).

\bibitem{Ali:2017jda}
A.~Ali, J.~S. Lange and S.~Stone,
\newblock Prog. Part. Nucl. Phys. {\bf 97}, 123 (2017).

\bibitem{Olsen:2017bmm}
S.~L. Olsen, T.~Skwarnicki and D.~Zieminska,
\newblock Rev. Mod. Phys. {\bf 90}, 015003 (2018).

\bibitem{Guo:2017jvc}
F.-K. Guo {\em et~al.},
\newblock Rev. Mod. Phys. {\bf 90}, 015004 (2018).

\bibitem{Liu:2019zoy}
Y.-R. Liu, H.-X. Chen, W.~Chen, X.~Liu and S.-L. Zhu,
\newblock Prog. Part. Nucl. Phys. {\bf 107}, 237 (2019).

\bibitem{Guo:2019twa}
F.-K. Guo, X.-H. Liu and S.~Sakai,
\newblock Prog. Part. Nucl. Phys. {\bf 112}, 103757 (2020).

\bibitem{Szczekowski:1988pq}
M.~Szczekowski,
\newblock Int. J. Mod. Phys. A {\bf 4}, 3985 (1989).

\bibitem{Anselmino:1989vt}
M.~Anselmino and E.~Predazzi, editors,
\newblock {\em {Diquarks. Proceedings, Workshop, Turin, Italy, October 24-26,
  1988}}, 1989.

\bibitem{Anselmino:1994uj}
M.~Anselmino and E.~Predazzi, editors,
\newblock {\em {Diquarks. Proceedings, 2nd Workshop, Villa Gualino, Turin,
  Italy, November 2-4, 1992}}, 1994.

\bibitem{Anselmino:1998rf}
M.~Anselmino and E.~Predazzi, editors,
\newblock {\em {Diquarks 3. Proceedings, 3rd Workshop, Torino, Italy, October
  28-30, 1996}}, 1998.

\bibitem{Skytt:1991rm}
B.-O. Skytt and S.~Fredriksson,
\newblock (1991),
\newblock {\emph{Diquark} '\emph{91}: \emph{A Literature Survey}}.

\bibitem{GellMann:1964nj}
M.~Gell-Mann,
\newblock Phys. Lett. {\bf 8}, 214 (1964).

\bibitem{Zweig:1981pd}
G.~Zweig,
\newblock (1964),
\newblock {\emph{An $SU(3)$ model for strong interaction symmetry and its
  breaking. Parts 1 and 2} (CERN Reports No.\ 8182/TH.\ 401 and No.\ 8419/TH.\
  412)}.

\bibitem{Ida:1966ev}
M.~Ida and R.~Kobayashi,
\newblock Prog. Theor. Phys. {\bf 36}, 846 (1966).

\bibitem{Lichtenberg:1967zz}
D.~B. Lichtenberg and L.~J. Tassie,
\newblock Phys. Rev. {\bf 155}, 1601 (1967).

\bibitem{Lichtenberg:1981pp}
D.~B. Lichtenberg,
\newblock Phys. Rev. {\bf 178}, 2197 (1969).

\bibitem{Neubert:1989fp}
M.~Neubert and B.~Stech,
\newblock Phys. Lett. B {\bf 231}, 477 (1989).

\bibitem{Close:1988br}
F.~E. Close and A.~W. Thomas,
\newblock Phys. Lett. {\bf B212}, 227 (1988).

\bibitem{Jaffe:2004ph}
R.~L. Jaffe,
\newblock Phys. Rept. {\bf 409}, 1 (2005).

\bibitem{Wilczek:2004im}
F.~Wilczek,
\newblock {Diquarks as inspiration and as objects},
\newblock in {\em {Deserfest: A Celebration of the Life and Works of Stanley
  Deser}}, pp. 322--338, 2004.

\bibitem{Selem:2006nd}
A.~Selem and F.~Wilczek,
\newblock {Hadron systematics and emergent diquarks -- hep-ph/0602128},
\newblock in {\em {Ringberg Workshop on New Trends in HERA Physics 2005}}, pp.
  337--356, 2006.

\bibitem{Martin:1985hw}
A.~Martin,
\newblock Z. Phys. C {\bf 32}, 315 (1986).

\bibitem{Johnson:1975sg}
K.~Johnson and C.~B. Thorn,
\newblock Phys. Rev. D {\bf 13}, 1934 (1976).

\bibitem{Santopinto:2004hw}
E.~Santopinto,
\newblock Phys. Rev. C {\bf 72}, 022201 (2005).

\bibitem{Galata:2012xt}
G.~Galata and E.~Santopinto,
\newblock Phys. Rev. C {\bf 86}, 045202 (2012).

\bibitem{Lichtenberg:1979de}
D.~Lichtenberg and R.~Johnson,
\newblock Hadronic J. {\bf 2}, 1 (1979).

\bibitem{Lichtenberg:1982jp}
D.~Lichtenberg, W.~Namgung, E.~Predazzi and J.~Wills,
\newblock Phys. Rev. Lett. {\bf 48}, 1653 (1982).

\bibitem{Ferretti:2011zz}
J.~Ferretti, A.~Vassallo and E.~Santopinto,
\newblock Phys. Rev. C {\bf 83}, 065204 (2011).

\bibitem{Santopinto:2014opa}
E.~Santopinto and J.~Ferretti,
\newblock Phys. Rev. C {\bf 92}, 025202 (2015).

\bibitem{Gutierrez:2014qpa}
C.~Gutierrez and M.~De~Sanctis,
\newblock Eur. Phys. J. A {\bf 50}, 169 (2014).

\bibitem{DeSanctis:2014ria}
M.~De~Sanctis, J.~Ferretti, E.~Santopinto and A.~Vassallo,
\newblock Eur. Phys. J. A {\bf 52}, 121 (2016).

\bibitem{Roncaglia:1995az}
R.~Roncaglia, D.~Lichtenberg and E.~Predazzi,
\newblock Phys. Rev. D {\bf 52}, 1722 (1995).

\bibitem{Ebert:1996ec}
D.~Ebert, R.~Faustov, V.~Galkin, A.~Martynenko and V.~Saleev,
\newblock Z. Phys. C {\bf 76}, 111 (1997).

\bibitem{Gershtein:2000nx}
S.~S. Gershtein, V.~V. Kiselev, A.~K. Likhoded and A.~I. Onishchenko,
\newblock Phys. Rev. D {\bf 62}, 054021 (2000).

\bibitem{Ebert:2002ig}
D.~Ebert, R.~N. Faustov, V.~O. Galkin and A.~P. Martynenko,
\newblock Phys. Rev. D {\bf 66}, 014008 (2002).

\bibitem{Giannuzzi:2009gh}
F.~Giannuzzi,
\newblock Phys. Rev. D {\bf 79}, 094002 (2009).

\bibitem{Ali:2017wsf}
A.~Ali {\em et~al.},
\newblock Eur. Phys. J. C {\bf 78}, 29 (2018).

\bibitem{Santopinto:2018ljf}
E.~Santopinto {\em et~al.},
\newblock Eur. Phys. J. C {\bf 79}, 1012 (2019).

\bibitem{Santopinto:2016fay}
E.~Santopinto and J.~Ferretti,
\newblock Few Body Syst. {\bf 57}, 1095 (2016).

\bibitem{Isgur:1978xj}
N.~Isgur and G.~Karl,
\newblock Phys. Rev. D {\bf 18}, 4187 (1978).

\bibitem{Capstick:1986bm}
S.~Capstick and N.~Isgur,
\newblock Phys. Rev. D {\bf 34}, 2809 (1986).

\bibitem{Capstick:1992th}
S.~Capstick and W.~Roberts,
\newblock Phys. Rev. D {\bf 47}, 1994 (1993).

\bibitem{Bijker:1994yr}
R.~Bijker, F.~Iachello and A.~Leviatan,
\newblock Annals Phys. {\bf 236}, 69 (1994).

\bibitem{Glozman:1995fu}
L.~{\relax Ya}. Glozman and D.~O. Riska,
\newblock Phys. Rept. {\bf 268}, 263 (1996).

\bibitem{Loring:2001kx}
U.~L{\"o}ring, B.~C. Metsch and H.~R. Petry,
\newblock Eur. Phys. J. A {\bf 10}, 395 (2001).

\bibitem{Giannini:2001kb}
M.~Giannini, E.~Santopinto and A.~Vassallo,
\newblock Eur. Phys. J. A {\bf 12}, 447 (2001).

\bibitem{Crede:2013sze}
V.~Crede and W.~Roberts,
\newblock Rept. Prog. Phys. {\bf 76}, 076301 (2013).

\bibitem{Mokeev:2015moa}
V.~I. Mokeev, I.~Aznauryan, V.~Burkert and R.~Gothe,
\newblock EPJ Web Conf. {\bf 113}, 01013 (2016).

\bibitem{Anisovich:2017pmi}
A.~V. Anisovich {\em et~al.},
\newblock Phys. Rev. Lett. {\bf 119}, 062004 (2017).

\bibitem{Jakob:1993th}
R.~Jakob, P.~Kroll, M.~Schurmann and W.~Schweiger,
\newblock Z. Phys. A {\bf 347}, 109 (1993).

\bibitem{Keiner:1995bu}
V.~Keiner,
\newblock Z. Phys. A {\bf 354}, 87 (1996).

\bibitem{Ma:2002ir}
B.-Q. Ma, D.~Qing and I.~Schmidt,
\newblock Phys. Rev. C {\bf 65}, 035205 (2002).

\bibitem{DeSanctis:2011zz}
M.~De~Sanctis, J.~Ferretti, E.~Santopinto and A.~Vassallo,
\newblock Phys. Rev. C {\bf 84}, 055201 (2011).

\bibitem{Lipkin:1981qb}
H.~J. Lipkin,
\newblock Phys. Rev. D {\bf 24}, 1437 (1981).

\bibitem{Weiss:1993kv}
C.~Weiss, A.~Buck, R.~Alkofer and H.~Reinhardt,
\newblock Phys. Lett. B {\bf 312}, 6 (1993).

\bibitem{Ramalho:2008ra}
G.~Ramalho, M.~T. Pena and F.~Gross,
\newblock Eur. Phys. J. A {\bf 36}, 329 (2008).

\bibitem{Jakob:1997wg}
R.~Jakob, P.~J. Mulders and J.~Rodrigues,
\newblock Nucl. Phys. A {\bf 626}, 937 (1997).

\bibitem{Brodsky:2002cx}
S.~J. Brodsky, D.~S. Hwang and I.~Schmidt,
\newblock Phys. Lett. B {\bf 530}, 99 (2002).

\bibitem{Gamberg:2003ey}
L.~P. Gamberg, G.~R. Goldstein and K.~A. Oganessyan,
\newblock Phys. Rev. D {\bf 67}, 071504 (2003).

\bibitem{Choi:2003ue}
S.~K. Choi {\em et~al.},
\newblock Phys. Rev. Lett. {\bf 91}, 262001 (2003).

\bibitem{Aubert:2004ns}
B.~Aubert {\em et~al.},
\newblock Phys. Rev. D {\bf 71}, 071103 (2005).

\bibitem{Acosta:2003zx}
D.~Acosta {\em et~al.},
\newblock Phys. Rev. Lett. {\bf 93}, 072001 (2004).

\bibitem{Abazov:2004kp}
V.~M. Abazov {\em et~al.},
\newblock Phys. Rev. Lett. {\bf 93}, 162002 (2004).

\bibitem{Aaltonen:2011at}
CDF, T.~Aaltonen {\em et~al.},
\newblock Mod. Phys. Lett. A {\bf 32}, 1750139 (2017).

\bibitem{Aaij:2016iza}
LHCb, R.~Aaij {\em et~al.},
\newblock Phys. Rev. Lett. {\bf 118}, 022003 (2017).

\bibitem{Aaij:2015tga}
R.~Aaij {\em et~al.},
\newblock Phys. Rev. Lett. {\bf 115}, 072001 (2015).

\bibitem{Aaij:2019vzc}
LHCb, R.~Aaij {\em et~al.},
\newblock Phys. Rev. Lett. {\bf 122}, 222001 (2019).

\bibitem{Berezhnoy:2011xn}
A.~V. Berezhnoy, A.~V. Luchinsky and A.~A. Novoselov,
\newblock Phys. Rev. D {\bf 86}, 034004 (2012).

\bibitem{Chen:2016jxd}
W.~Chen, H.-X. Chen, X.~Liu, T.~G. Steele and S.-L. Zhu,
\newblock Phys. Lett. B {\bf 773}, 247 (2017).

\bibitem{Karliner:2016zzc}
M.~Karliner, S.~Nussinov and J.~L. Rosner,
\newblock Phys. Rev. {\bf D95}, 034011 (2017).

\bibitem{Wang:2017jtz}
Z.-G. Wang,
\newblock Eur. Phys. J. C {\bf 77}, 432 (2017).

\bibitem{Anwar:2017toa}
M.~N. Anwar, J.~Ferretti, F.-K. Guo, E.~Santopinto and B.-S. Zou,
\newblock Eur. Phys. J. C {\bf 78}, 647 (2018).

\bibitem{Esposito:2018cwh}
A.~Esposito and A.~D. Polosa,
\newblock Eur. Phys. J. C {\bf 78}, 782 (2018).

\bibitem{Bedolla:2019zwg}
M.~A. Bedolla, J.~Ferretti, C.~D. Roberts and E.~Santopinto,
\newblock (2019),
\newblock {\emph{Spectrum of fully-heavy tetraquarks from a diquark+antidiquark
  perspective} -- arXiv:1911.00960 [hep-ph]}.

\bibitem{Wu:2016vtq}
J.~Wu, Y.-R. Liu, K.~Chen, X.~Liu and S.-L. Zhu,
\newblock Phys. Rev. D {\bf 97}, 094015 (2018).

\bibitem{Liu:2019zuc}
M.-S. Liu, Q.-F. L{\"u}, X.-H. Zhong and Q.~Zhao,
\newblock Phys. Rev. D {\bf 100}, 016006 (2019).

\bibitem{Eichten:2017ual}
E.~Eichten and Z.~Liu,
\newblock (2017),
\newblock {\emph{Would a Deeply Bound $b\bar b b\bar b$ Tetraquark Meson be
  Observed at the LHC?} -- arXiv:1709.09605 [hep-ph]}.

\bibitem{Aaij:2018zrb}
R.~Aaij {\em et~al.},
\newblock JHEP {\bf 10}, 086 (2018).

\bibitem{Aaij:2020fnh}
R.~Aaij {\em et~al.},
\newblock 2006.16957,
\newblock \emph{{Observation of structure in the $J/\psi$-pair mass spectrum}
  -- arXiv:2006.16957 [hep-ex]}.

\bibitem{Jaffe:1976ig}
R.~L. Jaffe,
\newblock Phys. Rev. D {\bf 15}, 267 (1977).

\bibitem{Lichtenberg:1996fi}
D.~Lichtenberg, R.~Roncaglia and E.~Predazzi,
\newblock {Diquark model of exotic mesons},
\newblock in {\em {3rd International Workshop on Diquarks and other Models of
  Compositeness (DIQUARKS III)}}, pp. 146--155, 1996.

\bibitem{Brink:1998as}
D.~M. Brink and F.~Stancu,
\newblock Phys. Rev. D {\bf 57}, 6778 (1998).

\bibitem{Maiani:2004vq}
L.~Maiani, F.~Piccinini, A.~D. Polosa and V.~Riquer,
\newblock Phys. Rev. D {\bf 71}, 014028 (2005).

\bibitem{Ebert:2008wm}
D.~Ebert, R.~Faustov and V.~Galkin,
\newblock Phys. Atom. Nucl. {\bf 72}, 184 (2009).

\bibitem{Deng:2014gqa}
C.~Deng, J.~Ping and F.~Wang,
\newblock Phys. Rev. D {\bf 90}, 054009 (2014).

\bibitem{Zhao:2014qva}
L.~Zhao, W.-Z. Deng and S.-L. Zhu,
\newblock Phys. Rev. {\bf D90}, 094031 (2014).

\bibitem{Bicudo:2015vta}
P.~Bicudo, K.~Cichy, A.~Peters, B.~Wagenbach and M.~Wagner,
\newblock Phys. Rev. D {\bf 92}, 014507 (2015).

\bibitem{Lu:2016cwr}
Q.-F. L{\"u} and Y.-B. Dong,
\newblock Phys. Rev. D {\bf 94}, 074007 (2016).

\bibitem{Anwar:2018sol}
M.~N. Anwar, J.~Ferretti and E.~Santopinto,
\newblock Phys. Rev. D {\bf 98}, 094015 (2018).

\bibitem{Yang:2019itm}
G.~Yang, J.~Ping and J.~Segovia,
\newblock Phys. Rev. D {\bf 101}, 014001 (2020).

\bibitem{Lichtenberg:1998dm}
D.~Lichtenberg,
\newblock J. Phys. G {\bf 24}, 2065 (1998).

\bibitem{Cheung:2003de}
K.~Cheung,
\newblock Phys. Rev. D {\bf 69}, 094029 (2004).

\bibitem{Shuryak:2003zi}
E.~Shuryak and I.~Zahed,
\newblock Phys. Lett. B {\bf 589}, 21 (2004).

\bibitem{Stewart:2004pd}
I.~W. Stewart, M.~E. Wessling and M.~B. Wise,
\newblock Phys. Lett. B {\bf 590}, 185 (2004).

\bibitem{Maiani:2015vwa}
L.~Maiani, A.~Polosa and V.~Riquer,
\newblock Phys. Lett. B {\bf 749}, 289 (2015).

\bibitem{Lebed:2015tna}
R.~F. Lebed,
\newblock Phys. Lett. B {\bf 749}, 454 (2015).

\bibitem{Li:2015gta}
G.-N. Li, X.-G. He and M.~He,
\newblock JHEP {\bf 12}, 128 (2015).

\bibitem{Hamermesh}
M.~Hamermesh,
\newblock {\em Group Theory and Its Application to Physical Problems} (Dover
  Publications Inc., New York, 1989).

\bibitem{DeRujula:1975qlm}
A.~De~Rujula, H.~Georgi and S.~L. Glashow,
\newblock Phys. Rev. D {\bf 12}, 147 (1975).

\bibitem{DeGrand:1975cf}
T.~A. DeGrand, R.~L. Jaffe, K.~Johnson and J.~Kiskis,
\newblock Phys. Rev. D {\bf 12}, 2060 (1975).

\bibitem{Ferretti:2015ada}
R.~Bijker, J.~Ferretti, G.~Galat{\`a}, H.~Garc{\'{\i}}a-Tecocoatzi and
  E.~Santopinto,
\newblock Phys. Rev. D {\bf 94}, 074040 (2016).

\bibitem{Santopinto:2016zkl}
E.~Santopinto,
\newblock {``Can Spectroscopy with Kaon Beams at JLab Discriminate between
  Quark Diquark and Three Quark Models?''},
\newblock in {\em {Proceedings: Workshop on Physics with Neutral Kaon Beam at
  JLab (KL2016): Newport News, VA, USA, February 1-3, 2016}}, pp. 127--142,
  2016.

\bibitem{Maiani:2004uc}
L.~Maiani, F.~Piccinini, A.~Polosa and V.~Riquer,
\newblock Phys. Rev. Lett. {\bf 93}, 212002 (2004).

\bibitem{Santopinto:2006my}
E.~Santopinto and G.~Gal{\`a}ta,
\newblock Phys. Rev. C {\bf 75}, 045206 (2007).

\bibitem{Ferretti:2019zyh}
J.~Ferretti,
\newblock Few Body Syst. {\bf 60}, 17 (2019).

\bibitem{Godfrey:1985xj}
S.~Godfrey and N.~Isgur,
\newblock Phys. Rev. D {\bf 32}, 189 (1985).

\bibitem{Tanabashi:2018oca}
M.~Tanabashi {\em et~al.},
\newblock Phys. Rev. D {\bf 98}, 030001 (2018).

\bibitem{Zhu:2015bba}
R.~Zhu and C.-F. Qiao,
\newblock Phys. Lett. B {\bf 756}, 259 (2016).

\bibitem{Ali:2016dkf}
A.~Ali, I.~Ahmed, M.~J. Aslam and A.~Rehman,
\newblock Phys. Rev. D {\bf 94}, 054001 (2016).

\bibitem{Ghosh:2017fwg}
R.~Ghosh, A.~Bhattacharya and B.~Chakrabarti,
\newblock Phys. Part. Nucl. Lett. {\bf 14}, 550 (2017).

\bibitem{Giannuzzi:2019esi}
F.~Giannuzzi,
\newblock Phys. Rev. D {\bf 99}, 094006 (2019).

\bibitem{Chang:2011vu}
L.~Chang, C.~D. Roberts and P.~C. Tandy,
\newblock Chin. J. Phys. {\bf 49}, 955 (2011).

\bibitem{Bashir:2012fs}
A.~Bashir {\em et~al.},
\newblock Commun. Theor. Phys. {\bf 58}, 79 (2012).

\bibitem{Fischer:2018sdj}
C.~S. Fischer,
\newblock Prog. Part. Nucl. Phys. {\bf 105}, 1 (2019).

\bibitem{Roberts:2016vyn}
C.~D. Roberts,
\newblock Few Body Syst. {\bf 58}, 5 (2017).

\bibitem{Salpeter:1951sz}
E.~E. Salpeter and H.~A. Bethe,
\newblock Phys. Rev. {\bf 84}, 1232 (1951).

\bibitem{Faddeev:1960su}
L.~D. Faddeev,
\newblock Sov. Phys. JETP {\bf 12}, 1014 (1961),
\newblock [Zh. Eksp. Teor. Fiz. \textbf{39} (1960) 1459].

\bibitem{Eichmann:2014xya}
G.~Eichmann, R.~Williams, R.~Alkofer and M.~Vujinovic,
\newblock Phys. Rev. D {\bf 89}, 105014 (2014).

\bibitem{Williams:2014iea}
R.~Williams,
\newblock Eur. Phys. J. A {\bf 51}, 57 (2015).

\bibitem{Cyrol:2014kca}
A.~K. Cyrol, M.~Q. Huber and L.~von Smekal,
\newblock Eur. Phys. J. C {\bf 75}, 102 (2015).

\bibitem{Aguilar:2015nqa}
A.~Aguilar, D.~Binosi and J.~Papavassiliou,
\newblock Phys. Rev. D {\bf 91}, 085014 (2015).

\bibitem{Williams:2015cvx}
R.~Williams, C.~S. Fischer and W.~Heupel,
\newblock Phys. Rev. D {\bf 93}, 034026 (2016).

\bibitem{Athenodorou:2016oyh}
A.~Athenodorou {\em et~al.},
\newblock Phys. Lett. B {\bf 761}, 444 (2016).

\bibitem{Binosi:2016wcx}
D.~Binosi, L.~Chang, J.~Papavassiliou, S.-X. Qin and C.~D. Roberts,
\newblock Phys. Rev. D {\bf 95}, 031501(R) (2017).

\bibitem{Aguilar:2019uob}
A.~Aguilar {\em et~al.},
\newblock Eur. Phys. J. C {\bf 80}, 154 (2020).

\bibitem{Huber:2020keu}
M.~Q. Huber,
\newblock Phys. Rev. D {\bf 101}, 11 (2020).

\bibitem{Mitter:2014wpa}
M.~Mitter, J.~M. Pawlowski and N.~Strodthoff,
\newblock Phys. Rev. D {\bf 91}, 054035 (2015).

\bibitem{Cyrol:2016tym}
A.~K. Cyrol, L.~Fister, M.~Mitter, J.~M. Pawlowski and N.~Strodthoff,
\newblock Phys. Rev. D {\bf 94}, 054005 (2016).

\bibitem{Cyrol:2017ewj}
A.~K. Cyrol, M.~Mitter, J.~M. Pawlowski and N.~Strodthoff,
\newblock Phys. Rev. D {\bf 97}, 054006 (2018).

\bibitem{Cucchieri:2007md}
A.~Cucchieri and T.~Mendes,
\newblock PoS {\bf LAT2007}, 297 (2007).

\bibitem{Cucchieri:2008qm}
A.~Cucchieri, A.~Maas and T.~Mendes,
\newblock Phys. Rev. D {\bf 77}, 094510 (2008).

\bibitem{Bogolubsky:2009dc}
I.~Bogolubsky, E.~Ilgenfritz, M.~Muller-Preussker and A.~Sternbeck,
\newblock Phys. Lett. B {\bf 676}, 69 (2009).

\bibitem{Maas:2011se}
A.~Maas,
\newblock Phys. Rept. {\bf 524}, 203 (2013).

\bibitem{Ayala:2012pb}
A.~Ayala, A.~Bashir, D.~Binosi, M.~Cristoforetti and J.~Rodr{\'i}guez-Quintero,
\newblock Phys. Rev. D {\bf 86}, 074512 (2012).

\bibitem{Duarte:2016iko}
A.~G. Duarte, O.~Oliveira and P.~J. Silva,
\newblock Phys. Rev. D {\bf 94}, 014502 (2016).

\bibitem{Binosi:2016xxu}
D.~Binosi, C.~D. Roberts and J.~Rodr{\'i}guez-Quintero,
\newblock Phys. Rev. D {\bf 95}, 114009 (2017).

\bibitem{Gao:2017uox}
F.~Gao, S.-X. Qin, C.~D. Roberts and J.~Rodr{\'{\i}}guez-Quintero,
\newblock Phys. Rev. D {\bf 97}, 034010 (2018).

\bibitem{Oliveira:2018lln}
O.~Oliveira, P.~J. Silva, J.-I. Skullerud and A.~Sternbeck,
\newblock Phys. Rev. D {\bf 99}, 094506 (2019).

\bibitem{Boucaud:2018xup}
P.~Boucaud, F.~De~Soto, K.~Raya, J.~Rodr{\'{\i}}guez-Quintero and
  S.~Zafeiropoulos,
\newblock Phys. Rev. D {\bf 98}, 114515 (2018).

\bibitem{Chang:2009zb}
L.~Chang and C.~D. Roberts,
\newblock Phys. Rev. Lett. {\bf 103}, 081601 (2009).

\bibitem{Chang:2010hb}
L.~Chang, Y.-X. Liu and C.~D. Roberts,
\newblock Phys. Rev. Lett. {\bf 106}, 072001 (2011).

\bibitem{Chang:2011ei}
L.~Chang and C.~D. Roberts,
\newblock Phys. Rev. C {\bf 85}, 052201(R) (2012).

\bibitem{Heupel:2014ina}
W.~Heupel, T.~Goecke and C.~S. Fischer,
\newblock Eur. Phys. J. A {\bf 50}, 85 (2014).

\bibitem{Sanchis-Alepuz:2015qra}
H.~Sanchis-Alepuz and R.~Williams,
\newblock Phys. Lett. B {\bf 749}, 592 (2015).

\bibitem{Binosi:2016rxz}
D.~Binosi, L.~Chang, S.-X. Qin, J.~Papavassiliou and C.~D. Roberts,
\newblock Phys. Rev. D {\bf 93}, 096010 (2016).

\bibitem{Yao:2020vef}
Z.-Q. Yao {\em et~al.},
\newblock Phys. Rev. D {\bf 102}, 014007 (2020).

\bibitem{Sanchis-Alepuz:2014wea}
H.~Sanchis-Alepuz, C.~S. Fischer and S.~Kubrak,
\newblock Phys. Lett. B {\bf 733}, 151 (2014).

\bibitem{Roberts:2019wov}
C.~D. Roberts,
\newblock {\emph{Resonance Electroproduction and the Origin of Mass} --
  arXiv:1909.11102 [nucl-th]},
\newblock in {\em {12th International Workshop on the Physics of Excited
  Nucleons (NSTAR 2019) Bonn, Germany, June 10-14, 2019}}, 2019.

\bibitem{Hecht:2002ej}
M.~B. Hecht {\em et~al.},
\newblock Phys. Rev. C {\bf 65}, 055204 (2002).

\bibitem{Bender:1996bb}
A.~Bender, C.~D. Roberts and L.~von Smekal,
\newblock Phys. Lett. B {\bf 380}, 7 (1996).

\bibitem{Bhagwat:2004hn}
M.~S. Bhagwat, A.~H{\"o}ll, A.~Krassnigg, C.~D. Roberts and P.~C. Tandy,
\newblock Phys. Rev. C {\bf 70}, 035205 (2004).

\bibitem{Reinhardt:1989rw}
H.~Reinhardt,
\newblock Phys. Lett. B {\bf 244}, 316 (1990).

\bibitem{Efimov:1990uz}
G.~V. Efimov, M.~A. Ivanov and V.~E. Lyubovitskij,
\newblock Z. Phys. C {\bf 47}, 583 (1990).

\bibitem{Burden:1988dt}
C.~J. Burden, R.~T. Cahill and J.~Praschifka,
\newblock Austral. J. Phys. {\bf 42}, 147 (1989).

\bibitem{Hellstern:1995ri}
G.~Hellstern and C.~Weiss,
\newblock Phys. Lett. B {\bf 351}, 64 (1995).

\bibitem{Hellstern:1997pg}
G.~Hellstern, R.~Alkofer, M.~Oettel and H.~Reinhardt,
\newblock Nucl. Phys. A {\bf 627}, 679 (1997).

\bibitem{Oettel:1999gc}
M.~Oettel, M.~Pichowsky and L.~von Smekal,
\newblock Eur. Phys. J. A {\bf 8}, 251 (2000).

\bibitem{AlvarengaNogueira:2019zcs}
J.~Alvarenga~Nogueira {\em et~al.},
\newblock Phys. Rev. D {\bf 100}, 016021 (2019).

\bibitem{Heupel:2012ua}
W.~Heupel, G.~Eichmann and C.~S. Fischer,
\newblock Phys. Lett. B {\bf 718}, 545 (2012).

\bibitem{Eichmann:2015cra}
G.~Eichmann, C.~S. Fischer and W.~Heupel,
\newblock Phys. Lett. B {\bf 753}, 282 (2016).

\bibitem{Wallbott:2019dng}
P.~C. Wallbott, G.~Eichmann and C.~S. Fischer,
\newblock Phys. Rev. D {\bf 100}, 014033 (2019).

\bibitem{Wallbott:2020jzh}
P.~C. Wallbott, G.~Eichmann and C.~S. Fischer,
\newblock (2020).

\bibitem{Xu:2019sns}
S.-S. Xu {\em et~al.},
\newblock Eur. Phys. J. A {\bf 55}, 113 (Lett.) (2019).

\bibitem{Chen:2012txa}
C.~Chen {\em et~al.},
\newblock Phys. Rev. C {\bf 87}, 045207 (2013).

\bibitem{Wilson:2011aa}
D.~J. Wilson, I.~C. Cloet, L.~Chang and C.~D. Roberts,
\newblock Phys. Rev. C {\bf 85}, 025205 (2012).

\bibitem{Chen:2012qr}
C.~Chen, L.~Chang, C.~D. Roberts, S.-L. Wan and D.~J. Wilson,
\newblock Few Body Syst. {\bf 53}, 293 (2012).

\bibitem{Buck:1992wz}
A.~Buck, R.~Alkofer and H.~Reinhardt,
\newblock Phys. Lett. B {\bf 286}, 29 (1992).

\bibitem{Xu:2015kta}
S.-S. Xu {\em et~al.},
\newblock Phys. Rev. D {\bf 92}, 114034 (2015).

\bibitem{Roberts:2011wy}
H.~L.~L. Roberts, A.~Bashir, L.~X. Guti{\'e}rrez-Guerrero, C.~D. Roberts and
  D.~J. Wilson,
\newblock Phys. Rev. C {\bf 83}, 065206 (2011).

\bibitem{Brown:2014ena}
Z.~S. Brown, W.~Detmold, S.~Meinel and K.~Orginos,
\newblock Phys. Rev. D {\bf 90}, 094507 (2014).

\bibitem{Mathur:2018epb}
N.~Mathur, M.~Padmanath and S.~Mondal,
\newblock Phys. Rev. Lett. {\bf 121}, 202002 (2018).

\bibitem{Gutierrez-Guerrero:2019uwa}
L.~Guti{\'e}rrez-Guerrero, A.~Bashir, M.~A. Bedolla and E.~Santopinto,
\newblock Phys. Rev. D {\bf 100}, 114032 (2019).

\bibitem{Segovia:2013rca}
J.~Segovia, C.~Chen, C.~D. Roberts and S.-L. Wan,
\newblock Phys. Rev. C {\bf 88}, 032201(R) (2013).

\bibitem{Segovia:2013uga}
J.~Segovia {\em et~al.},
\newblock Few Body Syst. {\bf 55}, 1 (2014).

\bibitem{Pitschmann:2014jxa}
M.~Pitschmann, C.-Y. Seng, C.~D. Roberts and S.~M. Schmidt,
\newblock Phys. Rev. D {\bf 91}, 074004 (2015).

\bibitem{Wang:2018kto}
Q.-W. Wang, S.-X. Qin, C.~D. Roberts and S.~M. Schmidt,
\newblock Phys. Rev. D {\bf 98}, 054019 (2018).

\bibitem{Roberts:2018hpf}
C.~D. Roberts,
\newblock Few Body Syst. {\bf 59}, 72 (2018).

\bibitem{Segovia:2014aza}
J.~Segovia, I.~C. Cloet, C.~D. Roberts and S.~M. Schmidt,
\newblock Few Body Syst. {\bf 55}, 1185 (2014).

\bibitem{Segovia:2016zyc}
J.~Segovia and C.~D. Roberts,
\newblock Phys. Rev. C {\bf 94}, 042201(R) (2016).

\bibitem{Chen:2018nsg}
C.~Chen {\em et~al.},
\newblock Phys. Rev. D {\bf 99}, 034013 (2019).

\bibitem{Cui:2020rmu}
Z.-F. Cui {\em et~al.},
\newblock Phys. Rev. D {\bf 102}, 014043 (2020).

\bibitem{Hecht:2000xa}
M.~B. Hecht, C.~D. Roberts and S.~M. Schmidt,
\newblock Phys. Rev. C {\bf 63}, 025213 (2001).

\bibitem{Jones:1999rz}
M.~K. Jones {\em et~al.},
\newblock Phys. Rev. Lett. {\bf 84}, 1398 (2000).

\bibitem{Gayou:2001qd}
O.~Gayou {\em et~al.},
\newblock Phys. Rev. Lett. {\bf 88}, 092301 (2002).

\bibitem{Punjabi:2005wq}
V.~Punjabi {\em et~al.},
\newblock Phys. Rev. C {\bf 71}, 055202 (2005),
\newblock [Erratum-ibid. C\,\textbf{71}, 069902 (2005)].

\bibitem{Puckett:2010ac}
A.~J.~R. Puckett {\em et~al.},
\newblock Phys. Rev. Lett. {\bf 104}, 242301 (2010).

\bibitem{Puckett:2011xg}
A.~J.~R. Puckett {\em et~al.},
\newblock Phys. Rev. C {\bf 85}, 045203 (2012).

\bibitem{Syritsyn:2017jrc}
S.~Syritsyn, A.~S. Gambhir, B.~Musch and K.~Orginos,
\newblock PoS {\bf LATTICE2016}, 176 (2017).

\bibitem{Kallidonis:2018cas}
C.~Kallidonis {\em et~al.},
\newblock PoS {\bf LATTICE2018}, 125 (2018).

\bibitem{Madey:2003av}
R.~Madey {\em et~al.},
\newblock Phys. Rev. Lett. {\bf 91}, 122002 (2003).

\bibitem{Riordan:2010id}
S.~Riordan {\em et~al.},
\newblock Phys. Rev. Lett. {\bf 105}, 262302 (2010).

\bibitem{Schlessinger:1966zz}
L.~Schlessinger and C.~Schwartz,
\newblock Phys. Rev. Lett. {\bf 16}, 1173 (1966).

\bibitem{PhysRev.167.1411}
L.~Schlessinger,
\newblock Phys. Rev. {\bf 167}, 1411 (1968).

\bibitem{Tripolt:2016cya}
R.~A. Tripolt, I.~Haritan, J.~Wambach and N.~Moiseyev,
\newblock Phys. Lett. B {\bf 774}, 411 (2017).

\bibitem{Binosi:2018rht}
D.~Binosi {\em et~al.},
\newblock Phys. Lett. B {\bf 790}, 257 (2019).

\bibitem{Alkofer:2004yf}
R.~Alkofer, A.~H{\"o}ll, M.~Kloker, A.~Krassnigg and C.~D. Roberts,
\newblock Few Body Syst. {\bf 37}, 1 (2005).

\bibitem{Cloet:2008re}
I.~C. Cloet, G.~Eichmann, B.~El-Bennich, T.~Kl{\"a}hn and C.~D. Roberts,
\newblock Few Body Syst. {\bf 46}, 1 (2009).

\bibitem{Cloet:2013gva}
I.~C. Cloet, C.~D. Roberts and A.~W. Thomas,
\newblock Phys. Rev. Lett. {\bf 111}, 101803 (2013).

\bibitem{Trivedi:2018rgo}
A.~Trivedi,
\newblock Few Body Syst. {\bf 60}, 5 (2019).

\bibitem{Burkert:2019opk}
V.~D. Burkert, V.~I. Mokeev and B.~S. Ishkhanov,
\newblock Moscow Univ. Phys. Bull. {\bf 74}, 243 (2019),
\newblock [Vestn. Mosk. Univ. Ser. III Fiz. Astron. \textbf{74} (2019) pp.\
  28-38].

\bibitem{Eichmann:2009zx}
G.~Eichmann,
\newblock (2009),
\newblock {\emph{Hadron Properties from QCD Bound-State Equations}, PhD thesis,
  Universit{\"a}t Graz -- arXiv:0909.0703 [hep-ph]}.

\bibitem{Sanchis-Alepuz:2017jjd}
H.~Sanchis-Alepuz and R.~Williams,
\newblock Comput. Phys. Commun. {\bf 232}, 1 (2018).

\bibitem{Qin:2019hgk}
S.-X. Qin, C.~D. Roberts and S.~M. Schmidt,
\newblock Few Body Syst. {\bf 60}, 26 (2019).

\bibitem{Eichmann:2007nn}
G.~Eichmann, A.~Krassnigg, M.~Schwinzerl and R.~Alkofer,
\newblock Annals Phys. {\bf 323}, 2505 (2008).

\bibitem{Maris:1997tm}
P.~Maris and C.~D. Roberts,
\newblock Phys. Rev. C {\bf 56}, 3369 (1997).

\bibitem{Maris:1999bh}
P.~Maris and P.~C. Tandy,
\newblock Phys. Rev. C {\bf 61}, 045202 (2000).

\bibitem{Qin:2011dd}
S.-X. Qin, L.~Chang, Y.-X. Liu, C.~D. Roberts and D.~J. Wilson,
\newblock Phys. Rev. C {\bf 84}, 042202(R) (2011).

\bibitem{Eichmann:2018adq}
G.~Eichmann and C.~S. Fischer,
\newblock Few Body Syst. {\bf 60}, 2 (2019).

\bibitem{Maris:1995ns}
P.~Maris,
\newblock Phys. Rev. D {\bf 52}, 6087 (1995).

\bibitem{Windisch:2012zd}
A.~Windisch, R.~Alkofer, G.~Haase and M.~Liebmann,
\newblock Comput. Phys. Commun. {\bf 184}, 109 (2013).

\bibitem{Windisch:2012sz}
A.~Windisch, M.~Q. Huber and R.~Alkofer,
\newblock Phys. Rev. D {\bf 87}, 065005 (2013).

\bibitem{Pawlowski:2017gxj}
J.~M. Pawlowski, N.~Strodthoff and N.~Wink,
\newblock Phys. Rev. D {\bf 98}, 074008 (2018).

\bibitem{Weil:2017knt}
E.~Weil, G.~Eichmann, C.~S. Fischer and R.~Williams,
\newblock Phys. Rev. D {\bf 96}, 014021 (2017).

\bibitem{Williams:2018adr}
R.~Williams,
\newblock Phys. Lett. B {\bf 798}, 134943 (2019).

\bibitem{Miramontes:2019mco}
{\'A}.~S. Miramontes and H.~Sanchis-Alepuz,
\newblock Eur. Phys. J. A {\bf 55}, 170 (2019).

\bibitem{Eichmann:2019dts}
G.~Eichmann, P.~Duarte, M.~Pe{\~n}a and A.~Stadler,
\newblock Phys. Rev. D {\bf 100}, 094001 (2019).

\bibitem{Nakanishi:1969ph}
N.~Nakanishi,
\newblock Prog. Theor. Phys. Suppl. {\bf 43}, 1 (1969).

\bibitem{Chang:2013nia}
L.~Chang, I.~C. Cloet, C.~D. Roberts, S.~M. Schmidt and P.~C. Tandy,
\newblock Phys. Rev. Lett. {\bf 111}, 141802 (2013).

\bibitem{Nicmorus:2008vb}
D.~Nicmorus, G.~Eichmann, A.~Krassnigg and R.~Alkofer,
\newblock Phys. Rev. D {\bf 80}, 054028 (2009).

\bibitem{Nicmorus:2010sd}
D.~Nicmorus, G.~Eichmann and R.~Alkofer,
\newblock Phys. Rev. D {\bf 82}, 114017 (2010).

\bibitem{Eichmann:2011aa}
G.~Eichmann and D.~Nicmorus,
\newblock Phys. Rev. D {\bf 85}, 093004 (2012).

\bibitem{Eichmann:2009qa}
G.~Eichmann, R.~Alkofer, A.~Krassnigg and D.~Nicmorus,
\newblock Phys. Rev. Lett. {\bf 104}, 201601 (2010).

\bibitem{Sanchis-Alepuz:2014sca}
H.~Sanchis-Alepuz and C.~S. Fischer,
\newblock Phys. Rev. D {\bf 90}, 096001 (2014).

\bibitem{Qin:2018dqp}
S.-X. Qin, C.~D. Roberts and S.~M. Schmidt,
\newblock Phys. Rev. D {\bf 97}, 114017 (2018).

\bibitem{Eichmann:2011vu}
G.~Eichmann,
\newblock Phys. Rev. D {\bf 84}, 014014 (2011).

\bibitem{Eichmann:2011pv}
G.~Eichmann and C.~S. Fischer,
\newblock Eur. Phys. J. A {\bf 48}, 9 (2012).

\bibitem{Sanchis-Alepuz:2015fcg}
H.~Sanchis-Alepuz and C.~S. Fischer,
\newblock Eur. Phys. J. A {\bf 52}, 34 (2016).

\bibitem{Eichmann:2016nsu}
G.~Eichmann,
\newblock Few Body Syst. {\bf 58}, 81 (2017).

\bibitem{Suzuki:2009nj}
N.~Suzuki {\em et~al.},
\newblock Phys. Rev. Lett. {\bf 104}, 042302 (2010).

\bibitem{Eichmann:2008ae}
G.~Eichmann, R.~Alkofer, I.~C. Cloet, A.~Krassnigg and C.~D. Roberts,
\newblock Phys. Rev. C {\bf 77}, 042202(R) (2008).

\bibitem{Pichowsky:1999mu}
M.~A. Pichowsky, S.~Walawalkar and S.~Capstick,
\newblock Phys. Rev. D {\bf 60}, 054030 (1999).

\bibitem{Lepage:1979zb}
G.~P. Lepage and S.~J. Brodsky,
\newblock Phys. Lett. B {\bf 87}, 359 (1979).

\bibitem{Efremov:1979qk}
A.~V. Efremov and A.~V. Radyushkin,
\newblock Phys. Lett. B {\bf 94}, 245 (1980).

\bibitem{Lepage:1980fj}
G.~P. Lepage and S.~J. Brodsky,
\newblock Phys. Rev. D {\bf 22}, 2157 (1980).

\bibitem{Chang:2013pq}
L.~Chang {\em et~al.},
\newblock Phys. Rev. Lett. {\bf 110}, 132001 (2013).

\bibitem{Chang:2013epa}
L.~Chang, C.~D. Roberts and S.~M. Schmidt,
\newblock Phys. Lett. B {\bf 727}, 255 (2013).

\bibitem{Segovia:2013eca}
J.~Segovia {\em et~al.},
\newblock Phys. Lett. B {\bf 731}, 13 (2014).

\bibitem{Gao:2014bca}
F.~Gao, L.~Chang, Y.-X. Liu, C.~D. Roberts and S.~M. Schmidt,
\newblock Phys. Rev. D {\bf 90}, 014011 (2014).

\bibitem{Mezrag:2014jka}
C.~Mezrag {\em et~al.},
\newblock Phys. Lett. B {\bf 741}, 190 (2015).

\bibitem{Shi:2015esa}
C.~Shi {\em et~al.},
\newblock Phys. Rev. D {\bf 92}, 014035 (2015).

\bibitem{Ding:2015rkn}
M.~Ding, F.~Gao, L.~Chang, Y.-X. Liu and C.~D. Roberts,
\newblock Phys. Lett. B {\bf 753}, 330 (2016).

\bibitem{Li:2016dzv}
B.~L. Li {\em et~al.},
\newblock Phys. Rev. D {\bf 93}, 114033 (2016).

\bibitem{Li:2016mah}
B.-L. Li, L.~Chang, M.~Ding, C.~D. Roberts and H.-S. Zong,
\newblock Phys. Rev. D {\bf 94}, 094014 (2016).

\bibitem{Ding:2018xwy}
M.~Ding {\em et~al.},
\newblock Phys. Rev. D {\bf 99}, 014014 (2019).

\bibitem{Cui:2020dlm}
Z.-F. Cui {\em et~al.},
\newblock (2020).

\bibitem{Cui:2020piK}
Z.-F. Cui {\em et~al.},
\newblock (2020),
\newblock {\emph{Kaon and pion parton distributions} -- in progress}.

\bibitem{Mezrag:2017znp}
C.~Mezrag, J.~Segovia, L.~Chang and C.~D. Roberts,
\newblock Phys. Lett. B {\bf 783}, 263 (2018).

\bibitem{Mezrag:2018hkk}
C.~Mezrag, J.~Segovia, M.~Ding, L.~Chang and C.~D. Roberts,
\newblock Springer Proc. Phys. {\bf 238}, 773 (2020).

\bibitem{Braun:2009jy}
V.~Braun {\em et~al.},
\newblock Phys. Rev. Lett. {\bf 103}, 072001 (2009).

\bibitem{Bali:2015ykx}
G.~S. Bali {\em et~al.},
\newblock JHEP {\bf 02}, 070 (2016).

\bibitem{Bali:2019ecy}
RQCD, G.~S. Bali {\em et~al.},
\newblock Eur. Phys. J. A {\bf 55}, 116 (2019).

\bibitem{Aoki:2010dy}
RBC, UKQCD, Y.~Aoki {\em et~al.},
\newblock Phys. Rev. D {\bf 83}, 074508 (2011).

\bibitem{Yang:2014sea}
Y.-B. Yang {\em et~al.},
\newblock Phys. Rev. D {\bf 92}, 034517 (2015).

\bibitem{Neuberger:1997fp}
H.~Neuberger,
\newblock Phys. Lett. B {\bf 417}, 141 (1998).

\bibitem{Chen:2007zzc}
Y.~Chen,
\newblock Mod. Phys. Lett. A {\bf 22}, 583 (2007).

\bibitem{Sun:2019aem}
M.~Sun {\em et~al.},
\newblock Phys. Rev. D {\bf 101}, 054511 (2020).

\bibitem{Chernyak:1983ej}
V.~L. Chernyak and A.~R. Zhitnitsky,
\newblock Phys. Rept. {\bf 112}, 173 (1984).

\bibitem{Kraenkl:2011qb}
S.~Krankl and A.~Manashov,
\newblock Phys. Lett. B {\bf 703}, 519 (2011).

\bibitem{Franklin:1968pn}
J.~Franklin,
\newblock Phys. Rev. {\bf 172}, 1807 (1968).

\bibitem{Braun:2000kw}
V.~Braun, R.~J. Fries, N.~Mahnke and E.~Stein,
\newblock Nucl. Phys. B {\bf 589}, 381 (2000),
\newblock [Erratum: Nucl. Phys. B\,\textbf{607}, 433 (2001)].

\bibitem{Wein:2015oqa}
P.~Wein and A.~Sch{\"a}fer,
\newblock JHEP {\bf 05}, 073 (2015).

\bibitem{Anikin:2013aka}
I.~V. Anikin, V.~M. Braun and N.~Offen,
\newblock Phys. Rev. D {\bf 88}, 114021 (2013).

\bibitem{Braun:2008ia}
V.~Braun, A.~Manashov and J.~Rohrwild,
\newblock Nucl. Phys. B {\bf 807}, 89 (2009).

\bibitem{Luscher:2011kk}
M.~Luscher and S.~Schaefer,
\newblock JHEP {\bf 07}, 036 (2011).

\bibitem{Luscher:2012av}
M.~Luscher and S.~Schaefer,
\newblock Comput. Phys. Commun. {\bf 184}, 519 (2013).

\bibitem{Bietenholz:2016mgn}
W.~Bietenholz,
\newblock Int. J. Mod. Phys. E {\bf 25}, 1642008 (2016).

\bibitem{Martinelli:1994ty}
G.~Martinelli, C.~Pittori, C.~T. Sachrajda, M.~Testa and A.~Vladikas,
\newblock Nucl. Phys. B {\bf 445}, 81 (1995).

\bibitem{Sturm:2009kb}
C.~Sturm {\em et~al.},
\newblock Phys. Rev. D {\bf 80}, 014501 (2009).

\bibitem{Kaltenbrunner:2008zz}
T.~Kaltenbrunner,
\newblock {\em {Renormalization of three-quark operators for the nucleon
  distribution amplitude}},
\newblock PhD thesis, Regensburg U., 2008.

\bibitem{Gockeler:2008we}
QCDSF, UKQCD, M.~Gockeler {\em et~al.},
\newblock Nucl. Phys. B {\bf 812}, 205 (2009).

\bibitem{Gruber:2017ozo}
M.~Gruber,
\newblock {\em {Renormalization of three-quark operators for baryon
  distribution amplitudes}},
\newblock PhD thesis, Regensburg U., 2017.

\bibitem{Wein:2011ix}
P.~Wein, P.~C. Bruns, T.~R. Hemmert and A.~Sch{\"a}fer,
\newblock Eur. Phys. J. A {\bf 47}, 149 (2011).

\bibitem{Chernyak:1987nu}
V.~Chernyak, A.~Ogloblin and I.~Zhitnitsky,
\newblock Yad. Fiz. {\bf 48}, 1410 (1988).

\bibitem{Braun:2014wpa}
V.~M. Braun {\em et~al.},
\newblock Phys. Rev. D {\bf 89}, 094511 (2014).

\bibitem{Chernyak:1984bm}
V.~L. Chernyak and I.~R. Zhitnitsky,
\newblock Nucl. Phys. B {\bf 246}, 52 (1984).

\bibitem{Hofstadter:1955ae}
R.~Hofstadter and R.~W. McAllister,
\newblock Phys. Rev. {\bf 98}, 217 (1955).

\bibitem{Punjabi:2015bba}
V.~Punjabi, C.~F. Perdrisat, M.~K. Jones, E.~J. Brash and C.~E. Carlson,
\newblock Eur. Phys. J. A {\bf 51}, 79 (2015).

\bibitem{Pacetti:2015iqa}
S.~Pacetti, R.~Baldini~Ferroli and E.~Tomasi-Gustafsson,
\newblock Phys. Rept. {\bf 550-551}, 1 (2015).

\bibitem{Rosenbluth:1950yq}
M.~Rosenbluth,
\newblock Phys. Rev. {\bf 79}, 615 (1950).

\bibitem{Bosted:1994tm}
P.~E. Bosted,
\newblock Phys. Rev. C {\bf 51}, 409 (1995).

\bibitem{Arnold:1986nq}
R.~G. Arnold {\em et~al.},
\newblock Phys. Rev. Lett. {\bf 57}, 174 (1986).

\bibitem{Sill:1992qw}
A.~F. Sill {\em et~al.},
\newblock Phys. Rev. D {\bf 48}, 29 (1993).

\bibitem{Lepage:1979za}
G.~Lepage and S.~J. Brodsky,
\newblock Phys. Rev. Lett. {\bf 43}, 545 (1979),
\newblock [Erratum: Phys.Rev.Lett. 43, 1625--1626 (1979)].

\bibitem{Perdrisat1989}
{Perdrisat, Ch and Punjabi, V. and others},
\newblock (1989),
\newblock \emph{Measurement of the Electric Form Factor of the Proton by Recoil
  Polarization}, approved Jefferson Lab experiment E89-014.

\bibitem{Scofield:1959zz}
J.~H. Scofield,
\newblock Phys. Rev. {\bf 113}, 1599 (1959).

\bibitem{Akhiezer:1968ek}
A.~Akhiezer and M.~Rekalo,
\newblock Sov. Phys. Dokl. {\bf 13}, 572 (1968).

\bibitem{Akhiezer:1974em}
A.~Akhiezer and M.~Rekalo,
\newblock Sov. J. Part. Nucl. {\bf 4}, 277 (1974),
\newblock [Fiz. Elem. Chast. Atom. Yadra \textbf{4}, 662 (1973)].

\bibitem{Arnold:1980zj}
R.~Arnold, C.~E. Carlson and F.~Gross,
\newblock Phys. Rev. C {\bf 23}, 363 (1981).

\bibitem{Gayou:2001qt}
O.~Gayou {\em et~al.},
\newblock Phys. Rev. C {\bf 64}, 038202 (2001).

\bibitem{Belitsky:2002kj}
A.~V. Belitsky, X.-d. Ji and F.~Yuan,
\newblock Phys. Rev. Lett. {\bf 91}, 092003 (2003).

\bibitem{Lachniet:2008qf}
J.~Lachniet {\em et~al.},
\newblock Phys. Rev. Lett. {\bf 102}, 192001 (2009).

\bibitem{Diehl:2013xca}
M.~Diehl and P.~Kroll,
\newblock Eur. Phys. J. C {\bf 73}, 2397 (2013).

\bibitem{Baillie:2011za}
N.~Baillie {\em et~al.},
\newblock Phys. Rev. Lett. {\bf 108}, 142001 (2012),
\newblock [Erratum: Phys.Rev.Lett. 108, 199902 (2012)].

\bibitem{Roberts:2013mja}
C.~D. Roberts, R.~J. Holt and S.~M. Schmidt,
\newblock Phys. Lett. B {\bf 727}, 249 (2013).

\bibitem{Chen:2020ijn}
X.~Chen, F.-K. Guo, C.~D. Roberts and R.~Wang,
\newblock (2020).

\bibitem{Accardi:2011fa}
A.~Accardi {\em et~al.},
\newblock Phys. Rev. D {\bf 84}, 014008 (2011).

\bibitem{Ablikim:2019eau}
M.~Ablikim {\em et~al.},
\newblock Phys. Rev. Lett. {\bf 124}, 042001 (2020).

\bibitem{CMD-3:2018kql}
R.~R. Akhmetshin {\em et~al.},
\newblock Phys. Lett. B {\bf 794}, 64 (2019).

\bibitem{Lees:2013ebn}
J.~P. Lees {\em et~al.},
\newblock Phys. Rev. D {\bf 87}, 092005 (2013).

\bibitem{Lees:2013uta}
J.~P. Lees {\em et~al.},
\newblock Phys. Rev. D {\bf 88}, 072009 (2013).

\bibitem{Kuraev:2011vq}
E.~Kuraev, E.~Tomasi-Gustafsson and A.~Dbeyssi,
\newblock Phys. Lett. B {\bf 712}, 240 (2012).

\bibitem{Kim:1988gg}
V.~Kim,
\newblock Mod. Phys. Lett. A {\bf 3}, 909 (1988).

\bibitem{Tiator:2003uu}
L.~Tiator {\em et~al.},
\newblock Eur. Phys. J. A {\bf 19}, 55 (2004),
\newblock [,55(2003)].

\bibitem{Mokeev:2015lda}
V.~I. Mokeev {\em et~al.},
\newblock Phys. Rev. C {\bf 93}, 025206 (2016).

\bibitem{Aznauryan:2011qj}
I.~Aznauryan and V.~Burkert,
\newblock Prog. Part. Nucl. Phys. {\bf 67}, 1 (2012).

\bibitem{Mokeev:2018zxt}
V.~I. Mokeev,
\newblock Few Body Syst. {\bf 59}, 46 (2018).

\bibitem{Isupov:2017lnd}
E.~L. Isupov {\em et~al.},
\newblock Phys. Rev. C {\bf 96}, 025209 (2017).

\bibitem{Mokeev:2012vsa}
V.~I. Mokeev {\em et~al.},
\newblock Phys. Rev. C {\bf 86}, 035203 (2012).

\bibitem{Markov:2019fjy}
N.~Markov {\em et~al.},
\newblock Phys. Rev. C {\bf 101}, 015208 (2020).

\bibitem{Aznauryan:2009mx}
I.~Aznauryan {\em et~al.},
\newblock Phys. Rev. C {\bf 80}, 055203 (2009).

\bibitem{Blin:2019fre}
A.~Hiller~Blin {\em et~al.},
\newblock Phys. Rev. C {\bf 100}, 035201 (2019).

\bibitem{Burkert:2018nvj}
V.~D. Burkert,
\newblock Ann. Rev. Nucl. Part. Sci. {\bf 68}, 405 (2018).

\bibitem{Dugger:2009pn}
M.~Dugger {\em et~al.},
\newblock Phys. Rev. C {\bf 79}, 065206 (2009).

\bibitem{Roberts:2016dnb}
C.~D. Roberts and J.~Segovia,
\newblock Few Body Syst. {\bf 57}, 1067 (2016).

\bibitem{Aznauryan:2016wwm}
I.~G. Aznauryan and V.~D. Burkert,
\newblock (2016),
\newblock {\emph{Configuration mixings and light-front relativistic quark model
  predictions for the electroexcitation of the \mbox{$\Delta(1232)3/2^+$},
  \mbox{$N(1440)1/2^+$}, and \mbox{$\Delta(1600)3/2^+$}}, arXiv:1603.06692
  [hep-ph]}.

\bibitem{JuliaDiaz:2006xt}
B.~Julia-Diaz, T.~S.~H. Lee, T.~Sato and L.~C. Smith,
\newblock Phys. Rev. C {\bf 75}, 015205 (2007).

\bibitem{Jones:1972ky}
H.~F. Jones and M.~D. Scadron,
\newblock Annals Phys. {\bf 81}, 1 (1973).

\bibitem{Ash1967165}
W.~Ash, K.~Berkelman, C.~Lichtenstein, A.~Ramanauskas and R.~Siemann,
\newblock Phys.\ Lett.\ B {\bf 24}, 165  (1967).

\bibitem{Carlson:1985mm}
C.~E. Carlson,
\newblock Phys. Rev. D {\bf 34}, 2704 (1986).

\bibitem{Capstick:1994ne}
S.~Capstick and B.~D. Keister,
\newblock Phys. Rev. D {\bf 51}, 3598 (1995).

\bibitem{Aznauryan:2015zta}
I.~G. Aznauryan and V.~D. Burkert,
\newblock Phys. Rev. C {\bf 92}, 035211 (2015).

\bibitem{Brodsky:2020vco}
S.~J. Brodsky {\em et~al.},
\newblock Intern. J. Mod. Phys. E , {\emph{in press}} (2020).

\bibitem{Carman:2020qmb}
D.~Carman, K.~Joo and V.~Mokeev,
\newblock Few Body Syst. {\bf 61}, 29 (2020).

\bibitem{MokeevPrivate2020}
V.~I. Mokeev,
\newblock (2020),
\newblock {\emph{private communication}}.

\bibitem{Mulders:1995dh}
P.~Mulders and R.~Tangerman,
\newblock Nucl.Phys. {\bf B461}, 197 (1996).

\bibitem{Boer:1997nt}
D.~Boer and P.~Mulders,
\newblock Phys. Rev. D {\bf 57}, 5780 (1998).

\bibitem{Mueller:1998fv}
D.~Mueller, D.~Robaschik, B.~Geyer, F.~M. Dittes and J.~Ho\v{r}ej\v{s}i,
\newblock Fortschr. Phys. {\bf 42}, 101 (1994).

\bibitem{Ji:1996nm}
X.-D. Ji,
\newblock Phys. Rev. D {\bf 55}, 7114 (1997).

\bibitem{Radyushkin:1997ki}
A.~V. Radyushkin,
\newblock Phys. Rev. D {\bf 56}, 5524 (1997), [hep-ph/9704207].

\bibitem{Ji:2003ak}
X.-d. Ji,
\newblock Phys. Rev. Lett. {\bf 91}, 062001 (2003).

\bibitem{Dudek:2012vr}
J.~Dudek {\em et~al.},
\newblock Eur.\ Phys.\ J. A {\bf 48}, 187 (2012).

\bibitem{Accardi:2011mz}
A.~Accardi, V.~Guzey, A.~Prokudin and C.~Weiss,
\newblock Eur. Phys. J. A {\bf 48}, 92 (2012).

\bibitem{Anselmino:2011ay}
M.~Anselmino {\em et~al.},
\newblock Eur. Phys. J. A {\bf 47}, 35 (2011).

\bibitem{Accardi:2012qut}
A.~Accardi {\em et~al.},
\newblock Eur. Phys. J. A {\bf 52}, 268 (2016).

\bibitem{Aschenauer:2014twa}
E.-C. Aschenauer {\em et~al.},
\newblock Eur. Phys. J. A {\bf 53}, 71 (2017).

\bibitem{EicCWP}
X.~Cao {\em et~al.},
\newblock Nucl. Tech. {\bf 43}, 020001 (2020).

\bibitem{EicCWPEL}
X.~Cao {\em et~al.},
\newblock Frontiers of Physics.  (2020),
\newblock {\emph{Electron Ion Collider in China (EicC)}}.

\bibitem{Bacchetta:2006tn}
A.~Bacchetta {\em et~al.},
\newblock JHEP {\bf 0702}, 093 (2007).

\bibitem{Collins:2011zzd}

\newblock J.~Collins{\em $,\,${Foundations of perturbative QCD}} Vol.~32
  (Cambridge University Press, 2013).

\bibitem{HarrisonThesis}
N.~A. Harrison,
\newblock {\em Exploring the Structure of the Proton via Semi-inclusive Pion
  Electroproduction},
\newblock PhD thesis, University of Connecticut, 2015.

\bibitem{Avakian:2003pk}
H.~Avakian {\em et~al.},
\newblock Phys. Rev. D {\bf 69}, 112004 (2004).

\bibitem{Aghasyan:2011ha}
M.~Aghasyan {\em et~al.},
\newblock Phys. Lett. B {\bf 704}, 397 (2011).

\bibitem{Gohn:2014zbz}
W.~Gohn {\em et~al.},
\newblock Phys. Rev. D {\bf 89}, 072011 (2014).

\bibitem{Pasquini:2008ax}
B.~Pasquini, S.~Cazzaniga and S.~Boffi,
\newblock Phys. Rev. D {\bf 78}, 034025 (2008).

\bibitem{Maji:2017bcz}
T.~Maji and D.~Chakrabarti,
\newblock Phys. Rev. D {\bf 95}, 074009 (2017).

\bibitem{Bacchetta:2008af}
A.~Bacchetta, F.~Conti and M.~Radici,
\newblock Phys. Rev. D {\bf 78}, 074010 (2008).

\bibitem{Avakian:2010br}
H.~Avakian, A.~Efremov, P.~Schweitzer and F.~Yuan,
\newblock Phys. Rev. D {\bf 81}, 074035 (2010).

\bibitem{Pasquini:2012jm}
B.~Pasquini and C.~Lorc{\'e},
\newblock Proc. Int. Sch. Phys. Fermi {\bf 180}, 197 (2012).

\bibitem{Mao:2013waa}
W.~Mao and Z.~Lu,
\newblock Eur. Phys. J. C {\bf 73}, 2557 (2013).

\bibitem{Maji:2016yqo}
T.~Maji and D.~Chakrabarti,
\newblock Phys. Rev. D {\bf 94}, 094020 (2016).

\bibitem{Anselmino:2005nn}
M.~Anselmino {\em et~al.},
\newblock Phys. Rev. D {\bf 71}, 074006 (2005).

\bibitem{Anselmino:2008sga}
M.~Anselmino {\em et~al.},
\newblock Eur. Phys. J. A {\bf 39}, 89 (2009).

\bibitem{Avakian:2015vha}
H.~Avakian, H.~Matevosyan, B.~Pasquini and P.~Schweitzer,
\newblock J. Phys. G {\bf 42}, 034015 (2015).

\bibitem{Kumericki:2016ehc}
K.~Kumericki, S.~Liuti and H.~Moutarde,
\newblock Eur. Phys. J. A {\bf 52}, 157 (2016).

\bibitem{Vanderhaeghen:1999xj}
M.~Vanderhaeghen, P.~A. Guichon and M.~Guidal,
\newblock Phys. Rev. D {\bf 60}, 094017 (1999).

\bibitem{Mukherjee:2002gb}
A.~Mukherjee, I.~Musatov, H.~Pauli and A.~Radyushkin,
\newblock Phys. Rev. D {\bf 67}, 073014 (2003).

\bibitem{Goloskokov:2005sd}
S.~Goloskokov and P.~Kroll,
\newblock Eur. Phys. J. C {\bf 42}, 281 (2005).

\bibitem{Kumericki:2009uq}
K.~Kumeri\v~cki and D.~Mueller,
\newblock Nucl. Phys. B {\bf 841}, 1 (2010).

\bibitem{Mezrag:2013mya}
C.~Mezrag, H.~Moutarde and F.~Sabati{\'e},
\newblock Phys. Rev. D {\bf 88}, 014001 (2013).

\bibitem{Mezrag:2016hnp}
C.~Mezrag, H.~Moutarde and J.~Rodr{\'i}guez-Quintero,
\newblock Few Body Syst. {\bf 57}, 729 (2016).

\bibitem{Moutarde:2018kwr}
H.~Moutarde, P.~Sznajder and J.~Wagner,
\newblock Eur. Phys. J. C {\bf 78}, 890 (2018).

\bibitem{Kumericki:2011rz}
K.~Kumericki, D.~Mueller and A.~Schafer,
\newblock JHEP {\bf 07}, 073 (2011).

\bibitem{Moutarde:2019tqa}
H.~Moutarde, P.~Sznajder and J.~Wagner,
\newblock Eur. Phys. J. C {\bf 79}, 614 (2019).

\bibitem{Chouika:2017dhe}
N.~Chouika, C.~Mezrag, H.~Moutarde and J.~Rodr{\'{\i}}guez-Quintero,
\newblock Eur. Phys. J. C {\bf 77}, 906 (2017).

\bibitem{Chouika:2017rzs}
N.~Chouika, C.~Mezrag, H.~Moutarde and J.~Rodr{\'{\i}}guez-Quintero,
\newblock Phys. Lett. B {\bf 780}, 287 (2018).

\bibitem{Ahmad:2006gn}
S.~Ahmad, H.~Honkanen, S.~Liuti and S.~K. Taneja,
\newblock Phys. Rev. D {\bf 75}, 094003 (2007).

\bibitem{Goldstein:2010gu}
G.~R. Goldstein, J.~Hernandez and S.~Liuti,
\newblock Phys. Rev. D {\bf 84}, 034007 (2011).

\bibitem{GonzalezHernandez:2012jv}
J.~Gonzalez-Hernandez, S.~Liuti, G.~R. Goldstein and K.~Kathuria,
\newblock Phys. Rev. C {\bf 88}, 065206 (2013).

\bibitem{Berthou:2015oaw}
B.~Berthou {\em et~al.},
\newblock Eur. Phys. J. C {\bf 78}, 478 (2018).

\bibitem{Moutarde:2013qs}
H.~Moutarde, B.~Pire, F.~Sabatie, L.~Szymanowski and J.~Wagner,
\newblock Phys. Rev. D {\bf 87}, 054029 (2013).

\bibitem{Defurne:2017paw}
M.~Defurne {\em et~al.},
\newblock Nature Commun. {\bf 8}, 1408 (2017).

\bibitem{Lane:1974he}
K.~D. Lane,
\newblock Phys. Rev. D {\bf 10}, 2605 (1974).

\bibitem{Politzer:1976tv}
H.~D. Politzer,
\newblock Nucl. Phys. B {\bf 117}, 397 (1976).

\bibitem{Qin:2014vya}
S.-X. Qin, C.~D. Roberts and S.~M. Schmidt,
\newblock Phys. Lett. B {\bf 733}, 202 (2014).

\bibitem{Badier:1983mj}
J.~Badier {\em et~al.},
\newblock Z. Phys. C {\bf 18}, 281 (1983).

\bibitem{Conway:1989fs}
J.~S. Conway {\em et~al.},
\newblock Phys. Rev. D {\bf 39}, 92 (1989).

\bibitem{Adloff:1998yg}
C.~Adloff {\em et~al.},
\newblock Eur. Phys. J. C {\bf 6}, 587 (1999).

\bibitem{Holt:2010vj}
R.~J. Holt and C.~D. Roberts,
\newblock Rev. Mod. Phys. {\bf 82}, 2991 (2010).

\bibitem{Wijesooriya:2005ir}
K.~Wijesooriya, P.~E. Reimer and R.~J. Holt,
\newblock Phys. Rev. C {\bf 72}, 065203 (2005).

\bibitem{Aicher:2010cb}
M.~Aicher, A.~Sch{\"a}fer and W.~Vogelsang,
\newblock Phys.\ Rev.\ Lett. {\bf 105}, 252003 (2010).

\bibitem{Ezawa:1974wm}
Z.~F. Ezawa,
\newblock Nuovo Cim. A {\bf 23}, 271 (1974).

\bibitem{Farrar:1975yb}
G.~R. Farrar and D.~R. Jackson,
\newblock Phys. Rev. Lett. {\bf 35}, 1416 (1975).

\bibitem{Berger:1979du}
E.~L. Berger and S.~J. Brodsky,
\newblock Phys. Rev. Lett. {\bf 42}, 940 (1979).

\bibitem{Ding:2019qlr}
M.~Ding {\em et~al.},
\newblock Chin. Phys. (Lett.) {\bf 44}, 031002 (2020).

\bibitem{Ding:2019lwe}
M.~Ding {\em et~al.},
\newblock Phys. Rev. D {\bf 101}, 054014 (2020).

\bibitem{JlabTDIS1}
{C. Keppel, B. Wojtsekhowski, P. King, D. Dutta, J. Annand, J. Zhang \emph{et
  al}.},
\newblock (2015),
\newblock approved Jefferson Lab experiment E12-15-006.

\bibitem{JlabTDIS2}
{K. Park, R. Montgomery, T. Horn \emph{et al}.},
\newblock (2015),
\newblock approved Jefferson Lab experiment C12-15-006A.

\bibitem{Peng:2016ebs}
J.-C. Peng and J.-W. Qiu,
\newblock The Universe {\bf 4}, 34 (2016).

\bibitem{Peng:2017ddf}
J.-C. Peng, W.-C. Chang, S.~Platchkov and T.~Sawada,
\newblock (2017),
\newblock {\emph{Valence Quark and Gluon Distributions of Kaon from J/Psi
  Production}, arXiv:1711.00839 [hep-ph]}.

\bibitem{Horn:2018fqr}
T.~Horn,
\newblock AIP Conf. Proc. {\bf 1970}, 030003 (2018).

\bibitem{Denisov:2018unj}
O.~Denisov {\em et~al.},
\newblock {\emph{Letter of Intent (Draft 2.0): A New QCD facility at the M2
  beam line of the CERN SPS} -- arXiv:1808.00848 [hep-ex]}.

\bibitem{Chang:2020rdy}
W.-C. Chang, J.-C. Peng, S.~Platchkov and T.~Sawada,
\newblock (2020),
\newblock {\emph{Constraining Gluon Density of Pions at Large $x$ by
  Pion-induced $J/\psi$ Production} -- arXiv:2006.06947 [hep-ph]}.

\bibitem{Sullivan:1971kd}
J.~D. Sullivan,
\newblock Phys. Rev. D {\bf 5}, 1732 (1972).

\bibitem{Qin:2017lcd}
S.-X. Qin, C.~Chen, C.~Mezrag and C.~D. Roberts,
\newblock Phys. Rev. C {\bf 97}, 015203 (2018).

\bibitem{Xu:2018eii}
S.-S. Xu, L.~Chang, C.~D. Roberts and H.-S. Zong,
\newblock Phys. Rev. D {\bf 97}, 094014 (2018).

\bibitem{Chen:2018fwa}
J.-H. Zhang {\em et~al.},
\newblock Phys. Rev. D {\bf 100}, 034505 (2019).

\bibitem{Karthik:2018wmj}
N.~Karthik {\em et~al.},
\newblock PoS {\bf LATTICE2018}, 109 (2018).

\bibitem{Sufian:2019bol}
R.~S. Sufian {\em et~al.},
\newblock Phys. Rev. D {\bf 99}, 074507 (2019).

\bibitem{PhysicsToday73}
T.~Feder,
\newblock Physics Today {\bf 73}, 22 (2020).

\bibitem{Aguilar:2019teb}
A.~C. Aguilar {\em et~al.},
\newblock Eur. Phys. J. A {\bf 55}, 190 (2019).

\bibitem{Gokhroo:2006bt}
G.~Gokhroo {\em et~al.},
\newblock Phys. Rev. Lett. {\bf 97}, 162002 (2006).

\bibitem{Aubert:2007rva}
B.~Aubert {\em et~al.},
\newblock Phys. Rev. D {\bf 77}, 011102 (2008).

\bibitem{Adachi:2008su}
T.~Aushev {\em et~al.},
\newblock Phys.Rev. {\bf D81}, 031103 (2010), [0810.0358].

\bibitem{Ablikim:2020xpq}
M.~Ablikim {\em et~al.},
\newblock Phys. Rev. Lett. {\bf 124}, 242001 (2020), [2001.01156].

\bibitem{Aaij:2011sn}
R.~Aaij {\em et~al.},
\newblock Eur. Phys. J. C {\bf 72}, 1972 (2012).

\bibitem{Chatrchyan:2013cld}
S.~Chatrchyan {\em et~al.},
\newblock JHEP {\bf 04}, 154 (2013).

\bibitem{CMS:2019vma}
{CMS Collaboration},
\newblock (2019),
\newblock {\emph{Evidence for $\chi_{c1}$(3872) in PbPb collisions and studies
  of its prompt production at
  $\sqrt{\smash[b]{s_{_{\mathrm{NN}}}}}=5.02\,$TeV}}.

\bibitem{Choi:2011fc}
S.-K. Choi {\em et~al.},
\newblock Phys. Rev. D {\bf 84}, 052004 (2011).

\bibitem{Aaij:2013zoa}
R.~Aaij {\em et~al.},
\newblock Phys. Rev. Lett. {\bf 110}, 222001 (2013).

\bibitem{Aaij:2015eva}
R.~Aaij {\em et~al.},
\newblock Phys. Rev. D {\bf 92}, 011102 (2015).

\bibitem{delAmoSanchez:2010jr}
P.~del Amo~Sanchez {\em et~al.},
\newblock Phys. Rev. D {\bf 82}, 011101 (2010).

\bibitem{Ablikim:2019zio}
M.~Ablikim {\em et~al.},
\newblock Phys. Rev. Lett. {\bf 122}, 232002 (2019).

\bibitem{Aghasyan:2017utv}
M.~Aghasyan {\em et~al.},
\newblock Phys. Lett. B {\bf 783}, 334 (2018).

\bibitem{Ablikim:2013mio}
M.~Ablikim {\em et~al.},
\newblock Phys. Rev. Lett. {\bf 110}, 252001 (2013).

\bibitem{Liu:2013dau}
Z.~Q. Liu {\em et~al.},
\newblock Phys. Rev. Lett. {\bf 110}, 252002 (2013).

\bibitem{Collaboration:2017njt}
M.~Ablikim {\em et~al.},
\newblock Phys.Rev.Lett. {\bf 119}, 072001 (2017).

\bibitem{Ablikim:2013xfr}
M.~Ablikim {\em et~al.},
\newblock Phys. Rev. Lett. {\bf 112}, 022001 (2014).

\bibitem{Ablikim:2015swa}
M.~Ablikim {\em et~al.},
\newblock Phys. Rev. D {\bf 92}, 092006 (2015).

\bibitem{Ablikim:2015tbp}
M.~Ablikim {\em et~al.},
\newblock Phys. Rev. Lett. {\bf 115}, 112003 (2015).

\bibitem{Ablikim:2015gda}
M.~Ablikim {\em et~al.},
\newblock Phys. Rev. Lett. {\bf 115}, 222002 (2015).

\bibitem{Ablikim:2013emm}
M.~Ablikim {\em et~al.},
\newblock Phys. Rev. Lett. {\bf 112}, 132001 (2014).

\bibitem{Ablikim:2013wzq}
M.~Ablikim {\em et~al.},
\newblock Phys. Rev. Lett. {\bf 111}, 242001 (2013).

\bibitem{Ablikim:2014dxl}
M.~Ablikim {\em et~al.},
\newblock (2014).

\bibitem{Ablikim:2015vvn}
M.~Ablikim {\em et~al.},
\newblock Phys. Rev. Lett. {\bf 115}, 182002 (2015).

\bibitem{Belle:2011aa}
A.~Bondar {\em et~al.},
\newblock Phys. Rev. Lett. {\bf 108}, 122001 (2012).

\bibitem{Krokovny:2013mgx}
P.~Krokovny {\em et~al.},
\newblock Phys. Rev. D {\bf 88}, 052016 (2013).

\bibitem{Garmash:2014dhx}
A.~Garmash {\em et~al.},
\newblock Phys. Rev. D {\bf 91}, 072003 (2015).

\bibitem{Garmash:2015rfd}
A.~Garmash {\em et~al.},
\newblock Phys. Rev. Lett. {\bf 116}, 212001 (2016).

\bibitem{Aubert:2005rm}
B.~Aubert {\em et~al.},
\newblock Phys. Rev. Lett. {\bf 95}, 142001 (2005).

\bibitem{He:2006kg}
Q.~He {\em et~al.},
\newblock Phys. Rev. D {\bf 74}, 091104 (2006).

\bibitem{Yuan:2007sj}
C.~Yuan {\em et~al.},
\newblock Phys. Rev. Lett. {\bf 99}, 182004 (2007).

\bibitem{Lees:2012pv}
J.~Lees {\em et~al.},
\newblock Phys. Rev. D {\bf 89}, 111103 (2014).

\bibitem{Ablikim:2016qzw}
M.~Ablikim {\em et~al.},
\newblock Phys. Rev. Lett. {\bf 118}, 092001 (2017).

\bibitem{BESIII:2016adj}
M.~Ablikim {\em et~al.},
\newblock Phys. Rev. Lett. {\bf 118}, 092002 (2017).

\bibitem{Ablikim:2017oaf}
M.~Ablikim {\em et~al.},
\newblock Phys. Rev. D {\bf 96}, 032004 (2017),
\newblock [Erratum: Phys. Rev. D \textbf{99}, 019903 (2019)].

\bibitem{Ablikim:2014qwy}
M.~Ablikim {\em et~al.},
\newblock Phys. Rev. Lett. {\bf 114}, 092003 (2015).

\bibitem{Ablikim:2018vxx}
M.~Ablikim {\em et~al.},
\newblock Phys. Rev. Lett. {\bf 122}, 102002 (2019).

\bibitem{Ablikim:2013dyn}
M.~Ablikim {\em et~al.},
\newblock Phys. Rev. Lett. {\bf 112}, 092001 (2014).

\bibitem{Pakhlova:2008vn}
G.~Pakhlova {\em et~al.},
\newblock Phys. Rev. Lett. {\bf 101}, 172001 (2008).

\bibitem{Cotugno:2009ys}
G.~Cotugno, R.~Faccini, A.~Polosa and C.~Sabelli,
\newblock Phys. Rev. Lett. {\bf 104}, 132005 (2010).

\bibitem{Abdesselam:2019gth}
R.~Mizuk {\em et~al.},
\newblock JHEP {\bf 10}, 220 (2019).

\bibitem{Ortega:2016hde}
P.~G. Ortega, J.~Segovia, D.~R. Entem and F.~Fern{\'a}ndez,
\newblock Phys. Rev. D {\bf 94}, 114018 (2016).

\bibitem{Aaij:2014jqa}
R.~Aaij {\em et~al.},
\newblock Phys. Rev. Lett. {\bf 112}, 222002 (2014).

\bibitem{Chilikin:2014bkk}
K.~Chilikin {\em et~al.},
\newblock Phys. Rev. D {\bf 90}, 112009 (2014).

\bibitem{Mizuk:2008me}
R.~Mizuk {\em et~al.},
\newblock Phys. Rev. D {\bf 78}, 072004 (2008).

\bibitem{Aaij:2018bla}
R.~Aaij {\em et~al.},
\newblock Eur. Phys. J. C {\bf 78}, 1019 (2018).

\bibitem{Aaij:2016nsc}
R.~Aaij {\em et~al.},
\newblock Phys. Rev. D {\bf 95}, 012002 (2017).

\bibitem{Rodas:2018owy}
A.~Rodas {\em et~al.},
\newblock Phys. Rev. Lett. {\bf 122}, 042002 (2019).

\bibitem{Szczepaniak:2015eza}
A.~P. Szczepaniak,
\newblock Phys. Lett. B {\bf 747}, 410 (2015).

\bibitem{Pilloni:2016obd}
A.~Pilloni {\em et~al.},
\newblock Phys. Lett. B {\bf 772}, 200 (2017).

\bibitem{Albaladejo:2015lob}
M.~Albaladejo, F.-K. Guo, C.~Hidalgo-Duque and J.~Nieves,
\newblock Phys. Lett. B {\bf 755}, 337 (2016).

\bibitem{Fernandez-Ramirez:2019koa}
C.~Fern{\'a}ndez-Ram{\'{\i}}rez {\em et~al.},
\newblock Phys. Rev. Lett. {\bf 123}, 092001 (2019).

\bibitem{Du:2019pij}
M.-L. Du {\em et~al.},
\newblock Phys.Rev.Lett. {\bf 124}, 072001 (2020).

\bibitem{Karliner:2015voa}
M.~Karliner and J.~L. Rosner,
\newblock Phys. Lett. B {\bf 752}, 329 (2016).

\bibitem{Kubarovsky:2015aaa}
V.~Kubarovsky and M.~Voloshin,
\newblock Phys. Rev. D {\bf 92}, 031502 (2015).

\bibitem{Blin:2016dlf}
A.~N. Hiller~Blin {\em et~al.},
\newblock Phys. Rev. D {\bf 94}, 034002 (2016).

\bibitem{Winney:2019edt}
D.~Winney {\em et~al.},
\newblock Phys. Rev. D {\bf 100}, 034019 (2019).

\bibitem{Ali:2019lzf}
A.~Ali {\em et~al.},
\newblock Phys. Rev. Lett. {\bf 123}, 072001 (2019).

\bibitem{Wu:2019adv}
J.-J. Wu, T.-S. Lee and B.-S. Zou,
\newblock Phys. Rev. C {\bf 100}, 035206 (2019).

\bibitem{Xu:2019ilh}
Y.-Z. Xu {\em et~al.},
\newblock Phys. Rev. D {\bf 100}, 114038 (2019).

\bibitem{Richard:2018yrm}
J.-M. Richard, A.~Valcarce and J.~Vijande,
\newblock Phys. Rev. C {\bf 97}, 035211 (2018).

\bibitem{Cardoso:2011fq}
N.~Cardoso, M.~Cardoso and P.~Bicudo,
\newblock Phys. Rev. D {\bf 84}, 054508 (2011).

\bibitem{Rossi:2016szw}
G.~Rossi and G.~Veneziano,
\newblock JHEP {\bf 06}, 041 (2016).

\bibitem{Maiani:2014aja}
L.~Maiani, F.~Piccinini, A.~Polosa and V.~Riquer,
\newblock Phys. Rev. D {\bf 89}, 114010 (2014), [1405.1551].

\bibitem{Faccini:2013lda}
L.~Maiani {\em et~al.},
\newblock Phys. Rev. D {\bf 87}, 111102 (2013).

\bibitem{Maiani:2016wlq}
L.~Maiani, A.~Polosa and V.~Riquer,
\newblock Phys. Rev. D {\bf 94}, 054026 (2016).

\bibitem{Ali:2019okl}
A.~Ali, L.~Maiani, A.~Y. Parkhomenko and W.~Wang,
\newblock Phys. Lett. B {\bf 802}, 135217 (2020).

\bibitem{Ali:2019clg}
A.~Ali, I.~Ahmed, M.~J. Aslam, A.~Y. Parkhomenko and A.~Rehman,
\newblock JHEP {\bf 10}, 256 (2019).

\bibitem{Francis:2016hui}
A.~Francis, R.~J. Hudspith, R.~Lewis and K.~Maltman,
\newblock Phys. Rev. Lett. {\bf 118}, 142001 (2017).

\bibitem{Karliner:2017qjm}
M.~Karliner and J.~L. Rosner,
\newblock Phys. Rev. Lett. {\bf 119}, 202001 (2017).

\bibitem{Eichten:2017ffp}
E.~J. Eichten and C.~Quigg,
\newblock Phys. Rev. Lett. {\bf 119}, 202002 (2017).

\bibitem{Carames:2011zz}
T.~Carames, A.~Valcarce and J.~Vijande,
\newblock Phys. Lett. B {\bf 699}, 291 (2011).

\bibitem{Esposito:2013fma}
A.~Esposito, M.~Papinutto, A.~Pilloni, A.~Polosa and N.~Tantalo,
\newblock Phys. Rev. D {\bf 88}, 054029 (2013).

\bibitem{Guerrieri:2014nxa}
A.~L. Guerrieri, M.~Papinutto, A.~Pilloni, A.~D. Polosa and N.~Tantalo,
\newblock PoS {\bf LATTICE2014}, 106 (2015).

\bibitem{Cheung:2017tnt}
G.~K.~C. Cheung, C.~E. Thomas, J.~J. Dudek and R.~G. Edwards,
\newblock JHEP {\bf 11}, 033 (2017).

\bibitem{Blitz:2015nra}
S.~H. Blitz and R.~F. Lebed,
\newblock Phys. Rev. D {\bf 91}, 094025 (2015).

\bibitem{Esposito:2016itg}
A.~Esposito, A.~Pilloni and A.~Polosa,
\newblock Phys. Lett. B {\bf 758}, 292 (2016).

\bibitem{Maiani:2017kyi}
L.~Maiani, A.~Polosa and V.~Riquer,
\newblock Phys. Lett. B {\bf 778}, 247 (2018).

\bibitem{Brodsky:2014xia}
S.~J. Brodsky, D.~S. Hwang and R.~F. Lebed,
\newblock Phys. Rev. Lett. {\bf 113}, 112001 (2014).

\bibitem{Giron:2019bcs}
J.~F. Giron, R.~F. Lebed and C.~T. Peterson,
\newblock JHEP {\bf 05}, 061 (2019).

\bibitem{Giron:2019cfc}
J.~F. Giron, R.~F. Lebed and C.~T. Peterson,
\newblock JHEP {\bf 01}, 124 (2020).

\bibitem{Maiani:2019cwl}
L.~Maiani, A.~D. Polosa and V.~Riquer,
\newblock Phys. Rev. D {\bf 100}, 014002 (2019), [1903.10253].

\bibitem{Maiani:2019lpu}
L.~Maiani, A.~D. Polosa and V.~Riquer,
\newblock Phys. Rev. D {\bf 100}, 074002 (2019).

\bibitem{Artoisenet:2010uu}
P.~Artoisenet and E.~Braaten,
\newblock Phys. Rev. D {\bf 83}, 014019 (2011).

\bibitem{Esposito:2013ada}
A.~Esposito, F.~Piccinini, A.~Pilloni and A.~Polosa,
\newblock J. Mod. Phys. {\bf 4}, 1569 (2013).

\bibitem{Guerrieri:2014gfa}
A.~L. Guerrieri, F.~Piccinini, A.~Pilloni and A.~D. Polosa,
\newblock Phys. Rev. D {\bf 90}, 034003 (2014).

\bibitem{Barabanov:2016jjv}
M.~Barabanov, A.~Vodopyanov, A.~Zinchenko and S.~Olsen,
\newblock Phys. Atom. Nucl. {\bf 79}, 126 (2016).

\bibitem{Albaladejo:2017blx}
M.~Albaladejo {\em et~al.},
\newblock Chin. Phys. C {\bf 41}, 121001 (2017).

\bibitem{Esposito:2017qef}
A.~Esposito {\em et~al.},
\newblock Chin. Phys. C {\bf 42}, 114107 (2018).

\bibitem{Wang:2017gay}
W.~Wang,
\newblock Chin. Phys. C {\bf 42}, 043103 (2018).

\bibitem{Braaten:2018eov}
E.~Braaten, L.-P. He and K.~Ingles,
\newblock Phys. Rev. D {\bf 100}, 094024 (2019).

\bibitem{Ortega:2009hj}
P.~Ortega, J.~Segovia, D.~Entem and F.~Fernandez,
\newblock Phys. Rev. D {\bf 81}, 054023 (2010).

\bibitem{Ferretti:2013faa}
J.~Ferretti, G.~Galat{\`a} and E.~Santopinto,
\newblock Phys. Rev. C {\bf 88}, 015207 (2013).

\bibitem{Ferretti:2014xqa}
J.~Ferretti, G.~Galat{\`a} and E.~Santopinto,
\newblock Phys. Rev. D {\bf 90}, 054010 (2014).

\bibitem{Ferretti:2018tco}
J.~Ferretti and E.~Santopinto,
\newblock Phys. Lett. B {\bf 789}, 550 (2019).

\bibitem{Esposito:2015fsa}
A.~Esposito {\em et~al.},
\newblock Phys. Rev. D {\bf 92}, 034028 (2015).

\bibitem{Cho:2013rpa}
S.~Cho and S.~H. Lee,
\newblock Phys. Rev. C {\bf 88}, 054901 (2013).

\bibitem{Sissakian:2009zza}
A.~Sissakian and A.~Sorin,
\newblock J. Phys. G {\bf 36}, 064069 (2009).

\bibitem{Barabanov:2019izj}
M.~Barabanov, A.~Vodopyanov and A.~Zinchenko,
\newblock Nuovo Cim. C {\bf 42}, 110 (2019).

\bibitem{Sjostrand:2014zea}
T.~Sj{\"o}strand {\em et~al.},
\newblock Comput. Phys. Commun. {\bf 191}, 159 (2015).

\bibitem{Li:2019kpj}
C.~Li and C.-Z. Yuan,
\newblock Phys. Rev. D {\bf 100}, 094003 (2019).

\bibitem{Puckett:2017flj}
A.~J.~R. Puckett {\em et~al.},
\newblock Phys. Rev. C {\bf 96}, 055203 (2017),
\newblock [erratum: Phys. Rev. C \textbf{98}, 019907 (2018)].

\bibitem{Meziane:2010xc}
M.~Meziane {\em et~al.},
\newblock Phys. Rev. Lett. {\bf 106}, 132501 (2011).

\bibitem{GEP5}
{E. Brash, E. Cisbani, M. K. Jones, M. Khandaker, L. P. Pentchev, C. F.
  Perdrisat, V. Punjabi, B. Wojtsekhowski, \emph{et al}.},
\newblock {Large Acceptance Proton Form Factor Ratio Measurements at 13 and 15
  GeV$^2$ Using Recoil Polarization Method},
\newblock 2007.

\bibitem{GEP5_PAC47}
{E. Cisbani, M. K. Jones, N. Liyanage, L. P. Pentchev, A. J. R. Puckett and B.
  Wojtsekhowski},
\newblock {Update on E12-07-109 to PAC 47: Large Acceptance Proton Form Factor
  Ratio Measurements at 13 and 15 GeV$^2$ Using Recoil Polarization Method},
\newblock 2019.

\bibitem{GMN}
{R. Gilman, B. Quinn, B. Wojtsekhowski \emph{et al}.},
\newblock {Precision Measurement of the Neutron Magnetic Form Factor up to $Q^2
  = 18.0$ (GeV/c)$^2$ by the Ratio Method},
\newblock 2007.

\bibitem{GEN2}
{G. Cates, S. Riordan, B. Wojtsekhowski \emph{et al}.},
\newblock {Measurement of the Neutron Electromagnetic Form Factor Ratio
  $G_E^n/G_M^n$ at High $Q^2$},
\newblock 2009.

\bibitem{Wojtsekhowski:2017kti}
B.~Wojtsekhowski,
\newblock {Prospect for Measuring ${G_E^n}$ at High Momentum Transfers --
  arXiv:1706.02747 [physics.ins-det]},
\newblock in {\em {Exclusive Processes at High Momentum Transfer}}, 2017.

\bibitem{Helium3article}
P.~A.~M. Dolph {\em et~al.},
\newblock Phys. Rev. C {\bf 84}, 065201 (2011).

\bibitem{Helium3_target2015}
J.~T. Singh {\em et~al.},
\newblock Phys. Rev. C {\bf 91}, 055205 (2015).

\bibitem{GENRP}
{J. R. M. Annand, V. Bellini, M. Kohl, N. Piskunov, B. Sawatzky, B.
  Wojtsekhowski \emph{et al}.},
\newblock { \emph{Measurement of the ratio $G_E^n/G_M^n$ by the
  double-polarized $^2H(\vec{e},e'\vec{n})$ reaction}, approved JLab experiment
  E12-17-004},
\newblock 2017.

\bibitem{Basilev:2019sno}
S.~Basilev {\em et~al.},
\newblock Eur. Phys. J. A {\bf 56}, 26 (2020).

\bibitem{SBS_SIDIS}
{G. Cates, E. Cisbani, G. Franklin, A. J. R. Puckett, B. Wojtsekhowski \emph{et
  al}.},
\newblock {Target Single-Spin Asymmetries in Semi-Inclusive Pion and Kaon
  Electroproduction on a Transversely Polarized $^3$He Target using Super
  BigBite and BigBite in Hall A},
\newblock 2011.

\bibitem{E12-07-108}
{J. Arrington, M. Christy, S. Gilad, V. Sulkosky, B. Wojtsekhowski \emph{et
  al}.},
\newblock Precision measurement of the proton elastic cross section at high
  $q^2$,
\newblock {A}pproved Jefferson Lab 12 GeV Experiment: E12-07-108, 2007.

\bibitem{Kelly:2004hm}
J.~J. Kelly,
\newblock Phys. Rev. C {\bf 70}, 068202 (2004).

\bibitem{Diehl:2004cx}
M.~Diehl, T.~Feldmann, R.~Jakob and P.~Kroll,
\newblock Eur. Phys. J. C {\bf 39}, 1 (2005).

\bibitem{Lomon:2006xb}
E.~L. Lomon,
\newblock (2006),
\newblock {\emph{Effect of revised R(n) measurements on extended
  Gari-Krumpelmann model fits to nucleon electromagnetic form factors} --
  nucl-th/0609020}.

\bibitem{Lomon:2012pn}
E.~L. Lomon and S.~Pacetti,
\newblock Phys. Rev. D {\bf 85}, 113004 (2012),
\newblock [Erratum: Phys. Rev. D \textbf{86}, 039901 (2012)].

\bibitem{Gross:2006fg}
F.~Gross, G.~Ramalho and M.~Pena,
\newblock Phys. Rev. C {\bf 77}, 015202 (2008).

\bibitem{Cloet:2012cy}
I.~C. Cloet and G.~A. Miller,
\newblock Phys. Rev. C {\bf 86}, 015208 (2012).

\bibitem{Miller:2002ig}
G.~A. Miller,
\newblock Phys. Rev. C {\bf 66}, 032201 (2002).

\bibitem{Sauli:1997qp}
F.~Sauli,
\newblock Nucl. Instrum. Meth. A {\bf 386}, 531 (1997).

\bibitem{Gnanvo:2014hpa}
K.~Gnanvo {\em et~al.},
\newblock Nucl. Instrum. Meth. A {\bf 782}, 77 (2015).

\bibitem{RiordanThesis}
S.~Riordan,
\newblock {\em {Measurements of the Electric Form Factor of the Neutron at $Q^2
  = 1.7 \text{ and }3.5\text{ GeV}^2$}},
\newblock PhD thesis, Carnegie Mellon University, 2008.

\bibitem{ObrechtThesis}
R.~F. Obrecht,
\newblock {\em {Electric Form Factor of the Neutron from Asymmetry
  Measurements}},
\newblock PhD thesis, University of Connecticut, 2019.

\bibitem{ECAL_NIM}
B.~Wojtsekhowski {\em et~al.},
\newblock Lead glass calorimeter with continuous thermal annealing,
\newblock In preparation.

\bibitem{Andersson:1983ia}
B.~Andersson, G.~Gustafson, G.~Ingelman and T.~Sjostrand,
\newblock Phys. Rept. {\bf 97}, 31 (1983).

\bibitem{Sjostrand:2006za}
T.~Sjostrand, S.~Mrenna and P.~Z. Skands,
\newblock JHEP {\bf 05}, 026 (2006).

\bibitem{Bellm_2016}
J.~Bellm {\em et~al.},
\newblock Eur. Phys. J. C {\bf 76}, 196 (2016).

\bibitem{Airapetian:2000ks}
A.~Airapetian {\em et~al.},
\newblock Eur. Phys. J. C , 479 (2001).

\bibitem{Airapetian:2003mi}
A.~Airapetian {\em et~al.},
\newblock Phys. Lett. B {\bf 577}, 37 (2003).

\bibitem{Airapetian:2005yh}
A.~Airapetian {\em et~al.},
\newblock Phys. Rev. Lett. {\bf 96}, 162301 (2006).

\bibitem{Airapetian:2007vu}
A.~Airapetian {\em et~al.},
\newblock Nucl. Phys. B {\bf 780}, 1 (2007).

\bibitem{Airapetian:2009jy}
A.~Airapetian {\em et~al.},
\newblock Phys. Lett. B {\bf 684}, 114 (2010).

\bibitem{Airapetian:2011jp}
A.~Airapetian {\em et~al.},
\newblock Eur. Phys. J. A {\bf 47}, 113 (2011).

\bibitem{Ashman:1991cx}
J.~Ashman {\em et~al.},
\newblock Z. Phys. C {\bf 52}, 1 (1991).

\bibitem{Accardi:2009qv}
A.~Accardi, F.~Arleo, W.~K. Brooks, D.~D'Enterria and V.~Muccifora,
\newblock Riv. Nuovo Cim. {\bf 32}, 439 (2010).

\bibitem{DelDuca:1992ru}
V.~Del~Duca, S.~J. Brodsky and P.~Hoyer,
\newblock Phys. Rev. D {\bf 46}, 931 (1992).

\bibitem{BL1}
W.~K. Brooks and J.~Lopez,
\newblock (in preparation).

\bibitem{Brooks_2019}
W.~K. Brooks and J.~A. L{\'o}pez,
\newblock Proceedings of the 8th International Conference on Quarks and Nuclear
  Physics (QNP2018)  (2019).

\bibitem{Ferreres-Sole:2018vgo}
S.~Ferreres-Sol{\'e} and T.~Sj{\"o}strand,
\newblock Eur. Phys. J. C {\bf 78}, 983 (2018).

\bibitem{Dok_1991}
Y.~L. Dokshitzer, V.~A. Khoze, A.~H. Mueller and S.~I. Troyan,
\newblock {\em Basics of Perturbative QCD} (Editions Fronti{\`e}res, Singapore,
  1991).

\bibitem{Collins:1997nm}
J.~C. Collins,
\newblock (1997),
\newblock {Light cone variables, rapidity and all that -- hep-ph/9705393}.

\bibitem{Breakstone:1985ef}
A.~Breakstone {\em et~al.},
\newblock Z. Phys. C {\bf 28}, 335 (1985).

\bibitem{E12-06-117}
{W. K. Brooks, G. Gilfoyle, H. Hakobyan, K. Hicks, M. Holtrop, K. Joo, G.
  Niculescu, I. Niculescu, L. B. Weinstein, M. Wood \emph{et al}},
\newblock (2006),
\newblock {\emph{Quark Propagation and Hadron Formation}, approved Jefferson
  Lab experiment E12-06-117}.

\bibitem{Chetry_2019}
T.~Chetry and L.~E. Fassi,
\newblock {Study of $\Lambda$ Hyperon Fragmentation in Current and Target
  Regions using CLAS}.

\bibitem{Jaffe:2005zz}
R.~Jaffe,
\newblock Nucl. Phys. B Proc. Suppl. {\bf 142}, 343 (2005).

\bibitem{Niiyama:2017wpp}
M.~Niiyama {\em et~al.},
\newblock Phys. Rev. D {\bf 97}, 072005 (2018).

\bibitem{Brodzicka:2012jm}
J.~Brodzicka {\em et~al.},
\newblock PTEP {\bf 2012}, 04D001 (2012).

\bibitem{Golonka:2005pn}
P.~Golonka and Z.~Was,
\newblock Eur. Phys. J. C {\bf 45}, 97 (2006).

\bibitem{Mizuk:2004yu}
R.~Mizuk {\em et~al.},
\newblock Phys. Rev. Lett. {\bf 94}, 122002 (2005).

\bibitem{Chambers:2017tuf}
A.~J. Chambers {\em et~al.},
\newblock Phys. Rev. D {\bf 96}, 114509 (2017).

\bibitem{Lepage:1989hd}
G.~Lepage,
\newblock {The Analysis of Algorithms for Lattice Field Theory},
\newblock in {\em {Theoretical Advanced Study Institute in Elementary Particle
  Physics}}, pp. 97--120, 1989.

\bibitem{Bali:2016lva}
G.~S. Bali, B.~Lang, B.~U. Musch and A.~Sch{\"a}fer,
\newblock Phys. Rev. D {\bf 93}, 094515 (2016).

\bibitem{Alberico:2008sz}
W.~Alberico, S.~Bilenky, C.~Giunti and K.~Graczyk,
\newblock Phys. Rev. C {\bf 79}, 065204 (2009), [0812.3539].

\bibitem{Green:2015wqa}
J.~Green {\em et~al.},
\newblock Phys. Rev. D {\bf 92}, 031501 (2015).

\end{thebibliography}

\end{document}